\documentclass[12pt, a4paper]{article}
\usepackage{natbib}
\setcitestyle{authoryear,open={(},close={)},citesep={;}}
\usepackage[]{hyperref}
\usepackage{xcolor}
\hypersetup{colorlinks = true, citecolor = blue, linkcolor=blue, linkbordercolor = {white}, urlcolor=blue}
\usepackage{amsmath}
\usepackage{mathrsfs}
\usepackage{amsthm}
\usepackage{amssymb}
\usepackage{mathtools}
\usepackage{bm}
\usepackage{graphicx}
\usepackage{caption}
\usepackage{subcaption}
\usepackage{multirow}
\usepackage{booktabs}
\usepackage{dsfont}
\usepackage{cleveref}
\usepackage{authblk}

\usepackage
[
        a4paper,
        left=2.5cm, 
        right=2.5cm,
        top=2.5cm,
        bottom=2.5cm,
]
{geometry}

\newtheorem{theorem}{Theorem}[section]
\newtheorem{lemma}{Lemma}[section]
\newtheorem{corollary}{Corollary}[section]
\newtheorem{proposition}{Proposition}[section]
\newtheorem{definition}{Definition}[section]

\newtheorem{remark}{Remark}[section]
\newtheorem{condition}{Condition}[section]

\newenvironment{customcon}[1]
{\renewcommand\thecondition{#1}\condition}
{\endcondition}

\newcommand{\dd}{\mathrm{d}}

\renewcommand{\P}{{\mathbb{P}}}
\newcommand{\E}{{\mathbb{E}}}

\newcommand{\MISE}{{\mathrm{MISE}}}

\newcommand{\tr}{\operatorname{tr}}
\newcommand{\CV}{\operatorname{CV}}
\newcommand{\Unif}{\operatorname{Unif}}

\newcommand{\mw}{w}
\newcommand{\mx}{x}

\newcommand{\mz}{z}

\newcommand{\mI}{\mathbf{I}}
\newcommand{\mX}{X}

\DeclareMathOperator*{\argmin}{\arg\min}

\newcommand{\Exp}{{\rm Exp}}
\newcommand{\Log}{{\rm Log}}
\newcommand{\ISE}{{\rm ISE}}

\renewcommand{\theequation}{\thesection.\arabic{equation}}

\begin{document}

\title{Local Fr\'echet Regression with Riemannian Predictors}

\author[1]{Chang Jun Im}
\author[2]{Jeong Min Jeon\thanks{Corresponding author.}}
\affil[1]{The Institute for Data Innovation in Science, Seoul National University, South Korea}
\affil[2]{Department of Statistics and School of Transdisciplinary Innovations, Seoul National University, South Korea}

\maketitle

\begin{abstract}
Fr\'echet regression is well developed for Euclidean predictors, but local linear methods remain limited for general manifold-valued predictors. We propose local constant and local linear estimators for predictors lying on a general Riemannian manifold and responses taking values in a general metric space. The proposed local linear estimator is the first local linear Fr\'echet regression method in this setting. Our construction uses geodesic neighborhoods, logarithmic-map coordinates, volume-density correction, and frame-invariant scalar equivalent weights. For both estimators, we establish not only pointwise consistency and convergence rates but also uniform consistency and convergence rates. Simulations and real data applications demonstrate the finite-sample performance and practical applicability of the proposed methods across diverse predictor and response geometries.
\end{abstract}

\small
\begin{quotation}
\end{quotation}
\normalsize

\section{Introduction}
\label{sec:introduction}

Modern statistical analysis increasingly encounters data objects with non-Euclidean geometric structure. Examples include directional observations on spheres, shape configurations in Kendall's shape spaces, covariance and diffusion tensor data represented by symmetric positive definite matrices, and probability distributions viewed as elements of Wasserstein spaces. Such objects do not naturally reside in a single global vector space, and applying Euclidean regression methods after an arbitrary coordinate representation may obscure their intrinsic geometry. This has motivated the development of statistical methods for random objects in metric and manifold-valued spaces.

The Fr\'echet regression framework of \cite{Petersen and Muller (2019)} provides a general approach to regression with responses taking values in a metric space $(\mathbb M,d_{\mathbb M})$. Their local constant and local linear estimators, together with the uniform theory of \cite{Chen and Muller (2022)}, form the principal foundation for local Fr\'echet smoothing with Euclidean predictors. Subsequent work has developed regularized, network, single-index, and variable-selection extensions \citep{Lin and Muller (2021), Zhou and Muller (2022), Bhattacharjee and Muller (2023), Tucker et al. (2023)}. Random-forest constructions provide further alternatives: \cite{Capitaine et al. (2024)} allow both predictors and responses to lie in general metric spaces, while \cite{Qiu et al. (2024)} use forest weights in local constant and local linear Fr\'echet procedures with Euclidean predictors.

There is also a substantial literature on smoothing with manifold-valued predictors or responses. Manifold kernel density estimation and scalar-response regression with volume-density correction were studied by \cite{Pelletier (2005)} and \cite{Pelletier (2006)}. Local polynomial regression with scalar or functional responses and predictors supported on a manifold was considered by \cite{Cheng and Wu (2013)} and \cite{Lin and Yao (2021)}, while \cite{Yuan et al. (2012)} developed intrinsic local polynomial regression for SPD-valued responses with Euclidean covariates. Fr\'echet regression with non-Euclidean predictors has been developed for particular predictor spaces, notably spheres and tori \citep{Im et al. (2025), Im and Jeon (2026)}. These space-specific procedures are designed to respect the intrinsic geometries of their respective predictor spaces.

Beyond Riemannian predictor manifolds, \cite{Tucker and Wu (2025)} provide, for predictors in a general metric space, a local constant construction based on metric neighborhoods and H\"older regularity, but not the intrinsic local linear construction or moving-frame uniform theory considered here. Recent work by \cite{Ruiz-Medina and Torres-Signes (2025)} develops local linear Fr\'echet curve regression for time-correlated manifold-valued functional predictors and responses, an infinite-dimensional curve-regression setting distinct from the present problem of a finite-dimensional Riemannian predictor, a general metric-space-valued response, and uniform estimation over predictor regions.

The existing literature leaves open the development of local linear Fr\'echet regression for general finite-dimensional Riemannian predictors and general metric-space-valued responses. The proposed local linear estimator is the first such method in this setting. For both the local constant and local linear estimators, we establish not only pointwise consistency and convergence rates but also uniform consistency and uniform convergence rates over compact predictor regions, while allowing tangent coordinates and local frames to vary with the evaluation point.

We develop local constant and local linear Fr\'echet regression for predictors on a finite-dimensional Riemannian manifold $\mathcal M$ and responses in a general metric space $\mathbb M$. The estimators are defined directly through metric-space squared loss and are built from geodesic neighborhoods, logarithmic-map coordinates, and volume-density correction. The local constant estimator is a volume-normalized manifold kernel smoother, while the local linear estimator uses tangent-space first and second moments to construct signed scalar equivalent weights; although the intermediate coordinates depend on a chosen local ordered orthonormal frame, the resulting scalar weights and fitted Fr\'echet criterion are invariant to that choice.

The main contribution is threefold. First, we provide a common intrinsic construction of local constant and local linear Fr\'echet regression on general finite-dimensional Riemannian predictor manifolds, including settings without a global coordinate chart, a global orthonormal frame, or a canonical ambient representation; the volume-density factor removes the Riemannian Jacobian from the leading normal-coordinate moments and yields a coordinate-invariant local-design formulation. Second, we establish pointwise and uniform consistency and convergence rates under explicit geometric, design, response-side, empirical-process, and Fr\'echet-margin conditions; under a quadratic Fr\'echet margin and the corresponding bandwidth choices, the pointwise and uniform distance upper rates are $O_{\mathbb P}\{n^{-2/(d+4)}\}$ and $O_{\mathbb P}\{(\log n/n)^{2/(d+4)}\}$, the standard twice-smooth Euclidean upper-bound orders with the Euclidean predictor dimension replaced by the intrinsic manifold dimension $d$. We do not establish a matching lower bound and therefore make no minimax-optimality claim. Third, the uniform theory separates the empirical-process requirements of the two estimators, with a zeroth-order kernel-window class for local constant smoothing and additional first- and second-order logarithmic-coordinate multiplier classes for local linear smoothing; we verify these requirements for standard compactly supported piecewise-polynomial kernels, including nonsmooth indicator-type profiles, on compact evaluation regions of Euclidean spaces, spheres, finite products of spheres, flat tori, and symmetric positive-definite manifolds equipped with the affine-invariant Riemannian metric.

We complement the theory with two simulation studies and two real data analyses. The simulations consider spherical predictors, for which the general-manifold construction is compared with the sphere-specific estimators under their respective conventional kernels, and symmetric positive-definite predictors equipped with the affine-invariant Riemannian metric. The real data analyses involve OASIS--3 diffusion-tensor data with an $\mathcal S_{++}^{3}$ predictor and a Wasserstein distributional response, and a head-and-gaze analysis with an $\mathrm{SO}(3)$ predictor and an $\mathbb S^{2}$ response; both predictor manifolds lie outside the scope of the earlier sphere- and torus-specific procedures. The proposed local linear estimator attains the lowest aggregate error in the reported comparisons.

The rest of the paper is organized as follows. \Cref{sec:riemannian_geometry} reviews the required Riemannian-geometric tools. \Cref{sec:estimation} introduces the problem setting and constructs the proposed estimators. \Cref{sec:asymptotic_theory} develops the pointwise and uniform theory. \Cref{sec:simulation} and \Cref{sec:real_data} present the simulation studies and real data analyses, respectively. Concluding remarks are given in \Cref{sec:discussion}, and all technical proofs are provided in the appendices.

\section{Preliminaries on Riemannian Geometry} \label{sec:riemannian_geometry}

In this section, we briefly review the basic Riemannian-geometric notions used throughout the paper. We focus on the local geometric properties needed for constructing and analyzing the proposed estimators, namely those that allow us to represent small neighborhoods of a manifold point in its tangent space. For a comprehensive geometric treatment, including the Levi--Civita connection, covariant derivatives, geodesics, the Hopf--Rinow theorem, and normal coordinates, we refer to \cite{Do Carmo (1992)} and \cite{Chavel (2006)}. Throughout this section, let $(\mathcal{M},g)$ be a connected $d$-dimensional complete Riemannian manifold without boundary. We denote by $d_{\mathcal M}$ the Riemannian geodesic distance induced by $g$. Let $x\in\mathcal M$ be a fixed point.

\subsection{Riemannian Metric and Geodesics}

The tangent space at $\mx$, denoted by $T_{\mx}\mathcal{M}$, is a $d$-dimensional vector space equipped with the inner product $\langle\cdot,\cdot\rangle_{\mx}$ induced by the Riemannian metric $g$. The corresponding norm is
\begin{align*}
    \|\mathbf{v}\|_{\mx}:=\sqrt{\langle\mathbf{v},\mathbf{v}\rangle_{\mx}}, \quad \mathbf{v}\in T_{\mx}\mathcal{M}.
\end{align*}

A smooth curve $\gamma:I\to\mathcal{M}$, defined on an interval $I\subset\mathbb{R}$, is called a \textit{geodesic} if its covariant acceleration vanishes:
\begin{align*}
    \nabla^{\mathcal{M}}_{\dot{\gamma}(t)}\dot{\gamma}(t)=\mathbf{0}_{\gamma(t)}, \quad t\in I.
\end{align*}
Here, $\nabla^{\mathcal{M}}$ denotes the Levi--Civita connection associated with the Riemannian metric $g$, and $\mathbf{0}_{\mz}$ denotes the zero vector in $T_{\mz}\mathcal{M}$ for each $\mz\in\mathcal{M}$. For a tangent vector $\mathbf{u}\in T_{\mx}\mathcal{M}$ and a smooth vector field $\mathbf{V}$ defined near $\mx$, $\nabla^{\mathcal{M}}_{\mathbf{u}}\mathbf{V}$ denotes the intrinsic directional derivative of $\mathbf{V}$ along $\mathbf{u}$, evaluated as a tangent vector at $\mx$. Geodesics are the Riemannian analogues of straight lines in Euclidean space; in particular, sufficiently short geodesic segments are locally length-minimizing.

\subsection{Exponential Map, Injectivity Radius, and Volume Density}

For any $\mathbf{v}\in T_{\mx}\mathcal{M}$, let $\gamma_{\mathbf{v}}$ be the geodesic satisfying $\gamma_{\mathbf{v}}(0)=\mx$ and $\dot{\gamma}_{\mathbf{v}}(0)=\mathbf{v}$. Since $\mathcal{M}$ is complete, the Hopf--Rinow theorem ensures that $\gamma_{\mathbf{v}}$ is defined on the whole real line. The \textit{exponential map} at $\mx$ is defined by
\begin{align*}
    \Exp_{\mx}(\mathbf{v}):=\gamma_{\mathbf{v}}(1), \quad \mathbf{v}\in T_{\mx}\mathcal{M}.
\end{align*}
Equivalently, $\gamma_{\mathbf{v}}(t)=\Exp_{\mx}(t\mathbf{v})$ for any $\mathbf{v}\in T_{\mx}\mathcal{M}$ and $t\in\mathbb{R}$.

The \textit{injectivity radius} at $\mx$, denoted by $i(\mx)$, is defined as
\begin{align*}
    i(\mx):=\sup\left\{r>0:\Exp_{\mx}\big|_{B_{\|\cdot\|_{\mx}}(\mathbf{0}_{\mx},r)}\text{ is a diffeomorphism onto its image}\right\}.
\end{align*}
By the inverse function theorem, $i(\mx)>0$ for every $\mx\in\mathcal{M}$. Hence, whenever $d_{\mathcal{M}}(\mx,\mz)<i(\mx)$, the \textit{logarithmic map}
\begin{align*}
    \Log_{\mx}(\mz):=\Exp_{\mx}^{-1}(\mz), \quad \mz\in B_{\mathcal{M}}(\mx,i(\mx)),
\end{align*}
is uniquely defined and smooth. In this normal neighborhood,
\begin{align*}
    d_{\mathcal{M}}(\mx,\mz)=\|\Log_{\mx}(\mz)\|_{\mx}, \quad \mz\in B_{\mathcal{M}}(\mx,i(\mx)).
\end{align*}

For later uniform arguments, we also recall the following standard fact. If $\mathcal{K}\subset\mathcal{M}$ is compact, then, by the positivity and continuity of the injectivity-radius function on a complete Riemannian manifold,
\begin{align*}
    i(\mathcal{K}):=\inf_{\mx\in\mathcal{K}}i(\mx)>0.
\end{align*}
Thus, for any fixed $\rho\in(0,i(\mathcal{K}))$, the logarithmic maps $\Log_{\mx}(\mz)$ are well-defined for all $\mx\in\mathcal{K}$ and $\mz\in B_{\mathcal{M}}(\mx,\rho)$. Moreover, the closed geodesic tube
\begin{align*}
    \mathcal{K}^{\rho}:=\{\mz\in\mathcal{M}:d_{\mathcal{M}}(\mz,\mathcal{K})\leq\rho\}
\end{align*}
is compact by the Hopf--Rinow theorem.

To express integration over $\mathcal{M}$ in normal coordinates, fix an \textit{ordered orthonormal basis}
\begin{align*}
    \mathbf{E}_{\mx}:=(\mathbf{E}_{\mx,1},\ldots,\mathbf{E}_{\mx,d})\in\mathcal{E}_{\mx},
\end{align*}
where $\mathcal{E}_{\mx}$ denotes the collection of all ordered orthonormal bases of $T_{\mx}\mathcal{M}$. For $\mz\in B_{\mathcal{M}}(\mx,i(\mx))$, write
\begin{align*}
    \mathbf{v}:=\Log_{\mx}(\mz)=\sum_{j=1}^{d}v_j\mathbf{E}_{\mx,j},
    \quad
    \bm{v}:=(v_1,\ldots,v_d)^\top\in\mathbb{R}^d.
\end{align*}
Let
\begin{align*}
    (d\Exp_{\mx})_{\mathbf{v}}:T_{\mathbf{v}}(T_{\mx}\mathcal{M})\to T_{\mz}\mathcal{M}, \quad \mathbf{v}=\Log_{\mx}(\mz),
\end{align*}
denote the differential of the exponential map at $\mathbf{v}$. Since $T_{\mx}\mathcal{M}$ is a vector space, we naturally identify $T_{\mathbf{v}}(T_{\mx}\mathcal{M})$ with $T_{\mx}\mathcal{M}$. Define the \textit{metric tensor matrix} in these normal coordinates by
\begin{align*}
    \left[\mathbf{G}_{\mathbf{E}_{\mx}}(\mz)\right]_{kl}:=\left\langle(d\Exp_{\mx})_{\mathbf{v}}(\mathbf{E}_{\mx,k}),(d\Exp_{\mx})_{\mathbf{v}}(\mathbf{E}_{\mx,l})\right\rangle_{\mz}, \quad 1\leq k,l\leq d.
\end{align*}
Since $(d\Exp_{\mx})_{\mathbf{0}_{\mx}}$ is the identity map,
\begin{align*}
    \mathbf{G}_{\mathbf{E}_{\mx}}(\mx)=\mI_d.
\end{align*}
The \textit{volume density function} is defined by
\begin{align*}
    \theta_{\mx}(\mz):=\sqrt{\det\left(\mathbf{G}_{\mathbf{E}_{\mx}}(\mz)\right)}, \quad \mz\in B_{\mathcal{M}}(\mx,i(\mx)).
\end{align*}
Although $\mathbf{G}_{\mathbf{E}_{\mx}}(\mz)$ depends on the chosen ordered orthonormal basis, its determinant does not; see \Cref{lemma:A.invariance_volume_density}. Hence $\theta_{\mx}(\mz)$ is intrinsically well-defined. Moreover, $\theta_{\mx}$ is smooth and strictly positive on $B_{\mathcal{M}}(\mx,i(\mx))$, and $\theta_{\mx}(\mx)=1$.

Under the change of variables $\mz=\Exp_{\mx}(\mathbf{v})$, the Riemannian volume measure satisfies
\begin{align*}
    \dd v_g(\mz)=\theta_{\mx}(\Exp_{\mx}(\mathbf{v}))\,\dd\mathbf{v}, \quad \mathbf{v}\in B_{\|\cdot\|_{\mx}}(\mathbf{0}_{\mx},i(\mx)),
\end{align*}
where $\dd\mathbf{v}$ denotes the Lebesgue measure on $T_{\mx}\mathcal{M}$ induced by the ordered orthonormal basis $\mathbf{E}_{\mx}$. Equivalently, writing $\mz=\Exp_{\mx}(\mathbf{v})$,
\begin{align*}
    \dd v_g(\mz)=\theta_{\mx}(\mz)\,\dd\mathbf{v}, \quad \mz\in B_{\mathcal{M}}(\mx,i(\mx)).
\end{align*}

\subsection{Gradient, Hessian, and Taylor Expansion}

For a smooth function $\phi:\mathcal{M}\to\mathbb{R}$, the \textit{Riemannian gradient} $\nabla\phi(\mx)\in T_{\mx}\mathcal{M}$ is defined by
\begin{align*}
    \left\langle\nabla\phi(\mx),\mathbf{v}\right\rangle_{\mx}=\frac{d}{dt}\phi(\Exp_{\mx}(t\mathbf{v}))\Big|_{t=0}, \quad \mathbf{v}\in T_{\mx}\mathcal{M}.
\end{align*}
The \textit{Riemannian Hessian} $\nabla^2\phi(\mx)$ is the symmetric bilinear form on $T_{\mx}\mathcal{M}$ defined by
\begin{align*}
    \nabla^2\phi(\mx)(\mathbf{u},\mathbf{v})=\left\langle\nabla^{\mathcal{M}}_{\mathbf{u}}\nabla\phi,\mathbf{v}\right\rangle_{\mx}, \quad \mathbf{u},\mathbf{v}\in T_{\mx}\mathcal{M}.
\end{align*}
In particular, along the geodesic $t\mapsto\Exp_{\mx}(t\mathbf{v})$,
\begin{align*}
    \nabla^2\phi(\mx)(\mathbf{v},\mathbf{v})=\frac{d^2}{dt^2}\phi(\Exp_{\mx}(t\mathbf{v}))\Big|_{t=0}, \quad \mathbf{v}\in T_{\mx}\mathcal{M}.
\end{align*}
In normal coordinates centered at $\mx$, the Christoffel symbols vanish at the origin. Therefore, for a sufficiently smooth $\phi$, the covariant Taylor expansion around $\mx$ takes the form
\begin{align*}
    \phi(\mz)=\phi(\mx)+\left\langle\nabla\phi(\mx),\Log_{\mx}(\mz)\right\rangle_{\mx}+\frac{1}{2}\nabla^2\phi(\mx)\left(\Log_{\mx}(\mz),\Log_{\mx}(\mz)\right)+R_{\mx}(\mz), \quad \mz\in B_{\mathcal{M}}(\mx,i(\mx)).
\end{align*}
Here, the remainder satisfies
\begin{align*}
    R_{\mx}(\mz)=o\left(d_{\mathcal{M}}^2(\mx,\mz)\right), \quad \text{as }\mz\to\mx.
\end{align*}
If $\phi$ is three times continuously differentiable in a neighborhood of $\mx$, the remainder can be strengthened to
\begin{align*}
    R_{\mx}(\mz)=O\left(d_{\mathcal{M}}^3(\mx,\mz)\right), \quad \text{as }\mz\to\mx.
\end{align*}

\section{Problem Setting and Estimators} \label{sec:estimation}
\setcounter{equation}{0}

\subsection{Problem Setting}

Let $(\Omega,\mathcal{F},\P)$ be an underlying probability space. Let $(\mathcal{M},g)$ be a connected $d$-dimensional complete Riemannian manifold without boundary, as introduced in \Cref{sec:riemannian_geometry}, and let $(\mathbb{M},d_{\mathbb{M}})$ be a general metric space. Let $\mX:\Omega\to\mathcal{M}$ be a manifold-valued predictor and let $Y:\Omega\to\mathbb{M}$ be the corresponding metric-space-valued response. Throughout the estimator construction, $(\mX^{(i)},Y^{(i)})$, $i=1,\ldots,n$, denote independent copies of $(\mX,Y)$.

Define the conditional Fr\'echet function by
\begin{align*}
    M_{\oplus}(\mx,y):=\E\left[d_{\mathbb{M}}^2(Y,y)\mid \mX=\mx\right], \quad (\mx,y)\in\mathcal{M}\times\mathbb{M}.
\end{align*}
When this minimizer exists and is unique, the target regression function is the conditional Fr\'echet mean
\begin{align*}
    m_{\oplus}(\mx):=\argmin_{y\in\mathbb{M}}M_{\oplus}(\mx,y), \quad \mx\in\mathcal{M}.
\end{align*}
When $\mathbb{M}=\mathbb{R}$, this definition reduces to the ordinary conditional mean because $\E[Y\mid \mX=\mx]=\argmin_{y\in\mathbb{R}}\E[(Y-y)^2\mid \mX=\mx]$. Thus, $m_{\oplus}$ is the natural Fr\'echet-regression analogue of the usual regression function \citep{Petersen and Muller (2019)}.

In the estimator construction below, we fix an evaluation point $\mx \in \mathcal{M}$. The predictor space $\mathcal{M}$ is not Euclidean, so local smoothing must account for both geodesic geometry and the relation between Riemannian volume and Euclidean volume in normal coordinates. Therefore, we use $d_{\mathcal{M}}$ to measure local proximity, $\Log_{\mx}$ to represent nearby predictors in $T_{\mx}\mathcal{M}$, and $\theta_{\mx}$ to correct the Riemannian volume measure in normal coordinates. The construction combines the normal-coordinate kernel normalization of \cite{Pelletier (2006)} with the local Fr\'echet equivalent-weight principle of \cite{Petersen and Muller (2019)}. It shares a first-order tangent-coordinate principle with the sphere- and torus-specific procedures of \cite{Im et al. (2025)} and \cite{Im and Jeon (2026)}, but differs in its kernel construction and, in the toroidal case, its bandwidth structure. Consequently, the resulting estimators are generally distinct.

\subsection{Local Constant Estimator} \label{sec:local_constant}

Let $K:[0,\infty)\to[0,\infty)$ be a nonnegative kernel function and let $h>0$ be a bandwidth. For $\mx\in\mathcal{M}$, define the volume-corrected manifold kernel weight by
\begin{align}
    \mathcal{L}_{\mx,h}(\mz):=
    \begin{cases}
    \theta_{\mx}(\mz)^{-1}K\left(\dfrac{d_{\mathcal{M}}(\mx,\mz)}{h}\right), & \mz\in B_{\mathcal{M}}(\mx,i(\mx)), \\
    0, & \mz\notin B_{\mathcal{M}}(\mx,i(\mx)).
    \end{cases} \label{eq:sec3.volume_corrected_kernel}
\end{align}
We omit the common factor $h^{-d}$ from $\mathcal{L}_{\mx,h}$ because it cancels from the local constant objective and from the local linear equivalent weights. The resulting estimators are therefore identical to those obtained by replacing $\mathcal{L}_{\mx,h}$ everywhere with $h^{-d}\mathcal{L}_{\mx,h}$, that is, by using
\begin{align*}
    h^{-d}\theta_{\mx}(\mz)^{-1}K\left(\dfrac{d_{\mathcal{M}}(\mx,\mz)}{h}\right),
    \quad \mz\in B_{\mathcal{M}}(\mx,i(\mx)),
\end{align*}
with value zero outside $B_{\mathcal{M}}(\mx,i(\mx))$. When $K$ is supported on $[0,1]$ and $h<i(\mx)$, only observations $\mz\in\mathcal{M}$ satisfying $d_{\mathcal{M}}(\mx,\mz)\leq h$ can receive nonzero weight, and all such observations lie inside the normal neighborhood $B_{\mathcal{M}}(\mx,i(\mx))$ where $\theta_{\mx}(\cdot)$ is well-defined. The factor $\theta_{\mx}(\cdot)^{-1}$ converts integration with respect to the Riemannian volume measure into integration with respect to Lebesgue measure on $T_{\mx}\mathcal{M}$ in normal coordinates. On compact normal-coordinate neighborhoods, $\theta_{\mx}(\mz)=1+O\{d_{\mathcal M}^2(\mx,\mz)\}$ as $\mz\to\mx$, so the correction can be numerically small for small bandwidths or weakly curved local regions. The structural role of this correction in the local moments and bias expansions is discussed in \Cref{rem:volume_density_role_theory}.

To motivate the construction, first suppose that $\mathbb{M}=\mathbb{R}$ and write $m(\mx):=\E[Y\mid \mX=\mx]$ for the usual scalar regression function. The local constant estimator of $m(\mx)$ is obtained from the locally weighted least-squares problem
\begin{align}
    \hat{\alpha}_{h,0}(\mx)
    &:=
    \argmin_{\alpha\in\mathbb{R}}
    \sum_{i=1}^{n}
    \mathcal{L}_{\mx,h}\left(\mX^{(i)}\right)
    \left(Y^{(i)}-\alpha\right)^2. \label{eq:sec3.optim_scalar_local_constant}
\end{align}
Define the zeroth local sample moment by
\begin{align*}
    \hat{\mu}_{h,0}(\mx)
    &:=
    n^{-1}\sum_{i=1}^{n}
    \mathcal{L}_{\mx,h}\left(\mX^{(i)}\right).
\end{align*}
\Cref{lemma:B.pointwise_empirical_moments} implies that $\hat{\mu}_{h,0}(\mx)>0$ with probability tending to one. On the event that $\hat{\mu}_{h,0}(\mx)>0$, define the empirical local constant equivalent-weight function by
\begin{align} \label{eq:sec3.empirical_equiv_weight_lc}
    \hat{W}_{\mx,h,0}(\mz)
    &:=
    \frac{\mathcal{L}_{\mx,h}(\mz)}{\hat{\mu}_{h,0}(\mx)},
    \quad \mz\in\mathcal{M}.
\end{align}
Then the solution of \eqref{eq:sec3.optim_scalar_local_constant} is
\begin{align*}
    \hat{\alpha}_{h,0}(\mx)
    =
    n^{-1}\sum_{i=1}^{n}
    \hat{W}_{\mx,h,0}\left(\mX^{(i)}\right)Y^{(i)}.
\end{align*}
The equivalent weights satisfy the exact normalization identity $n^{-1}\sum_{i=1}^{n}\hat{W}_{\mx,h,0}(\mX^{(i)})=1$ whenever $\hat{\mu}_{h,0}(\mx)$ is positive. Therefore, $\hat{\alpha}_{h,0}(\mx)$ can be written as the minimizer
\begin{align}
    \hat{\alpha}_{h,0}(\mx)
    =
    \argmin_{y\in\mathbb{R}}
    n^{-1}\sum_{i=1}^{n}
    \hat{W}_{\mx,h,0}\left(\mX^{(i)}\right)
    \left(Y^{(i)}-y\right)^2. \label{eq:sec3.equiv_kernel_form_lc}
\end{align}

Equation \eqref{eq:sec3.equiv_kernel_form_lc} provides the bridge to general metric-space-valued responses. Replacing the Euclidean squared loss with $d_{\mathbb{M}}^2$ leads to the local constant empirical Fr\'echet objective
\begin{align*}
    \hat{M}_{h,0}(\mx,y)
    &:=
    n^{-1}\sum_{i=1}^{n}
    \hat{W}_{\mx,h,0}\left(\mX^{(i)}\right)
    d_{\mathbb{M}}^2\left(Y^{(i)},y\right), 
    \quad y\in\mathbb{M}.
\end{align*}
The objective is understood on the event that $\hat{\mu}_{h,0}(\mx)>0$; under the regularity conditions below, this event has probability tending to one by \Cref{lemma:B.pointwise_empirical_moments}. The local constant Fr\'echet regression estimator is then
\begin{align*}
    \hat{m}_{h,0}(\mx)
    \in
    \argmin_{y\in\mathbb{M}}
    \hat{M}_{h,0}(\mx,y).
\end{align*}

\subsection{Local Linear Estimator} \label{sec:local_linear}

We next introduce the local linear estimator. The key idea is to perform a local linear approximation in the tangent space $T_{\mx}\mathcal{M}$, while keeping the response space $\mathbb{M}$ purely metric. To motivate the construction, first suppose that $\mathbb{M}=\mathbb{R}$ and write $m(\mx):=\E[Y\mid\mX=\mx]$ for the usual scalar regression function.

Let $m:\mathcal{M}\to\mathbb{R}$ be sufficiently smooth. For a nearby point $\mz\in B_{\mathcal{M}}(\mx,i(\mx))$, the Riemannian Taylor expansion in \Cref{sec:riemannian_geometry} gives the first-order approximation
\begin{align}
    m(\mz)\approx m(\mx)+\left\langle\nabla m(\mx),\Log_{\mx}(\mz)\right\rangle_{\mx}, \quad \mz\in B_{\mathcal{M}}(\mx,i(\mx)). \label{eq:sec3.first_order_taylor_manifold}
\end{align}
For an ordered orthonormal basis $\mathbf{E}_{\mx}=(\mathbf{E}_{\mx,1},\ldots,\mathbf{E}_{\mx,d})\in\mathcal{E}_{\mx}$, define the coordinate isomorphism $\bm{\Phi}_{\mathbf{E}_{\mx}}:T_{\mx}\mathcal{M}\to\mathbb{R}^d$ by
\begin{align*}
    \bm{\Phi}_{\mathbf{E}_{\mx}}(\mathbf{u}):=(u_1,\ldots,u_d)^\top,
    \quad
    \mathbf{u}=\sum_{j=1}^{d}u_j\mathbf{E}_{\mx,j}\in T_{\mx}\mathcal{M}.
\end{align*}
For $\mz\in\mathcal{M}$, define the basis-dependent tangent-coordinate map $\mathbf{v}_{\mx}^{\mathbf{E}_{\mx}}:\mathcal{M}\to\mathbb{R}^d$ by
\begin{align} \label{eq:sec3.tangent_coordinate_map}
    \mathbf{v}_{\mx}^{\mathbf{E}_{\mx}}(\mz):=
    \begin{cases}
    \bm{\Phi}_{\mathbf{E}_{\mx}}\left(\Log_{\mx}(\mz)\right), & \mz\in B_{\mathcal{M}}(\mx,i(\mx)), \\
    \mathbf{0}_d, & \mz\notin B_{\mathcal{M}}(\mx,i(\mx)).
    \end{cases}
\end{align}
This convention has no effect on the local criterion for all sufficiently small $h<i(\mx)$, because $\mathcal{L}_{\mx,h}(\mz)=0$ unless $d_{\mathcal{M}}(\mx,\mz)\leq h$. If $\bm{\beta}:=\bm{\Phi}_{\mathbf{E}_{\mx}}(\nabla m(\mx))\in\mathbb{R}^d$ and $\mathbf{v}:=\bm{\Phi}_{\mathbf{E}_{\mx}}(\Log_{\mx}(\mz))\in\mathbb{R}^d$, then the orthonormality of $\mathbf{E}_{\mx}$ gives
\begin{align*}
    \left\langle\nabla m(\mx),\Log_{\mx}(\mz)\right\rangle_{\mx}=\bm{\beta}^{\top}\mathbf{v}, \quad \mz\in B_{\mathcal{M}}(\mx,i(\mx)).
\end{align*}
Motivated by the first-order approximation in \eqref{eq:sec3.first_order_taylor_manifold} and its coordinate representation above, the scalar-response local linear estimator of $m(\mx)$ is obtained from the locally weighted least-squares problem
\begin{align}
    \left(\hat{\alpha}_{h,1}(\mx),\hat{\bm{\beta}}_{h,1}(\mx)\right):=\argmin_{\alpha\in\mathbb{R},\,\bm{\beta}\in\mathbb{R}^d} n^{-1}\sum_{i=1}^{n}\mathcal{L}_{\mx,h}\left(\mX^{(i)}\right)\left(Y^{(i)}-\alpha-\bm{\beta}^{\top}\mathbf{v}_{\mx}^{\mathbf{E}_{\mx}}\left(\mX^{(i)}\right)\right)^{2}. \label{eq:sec3.optim_scalar_local_linear}
\end{align}
For $\mathbf{v}\in\mathbb{R}^d$, define $\mathbf{v}^{\otimes0}:=1$, $\mathbf{v}^{\otimes1}:=\mathbf{v}$, and $\mathbf{v}^{\otimes2}:=\mathbf{v}\mathbf{v}^{\top}$. Together with the zeroth local sample moment $\hat{\mu}_{h,0}(\mx)$ defined in \Cref{sec:local_constant}, the first and second local sample moments associated with $\mathbf{E}_{\mx}$ are
\begin{align*}
    \bm{\hat{\mu}}_{h,1}(\mx,\mathbf{E}_{\mx})&:=n^{-1}\sum_{i=1}^{n}\mathcal{L}_{\mx,h}\left(\mX^{(i)}\right)\mathbf{v}_{\mx}^{\mathbf{E}_{\mx}}\left(\mX^{(i)}\right), \\
    \bm{\hat{\mu}}_{h,2}(\mx,\mathbf{E}_{\mx})&:=n^{-1}\sum_{i=1}^{n}\mathcal{L}_{\mx,h}\left(\mX^{(i)}\right)\mathbf{v}_{\mx}^{\mathbf{E}_{\mx}}\left(\mX^{(i)}\right)\left(\mathbf{v}_{\mx}^{\mathbf{E}_{\mx}}\left(\mX^{(i)}\right)\right)^{\top}.
\end{align*}
The normal equation corresponding to \eqref{eq:sec3.optim_scalar_local_linear} is
\begin{align}
    \begin{bmatrix}
        n^{-1}\sum_{i=1}^{n}\mathcal{L}_{\mx,h}\left(\mX^{(i)}\right)Y^{(i)} \\
        n^{-1}\sum_{i=1}^{n}\mathcal{L}_{\mx,h}\left(\mX^{(i)}\right)\mathbf{v}_{\mx}^{\mathbf{E}_{\mx}}\left(\mX^{(i)}\right)Y^{(i)}
    \end{bmatrix}
    =
    \begin{bmatrix}
        \hat{\mu}_{h,0}(\mx) & \bm{\hat{\mu}}_{h,1}(\mx,\mathbf{E}_{\mx})^{\top} \\
        \bm{\hat{\mu}}_{h,1}(\mx,\mathbf{E}_{\mx}) & \bm{\hat{\mu}}_{h,2}(\mx,\mathbf{E}_{\mx})
    \end{bmatrix}
    \begin{bmatrix}
        \hat{\alpha}_{h,1}(\mx) \\
        \hat{\bm{\beta}}_{h,1}(\mx)
    \end{bmatrix}. \label{eq:sec3.normal_equation_local_linear}
\end{align}
\Cref{lemma:B.pointwise_empirical_moments} implies the invertibility of $\bm{\hat{\mu}}_{h,2}(\mx,\mathbf{E}_{\mx})$ with probability tending to one as $n \to \infty$. Although $\bm{\hat{\mu}}_{h,1}(\mx,\mathbf{E}_{\mx})$, $\bm{\hat{\mu}}_{h,2}(\mx,\mathbf{E}_{\mx})$, and $\mathbf{v}_{\mx}^{\mathbf{E}_{\mx}}$ depend on the chosen ordered orthonormal basis, \Cref{lemma:A.invariance_local_linear_weights} implies that the scalar quantities
\begin{align*}
    \bm{\hat{\mu}}_{h,1}(\mx,\mathbf{E}_{\mx})^{\top}
    \bm{\hat{\mu}}_{h,2}(\mx,\mathbf{E}_{\mx})^{-1}
    \bm{\hat{\mu}}_{h,1}(\mx,\mathbf{E}_{\mx}), \quad \bm{\hat{\mu}}_{h,1}(\mx,\mathbf{E}_{\mx})^{\top}
    \bm{\hat{\mu}}_{h,2}(\mx,\mathbf{E}_{\mx})^{-1}
    \mathbf{v}_{\mx}^{\mathbf{E}_{\mx}}\left(\mX^{(i)}\right)
\end{align*}
are invariant under a change of basis. Hence, we define the basis-independent normalization factor and empirical equivalent-weight function by
\begin{align} \label{eq:sec3.empirical_equiv_weight_ll}
\begin{split} 
    \hat{\sigma}_{h}(\mx)
    &:=
    \hat{\mu}_{h,0}(\mx)
    -
    \bm{\hat{\mu}}_{h,1}(\mx,\mathbf{E}_{\mx})^{\top}
    \bm{\hat{\mu}}_{h,2}(\mx,\mathbf{E}_{\mx})^{-1}
    \bm{\hat{\mu}}_{h,1}(\mx,\mathbf{E}_{\mx}), \\
    \hat{W}_{\mx,h,1}(\mz)
    &:=
    \frac{\mathcal{L}_{\mx,h}(\mz)}{\hat{\sigma}_{h}(\mx)}
    \left[
    1-
    \bm{\hat{\mu}}_{h,1}(\mx,\mathbf{E}_{\mx})^{\top}
    \bm{\hat{\mu}}_{h,2}(\mx,\mathbf{E}_{\mx})^{-1}
    \mathbf{v}_{\mx}^{\mathbf{E}_{\mx}}(\mz)
    \right], \quad \mz\in\mathcal{M}.
\end{split}
\end{align}
Also, \Cref{lemma:B.pointwise_denominators} implies that $\hat{\sigma}_{h}(\mx)>0$ with probability tending to one. On the event that $\bm{\hat{\mu}}_{h,2}(\mx,\mathbf{E}_{\mx})$ is nonsingular and $\hat{\sigma}_{h}(\mx)>0$, block matrix inversion of \eqref{eq:sec3.normal_equation_local_linear} gives
\begin{align*}
    \hat{\alpha}_{h,1}(\mx)=n^{-1}\sum_{i=1}^{n}\hat{W}_{\mx,h,1}(\mX^{(i)})Y^{(i)}.
\end{align*}
The equivalent weights satisfy the exact normalization identity $n^{-1}\sum_{i=1}^{n}\hat{W}_{\mx,h,1}(\mX^{(i)})=1$ whenever $\bm{\hat{\mu}}_{h,2}(\mx,\mathbf{E}_{\mx})$ is nonsingular and $\hat{\sigma}_{h}(\mx)\neq0$. Therefore, $\hat{\alpha}_{h,1}(\mx)$ can be written as the minimizer
\begin{align}
    \hat{\alpha}_{h,1}(\mx)=\argmin_{y\in\mathbb{R}} n^{-1}\sum_{i=1}^{n}\hat{W}_{\mx,h,1}(\mX^{(i)})(Y^{(i)}-y)^2. \label{eq:sec3.equiv_kernel_form_ll}
\end{align}

Equation \eqref{eq:sec3.equiv_kernel_form_ll} provides the bridge to general metric-space-valued responses. Replacing the Euclidean squared loss with $d_{\mathbb{M}}^2$ leads to the local linear empirical Fr\'echet objective
\begin{align*}
    \hat{M}_{h,1}(\mx,y):=n^{-1}\sum_{i=1}^{n}\hat{W}_{\mx,h,1}(\mX^{(i)})d_{\mathbb{M}}^2(Y^{(i)},y), \quad y\in\mathbb{M}.
\end{align*}
As in ordinary local linear smoothing, the equivalent weights may be negative. For Euclidean or Hilbert responses, the weighted squared-loss representation remains algebraically transparent. For a general metric response space, however, signed weights do not automatically preserve convexity, uniqueness, or stable measurable selection of the empirical Fr\'echet minimizer. The asymptotic theory below therefore imposes existence, uniqueness, separation, and local margin conditions for the relevant population, localized, and empirical Fr\'echet objectives. These are high-level sufficient conditions. For signed local linear objectives, their verification is response-space- and model-specific and is not implied by compactness or by ordinary Fr\'echet-mean assumptions alone. Recent work gives explicit existence and optimization conditions for signed Fr\'echet objectives on bounded-curvature response manifolds \citep{Zhou and Uribe (2026)}; the present assumptions retain a broader metric-response formulation and do not rely on those manifold-specific conditions. Natural settings in which the assumptions may be verified include compact Euclidean response sets with a nonsingular conditional second moment structure, compact geodesically convex regions of Riemannian response manifolds away from cut loci, and bounded one-dimensional Wasserstein classes. The local linear Fr\'echet regression estimator is then
\begin{align*}
    \hat{m}_{h,1}(\mx)\in\argmin_{y\in\mathbb{M}}\hat{M}_{h,1}(\mx,y).
\end{align*}

To regard the estimators as fully defined random elements, fix once and for all a reference point $y_{\circ}\in\mathbb M$. At any evaluation point $\mx$ and on any sample outcome for which a required local moment, denominator, or measurable minimizer is not well defined, set $\hat m_{h,s}(\mx):=y_{\circ}$, $s\in\{0,1\}$. The same pointwise convention defines the fitted map over a compact evaluation set. Under the pointwise or uniform good events established below, this convention is inactive; since the exceptional probabilities tend to zero, it does not alter any consistency or rate conclusion.

\begin{remark}[Euclidean specialization]\label{rem:euclidean_specialization}
When $\mathcal M=\mathbb R^d$ with its Euclidean metric, $\theta_{\mx}\equiv1$, $\Log_{\mx}(\mz)=\mz-\mx$, and a single fixed orthonormal frame may be used. The equivalent-weight functions are then normalized to have unit sample average, and the effective observation weights $n^{-1}\hat W_{\mx,h,0}$ and $n^{-1}\hat W_{\mx,h,1}$ coincide with the usual normalized radial Nadaraya--Watson and multivariate local linear equivalent weights, respectively. In particular, when $d=1$, the corresponding metric-response estimators coincide with the local Fr\'echet estimators of \cite{Petersen and Muller (2019)}; for $d>1$, the formulas give their standard multivariate local linear extension.
\end{remark}

The preceding construction also explains why the uniform theory for the local linear estimator is more demanding than that for the local constant estimator. Local constant smoothing only uses empirical-process control of the zeroth-order kernel-window class generated by $\mathcal L_{\mx,h}$. Local linear smoothing additionally uses tangent-coordinate multiplier classes involving $\mathcal L_{\mx,h}\{\mathbf v_{\mx}^{\mathbf E_{\mx}}\}^{\otimes j}$ for $j=1,2$, because these terms determine the empirical first moment, second moment, inverse moment matrix, and equivalent weights. In \Cref{sec:asymptotic_theory}, U-K1 records the zeroth-order input and U-K2 records the additional first- and second-order multiplier input. On the tame manifolds treated in \Cref{app:condition_verification}, both conditions are verified jointly. Their separation therefore identifies the minimal empirical-process complexity used by each estimator; it is not a claim that U-K1 and U-K2 are inequivalent on every manifold of interest.

\begin{remark}[Scope of predictor manifolds]
The construction above is not tied to a particular predictor manifold. It applies to any finite-dimensional Riemannian predictor manifold on which the logarithmic map, normal-coordinate volume density, and local orthonormal frames are available on the relevant evaluation region. In the pointwise theory this region is a fixed normal neighborhood of the evaluation point, whereas in the uniform theory it is a compact subset admitting a finite smooth local-frame cover. This finite-cover formulation is essential on a general manifold, where no single global frame or global Euclidean coordinate representation need be available. Thus the framework is designed to cover compact manifolds such as circles, spheres, products of spheres, and tori, as well as noncompact manifolds such as symmetric positive definite matrix manifolds after restricting attention to compact evaluation regions with eigenvalues bounded away from zero and infinity.

The formal asymptotic results additionally require the corresponding design, Fr\'echet-margin, and empirical-process conditions stated in \Cref{sec:asymptotic_theory}. In particular, the uniform kernel conditions U-K1 and U-K2 should be read as sufficient empirical-process assumptions on the local-design classes induced jointly by the kernel, the distance function, the logarithmic map, the volume density, and the chosen finite frame cover. They are weaker in spirit than imposing global Lipschitz structure on every sample path of the estimator, and they allow standard compactly supported kernels when the induced classes are of VC type. At the same time, for a general Riemannian predictor manifold these conditions are not automatic consequences of smoothness alone; they are verified through either general entropy arguments for the specific manifold and kernel under consideration or through the sufficient examples discussed with the uniform theory.
\end{remark}

\begin{remark}[Relation to the sphere- and torus-specific procedures]
\label{rem:sphere_torus_relation}
Within their respective normal neighborhoods, the tangent regressors of the spherical procedure of \cite{Im et al. (2025)} and the signed-angle coordinates of the toroidal procedure of \cite{Im and Jeon (2026)} are coordinate representations of the corresponding logarithmic-map displacements, so all three constructions share an intrinsic first-order local-coordinate principle. Their complete weighting schemes nevertheless differ. The spherical procedure uses directional kernel weights based on $L\{(1-\mz^{\top}\mx)/h^{2}\}$, and the toroidal procedure uses the product kernel $\prod_{\ell=1}^{d}L\{(1-\mz_{\ell}^{\top}\mx_{\ell})/h_{\ell}^{2}\}$. These inner-product kernels are globally defined on their compact predictor spaces, their profiles need not be compactly supported, and the toroidal formulation permits coordinate-specific bandwidths without bounded-ratio restrictions. By contrast, the present construction uses the volume-normalized radial weight $\theta_{\mx}(\mz)^{-1}K\{d_{\mathcal M}(\mx,\mz)/h\}$ with a scalar bandwidth and a compactly supported profile, so that all observations receiving nonzero weight remain in a uniform normal neighborhood on which the logarithmic map is single-valued and smooth. On a sphere, the volume-density factor is nontrivial and the radial geodesic kernel differs from the inner-product kernel; on a flat torus, $\theta_{\mx}(\mz)\equiv1$, but the scalar radial kernel still differs from the product kernel with coordinate-specific bandwidths. Consequently, applying the present framework to a sphere or a flat torus does not in general reproduce the space-specific estimators. The special-space methods gain kernel-support flexibility and, for tori, bandwidth anisotropy from their canonical global representations, whereas the present framework addresses the geometric and analytical difficulties, notably finite smooth local-frame covers and frame invariance, that arise without such a representation, and it also covers noncompact manifolds when uniform estimation is restricted to a compact evaluation region.
\end{remark}

\section{Asymptotic Theory} \label{sec:asymptotic_theory}
\setcounter{equation}{0}

In this section, we establish pointwise and uniform asymptotic properties of the proposed estimators. Throughout this section, $\mathcal{M}$ is a connected $d$-dimensional complete Riemannian manifold without boundary, $\dd v_g$ denotes its Riemannian volume measure, and $f$ denotes the density of $\mX$ with respect to $\dd v_g$. Let $P$ denote the joint distribution of $(\mX,Y)$, and $P_{\mX}$, $P_Y$ be the marginal distributions of $\mX$, $Y$, respectively. For each predictor value $\mz\in\mathcal M$, let $P_{Y|\mX=\mz}$ be a regular conditional distribution of $Y$ given $\mX=\mz$. We assume that $P_{Y|\mX=\mz}$ is absolutely continuous with respect to $P_Y$ for the predictor values considered below, and fix a jointly measurable nonnegative version of the Radon--Nikodym derivative
\begin{align*}
    g_{\omega}(\mz)
    :=
    \frac{\dd P_{Y|\mX=\mz}}{\dd P_Y}(\omega),
    \quad
    (\mz,\omega)\in\mathcal{M}\times\mathbb{M}.
\end{align*}
All pointwise and uniform conditions involving $g_\omega$ below are imposed on this fixed jointly measurable nonnegative version. The domination representation is a sufficient analytic device for controlling the conditional Fr\'echet objective; it is not required to define the estimators and does not cover every metric-space regression model, including some deterministic or otherwise singular conditional laws. Here and below, $y$ denotes a candidate point in the Fr\'echet objective, whereas $\omega$ denotes a generic response-space argument in the Radon--Nikodym derivative $g_{\omega}$. 

We use the prefix ``P'' for pointwise assumptions and the prefix ``U'' for uniform assumptions. The letters K, B, D, and M refer respectively to kernel, bandwidth, design smoothness, and Fr\'echet/metric-space conditions.

The assumptions below are organized to separate pointwise from uniform arguments and local constant from local linear smoothing. The pointwise theory is local at a fixed predictor value and does not require a finite moving-frame cover or uniform empirical-process entropy conditions. The uniform theory over a compact set $\mathcal K$ requires empirical-process control of local-design classes. For local constant smoothing, only the zeroth-order kernel-window class is needed; for local linear smoothing, the first- and second-order tangent-coordinate multiplier classes are also required. \Cref{tab:assumption_dependency} summarizes the main dependency structure. The table is only a guide to the assumptions used in the theorem statements and is not an additional condition.

\begin{table}[!ht]
\centering
\caption{
Main assumption dependencies for the asymptotic theory. The notation ``previous'' means the assumptions in the corresponding consistency result.
}
\label{tab:assumption_dependency}
\small
\begin{tabular}{lll}
\toprule
Result & Estimator & Main assumptions \\
\midrule
Pointwise consistency
& LC, LL
& P-K1, P-B1, P-D1--P-D2, M1, P-M2 \\
Pointwise rate
& LC, LL
& previous $+$ P-D3--P-D4, P-M3--P-M4 \\
Uniform consistency
& LC
& U-K1, U-B1, U-D1--U-D2, M1, U-M2 \\
Uniform consistency
& LL
& U-K1, U-K2, U-B1, U-D1--U-D2, M1, U-M2 \\
Uniform rate
& LC
& uniform consistency assumptions $+$ U-D3--U-D4, U-M3--U-M4 \\
Uniform rate
& LL
& uniform consistency assumptions $+$ U-D3--U-D4, U-M3--U-M4 \\
\bottomrule
\end{tabular}
\end{table}

The predictor-side smoothness conditions involving $f$ and $g_\omega$ are used to control deterministic local-objective bias. The Fr\'echet-side conditions have a different role: existence, uniqueness, and separation identify the target and localized minimizers; margin conditions convert objective-level error into metric error; and response-space entropy conditions control stochastic fluctuations over candidate response values. For the local linear estimator, the empirical equivalent weights can be signed, so well-posedness of the resulting empirical Fr\'echet minimizer is not automatic in a general metric space and is stated explicitly.

\subsection{Pointwise Consistency}

Fix $\mx\in\mathcal{M}$. Choose $\rho_{\mx}\in(0,i(\mx))$ and, throughout the pointwise analysis, use this radius to define the local normal neighborhood $B_{\mathcal{M}}(\mx,\rho_{\mx})$. The radius $\rho_{\mx}$ is always understood to be chosen small enough for the local pointwise conditions imposed below; when several local radii are available, we replace $\rho_{\mx}$ by their minimum and relabel it as $\rho_{\mx}$. Since $h\to0$ as $n\to\infty$, we have $h<\rho_{\mx}$ for all sufficiently large $n$, and hence all observations receiving nonzero kernel weight lie inside this normal neighborhood.

\begin{customcon}{P-K1} \label{con:P-K1}
The kernel $K:[0,\infty)\to[0,\infty)$ is bounded, Lebesgue measurable, compactly supported on $[0,1]$, and nonnegative. Moreover, $\lambda_1\{t\in[0,1]:K(t)>0\}>0$, where $\lambda_1$ denotes one-dimensional Lebesgue measure.
\end{customcon}

Condition~\ref{con:P-K1} is the standard compactly supported radial-kernel condition used for local smoothing. Compact support ensures that the estimator only uses observations in the normal neighborhood $B_{\mathcal{M}}(\mx,\rho_{\mx})$ for all sufficiently large $n$, while nonnegativity and nondegeneracy guarantee a positive leading local design mass and a nonsingular leading second-moment matrix. The volume correction in \eqref{eq:sec3.volume_corrected_kernel} follows the manifold smoothing construction of \cite{Pelletier (2006)}.

\begin{customcon}{P-B1} \label{con:P-B1}
The bandwidth satisfies $h\to0$ and $nh^d\to\infty$ as $n\to\infty$.
\end{customcon}

Condition~\ref{con:P-B1} is the usual pointwise bandwidth condition for a $d$-dimensional local smoothing problem. The requirement $h\to0$ controls bias, while $nh^d\to\infty$ guarantees that the effective local sample size diverges.

\begin{customcon}{P-D1} \label{con:P-D1}
The predictor density satisfies $f(\mx)>0$ and is continuous at $\mx$. Moreover,
\begin{align*}
    \sup_{\mz\in B_{\mathcal{M}}(\mx,\rho_{\mx})}
    f(\mz)
    <\infty.
\end{align*}
\end{customcon}

The essential local requirements in Condition~\ref{con:P-D1} are positivity and continuity of $f$ at the evaluation point $\mx$. The displayed upper bound is imposed as a local envelope condition used in the moment bounds. It is automatic, after possibly decreasing $\rho_{\mx}$, if $f$ is continuous on a neighborhood of $\mx$. Positivity and continuity also imply that $f$ is bounded away from zero on a sufficiently small neighborhood of $\mx$.

\begin{customcon}{P-D2} \label{con:P-D2}
The family $\{g_{\omega}:\omega\in\mathbb{M}\}$ is equicontinuous at $\mx$. Moreover,
\begin{align*}
    \sup_{\omega\in\mathbb{M}}
    \sup_{\mz\in B_{\mathcal{M}}(\mx,\rho_{\mx})}
    g_{\omega}(\mz)
    <\infty.
\end{align*}
\end{customcon}

Condition~\ref{con:P-D2} is the corresponding local envelope and equicontinuity condition for the Radon--Nikodym density-ratio family. Since the version of $g_{\omega}$ fixed above is nonnegative, no absolute value is needed in the displayed envelope bound. This condition should be read as a sufficient local regularity assumption for Fr\'echet-objective approximation, not as an automatic consequence of the metric response space. A primitive sufficient setting is the following dominated model: suppose that $P_{Y|\mX=\mz}$ admits a density $p(\omega|\mz)$ with respect to a common measure $\nu$, that $P_Y$ has density $p_Y(\omega)$ with $p_Y(\omega)>0$ on the relevant support, and that $p(\omega|\mz)/p_Y(\omega)$ is locally equicontinuous in $\mz$ with a uniform local envelope over $\omega$. Then the fixed density-ratio version $g_\omega(\mz)=p(\omega|\mz)/p_Y(\omega)$ satisfies Condition~\ref{con:P-D2}. Standard finite-dimensional examples include compactly localized Gaussian location models with variance bounded away from zero and smooth mean map, and von Mises--Fisher-type models on a sphere with concentration bounded on compact parameter ranges and smooth predictor-dependent mean direction.

To state the metric-space and Fr\'echet-objective conditions, we first introduce the population versions of the local objectives. Let $\mathbf{E}_{\mx}\in\mathcal{E}_{\mx}$ be an arbitrary ordered orthonormal basis. We use the tangent-coordinate map $\mathbf{v}_{\mx}^{\mathbf{E}_{\mx}}:\mathcal{M}\to\mathbb{R}^d$ defined in \Cref{sec:local_linear}. For $j=0,1,2$, define the population local moments by
\begin{align*}
    \tilde{\mu}_{h,0}(\mx)&:=\E\left[\mathcal{L}_{\mx,h}\left(\mX\right)\right], \\
    \bm{\tilde{\mu}}_{h,1}(\mx,\mathbf{E}_{\mx})&:=\E\left[\mathcal{L}_{\mx,h}\left(\mX\right)\mathbf{v}_{\mx}^{\mathbf{E}_{\mx}}(\mX)\right], \\
    \bm{\tilde{\mu}}_{h,2}(\mx,\mathbf{E}_{\mx})&:=\E\left[\mathcal{L}_{\mx,h}\left(\mX\right)\mathbf{v}_{\mx}^{\mathbf{E}_{\mx}}(\mX)\mathbf{v}_{\mx}^{\mathbf{E}_{\mx}}(\mX)^{\top}\right].
\end{align*}
\Cref{lemma:B.pointwise_moment_orders} implies the invertibility of $\bm{\tilde{\mu}}_{h,2}(\mx,\mathbf{E}_{\mx})$ for small $h$. Although $\bm{\tilde{\mu}}_{h,1}(\mx,\mathbf{E}_{\mx})$, $\bm{\tilde{\mu}}_{h,2}(\mx,\mathbf{E}_{\mx})$, and $\mathbf{v}_{\mx}^{\mathbf{E}_{\mx}}$ depend on the chosen ordered orthonormal basis, \Cref{lemma:A.invariance_local_linear_weights} implies that the scalar quantities
\begin{align*}
    \bm{\tilde{\mu}}_{h,1}(\mx,\mathbf{E}_{\mx})^{\top}
    \bm{\tilde{\mu}}_{h,2}(\mx,\mathbf{E}_{\mx})^{-1}
    \bm{\tilde{\mu}}_{h,1}(\mx,\mathbf{E}_{\mx}),
    \quad
    \bm{\tilde{\mu}}_{h,1}(\mx,\mathbf{E}_{\mx})^{\top}
    \bm{\tilde{\mu}}_{h,2}(\mx,\mathbf{E}_{\mx})^{-1}
    \mathbf{v}_{\mx}^{\mathbf{E}_{\mx}}(\mX)
\end{align*}
are invariant under a change of basis, analogously to the empirical case. By \Cref{lemma:B.pointwise_moment_orders,lemma:B.pointwise_denominators}, $\tilde{\mu}_{h,0}(\mx)>0$ and $\tilde{\sigma}_{h}(\mx)>0$ for all sufficiently small $h$. Hence the following population normalizing factors and equivalent-weight functions are well-defined:
\begin{align}
\begin{split} \label{eq:B.population_equiv_weights}
    \tilde{W}_{\mx,h,0}(\mz)
    &:=
    \frac{\mathcal{L}_{\mx,h}(\mz)}{\tilde{\mu}_{h,0}(\mx)}, \\
    \tilde{\sigma}_{h}(\mx)
    &:=
    \tilde{\mu}_{h,0}(\mx)
    -
    \bm{\tilde{\mu}}_{h,1}(\mx,\mathbf{E}_{\mx})^{\top}
    \bm{\tilde{\mu}}_{h,2}(\mx,\mathbf{E}_{\mx})^{-1}
    \bm{\tilde{\mu}}_{h,1}(\mx,\mathbf{E}_{\mx}), \\
    \tilde{W}_{\mx,h,1}(\mz)
    &:=
    \frac{\mathcal{L}_{\mx,h}(\mz)}{\tilde{\sigma}_{h}(\mx)}
    \left[
    1-
    \bm{\tilde{\mu}}_{h,1}(\mx,\mathbf{E}_{\mx})^{\top}
    \bm{\tilde{\mu}}_{h,2}(\mx,\mathbf{E}_{\mx})^{-1}
    \mathbf{v}_{\mx}^{\mathbf{E}_{\mx}}(\mz)
    \right], \quad \mz\in\mathcal{M}.
\end{split}
\end{align}
Similarly, for $s\in\{0,1\}$, we define the population version of $\hat{M}_{h,s}(\mx,y)$ as $\tilde{M}_{h,s}(\mx,y)$ by
\begin{align*}
    \tilde{M}_{h,s}(\mx,y)&:=\E\left[\tilde{W}_{\mx,h,s}(\mX)d_{\mathbb{M}}^2(Y,y)\right], \quad y\in\mathbb{M}.
\end{align*}
Whenever the minimizer exists, the corresponding population local minimizer is denoted by
\begin{align*}
    \tilde{m}_{h,s}(\mx)\in\argmin_{y\in\mathbb{M}}\tilde{M}_{h,s}(\mx,y), \quad s\in\{0,1\}.
\end{align*}

\begin{customcon}{M1} \label{con:M1}
The metric space $(\mathbb{M},d_{\mathbb{M}})$ is totally bounded.
\end{customcon}

Condition~\ref{con:M1} is common to both pointwise and uniform theory. It gives stochastic equicontinuity for Fr\'echet objectives indexed by $y\in\mathbb{M}$ and ensures that the squared metric loss is uniformly bounded.

\begin{customcon}{P-M2} \label{con:P-M2}
For $s\in\{0,1\}$, the following statements hold. First, $m_{\oplus}(\mx)$ and $\tilde{m}_{h,s}(\mx)$ exist and are unique for all sufficiently small $h$, and a measurable minimizer $\hat{m}_{h,s}(\mx)\in\argmin_{y\in\mathbb{M}}\hat{M}_{h,s}(\mx,y)$ exists with probability tending to one. Second, for any $\epsilon>0$,
\begin{align*}
    \inf_{y\in\mathbb{M}:d_{\mathbb{M}}(y,m_{\oplus}(\mx))>\epsilon}
    \left[
    M_{\oplus}(\mx,y)-M_{\oplus}(\mx,m_{\oplus}(\mx))
    \right]>0,
\end{align*}
and
\begin{align*}
    \liminf_{h\downarrow0}
    \inf_{s\in\{0,1\}}
    \inf_{y\in\mathbb{M}:d_{\mathbb{M}}(y,\tilde{m}_{h,s}(\mx))>\epsilon}
    \left[
    \tilde{M}_{h,s}(\mx,y)-\tilde{M}_{h,s}(\mx,\tilde{m}_{h,s}(\mx))
    \right]>0.
\end{align*}
\end{customcon}

Condition~\ref{con:P-M2} is the pointwise identifiability and separation condition for the population and localized Fr\'echet objectives. This type of condition is standard for consistency of Fr\'echet-type M-estimators \citep{Petersen and Muller (2019), Im et al. (2025)}. The stochastic convergence of the empirical local objectives to their population counterparts is established in \Cref{lemma:B.pointwise_empirical_objective}.

\begin{theorem}[Pointwise consistency] \label{thm:pointwise_consistency}
Assume Conditions~\ref{con:P-K1}, \ref{con:P-B1}, \ref{con:P-D1}, \ref{con:P-D2}, \ref{con:M1}, and \ref{con:P-M2}. For a fixed $\mx\in\mathcal{M}$ and $s\in\{0,1\}$,
\begin{align*}
    d_{\mathbb{M}}(\hat{m}_{h,s}(\mx),m_{\oplus}(\mx))=o_{\P}(1).
\end{align*}
\end{theorem}

\subsection{Pointwise Convergence Rate}

To derive a pointwise convergence rate, we strengthen the local design conditions to second-order smoothness conditions on the same normal neighborhood $B_{\mathcal{M}}(\mx,\rho_{\mx})$. These assumptions are used to obtain the $O(h^2)$ local smoothing bias after the volume-density correction.

\begin{customcon}{P-D3} \label{con:P-D3}
The density $f$ is $C^2$ on $B_{\mathcal{M}}(\mx,\rho_{\mx})$. Moreover,
\begin{align*}
    \sup_{\mz\in B_{\mathcal{M}}(\mx,\rho_{\mx})}
    \|\nabla f(\mz)\|_{\mz}<\infty,
    \quad
    \sup_{\mz\in B_{\mathcal{M}}(\mx,\rho_{\mx})}
    \|\nabla^2 f(\mz)\|_{\mathrm{op}}<\infty.
\end{align*}
\end{customcon}

Condition~\ref{con:P-D3} is used to obtain second-order Taylor bounds in normal coordinates and to control the resulting remainders uniformly over local kernel neighborhoods. Since $h\downarrow0$, for every fixed $r\in(0,\rho_{\mx})$ all observations receiving nonzero kernel weight lie in $B_{\mathcal{M}}(\mx,r)$ for all sufficiently large $n$. Thus the pointwise rate arguments may be carried out on compact subballs strictly contained in $B_{\mathcal{M}}(\mx,\rho_{\mx})$, while the displayed envelope bounds provide a convenient uniform control on the fixed working neighborhood.

\begin{customcon}{P-D4} \label{con:P-D4}
For every $\omega\in\mathbb{M}$, the scalar function $g_{\omega}:\mathcal{M}\to\mathbb{R}$ is twice covariantly differentiable on $B_{\mathcal{M}}(\mx,\rho_{\mx})$. Moreover,
\begin{align*}
    \sup_{\omega\in\mathbb{M}}
    \sup_{\mz\in B_{\mathcal{M}}(\mx,\rho_{\mx})}
    \|\nabla g_{\omega}(\mz)\|_{\mz}
    <\infty,
    \quad
    \sup_{\omega\in\mathbb{M}}
    \sup_{\mz\in B_{\mathcal{M}}(\mx,\rho_{\mx})}
    \|\nabla^2 g_{\omega}(\mz)\|_{\mathrm{op}}
    <\infty.
\end{align*}
\end{customcon}

Conditions~\ref{con:P-D3} and~\ref{con:P-D4} are the Riemannian analogues of the second-order design smoothness assumptions used in Euclidean local linear Fr\'echet regression. The Taylor bounds in normal coordinates, together with the volume correction in \eqref{eq:sec3.volume_corrected_kernel} and the cancellation identity in \Cref{lemma:A.pointwise_coord_cancellation}, yield the same second-order bias order as in Euclidean local smoothing. Condition~\ref{con:P-D4} imposes uniform local bounds over $\omega\in\mathbb M$ for the first and second covariant derivatives of the conditional density-ratio family; these bounds are not automatic from pointwise twice differentiability of each $g_\omega$.

\begin{remark}[Role of the volume-density correction]
\label{rem:volume_density_role_theory}
In normal coordinates at $\mx$, Riemannian integration has the form
\begin{align*}
    \int_{\mathcal M} \varphi(\mz)\,\dd v_g(\mz)
    =
    \int_{T_{\mx}\mathcal M}
    \varphi\left(\Exp_{\mx}(\mathbf u)\right)
    \theta_{\mx}\left(\Exp_{\mx}(\mathbf u)\right)
    \dd\mathbf u
\end{align*}
on the normal neighborhood. The factor $\theta_{\mx}(\mz)^{-1}$ in $\mathcal L_{\mx,h}$ cancels this Jacobian factor and makes the leading local moments have the same radial Euclidean form as in ordinary kernel smoothing. Since $\theta_{\mx}(\Exp_{\mx}(\mathbf u))=1+O(\|\mathbf u\|_{\mx}^2)$ locally, omitting the correction would not necessarily change the $O(h^2)$ bias order under sufficient smoothness, but it would introduce curvature-dependent terms into the local moment expansions. We include the correction because it is the intrinsic normal-coordinate construction and yields cleaner, geometrically coherent bias calculations. Its finite-sample numerical effect may be small when the selected bandwidths are small or the predictor region is close to Euclidean.
\end{remark}

\begin{remark}[Local nature of the pointwise predictor assumptions]
The pointwise predictor-side assumptions are local at the fixed evaluation point $\mx$. Once a normal-neighborhood radius $\rho_{\mx}\in(0,i(\mx))$ and an ordered orthonormal basis of $T_{\mx}\mathcal M$ are fixed, the pointwise arguments use only the compact support and boundedness properties of the radial kernel in Condition~\ref{con:P-K1}, the pointwise bandwidth condition in Condition~\ref{con:P-B1}, and the local regularity of the design density and conditional density-ratio functions on $B_{\mathcal M}(\mx,\rho_{\mx})$ in Conditions~\ref{con:P-D1}--\ref{con:P-D4}. In particular, no finite moving-frame cover and no uniform VC-type entropy condition over evaluation points are required for the pointwise theory. These uniform empirical-process requirements enter only in the uniform theory over a compact set $\mathcal K$.
\end{remark}

\begin{customcon}{P-M3} \label{con:P-M3}
There exist constants $h_{\oplus,\mx}>0$, $\eta_{\oplus,\mx}>0$, $C_{\oplus,\mx}>0$, and $\beta_{\oplus,\mx}\in(1,\infty)$ such that, for all $y\in\mathbb{M}$ satisfying $d_{\mathbb{M}}(y,m_{\oplus}(\mx))<\eta_{\oplus,\mx}$,
\begin{align*}
    M_{\oplus}(\mx,y)-M_{\oplus}(\mx,m_{\oplus}(\mx))
    \geq
    C_{\oplus,\mx}
    d_{\mathbb{M}}(y,m_{\oplus}(\mx))^{\beta_{\oplus,\mx}},
\end{align*}
and, for $s\in\{0,1\}$, all $h<h_{\oplus,\mx}$, and all $y\in\mathbb{M}$ satisfying $d_{\mathbb{M}}(y,\tilde{m}_{h,s}(\mx))<\eta_{\oplus,\mx}$,
\begin{align*}
    \tilde{M}_{h,s}(\mx,y)-\tilde{M}_{h,s}(\mx,\tilde{m}_{h,s}(\mx))
    \geq
    C_{\oplus,\mx}
    d_{\mathbb{M}}(y,\tilde{m}_{h,s}(\mx))^{\beta_{\oplus,\mx}}.
\end{align*}
\end{customcon}

\begin{customcon}{P-M4} \label{con:P-M4}
Let $N(r,B_{\mathbb{M}}(y,\delta),d_{\mathbb{M}})$ denote the $r$-covering number of $B_{\mathbb{M}}(y,\delta)$ under $d_{\mathbb{M}}$. There exists a constant $r_{\mathbb{M},\mx}>0$ such that
\begin{align*}
    \sup_{y\in\mathbb{M}:d_{\mathbb{M}}\left(y,m_{\oplus}(\mx)\right)<r_{\mathbb{M},\mx}}
    \int_{0}^{1/2}
    \sqrt{
    1+
    \log N\left(\delta\epsilon,B_{\mathbb{M}}(y,\delta),d_{\mathbb{M}}\right)
    }
    \,\dd\epsilon
    =
    O(1),
    \quad
    \delta\downarrow0.
\end{align*}
\end{customcon}

Condition~\ref{con:P-M3} is a local curvature, or margin, condition for the Fr\'echet objective around $m_{\oplus}(\mx)$, while Condition~\ref{con:P-M4} is a localized entropy condition for small response-space balls whose centers remain in a fixed neighborhood of $m_{\oplus}(\mx)$. Together, these conditions convert localized objective-level deterministic and stochastic bounds into a rate for the corresponding minimizers; see \cite{Petersen and Muller (2019)} and \cite{Im et al. (2025)} for closely related Fr\'echet-regression arguments.

\begin{remark}[Examples and interpretation of the pointwise response-space conditions]
\label{rem:pointwise_response_space_examples}
Conditions~\ref{con:M1} and~\ref{con:P-M2}--\ref{con:P-M4} are high-level Fr\'echet regularity conditions on the response space and on the local Fr\'echet objectives at the fixed predictor value $\mx$. They are not intended to be automatic consequences of the Riemannian structure of the predictor manifold. Rather, they play the same role as the compactness, uniqueness, separation, margin, and local entropy conditions commonly imposed in Fr\'echet regression with general metric-space-valued responses.

Several standard response spaces satisfy the compactness and local entropy parts after the usual localization. If $\mathbb M$ is a compact subset of a finite-dimensional Euclidean space, or more generally a compact finite-dimensional Riemannian manifold equipped with its geodesic distance, then local covering numbers are polynomial in the covering radius, and Condition~\ref{con:P-M4} follows from the usual finite-dimensional entropy bound. The same conclusion applies to compact subsets of finite-dimensional normed spaces, such as bounded graph-Laplacian representations of weighted networks with a fixed number of nodes, and to compact subsets of the symmetric positive definite cone under a metric that is isometric to a finite-dimensional Euclidean representation, such as the log-Euclidean metric. For one-dimensional Wasserstein responses, the condition is natural on $\mathcal W_2([a,b])$ with $-\infty<a<b<\infty$, or on other totally bounded subclasses of $\mathcal W_2(\mathbb R)$; it should not be read as applying to the unrestricted space $\mathcal W_2(\mathbb R)$ without additional localization or tightness restrictions.

The remaining conditions are Fr\'echet-objective conditions rather than purely metric entropy conditions. Condition~\ref{con:P-M2} imposes existence, uniqueness, and separation of the relevant pointwise minimizers, while Condition~\ref{con:P-M3} imposes a local margin around the population target. In Euclidean or Hilbert-valued settings these conditions reduce to familiar convexity and nondegeneracy requirements. For manifold-valued responses they may be verified under localization in a geodesically convex region, for instance away from cut-locus and antipodal ambiguities on positively curved spaces. For the local linear estimator, the corresponding oracle objective may involve signed equivalent weights, so well-posedness is stated explicitly rather than derived from global nonpositive curvature or convexity alone.
\end{remark}

\begin{theorem}[Pointwise convergence rate] \label{thm:pointwise_rate}
Assume the conditions of \Cref{thm:pointwise_consistency} and Conditions~\ref{con:P-D3}, \ref{con:P-D4}, \ref{con:P-M3}, and~\ref{con:P-M4}. For a fixed $\mx\in\mathcal{M}$ and $s\in\{0,1\}$,
\begin{align*}
    d_{\mathbb{M}}(\hat{m}_{h,s}(\mx),m_{\oplus}(\mx))
    =
    O\left(h^{2/(\beta_{\oplus,\mx}-1)}\right)
    +
    O_{\P}\left((nh^d)^{-1/(2\beta_{\oplus,\mx}-2)}\right).
\end{align*}
\end{theorem}

When $\beta_{\oplus,\mx}=2$ and $h\asymp n^{-1/(d+4)}$, \Cref{thm:pointwise_rate} yields
\begin{align*}
    d_{\mathbb{M}}(\hat{m}_{h,s}(\mx),m_{\oplus}(\mx))
    =
    O_{\P}\left(n^{-2/(d+4)}\right),
    \quad
    s\in\{0,1\}.
\end{align*}
Thus the pointwise rate is of the same order as the standard twice-smooth Euclidean upper-bound rate in intrinsic dimension $d$. The theoretical contribution of the local linear analysis is therefore not a faster interior rate, but the construction and control of intrinsic signed local linear Fr\'echet weights on moving tangent spaces and the separation of the zeroth-order and multiplier empirical-process requirements. A sharper leading-bias-constant comparison between the local constant and local linear Fr\'echet estimators is not established here and is left as a useful direction for further refinement.

\subsection{Uniform Consistency}

For the uniform theory, let $\mathcal K\subset\mathcal M$ be a fixed compact set and choose a radius $\rho\in(0,i(\mathcal K))$, where
\begin{align*}
    i(\mathcal K)
    &:=
    \inf_{\mx\in\mathcal K}i(\mx)
    >0.
\end{align*}
Define the closed geodesic tube
\begin{align*}
    \mathcal K^\rho
    &:=
    \left\{
    \mz\in\mathcal M:
    d_{\mathcal M}(\mz,\mathcal K)\leq\rho
    \right\}.
\end{align*}
The radius $\rho$ is fixed throughout the uniform arguments and is understood to be chosen small enough for the uniform local conditions imposed below. Since $h\to0$ as $n\to\infty$, we have $h<\rho$ for all sufficiently large $n$. Hence, for every $\mx\in\mathcal K$, all observations receiving nonzero kernel weight lie in $B_{\mathcal M}(\mx,\rho)$ and therefore in $\mathcal K^\rho$. Since $\mathcal M$ is complete, the Hopf--Rinow theorem implies that $\mathcal K^\rho$ is compact. Moreover, by \Cref{lemma:A.uniform_normal_neighborhoods}, the logarithmic maps and volume-density functions are uniformly well-behaved on the relevant normal neighborhoods.

All empirical-process suprema appearing below are assumed to be measurable. If measurability is not verified directly, the corresponding probability statements may instead be interpreted in the outer-probability sense.

Before stating the uniform kernel conditions, we choose and fix a finite local-frame system used only to express tangent-coordinate components in the uniform empirical-process arguments. Since smooth local orthonormal frames exist locally and $\mathcal K^\rho$ is compact, there exist open sets $\mathcal O^1,\ldots,\mathcal O^{N_{\mathcal K}}$ in $\mathcal M$, where $N_{\mathcal K}\in\mathbb N$, such that
\begin{align}
    \mathcal K^\rho
    \subset
    \bigcup_{\alpha=1}^{N_{\mathcal K}}\mathcal O^\alpha,
    \label{eq:U.fixed_frame_cover}
\end{align}
and each $\mathcal O^\alpha$ admits a smooth ordered orthonormal frame. For each $\alpha\in\{1,\ldots,N_{\mathcal K}\}$, fix one such frame and write
\begin{align*}
    \mathbf E^\alpha
    &:=
    \left(
    \mathbf E^\alpha_1,
    \ldots,
    \mathbf E^\alpha_d
    \right),
\end{align*}
where each $\mathbf E^\alpha_r$ is a smooth vector field on $\mathcal O^\alpha$. Thus, for every $\mz\in\mathcal O^\alpha$,
\begin{align*}
    \mathbf E^\alpha_{\mz,r}
    &:=
    \mathbf E^\alpha_r(\mz)
    \in T_{\mz}\mathcal M,
    \quad r=1,\ldots,d,
\end{align*}
and
\begin{align*}
    \mathbf E^\alpha_{\mz}
    &:=
    \left(
    \mathbf E^\alpha_{\mz,1},
    \ldots,
    \mathbf E^\alpha_{\mz,d}
    \right)
    \in\mathcal E_{\mz}.
\end{align*}
Equivalently, $\mathbf E^\alpha_{\mz}$ is an ordered orthonormal basis of $T_{\mz}\mathcal M$ for every $\mz\in\mathcal O^\alpha$, and the basis vectors vary smoothly with $\mz$.

For $\mx\in\mathcal O^\alpha$, define the frame-induced coordinate isomorphism
\begin{align*}
    \bm{\Phi}_{\mathbf E^\alpha_{\mx}}
    :
    T_{\mx}\mathcal M
    \to
    \mathbb R^d
\end{align*}
by
\begin{align*}
    \bm{\Phi}_{\mathbf E^\alpha_{\mx}}(\mathbf u)
    &:=
    (u_1,\ldots,u_d)^\top,
    \quad
    \mathbf u
    =
    \sum_{r=1}^d
    u_r\mathbf E^\alpha_{\mx,r}
    \in T_{\mx}\mathcal M.
\end{align*}
For $\mz\in\mathcal M$, define
\begin{align*}
    \mathbf v_{\mx}^{\alpha}(\mz)
    &:=
    \begin{cases}
    \bm{\Phi}_{\mathbf E^\alpha_{\mx}}
    \left(
    \Log_{\mx}(\mz)
    \right),
    & \mz\in B_{\mathcal M}(\mx,i(\mx)), \\
    \mathbf 0_d,
    & \mz\notin B_{\mathcal M}(\mx,i(\mx)).
    \end{cases}
\end{align*}
This finite frame system is fixed once and for all throughout the uniform analysis. The empirical-process conditions below are imposed relative to this fixed cover, and their constants may depend on the cover. No uniformity over all possible finite frame covers is required.

No unique frame is assigned on overlaps. If
$\mx\in\mathcal O^\alpha\cap\mathcal O^\beta$, the two frames are related by an orthogonal change of basis, which leaves the scalar local linear equivalent weights unchanged by
\Cref{lemma:A.invariance_local_linear_weights}. The required componentwise bounds are therefore established separately on each cover element and combined by taking the maximum over the finite cover. No measurable frame selection or partition of unity is needed.

\begin{customcon}{U-K1} \label{con:U-K1}
Condition~\ref{con:P-K1} holds. In addition, for some $h_0\in(0,\rho)$ and each $\alpha\in\{1,\ldots,N_{\mathcal K}\}$, define
\begin{align*}
    \mathcal F_{\alpha,0}
    &:=
    \left\{
    \mz\mapsto
    \mathcal L_{\mx,h}(\mz):
    \mx\in\mathcal K\cap\mathcal O^\alpha,\ 
    0<h<h_0
    \right\}.
\end{align*}
There exist constants $A_0<\infty$, $v_0<\infty$, and $C_0<\infty$, independent of $\alpha$, such that $\mathcal F_{\alpha,0}$ has envelope bounded by $C_0$ and, for every finitely discrete probability measure $Q$ on $\mathcal M$ and every $\epsilon\in(0,1)$,
\begin{align*}
    N\left(
    \epsilon C_0,
    \mathcal F_{\alpha,0},
    L_2(Q)
    \right)
    \leq
    \left(
    \frac{A_0}{\epsilon}
    \right)^{v_0}.
\end{align*}
\end{customcon}

\begin{customcon}{U-K2} \label{con:U-K2}
Let $h_0\in(0,\rho)$ be as in Condition~\ref{con:U-K1}. For
$\alpha\in\{1,\ldots,N_{\mathcal K}\}$ and $r,s\in\{1,\ldots,d\}$, define
the first- and second-order multiplier-augmented local-design classes by
\begin{align*}
    \mathcal F_{\alpha,1,r}
    &:=
    \left\{
    \mz\mapsto
    \mathcal L_{\mx,h}(\mz)
    h^{-1}
    \left[
    \mathbf v^\alpha_{\mx}(\mz)
    \right]_r:
    \mx\in\mathcal K\cap\mathcal O^\alpha,\ 
    0<h<h_0
    \right\}, \\
    \mathcal F_{\alpha,2,r,s}
    &:=
    \left\{
    \mz\mapsto
    \mathcal L_{\mx,h}(\mz)
    h^{-2}
    \left[
    \mathbf v^\alpha_{\mx}(\mz)
    \right]_r
    \left[
    \mathbf v^\alpha_{\mx}(\mz)
    \right]_s:
    \mx\in\mathcal K\cap\mathcal O^\alpha,\ 
    0<h<h_0
    \right\}.
\end{align*}
For each $j\in\{1,2\}$, there exist constants
$A_j<\infty$, $v_j<\infty$, and $C_j<\infty$, independent of
$\alpha$, $r$, and $s$, such that every class of order $j$ has envelope
bounded by $C_j$ and, for every finitely discrete probability measure $Q$
on $\mathcal M$ and every $\epsilon\in(0,1)$,
\begin{align*}
    N\left(
    \epsilon C_1,
    \mathcal F_{\alpha,1,r},
    L_2(Q)
    \right)
    \leq
    \left(
    \frac{A_1}{\epsilon}
    \right)^{v_1}, \quad
    N\left(
    \epsilon C_2,
    \mathcal F_{\alpha,2,r,s},
    L_2(Q)
    \right)
    \leq
    \left(
    \frac{A_2}{\epsilon}
    \right)^{v_2}.
\end{align*}
\end{customcon}

Condition~\ref{con:U-K1} controls the zeroth-order local-design class needed for the local constant estimator. Condition~\ref{con:U-K2} controls the first- and second-order tangent-coordinate multiplier classes needed for the empirical first moment, second moment matrix, and denominator of the local linear estimator. On the kernel support,
\begin{align*}
    \left|
    h^{-1}
    \left[
    \mathbf v^\alpha_{\mx}(\mz)
    \right]_r
    \right|
    \leq1, \quad
    \left|
    h^{-2}
    \left[
    \mathbf v^\alpha_{\mx}(\mz)
    \right]_r
    \left[
    \mathbf v^\alpha_{\mx}(\mz)
    \right]_s
    \right|
    \leq1,
\end{align*}
for every $r,s\in\{1,\ldots,d\}$. Together with the uniform bound for $\theta_{\mx}(\mz)^{-1}$ in \Cref{lemma:A.uniform_normal_neighborhoods}, this shows that the envelope requirements follow from the compactly supported local geometry. The substantive content of Conditions~\ref{con:U-K1} and~\ref{con:U-K2} is therefore the uniform polynomial covering-number bound. In particular, these are entropy conditions on the actual moving local-design classes and not Lipschitz conditions on the radial kernel profile.

The separate formulation of Condition~\ref{con:U-K2} reflects the additional moving-anchor complexity of local linear smoothing. Unlike the zeroth-order class, its functions involve the logarithmic map, its coordinates under the fixed finite frame cover, and first- or second-order tangent-coordinate multipliers. This formulation requires neither a global coordinate chart nor a global frame.

\begin{remark}[Verification of Conditions~\ref{con:U-K1} and~\ref{con:U-K2}]
The Euclidean verification is given in
\Cref{lem:app_UK1_implies_UK2_euclidean,lem:app_standard_kernels_UK}. More generally,
\Cref{prop:F.tame_UK_verification} shows that finitely piecewise-polynomial compactly supported kernels satisfy Conditions~\ref{con:U-K1} and~\ref{con:U-K2} under the tame-local-geometry condition of \Cref{app:condition_verification}. By
\Cref{prop:F.analytic_implies_tame,cor:F.standard_analytic_manifolds}, this includes compact evaluation regions of Euclidean spaces, spheres, finite products of spheres, flat tori, and SPD manifolds equipped with the affine-invariant Riemannian metric. Consequently, the uniform, triangular, Epanechnikov, biweight, and triweight kernels satisfy the two conditions in these settings. For a general smooth Riemannian predictor manifold outside this sufficient class, Conditions~\ref{con:U-K1} and~\ref{con:U-K2} remain explicit high-level empirical-process assumptions.
\end{remark}

\begin{customcon}{U-B1} \label{con:U-B1}
The bandwidth satisfies $h\to0$ and $nh^d/\log n\to\infty$ as $n\to\infty$.
\end{customcon}

Condition~\ref{con:U-B1} is the uniform analogue of Condition~\ref{con:P-B1}. The requirement $h\to0$ ensures that all kernel neighborhoods eventually lie in the fixed tube $\mathcal K^\rho$, while the additional logarithmic factor accounts for taking suprema over $\mx\in\mathcal K$.

\begin{customcon}{U-D1} \label{con:U-D1}
The predictor density $f$ is continuous on $\mathcal K^\rho$ and satisfies
\begin{align*}
    c_{\mathcal K}:=\inf_{\mx\in\mathcal K} f(\mx)>0.
\end{align*}
\end{customcon}

\begin{customcon}{U-D2} \label{con:U-D2}
The family $\{g_{\omega}:\omega\in\mathbb M\}$ is equicontinuous on $\mathcal K^\rho$, in the sense that
\begin{align*}
    \lim_{r\downarrow0}
    \sup_{\omega\in\mathbb M}
    \sup_{\substack{\mz,\mw\in\mathcal K^\rho \\
    d_{\mathcal M}(\mz,\mw)\leq r}}
    \left|
    g_{\omega}(\mz)-g_{\omega}(\mw)
    \right|
    =
    0.
\end{align*}
Moreover,
\begin{align*}
    \sup_{\omega\in\mathbb M}
    \sup_{\mz\in\mathcal K^\rho}
    g_{\omega}(\mz)
    <\infty.
\end{align*}
\end{customcon}

Under Condition~\ref{con:U-D1}, we use the notation
\begin{align} \label{eq:U.design_density_bounds_on_tube}
    C_{\mathcal K,\rho}
    :=
    \sup_{\mz\in\mathcal K^\rho}f(\mz)<\infty.
\end{align}
The lower bound is imposed only on the center set $\mathcal K$, while the compact tube $\mathcal K^\rho$ is used to obtain upper bounds and uniform continuity. In particular, the proofs below do not require $\inf_{\mz\in\mathcal K^\rho}f(\mz)>0$.

Conditions~\ref{con:U-D1} and~\ref{con:U-D2} are the uniform counterparts of Conditions~\ref{con:P-D1} and~\ref{con:P-D2}. They impose local positivity at the kernel centers, continuity, equicontinuity, and envelope requirements on the fixed tube $\mathcal K^\rho$. Since $\mathcal K^\rho$ is compact, Condition~\ref{con:U-D1} implies that $f$ is bounded above and uniformly continuous on $\mathcal K^\rho$, while the positive lower bound needed for denominator arguments is provided by $c_{\mathcal K}>0$ on the center set $\mathcal K$. Condition~\ref{con:U-D2} states the corresponding uniform equicontinuity and envelope requirements for the conditional density-ratio family $\{g_\omega:\omega\in\mathbb M\}$.

\begin{remark}[On the density-ratio regularity conditions]
The conditions on $\{g_\omega:\omega\in\mathbb M\}$ are sufficient regularity assumptions used to compare localized weighted objectives with the target Fr\'echet objective. They should not be read as automatic consequences of the metric-space response structure. One primitive setting in which such conditions can be checked is the following: the conditional law of $Y$ given $\mX=\mz$ is dominated by a common measure $\lambda$ with density $p(\omega\mid\mz)$, the marginal density $p_Y(\omega)$ is bounded away from zero on the relevant response support, and $p(\omega\mid\mz)/p_Y(\omega)$ is uniformly bounded and uniformly continuous in $\mz$ over $\omega$ on $\mathcal K^\rho$. The second-order condition below is verified similarly if the first and second covariant derivatives in $\mz$ of $p(\omega\mid\mz)/p_Y(\omega)$ are uniformly bounded. These conditions are therefore best viewed as objective-level smoothness assumptions on the conditional law, imposed to obtain second-order bias bounds.
\end{remark}

\begin{customcon}{U-M2} \label{con:U-M2}
For each $s\in\{0,1\}$, the following statements hold. First, $m_{\oplus}(\mx)$ and $\tilde{m}_{h,s}(\mx)$ exist and are unique for all $\mx\in\mathcal K$ and all sufficiently small $h$, and a measurable version of
\begin{align*}
    \hat{m}_{h,s}(\mx)
    \in
    \argmin_{y\in\mathbb M}
    \hat{M}_{h,s}(\mx,y)
\end{align*}
exists on $\mathcal K$ with probability tending to one. Moreover, for any $\epsilon>0$,
\begin{align*}
    \inf_{\mx\in\mathcal K}
    \inf_{y\in\mathbb M:d_{\mathbb M}(y,m_{\oplus}(\mx))>\epsilon}
    \left[
    M_{\oplus}(\mx,y)-M_{\oplus}(\mx,m_{\oplus}(\mx))
    \right]
    >
    0,
\end{align*}
and
\begin{align*}
    \liminf_{h\downarrow0}
    \inf_{\mx\in\mathcal K}
    \inf_{y\in\mathbb M:d_{\mathbb M}(y,\tilde{m}_{h,s}(\mx))>\epsilon}
    \left[
    \tilde{M}_{h,s}(\mx,y)-\tilde{M}_{h,s}(\mx,\tilde{m}_{h,s}(\mx))
    \right]
    >
    0.
\end{align*}
\end{customcon}

Condition~\ref{con:U-M2} is the uniform identifiability and well-posedness condition. It prevents the population and localized Fr\'echet objectives from having nearly tied minimizers uniformly over $\mx\in\mathcal K$.

\begin{remark}[On Fr\'echet-side well-posedness]
Condition~\ref{con:U-M2} collects assumptions that are standard but important in Fr\'echet regression: existence, uniqueness, separation of the minimizer, and measurability of empirical minimizers. These requirements are mild for compact Euclidean response sets with strictly convex quadratic objectives, and they can be verified for many manifold-valued responses when the relevant conditional distributions are concentrated in strongly convex geodesic balls away from cut loci. For local linear Fr\'echet regression, however, the equivalent weights may be signed, so existence and uniqueness of $\tilde m_{h,1}(\mx)$ and $\hat m_{h,1}(\mx)$ are not automatic in a general metric space. Condition~\ref{con:U-M2} makes this cost explicit. In applications with compact response spaces, measurable selection can typically be obtained by combining continuity of the empirical objective in $y$ with separability or compactness of the response space.
\end{remark}

\begin{theorem}[Uniform consistency] \label{thm:uniform_consistency}
Assume Conditions~\ref{con:U-K1}, \ref{con:U-B1}, \ref{con:U-D1}, \ref{con:U-D2}, \ref{con:M1}, and~\ref{con:U-M2}. Then the local constant estimator satisfies
\begin{align*}
    \sup_{\mx\in\mathcal K}
    d_{\mathbb M}\left(\hat{m}_{h,0}(\mx),m_{\oplus}(\mx)\right)
    =
    o_{\P}(1).
\end{align*}
If, in addition, Condition~\ref{con:U-K2} holds, then the local linear estimator satisfies
\begin{align*}
    \sup_{\mx\in\mathcal K}
    d_{\mathbb M}\left(\hat{m}_{h,1}(\mx),m_{\oplus}(\mx)\right)
    =
    o_{\P}(1).
\end{align*}
\end{theorem}

\subsection{Uniform Convergence Rate}

Finally, we establish the uniform convergence rate over $\mathcal K$. The radius $\rho\in(0,i(\mathcal K))$ fixed above remains fixed throughout this subsection. Since the kernel has compact support, the estimators depend only on observations inside $\mathcal K^\rho$ for all sufficiently small $h$. Thus, the second-order smoothness conditions are imposed on an open neighborhood of this fixed compact tube.

\begin{customcon}{U-D3} \label{con:U-D3}
The density $f$ is $C^2$ on an open neighborhood of $\mathcal K^\rho$.
\end{customcon}

\begin{customcon}{U-D4} \label{con:U-D4}
The family $\{g_{\omega}:\omega\in\mathbb M\}$ is defined so that, for every $\omega\in\mathbb M$, the scalar function $g_{\omega}:\mathcal M\to\mathbb R$ is twice covariantly differentiable on a common open neighborhood of $\mathcal K^\rho$. Moreover,
\begin{align*}
    \sup_{\omega\in\mathbb M}
    \sup_{\mz\in\mathcal K^\rho}
    \|\nabla g_{\omega}(\mz)\|_{\mz}
    <\infty,
    \quad
    \sup_{\omega\in\mathbb M}
    \sup_{\mz\in\mathcal K^\rho}
    \|\nabla^2 g_{\omega}(\mz)\|_{\mathrm{op}}
    <\infty.
\end{align*}
\end{customcon}

Condition~\ref{con:U-D3} gives the uniform second-order smoothness of the predictor density. Since $\mathcal K^\rho$ is compact, the boundedness and uniform continuity of $\nabla f$ and $\nabla^2 f$ on $\mathcal K^\rho$ follow automatically from Condition~\ref{con:U-D3}. Condition~\ref{con:U-D4} imposes the corresponding uniform second-order boundedness requirements on the conditional density-ratio family. Unlike the density condition, the displayed derivative bounds are not automatic from pointwise twice differentiability of each $g_{\omega}$, because the bounds must hold uniformly over $\omega\in\mathbb M$.

\begin{customcon}{U-M3} \label{con:U-M3}
There exist constants $h_{\oplus,\mathcal K}>0$, $\eta_{\oplus,\mathcal K}>0$, $C_{\oplus,\mathcal K}>0$, and $\beta_{\oplus,\mathcal K}\in(1,\infty)$ such that, for all $\mx\in\mathcal K$ and all $y\in\mathbb M$ satisfying
\begin{align*}
    d_{\mathbb M}(y,m_{\oplus}(\mx))<\eta_{\oplus,\mathcal K},
\end{align*}
we have
\begin{align*}
    M_{\oplus}(\mx,y)-M_{\oplus}(\mx,m_{\oplus}(\mx))
    \geq
    C_{\oplus,\mathcal K}
    d_{\mathbb M}(y,m_{\oplus}(\mx))^{\beta_{\oplus,\mathcal K}}.
\end{align*}
Moreover, for each $s\in\{0,1\}$, all $h<h_{\oplus,\mathcal K}$, all $\mx\in\mathcal K$, and all $y\in\mathbb M$ satisfying
\begin{align*}
    d_{\mathbb M}(y,\tilde{m}_{h,s}(\mx))<\eta_{\oplus,\mathcal K},
\end{align*}
we have
\begin{align*}
    \tilde{M}_{h,s}(\mx,y)-\tilde{M}_{h,s}(\mx,\tilde{m}_{h,s}(\mx))
    \geq
    C_{\oplus,\mathcal K}
    d_{\mathbb M}(y,\tilde{m}_{h,s}(\mx))^{\beta_{\oplus,\mathcal K}}.
\end{align*}
\end{customcon}

\begin{remark}
Condition~\ref{con:U-M3} is a local curvature condition on the population and localized Fr\'echet objectives. The first part is the usual uniform margin condition around $m_{\oplus}(\mx)$. The second part imposes the same type of margin on the localized oracle objectives; it is used to convert objective-level stochastic and bias bounds into distance bounds for $\tilde m_{h,s}$ and $\hat m_{h,s}$. In settings where the population objective is uniformly strongly convex in a neighborhood of its minimizer and the localized objectives converge smoothly to the population objective, this localized margin can often be derived. In the present general metric-space formulation it is imposed as a sufficient high-level condition.
\end{remark}

\begin{customcon}{U-M4} \label{con:U-M4} 
There exists a constant $r_{\mathbb M,\mathcal K}>0$ such that
\begin{align*}
    \sup_{\mx\in\mathcal K}
    \sup_{y\in\mathbb M:d_{\mathbb M}\left(y,m_{\oplus}(\mx)\right)<r_{\mathbb M,\mathcal K}}
    \int_{0}^{1/2}
    \sqrt{
    1+
    \log N\left(
    \delta\epsilon,
    B_{\mathbb M}(y,\delta),
    d_{\mathbb M}
    \right)
    }
    \,\dd\epsilon
    =
    O(1),
    \quad
    \delta\downarrow0.
\end{align*}
\end{customcon}

Condition~\ref{con:U-M4} is the uniform counterpart of Condition~\ref{con:P-M4}. It is needed because the local target $m_{\oplus}(\mx)$ varies with $\mx\in\mathcal K$.

\begin{remark}[Uniform response-space conditions]
\label{rem:uniform_response_space_examples}
Conditions~\ref{con:U-M2}--\ref{con:U-M4} are uniform versions of the preceding pointwise Fr\'echet regularity conditions over the compact predictor region $\mathcal K$. They require the same type of response-space compactness, local entropy, uniqueness, separation, and margin behavior to hold uniformly for $\mx\in\mathcal K$ and, for the oracle local objectives, for all sufficiently small bandwidths. Thus the uniform conditions are stronger than imposing the pointwise conditions separately at each $\mx$.

These assumptions are reasonable in finite-dimensional compact response settings when the population regression image remains in a region where the Fr\'echet objective is uniformly well behaved. Examples include compact subsets of Euclidean spaces, compact finite-dimensional Riemannian manifolds subject to uniform localization away from nonunique Fr\'echet means, bounded network spaces represented by graph Laplacians with a fixed number of nodes and uniformly bounded edge weights, and compact subsets of symmetric positive definite matrices under a finite-dimensional metric representation. For Wasserstein responses, a typical admissible setting is $\mathcal W_2([a,b])$ or a uniformly totally bounded subclass of one-dimensional distributions. The unrestricted space $\mathcal W_2(\mathbb R)$ is not covered by Condition~\ref{con:M1} without additional restrictions.

The uniform margin and minimizer assumptions are substantive. They ensure that the population and oracle local Fr\'echet objectives have uniformly identifiable minimizers and that stochastic perturbations of the objectives can be converted into uniform distance bounds. In particular, for local linear smoothing, the signed nature of the equivalent weights makes it important to impose uniform well-posedness of the oracle and empirical minimization problems. These assumptions are standard in general Fr\'echet regression theory, but their verification is response-space and model dependent.
\end{remark}

\begin{theorem}[Uniform convergence rate] \label{thm:uniform_rate}
Assume Conditions~\ref{con:U-K1}, \ref{con:U-B1}, \ref{con:U-D1}, \ref{con:U-D2}, \ref{con:M1}, \ref{con:U-M2}, \ref{con:U-D3}, \ref{con:U-D4}, \ref{con:U-M3}, and~\ref{con:U-M4}. Then
\begin{align*}
    \sup_{\mx\in\mathcal K}
    d_{\mathbb M}(\hat{m}_{h,0}(\mx),m_{\oplus}(\mx))
    =
    O\left(h^{2/(\beta_{\oplus,\mathcal K}-1)}\right)
    +
    O_{\P}\left(
    \left(\frac{\log n}{nh^d}\right)^{1/(2\beta_{\oplus,\mathcal K}-2)}
    \right).
\end{align*}
If, in addition, Condition~\ref{con:U-K2} holds, then
\begin{align*}
    \sup_{\mx\in\mathcal K}
    d_{\mathbb M}(\hat{m}_{h,1}(\mx),m_{\oplus}(\mx))
    =
    O\left(h^{2/(\beta_{\oplus,\mathcal K}-1)}\right)
    +
    O_{\P}\left(
    \left(\frac{\log n}{nh^d}\right)^{1/(2\beta_{\oplus,\mathcal K}-2)}
    \right).
\end{align*}
\end{theorem}

Under a quadratic uniform Fr\'echet margin, $\beta_{\oplus,\mathcal K}=2$, and the bandwidth choice $h\asymp(\log n/n)^{1/(d+4)}$, the uniform rate becomes
\begin{align*}
    \sup_{\mx\in\mathcal K}
    d_{\mathbb{M}}(\hat{m}_{h,s}(\mx),m_{\oplus}(\mx))
    =
    O_{\P}\left(\left(\frac{\log n}{n}\right)^{2/(d+4)}\right),
    \quad
    s\in\{0,1\}.
\end{align*}
This is of the same order as the usual logarithmic uniform upper-bound rate for twice-smooth Euclidean nonparametric regression in intrinsic dimension $d$. The theoretical contribution of the local linear analysis is therefore not a faster interior rate, but the construction and control of intrinsic signed local linear Fr\'echet weights on moving tangent spaces and the separation of the zeroth-order and multiplier empirical-process requirements.

\section{Simulation Studies}
\label{sec:simulation}
\setcounter{equation}{0}

We conduct two simulation studies to examine the finite-sample behavior of local Fr\'echet regression with Riemannian manifold predictors. The simulations are intended as controlled numerical illustrations rather than as exhaustive empirical validation over all possible Riemannian predictor manifolds. Simulation A considers spherical predictors and spherical responses and compares the proposed general Riemannian construction with the sphere-specific estimators of \cite{Im et al. (2025)} and a Tucker--Wu-type local constant metric-predictor smoother. Simulation B considers symmetric positive-definite predictors under the affine-invariant Riemannian metric and spherical responses, illustrating that the proposed intrinsic construction also applies beyond compact constant-curvature predictor manifolds.

The two simulations play complementary roles. Simulation A directly compares the proposed estimator with methods specifically constructed for spherical predictors. Even when the predictor manifold is $\mathbb S^2$, the proposed construction differs by incorporating the Riemannian normal-coordinate volume-density correction and by defining the local linear weights through a general frame-invariant tangent-space formulation. Simulation B investigates the same construction on $\mathcal S_{++}^2$ under the affine-invariant metric using a nonidentity affine-invariant normal-coordinate design. Since the two estimators have the same interior rate order under the assumptions of \Cref{sec:asymptotic_theory}, the numerical experiments examine their finite-sample behavior and leading-error differences rather than rate superiority.

Bandwidths are selected separately for each method and each Monte Carlo replication by five-fold cross-validation. For a candidate bandwidth $h$, the cross-validation criterion is
\begin{align*}
    \CV(h)
    &:=
    \frac{1}{n}
    \sum_{i=1}^n
    d_{\mathbb M}^2
    \left(
    Y^{(i)},
    \hat m_{-k(i),h}
    \left(
    X^{(i)}
    \right)
    \right),
\end{align*}
where $k(i)$ denotes the validation fold containing observation $i$, and $\hat m_{-k(i),h}$ is fitted without the observations in that fold. The same folds are used for all methods within each replication. Candidate bandwidths producing non-finite validation predictions or numerical failures are assigned an infinite cross-validation score.

For each configuration, we use $R=100$ Monte Carlo replications. In Simulation A, a replication is accepted only when all five competing methods produce finite final predictions, so the methods are evaluated on the same generated data within every accepted replication. All Simulation A replications were accepted on their first attempt, and every method produced a finite integrated squared error. Simulation B uses $100$ fixed replications for each configuration without resampling; all reported full-region integrated squared errors were finite.

Let $\nu$ denote the evaluation measure on the predictor region. For the estimator obtained in the $r$th replication with selected bandwidth $\hat h^{(r)}$, define the integrated squared error by
\begin{align*}
    \ISE^{(r)}
    &:=
    \int_{\mathcal M}
    d_{\mathbb M}^2
    \left(
    \hat m_{\hat h^{(r)}}^{(r)}(x),
    m_{\oplus}(x)
    \right)
    \dd\nu(x).
\end{align*}
The corresponding mean integrated squared error is
\begin{align*}
    \MISE
    &:=
    \E
    \left[
    \ISE^{(r)}
    \right],
\end{align*}
where the expectation is taken over the training sample and response noise and, in Simulation B, over the independently generated evaluation sample used to approximate integration over the predictor region.

This quantity is approximated using $N_{\mathrm{eval}}=800$ evaluation points and $R=100$ Monte Carlo replications. Let $\{X_{\mathrm{eval}}^{(r,j)}\}_{j=1}^{N_{\mathrm{eval}}}$ denote the evaluation set used in the $r$th replication. In Simulation A, the same fixed approximately uniform Fibonacci grid on $\mathbb S^2$ is used in every replication, so $X_{\mathrm{eval}}^{(r,j)}$ does not depend on $r$. In Simulation B, a new evaluation sample is generated independently in each replication over the full normal-coordinate region $[-0.9,0.9]^3$, and the resulting evaluation predictors are shared by all methods within that replication. The Monte Carlo approximation to the MISE is
\begin{align*}
    \widehat{\MISE}
    &:=
    \frac{1}{R}
    \sum_{r=1}^R
    \frac{1}{N_{\mathrm{eval}}}
    \sum_{j=1}^{N_{\mathrm{eval}}}
    d_{\mathbb M}^2
    \left(
    \hat m_{\hat h^{(r)}}^{(r)}
    \left(
    X_{\mathrm{eval}}^{(r,j)}
    \right),
    m_{\oplus}
    \left(
    X_{\mathrm{eval}}^{(r,j)}
    \right)
    \right).
\end{align*}
For simplicity, we refer to $\widehat{\MISE}$ as MISE in the numerical results.

\subsection{Simulation A: Spherical predictor benchmark}
\label{subsec:simulation_A}

Simulation A considers predictors and responses on the unit sphere $\mathbb S^2$. The predictors are generated as $X^{(i)}\sim\Unif(\mathbb S^2)$, $i=1,\ldots,n$. To avoid tying the data-generating process to longitude--latitude coordinates, we define the regression function by normalizing a smooth ambient-coordinate map. For $x=(x_1,x_2,x_3)^\top\in\mathbb S^2$, define
\begin{align*}
    m_{\oplus}(x)
    &:=
    \frac{
    \left(
    m_{\oplus,1}(x),
    m_{\oplus,2}(x),
    m_{\oplus,3}(x)
    \right)^\top
    }{
    \left\|
    \left(
    m_{\oplus,1}(x),
    m_{\oplus,2}(x),
    m_{\oplus,3}(x)
    \right)^\top
    \right\|_2
    }
    \in\mathbb S^2,
\end{align*}
where
\begin{align*}
    m_{\oplus,1}(x)
    &:=
    1.2+0.6x_2+0.8x_3, \\
    m_{\oplus,2}(x)
    &:=
    -0.8+0.5x_1, \\
    m_{\oplus,3}(x)
    &:=
    2.0+0.4x_1+0.3x_2.
\end{align*}
This construction defines a smooth nonlinear map $m_{\oplus}:\mathbb S^2\to\mathbb S^2$. Moreover,
\begin{align*}
    m_{\oplus,3}(x)
    &\geq
    2-\sqrt{0.4^2+0.3^2}
    =
    1.5,
    \quad
    x\in\mathbb S^2,
\end{align*}
so the normalization is uniformly well defined.

Responses are generated according to
\begin{align*}
    Y^{(i)}
    &:=
    \Exp_{m_{\oplus}\left(X^{(i)}\right)}^{\mathbb S^2}
    \left(
    \bm{\varepsilon}^{(i)}
    \right),
    \quad
    i=1,\ldots,n,
\end{align*}
where, conditionally on $X^{(i)}$, $\bm{\varepsilon}^{(i)}$ is isotropic Gaussian noise in the two-dimensional tangent space
$T_{m_{\oplus}(X^{(i)})}\mathbb S^2$ with componentwise standard deviation $\sigma$. Its tangent norm is truncated at $2.5\sigma$ to keep the generated responses within a numerically stable geodesic neighborhood. Since $\sigma\leq0.25$, the conditional response is supported in the geodesic ball of radius $0.625<\pi/4$ centered at $m_{\oplus}(X^{(i)})$. The support-radius condition gives uniqueness of the intrinsic conditional Fr\'echet mean \citep{Afsari (2011)}. Rotational symmetry about the displayed center then identifies this unique minimizer as $m_{\oplus}(X^{(i)})$. We consider $n\in\{100,200,400\}$ and $\sigma\in\{0.10,0.25\}$. The evaluation points are given by a fixed approximately uniform Fibonacci grid of size $N_{\mathrm{eval},A}=800$ on $\mathbb S^2$.

For the proposed estimators, the local kernel is constructed from the geodesic distance on $\mathbb S^2$. Away from the cut locus of $x$, the normal-coordinate volume density is
\begin{align*}
    \theta_x(z)
    &:=
    \frac{
    \sin\left(
    d_{\mathbb S^2}(x,z)
    \right)
    }{
    d_{\mathbb S^2}(x,z)
    },
    \quad
    z\in
    \mathbb S^2\setminus\{-x\},
\end{align*}
where the ratio is interpreted as one when
$d_{\mathbb S^2}(x,z)=0$. Hence the proposed local kernel uses the correction factor
\begin{align*}
    \theta_x(z)^{-1}
    &:=
    \frac{
    d_{\mathbb S^2}(x,z)
    }{
    \sin\left(
    d_{\mathbb S^2}(x,z)
    \right)
    },
    \quad
    z\in
    \mathbb S^2\setminus\{-x\}.
\end{align*}
The largest candidate bandwidth is $1.50<\pi$, so the compactly supported proposed kernels do not place positive weight at the cut locus.

We compare the following five estimators:
\begin{itemize}
    \item Proposed LC: The proposed volume-corrected local constant Fr\'echet estimator.
    \item Proposed LL: The proposed volume-corrected local linear Fr\'echet estimator.
    \item Spherical LC: The sphere-specific local constant estimator of \cite{Im et al. (2025)}.
    \item Spherical LL: The sphere-specific local linear estimator of \cite{Im et al. (2025)}.
    \item TW LC: A generic metric-predictor local constant Fr\'echet smoother motivated by the local smoothing construction in \cite{Tucker and Wu (2025)}.
\end{itemize}
The TW LC estimator is included as a generic metric-predictor local constant benchmark. The Spherical LC and Spherical LL estimators provide the most relevant specialized comparison because they were constructed specifically for spherical predictors.

The proposed estimators and TW LC use the bandwidth grid $\{0.25+0.05\ell:\ell=1,\ldots,25\}=\{0.30,0.35,\ldots,1.50\}$, whereas Spherical LC and Spherical LL use their method-specific bandwidth scale $\{0.10+0.02\ell:\ell=1,\ldots,25\}=\{0.12,0.14,\ldots,0.60\}$. Each grid contains $25$ candidate values. The proposed estimators and TW LC use the same compactly supported Epanechnikov kernel profile
\begin{align*}
    K(t)
    &:=
    \frac{3}{4}
    \left(1-t^2\right)
    \mathbf{1}\{0\leq t\leq1\},
    \quad
    t\in[0,\infty).
\end{align*}
The proposed estimators additionally incorporate the Riemannian volume-density correction, while Proposed LL further uses the tangent-space local linear adjustment. Spherical LC and Spherical LL use the sphere-specific exponential kernel
\begin{align*}
    K_{\mathrm{sph},h}(x,z)
    &:=
    \exp\left(
    -\frac{
    1-x^\top z
    }{
    h^2
    }
    \right),
    \quad
    x,z\in\mathbb S^2,
\end{align*}
which is proportional, for fixed $h$, to a von Mises--Fisher kernel with concentration parameter $h^{-2}$. The numerical bandwidth values are not directly comparable across these kernel parameterizations. Each grid is chosen on the scale convention used by the corresponding method and is intended to cover a suitable range of effective neighborhoods. Accordingly, the comparison with Spherical LC and Spherical LL is a comparison of the complete conventional smoothing procedures, not a kernel-controlled attribution of small differences to the general-manifold construction alone.

\Cref{tab:simA_sphere_mise} reports $10^3\times\MISE$ over the $R=100$ Monte Carlo replications. Proposed LL has the smallest reported Monte Carlo mean in every configuration, although its difference from the specialized Spherical LL estimator is modest. Under $\sigma=0.10$, both local linear procedures have lower mean MISE than the local constant procedures, while under $\sigma=0.25$ the differences are smaller. Proposed LC and TW LC are close across the reported configurations when their distance and kernel profiles are held fixed, with the largest gap occurring at $(n,\sigma)=(100,0.25)$. This pattern indicates that the finite-sample effect of the spherical volume-density correction is small in this design.

\begin{table}[!ht]
\centering
\caption{
Simulation A: $10^3\times\MISE$ for the spherical-predictor benchmark, based on $100$ Monte Carlo replications.
}
\label{tab:simA_sphere_mise}
\small
\begin{tabular}{lcccccc}
\toprule
& \multicolumn{3}{c}{$\sigma=0.10$}
& \multicolumn{3}{c}{$\sigma=0.25$} \\
\cmidrule(lr){2-4}
\cmidrule(lr){5-7}
Method
& $n=100$ & $n=200$ & $n=400$
& $n=100$ & $n=200$ & $n=400$ \\
\midrule
Proposed LC
& 3.538 & 2.065 & 1.194
& 10.329 & 6.211 & 3.672 \\
Proposed LL
& \textbf{2.843} & \textbf{1.748} & \textbf{1.024}
& \textbf{9.722} & \textbf{6.056} & \textbf{3.510} \\
Spherical LC
& 3.360 & 2.060 & 1.191
& 10.277 & 6.328 & 3.769 \\
Spherical LL
& 2.868 & 1.819 & 1.072
& 10.009 & 6.174 & 3.675 \\
TW LC
& 3.511 & 2.064 & 1.189
& 10.049 & 6.199 & 3.644 \\
\bottomrule
\end{tabular}
\end{table}

\begin{figure}[!ht]
\centering
\includegraphics[width=\textwidth]{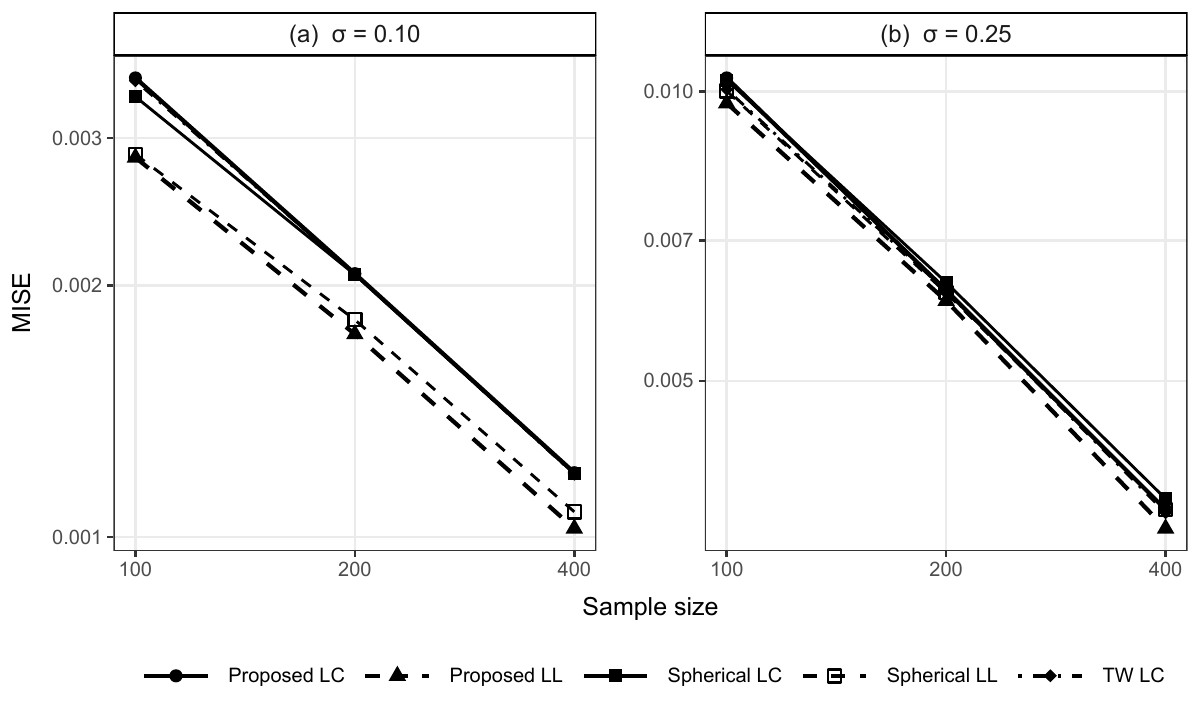}
\caption{
Simulation A: log-log plots of MISE against sample size for the spherical-predictor benchmark under (a) $\sigma=0.10$ and (b) $\sigma=0.25$.
}
\label{fig:simA_sphere_loglog_mise}
\end{figure}

\Cref{fig:simA_sphere_loglog_mise} displays the same MISE values on logarithmic axes. All methods improve steadily as the sample size increases. Proposed LL and Spherical LL remain below the local constant estimators under the lower noise level, while the method differences are less pronounced under the higher noise level. Proposed LL has the lowest MISE throughout, but its trajectory remains close to that of the specialized Spherical LL estimator.

\subsection{Simulation B: SPD predictor under the affine-invariant metric}
\label{subsec:simulation_B}

Simulation B considers the manifold $\mathcal S_{++}^2$ of $2\times2$ symmetric positive-definite matrices equipped with the affine-invariant Riemannian metric
\begin{align*}
    \left\langle
    \mathbf{A},
    \mathbf{B}
    \right\rangle_X
    &:=
    \tr\left(
    X^{-1}
    \mathbf{A}
    X^{-1}
    \mathbf{B}
    \right),
    \quad
    X\in\mathcal S_{++}^2,
    \quad
    \mathbf{A},\mathbf{B}\in T_X\mathcal S_{++}^2.
\end{align*}
The intrinsic dimension of $\mathcal S_{++}^2$ is $d=3$. The affine-invariant exponential map, logarithmic map, and geodesic distance are
\begin{align*}
    \Exp_X(\mathbf{A})
    &:=
    X^{1/2}
    \exp\left(
    X^{-1/2}
    \mathbf{A}
    X^{-1/2}
    \right)
    X^{1/2},
    \quad
    X\in\mathcal S_{++}^2,
    \quad
    \mathbf{A}\in T_X\mathcal S_{++}^2, \\
    \Log_X(Z)
    &:=
    X^{1/2}
    \log\left(
    X^{-1/2}
    Z
    X^{-1/2}
    \right)
    X^{1/2},
    \quad
    X,Z\in\mathcal S_{++}^2, \\
    d_{\mathcal S_{++}^2}(X,Z)
    &:=
    \left\|
    \log\left(
    X^{-1/2}
    Z
    X^{-1/2}
    \right)
    \right\|_F,
    \quad
    X,Z\in\mathcal S_{++}^2.
\end{align*}

To generate a nonidentity affine-invariant normal-coordinate design, let
\begin{align*}
    \mathbf{Q}
    &:=
    \frac{1}{\sqrt{2}}
    \begin{pmatrix}
        1 & -1 \\
        1 & 1
    \end{pmatrix},
    \qquad
    \mathbf{P}_0
    :=
    \mathbf{Q}
    \begin{pmatrix}
        4 & 0 \\
        0 & 0.25
    \end{pmatrix}
    \mathbf{Q}^\top
    =
    \begin{pmatrix}
        17/8 & 15/8 \\
        15/8 & 17/8
    \end{pmatrix}.
\end{align*}
Thus, $\mathbf{P}_0$ has eigenvalues $4$ and $0.25$ and condition number $16$. For $\mathbf{v}=(v_1,v_2,v_3)^\top\in\mathbb R^3$, define
\begin{align*}
    \mathbf{S}(\mathbf{v})
    :=
    \begin{pmatrix}
        v_1 & v_3/\sqrt{2} \\
        v_3/\sqrt{2} & v_2
    \end{pmatrix},
    \quad
    \mathbf{V}(\mathbf{v})
    :=
    \mathbf{P}_0^{1/2}
    \mathbf{S}(\mathbf{v})
    \mathbf{P}_0^{1/2}
    \in
    T_{\mathbf{P}_0}\mathcal S_{++}^2.
\end{align*}
The coordinates are orthonormal with respect to the affine-invariant metric at $\mathbf{P}_0$, since
\begin{align*}
    \left\|
    \mathbf{V}(\mathbf{v})
    \right\|_{\mathbf{P}_0}^2
    &:=
    \tr\left(
    \mathbf{P}_0^{-1}
    \mathbf{V}(\mathbf{v})
    \mathbf{P}_0^{-1}
    \mathbf{V}(\mathbf{v})
    \right)
    =
    \left\|
    \mathbf{S}(\mathbf{v})
    \right\|_F^2
    =
    \|\mathbf{v}\|_2^2.
\end{align*}

For each observation, we draw
\begin{align*}
    \mathbf{v}^{(i)}
    &:=
    \left(
    v_1^{(i)},
    v_2^{(i)},
    v_3^{(i)}
    \right)^\top
    \sim
    \Unif\left([-0.9,0.9]^3\right),
    \quad
    i=1,\ldots,n,
\end{align*}
and set
\begin{align*}
    X^{(i)}
    &:=
    \Exp_{\mathbf{P}_0}
    \left(
    \mathbf{V}\left(\mathbf{v}^{(i)}\right)
    \right)
    =
    \mathbf{P}_0^{1/2}
    \exp\left(
    \mathbf{S}\left(\mathbf{v}^{(i)}\right)
    \right)
    \mathbf{P}_0^{1/2},
    \quad
    i=1,\ldots,n.
\end{align*}
The bounded normal-coordinate region is mapped continuously to a compact subset of $\mathcal S_{++}^2$ whose eigenvalues are uniformly bounded away from zero and infinity. Unlike an identity-centered construction of the form $X=\exp(\mathbf{U})$, the resulting data-generating coordinates are intrinsic normal coordinates at the nonidentity base point $\mathbf{P}_0$.

Define
\begin{align*}
    \mathbf{A}_{X,Z}
    &:=
    \log\left(
    X^{-1/2}
    Z
    X^{-1/2}
    \right),
    \quad
    X,Z\in\mathcal S_{++}^2,
\end{align*}
and let
$\lambda_1\left(\mathbf{A}_{X,Z}\right)$ and
$\lambda_2\left(\mathbf{A}_{X,Z}\right)$
denote its eigenvalues. Under the affine-invariant metric, the normal-coordinate volume density is
\begin{align*}
    \theta_X(Z)
    &:=
    \frac{
    \sinh\left(
    \left|
    \lambda_1\left(
    \mathbf{A}_{X,Z}
    \right)
    -
    \lambda_2\left(
    \mathbf{A}_{X,Z}
    \right)
    \right|/2
    \right)
    }{
    \left|
    \lambda_1\left(
    \mathbf{A}_{X,Z}
    \right)
    -
    \lambda_2\left(
    \mathbf{A}_{X,Z}
    \right)
    \right|/2
    },
    \quad
    X,Z\in\mathcal S_{++}^2,
\end{align*}
where the ratio is interpreted as one when its argument is zero. Hence the proposed local kernel uses the correction factor
\begin{align*}
    \theta_X(Z)^{-1}
    &:=
    \frac{
    \left|
    \lambda_1\left(
    \mathbf{A}_{X,Z}
    \right)
    -
    \lambda_2\left(
    \mathbf{A}_{X,Z}
    \right)
    \right|/2
    }{
    \sinh\left(
    \left|
    \lambda_1\left(
    \mathbf{A}_{X,Z}
    \right)
    -
    \lambda_2\left(
    \mathbf{A}_{X,Z}
    \right)
    \right|/2
    \right)
    },
    \quad
    X,Z\in\mathcal S_{++}^2.
\end{align*}

The response space is $\mathbb S^2$, equipped with the geodesic distance
\begin{align*}
    d_{\mathbb S^2}(y,z)
    &:=
    \arccos\left(
    y^\top z
    \right),
    \quad
    y,z\in\mathbb S^2.
\end{align*}
Let $\bm{\mu}_0:=(0,0,1)^\top$. For $\mathbf{v}=(v_1,v_2,v_3)^\top\in[-0.9,0.9]^3$, define
\begin{align*}
    \bm{\eta}_B(\mathbf{v})
    &:=
    \begin{pmatrix}
        0.55\sin(1.5\pi v_1)
        +0.30v_2v_3
        +0.15v_1v_2
        \\
        0.45\sin(\pi v_2v_3)
        +0.25v_1^2
        +0.20\sin(\pi v_3)
        -0.10
        \\
        0
    \end{pmatrix}
    \in
    T_{\bm{\mu}_0}\mathbb S^2,
\end{align*}
and set
\begin{align*}
    m_{\oplus}
    \left(
    X(\mathbf{v})
    \right)
    &:=
    \Exp_{\bm{\mu}_0}^{\mathbb S^2}
    \left(
    \bm{\eta}_B(\mathbf{v})
    \right),
\end{align*}
where
\begin{align*}
    X(\mathbf{v})
    &:=
    \mathbf{P}_0^{1/2}
    \exp\left(
    \mathbf{S}(\mathbf{v})
    \right)
    \mathbf{P}_0^{1/2}.
\end{align*}
Responses are generated according to
\begin{align*}
    Y^{(i)}
    &:=
    \Exp_{m_{\oplus}\left(X^{(i)}\right)}^{\mathbb S^2}
    \left(
    \bm{\varepsilon}^{(i)}
    \right),
    \quad
    i=1,\ldots,n,
\end{align*}
where, conditionally on $X^{(i)}$, $\bm{\varepsilon}^{(i)}$ is isotropic Gaussian noise in
$T_{m_{\oplus}(X^{(i)})}\mathbb S^2$ with componentwise standard deviation $\sigma$. Its tangent norm is truncated at $2.5\sigma$ for numerical stability. As in Simulation A, the maximal conditional support radius is $0.625<\pi/4$. The support-radius condition gives uniqueness of the intrinsic conditional Fr\'echet mean \citep{Afsari (2011)}, and rotational symmetry identifies the displayed center $m_{\oplus}(X^{(i)})$ as that unique minimizer. We consider $n\in\{100,200,400\}$ and $\sigma\in\{0.10,0.25\}$.

For each replication, all methods are evaluated using
$N_{\mathrm{eval},B}=800$ predictors generated independently from the full normal-coordinate region $[-0.9,0.9]^3$ through the same affine-invariant exponential-map construction as the training predictors. The same evaluation sample is used for all methods within the replication. Thus, the reported MISE measures prediction accuracy over the entire predictor region, including points arbitrarily close to the boundary of the coordinate cube. Because the design density contains a support indicator at this boundary, the full-support experiment does not satisfy the interior smooth-density assumptions of the uniform rate theorem over the entire evaluation region. Simulation B is therefore a finite-sample boundary and design-adaptivity experiment, not a direct numerical verification of the uniform rate theorem. The data-generating regression function is deliberately smooth in AIRM normal coordinates at $\mathbf P_0$; the design assesses the estimator under a correctly specified affine-invariant geometry.

We report the following three estimators:
\begin{itemize}
    \item Proposed LC: The proposed affine-invariant, volume-corrected local constant Fr\'echet estimator.
    \item Proposed LL: The proposed affine-invariant, volume-corrected local linear Fr\'echet estimator.
    \item TW LC: A generic metric-predictor local constant Fr\'echet smoother motivated by the local smoothing construction in \cite{Tucker and Wu (2025)}.
\end{itemize}
The TW LC estimator is included as a generic metric-predictor local constant benchmark. All three procedures use the affine-invariant predictor distance, so the experiment holds the predictor geometry fixed and examines local linear versus local constant smoothing, apart from the volume-density normalization distinguishing Proposed LC from TW LC.

All three methods use the sample-size-dependent bandwidth grid $\{0.2\ell n^{-1/7}:\ell=1,\ldots,25\}$. The factor $n^{-1/7}$ reflects the intrinsic dimension $d=3$, for which the pointwise twice-smooth bandwidth order is $n^{-1/(d+4)}=n^{-1/7}$. Each grid contains $25$ candidate values. All estimators use the same compactly supported Epanechnikov kernel profile
\begin{align*}
    K(t)
    &:=
    \frac{3}{4}
    \left(1-t^2\right)
    \mathbf{1}\{0\leq t\leq1\},
    \quad
    t\in[0,\infty).
\end{align*}
The proposed estimators additionally incorporate the Riemannian volume-density correction, while Proposed LL further uses the tangent-space local linear adjustment.

\Cref{tab:simB_full_mise} reports the MISE values, multiplied by $10^2$ for readability, over the $R=100$ Monte Carlo replications. Proposed LL has the smallest reported Monte Carlo mean in every configuration. Its difference from the two local constant estimators is appreciable under both noise levels and becomes larger with sample size in this full-support AIRM-aligned design. Proposed LC and TW LC yield nearly identical values throughout. Since these two procedures use the same affine-invariant distance and Epanechnikov kernel and differ only through the volume-density normalization, their similarity indicates that the normalization has a small direct numerical effect here. The separation of Proposed LL is consistent with the tangent-space first-order correction, including its familiar design- and support-boundary adaptivity.

\begin{table}[!ht]
\centering
\caption{
Simulation B: $10^2\times\MISE$ over the full normal-coordinate region
$[-0.9,0.9]^3$, based on $100$ Monte Carlo replications.
}
\label{tab:simB_full_mise}
\small
\begin{tabular}{lcccccc}
\toprule
& \multicolumn{3}{c}{$\sigma=0.10$}
& \multicolumn{3}{c}{$\sigma=0.25$} \\
\cmidrule(lr){2-4}
\cmidrule(lr){5-7}
Method
& $n=100$ & $n=200$ & $n=400$
& $n=100$ & $n=200$ & $n=400$ \\
\midrule
Proposed LC
& 15.095 & 10.831 & 6.943
& 15.566 & 11.180 & 7.434 \\
Proposed LL
& \textbf{10.404} & \textbf{7.040} & \textbf{4.331}
& \textbf{12.681} & \textbf{8.701} & \textbf{5.368} \\
TW LC
& 15.183 & 10.877 & 6.964
& 15.653 & 11.223 & 7.454 \\
\bottomrule
\end{tabular}
\end{table}

\begin{figure}[!ht]
\centering
\includegraphics[width=\textwidth]{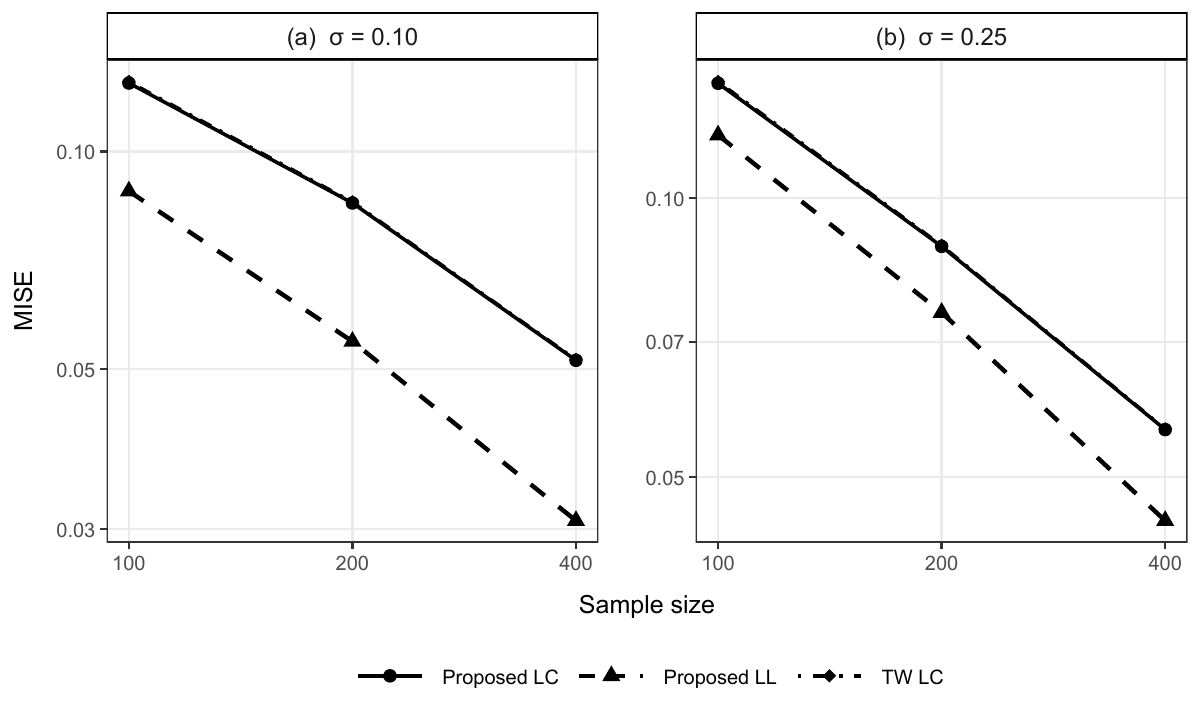}
\caption{
Simulation B: log-log plots of full-region MISE against sample size under (a) $\sigma=0.10$ and (b) $\sigma=0.25$.
}
\label{fig:simB_loglog_mise}
\end{figure}

\Cref{fig:simB_loglog_mise} displays the same full-region MISE values on logarithmic axes. Proposed LL remains below both local constant estimators across the considered sample sizes and noise levels, while Proposed LC and TW LC follow nearly identical trajectories.

\subsection{Summary of simulation findings}

The two experiments have different purposes. In the spherical benchmark, the general Riemannian local linear implementation is numerically competitive with the sphere-specific local linear procedure under each method's conventional kernel and tuning scale.

In the SPD experiment, the AIRM geometry is fixed and the data-generating relation is deliberately smooth in AIRM normal coordinates. Proposed LL has lower mean MISE than the two local constant procedures throughout the reported full-support design, while Proposed LC and TW LC again remain nearly identical. Since the evaluation region includes the support boundary, the observed separation is best read as a finite-sample illustration of local linear design and boundary adaptation under a correctly specified intrinsic geometry. The reported tables contain Monte Carlo averages and are used descriptively; small numerical differences are not interpreted as uncertainty-qualified superiority claims.

\section{Real Data Analysis}
\label{sec:real_data}
\setcounter{equation}{0}

\subsection{Diffusion tensor imaging data}
\label{subsec:real_data_dti}

We first analyze axial diffusion tensor imaging data from the OASIS--3 study \citep{LaMontagne et al. (2019)}, available at \url{https://www.oasis-brains.org/}. After preprocessing and quality control, the analysis includes 281 subjects. For each subject, the predictor $X\in\mathcal S_{++}^3$ is the mean diffusion tensor within the body of the corpus callosum. The response $Y\in\mathcal W_2(\mathbb R)$ is the empirical distribution of voxelwise fractional anisotropy values within the splenium of the corpus callosum. We represent each response by its empirical quantile function at the 99 probability levels $0.01,\ldots,0.99$ and approximate squared $2$-Wasserstein distance by the average squared difference between the corresponding quantile values.

We compare four estimators: Proposed LC and Proposed LL under the affine-invariant Riemannian metric, TW LC using the same affine-invariant distance without the volume-density normalization, and PM LL obtained by applying the Euclidean-predictor local linear Fr\'echet estimator of \cite{Petersen and Muller (2019)} to the Frobenius-scaled half-vectorization of the diffusion tensors. All four procedures use the Epanechnikov kernel. Prediction performance is evaluated by five repetitions of five-fold outer cross-validation. Within each outer training fold, the bandwidth is selected by inner five-fold cross-validation using identical subject folds across methods. The absolute bandwidth grids are $\{0.05k:k=1,\ldots,50\}$ for Proposed LC and TW LC, $\{0.10k:k=1,\ldots,50\}$ for Proposed LL, and $\{10^{-4}k:k=1,\ldots,50\}$ for PM LL.

The aggregate root mean squared Wasserstein prediction errors are $0.07389$ for Proposed LC, $\mathbf{0.07277}$ for Proposed LL, $0.07391$ for TW LC, and $0.07722$ for PM LL. The corresponding standard deviations across the five repeat-level summaries are $0.00083$, $0.00067$, $0.00083$, and $0.00363$, respectively. Among the four implemented procedures, Proposed LL therefore has the lowest aggregate error, although its difference from the two AIRM local constant procedures is modest. Proposed LC and TW LC are nearly indistinguishable, indicating a small direct numerical effect of the volume-density normalization in this data set. The raw Frobenius implementation is less accurate, but the comparator set is not intended as a comprehensive comparison of possible metrics or global representations on the SPD cone.

\begin{figure}[t]
    \centering
    \includegraphics[width=\linewidth]{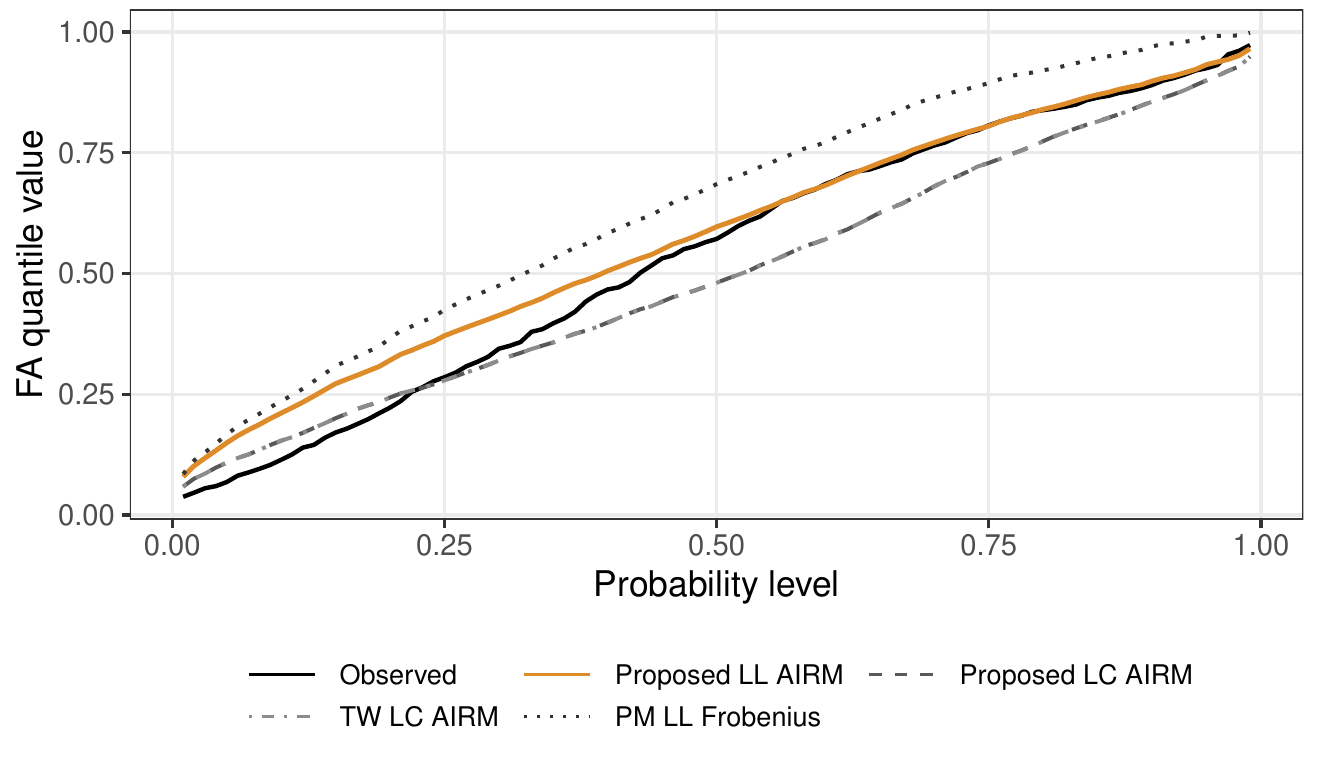}
    \caption{Observed and predicted quantile functions for an illustrative held-out FA distribution in the OASIS--3 analysis. The predictions are obtained from Proposed LL under AIRM, Proposed LC under AIRM, TW LC using the AIRM distance, and PM LL using the raw Frobenius representation. The aggregate repeated-cross-validation results, rather than this single illustration, are the basis for the method comparison.}
    \label{fig:real_data_dti_quantile}
\end{figure}

\Cref{fig:real_data_dti_quantile} illustrates the form of the distribution-valued predictions for one held-out response. Proposed LC and TW LC produce nearly overlapping curves. The displayed case is qualitative; the numerical comparison above is based on all held-out subjects and all outer folds.

\subsection{Gaze-direction data}
\label{subsec:real_data_gaze}

We next use the Head and Gaze VR Behavior Dataset of \cite{Jin et al. (2022)}, available at \url{https://cuhksz-inml.github.io/head_gaze_dataset/}. We analyze the recordings for video~11, which supplies a common stimulus across participants. The predictor $X\in\mathrm{SO}(3)$ is the head-to-world rotation matrix, and the response $Y\in\mathbb S^2$ is the binocular gaze direction expressed in world coordinates. The response is obtained by normalizing the average of the transformed left- and right-eye directions, with the available eye used when only one eye yields a valid direction. After quality control, 97 subjects are retained. The original $10$~Hz sequences are reduced to $2$~Hz for computation.

Subjects, rather than individual time points, are used as the sampling units for evaluation. We randomly assign 77 subjects to training and 20 subjects to testing. To prevent longer recordings from dominating the fit, the training observations are weighted so that each subject has the same total weight. We compare Proposed LC, Proposed LL, and TW LC. Bandwidths are selected separately by subject-level five-fold cross-validation within the training set over the common grid $\{0.02k:k=1,\ldots,50\}$. The primary criterion is the equal-subject mean squared geodesic error on $\mathbb S^2$.

In this split, the subject-equal angular root mean squared prediction errors are $19.05^{\circ}$ for Proposed LC, $\mathbf{15.57^{\circ}}$ for Proposed LL, and $19.04^{\circ}$ for TW LC. Proposed LL has the lowest error in the selected split, whereas the two local constant fits are nearly identical. The observations are longitudinal within subject, and the equal-subject fitting weights define a clustered, weighted implementation of the estimator. The independent-pair asymptotic theory in \Cref{sec:asymptotic_theory} does not directly cover this analysis; it is included as a prediction illustration with subject-level separation rather than as uncertainty-qualified evidence across repeated splits or videos.

\begin{figure}[t]
    \centering
    \includegraphics[width=\linewidth]{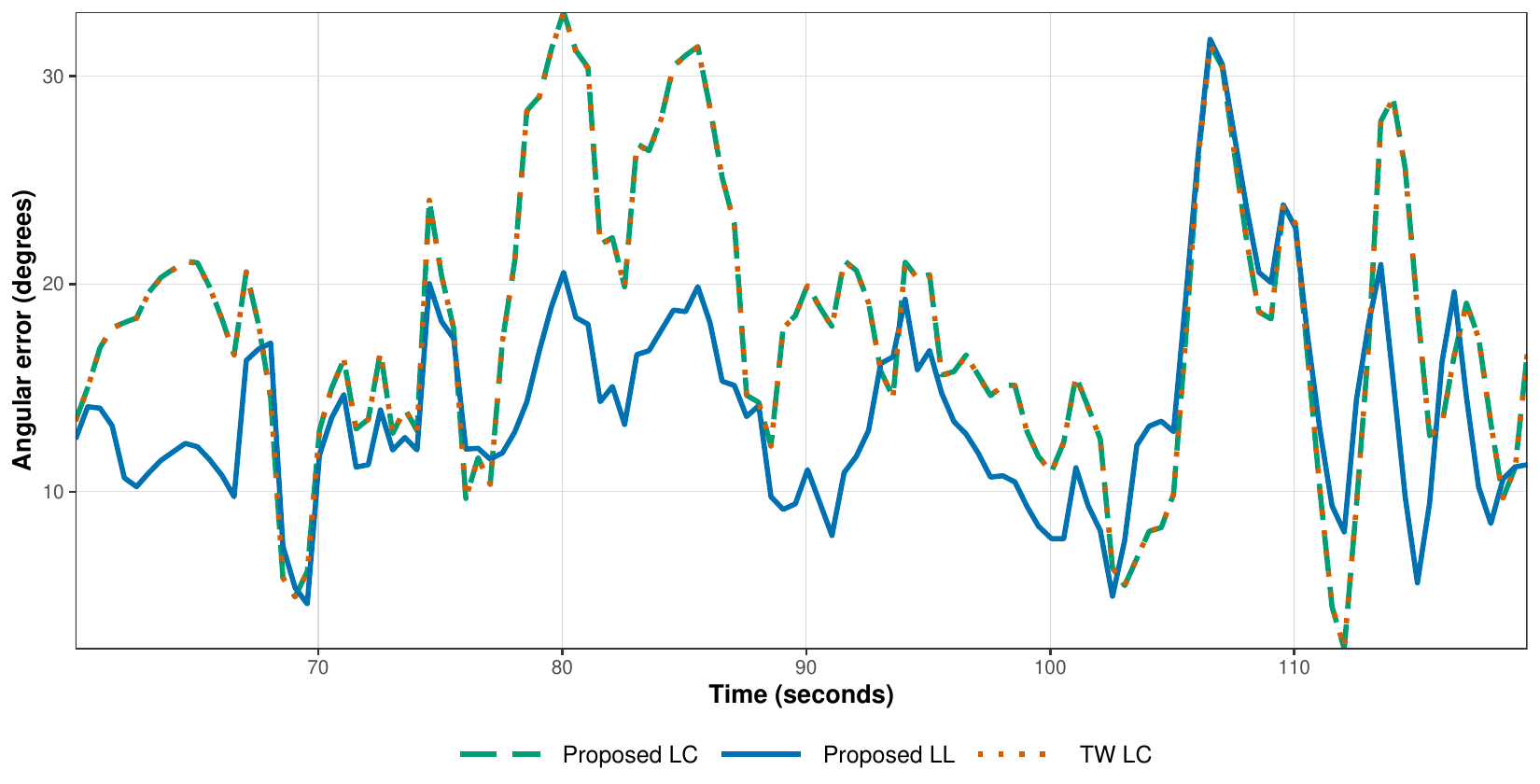}
    \caption{Angular prediction errors over time for a representative held-out subject in the gaze analysis. The subject is selected as the test subject whose Proposed LL root mean squared error is closest to the median across test subjects. The curves are smoothed by a one-second moving average for display.}
    \label{fig:real_data_gaze_error}
\end{figure}

\Cref{fig:real_data_gaze_error} shows the prediction errors for this representative test subject. The Proposed LL curve is lower than the local constant curves over much of the displayed interval, while Proposed LC and TW LC are nearly indistinguishable. Together, the two applications show that the construction can be implemented for distinct predictor--response geometries, including $\mathcal S_{++}^3\to\mathcal W_2(\mathbb R)$ and $\mathrm{SO}(3)\to\mathbb S^2$ regression, subject to the different validation designs and dependence structures described above.

\section{Discussion and Future Work}
\label{sec:discussion}

This paper develops local Fr\'echet regression for predictors on a finite-dimensional Riemannian manifold and metric-space-valued responses. The central methodological contribution is the intrinsic local linear construction on moving tangent spaces, with scalar equivalent weights that are invariant to local frame choice. Pointwise and uniform consistency and upper rates are established under the stated geometric, design, dominated conditional-law, empirical-process, and Fr\'echet-margin conditions. Under a quadratic margin, the rates are of the same order as standard twice-smooth Euclidean upper-bound rates in the intrinsic predictor dimension; no matching lower bound or optimality claim is made.

The separate conditions U-K1 and U-K2 identify the empirical-process input used by the two estimators. Local constant smoothing consumes only zeroth-order kernel-window complexity, while local linear smoothing also consumes first- and second-order tangent-coordinate multiplier complexity. On the tame analytic manifolds covered by \Cref{app:condition_verification}, both conditions are verified jointly. The value of the separation is therefore conceptual and technical: it records which parts of the moving-frame empirical process are needed by each estimator. The same verification also permits standard piecewise-polynomial compact kernels, including nonsmooth and indicator-type profiles.

The local linear estimator is an intrinsic first-order extension of local constant smoothing, not an estimator with a uniformly better interior rate. Its role is to correct local design moments in tangent coordinates. Proposed LL has the lowest aggregate error in the reported experiments, but the strength of the numerical separation varies: the difference in OASIS--3 is modest, Simulation~B includes support-boundary effects under an AIRM-aligned design, and the gaze result is based on one subject split. These findings illustrate the potential effect of first-order correction without establishing universal dominance or a leading-bias comparison.

The volume-density factor is primarily a geometric normalization. It removes the normal-coordinate Jacobian from the leading local moments and makes the bias calculations canonical. Since $\theta_x(z)=1+O\{d_{\mathcal M}^2(x,z)\}$ locally, its numerical effect can be small, as reflected by the near equality of Proposed LC and TW LC throughout the reported studies. It should therefore not be interpreted as a finite-sample improvement device.

Several assumptions on the response side are high-level sufficient conditions. In particular, signed local linear Fr\'echet objectives require model-specific existence, uniqueness, measurable selection, separation, and margin behavior. The domination formulation used for the conditional law is also sufficient rather than universal and excludes some singular conditional models. These restrictions delimit the rate theory but do not affect the definition of the estimators themselves.

The theory is finite dimensional and uses compact localization for uniform results. On noncompact manifolds such as the SPD cone, this requires predictor regions whose eigenvalues remain bounded away from zero and infinity. The intrinsic dimension of $\mathcal S_{++}^p$ is $p(p+1)/2$, so the usual curse of dimensionality remains. Intrinsic computation also requires repeated geodesic distances, logarithmic maps, local-moment inversion, and response-space Fr\'echet optimization. Euclideanized methods may be cheaper and competitive when a scientifically appropriate global representation is available. The metric is part of the predictor model because it determines both neighborhoods and tangent coordinates; scientific invariance considerations or nested validation over a prespecified metric collection can guide that choice.

Statistical inference for conditional Fr\'echet means with Riemannian predictors is deliberately outside the scope of this estimation-and-rates paper and is being developed separately, together with single-index and additive extensions. Further work also includes sharper bias expansions, theory for data-driven bandwidth selection, noncompact radial kernels, and dimension reduction for high-dimensional predictor manifolds.

\section*{Data Availability}

OASIS--3 data are available through the OASIS project at \url{https://www.oasis-brains.org/} under its Data Use Agreement. The Head and Gaze VR Behavior Dataset is available at \url{https://cuhksz-inml.github.io/head_gaze_dataset/} under the terms specified by the data providers. Both analyses use previously collected secondary data under the respective data-use conditions, and no attempt was made to identify participants.

\section*{Acknowledgements}

Chang Jun Im was supported by the National Research Foundation of Korea grant funded by the Korea government (MSIT) (No. RS-2025-00515381). Jeong Min Jeon was supported by the National Research Foundation of Korea grant funded by the Korea government (MSIT) (No. RS-2023-00211910). Data were provided in part by OASIS--3 (Principal Investigators: T. Benzinger, D. Marcus, and J. Morris; NIH grants P30 AG066444, P50 AG00561, P30 NS09857781, P01 AG026276, P01 AG003991, R01 AG043434, UL1 TR000448, and R01 EB009352).

\newpage

\appendix
\setcounter{equation}{0}
\renewcommand{\theequation}{App.\arabic{equation}}
\renewcommand{\theHequation}{App.\arabic{equation}}

{\centering \LARGE{\bf Appendix}}

Throughout the appendices, when the order $j\in\{0,1,2\}$ is clear from context, we use $\|\cdot\|_{\star}$ to denote absolute value for scalar quantities corresponding to $j=0$, the Euclidean norm for vector quantities corresponding to $j=1$, and the operator norm for matrix quantities corresponding to $j=2$. We use the tensor-power convention $\mathbf u^{\otimes0}=1$, $\mathbf u^{\otimes1}=\mathbf u$, and $\mathbf u^{\otimes2}=\mathbf u\mathbf u^{\top}$ whenever the order is clear. For $\mx\in\mathcal M$ and $\mathbf E_{\mx}\in\mathcal E_{\mx}$, we write
\begin{align}
    \Exp_{\mx}^{\mathbf E_{\mx}}(\mathbf u)
    &:=
    \Exp_{\mx}\left(\bm{\Phi}_{\mathbf E_{\mx}}^{-1}(\mathbf u)\right),
    \quad \mathbf u\in B_{\mathbb R^d}\left(\mathbf 0_d,i(\mx)\right).
    \label{eq:app.exp_coordinate_shorthand}
\end{align}
Throughout the appendices, the empirical-process suprema appearing below are assumed to be measurable. Otherwise, the corresponding expectations and probabilities may be interpreted in the outer sense.

\section{Auxiliary Geometric Results} \label{app:auxiliary_geometry}
\setcounter{equation}{0}
\renewcommand{\theequation}{A.\arabic{equation}}
\renewcommand{\theHequation}{A.\arabic{equation}}

This appendix collects the geometric facts used in the proofs. The definitions of the exponential map, logarithmic map, injectivity radius, normal coordinates, and volume density are given in \Cref{sec:riemannian_geometry}; see also \cite{Do Carmo (1992)}, \cite{Chavel (2006)}, and \cite{Pelletier (2006)} for standard references. Throughout this appendix, $(\mathcal{M},g)$ is a connected $d$-dimensional complete Riemannian manifold without boundary, and $\dd v_g$ denotes the Riemannian volume measure.

\subsection{Volume density and normal-coordinate neighborhoods}

\begin{lemma}[Basis-invariance of the volume density] \label{lemma:A.invariance_volume_density}
For any $\mx\in\mathcal{M}$ and any $\mz\in B_{\mathcal{M}}(\mx,i(\mx))$, the volume density
\begin{align*}
    \theta_{\mx}(\mz)=\sqrt{\det\left(\mathbf{G}_{\mathbf{E}_{\mx}}(\mz)\right)}, \quad \mz\in B_{\mathcal{M}}(\mx,i(\mx))
\end{align*}
is independent of the chosen ordered orthonormal basis $\mathbf{E}_{\mx}\in\mathcal{E}_{\mx}$.
\end{lemma}

\begin{proof}[Proof of \Cref{lemma:A.invariance_volume_density}]
Let $\mathbf{E}_{\mx}=(\mathbf{E}_{\mx,1},\ldots,\mathbf{E}_{\mx,d})$ and $\widetilde{\mathbf{E}}_{\mx}=(\widetilde{\mathbf{E}}_{\mx,1},\ldots,\widetilde{\mathbf{E}}_{\mx,d})$ be two ordered orthonormal bases of $T_{\mx}\mathcal{M}$. Then there exists an orthogonal matrix $\mathbf{Q}\in O(d)$ such that
\begin{align*}
    \widetilde{\mathbf{E}}_{\mx,j}=\sum_{k=1}^{d}\mathbf{Q}_{kj}\mathbf{E}_{\mx,k}, \quad j=1,\ldots,d.
\end{align*}
Fix $\mz\in B_{\mathcal{M}}(\mx,i(\mx))$ and set $\mathbf{v}:=\Log_{\mx}(\mz)\in T_{\mx}\mathcal{M}$. The differential $(d\Exp_{\mx})_{\mathbf{v}}$ is a linear map from $T_{\mathbf{v}}(T_{\mx}\mathcal{M})$ to $T_{\mz}\mathcal{M}$. Since $T_{\mx}\mathcal{M}$ is a vector space, we identify $T_{\mathbf{v}}(T_{\mx}\mathcal{M})$ naturally with $T_{\mx}\mathcal{M}$. Hence, by linearity of $(d\Exp_{\mx})_{\mathbf{v}}$,
\begin{align*}
    (d\Exp_{\mx})_{\mathbf{v}}(\widetilde{\mathbf{E}}_{\mx,j})
    =
    \sum_{k=1}^{d}\mathbf{Q}_{kj}(d\Exp_{\mx})_{\mathbf{v}}(\mathbf{E}_{\mx,k}), \quad j=1,\ldots,d.
\end{align*}
Therefore, for $1\leq j,\ell\leq d$,
\begin{align*}
    \left[\mathbf{G}_{\widetilde{\mathbf{E}}_{\mx}}(\mz)\right]_{j\ell}
    &=
    \left\langle
    (d\Exp_{\mx})_{\mathbf{v}}(\widetilde{\mathbf{E}}_{\mx,j}),
    (d\Exp_{\mx})_{\mathbf{v}}(\widetilde{\mathbf{E}}_{\mx,\ell})
    \right\rangle_{\mz} \\
    &=
    \left\langle
    \sum_{k=1}^{d}\mathbf{Q}_{kj}(d\Exp_{\mx})_{\mathbf{v}}(\mathbf{E}_{\mx,k}),
    \sum_{m=1}^{d}\mathbf{Q}_{m\ell}(d\Exp_{\mx})_{\mathbf{v}}(\mathbf{E}_{\mx,m})
    \right\rangle_{\mz} \\
    &=
    \sum_{k=1}^{d}\sum_{m=1}^{d}
    \mathbf{Q}_{kj}\mathbf{Q}_{m\ell}
    \left\langle
    (d\Exp_{\mx})_{\mathbf{v}}(\mathbf{E}_{\mx,k}),
    (d\Exp_{\mx})_{\mathbf{v}}(\mathbf{E}_{\mx,m})
    \right\rangle_{\mz} \\
    &=
    \sum_{k=1}^{d}\sum_{m=1}^{d}
    \mathbf{Q}_{kj}
    \left[\mathbf{G}_{\mathbf{E}_{\mx}}(\mz)\right]_{km}
    \mathbf{Q}_{m\ell}.
\end{align*}
Equivalently,
\begin{align*}
    \mathbf{G}_{\widetilde{\mathbf{E}}_{\mx}}(\mz)=\mathbf{Q}^{\top}\mathbf{G}_{\mathbf{E}_{\mx}}(\mz)\mathbf{Q}.
\end{align*}
Taking determinants gives
\begin{align*}
    \det\{\mathbf{G}_{\widetilde{\mathbf{E}}_{\mx}}(\mz)\}
    =
    \det(\mathbf{Q})^2\det\{\mathbf{G}_{\mathbf{E}_{\mx}}(\mz)\}
    =
    \det\{\mathbf{G}_{\mathbf{E}_{\mx}}(\mz)\},
\end{align*}
because $\mathbf{Q}$ is orthogonal and hence $\det(\mathbf{Q})^2=1$. Thus $\theta_{\mx}(\mz)$ is independent of the chosen ordered orthonormal basis.
\end{proof}

\begin{lemma}[Basis-invariance of local linear scalar weights] \label{lemma:A.invariance_local_linear_weights}
Fix $\mx\in\mathcal{M}$ and let $h<i(\mx)$. Let $\mathbf{E}_{\mx},\widetilde{\mathbf{E}}_{\mx}\in\mathcal{E}_{\mx}$ be two ordered orthonormal bases of $T_{\mx}\mathcal{M}$. Suppose that $\bm{\hat{\mu}}_{h,2}(\mx,\mathbf{E}_{\mx})$ is nonsingular. Then $\bm{\hat{\mu}}_{h,2}(\mx,\widetilde{\mathbf{E}}_{\mx})$ is nonsingular, and the scalar quantities
\begin{align*}
    \bm{\hat{\mu}}_{h,1}(\mx,\mathbf{E}_{\mx})^{\top}
    \bm{\hat{\mu}}_{h,2}(\mx,\mathbf{E}_{\mx})^{-1}
    \bm{\hat{\mu}}_{h,1}(\mx,\mathbf{E}_{\mx})
\end{align*}
and
\begin{align*}
    \bm{\hat{\mu}}_{h,1}(\mx,\mathbf{E}_{\mx})^{\top}
    \bm{\hat{\mu}}_{h,2}(\mx,\mathbf{E}_{\mx})^{-1}
    \mathbf{v}_{\mx}^{\mathbf{E}_{\mx}}(\mz),
    \quad \mz\in B_{\mathcal{M}}(\mx,i(\mx)),
\end{align*}
are invariant under replacing $\mathbf{E}_{\mx}$ by $\widetilde{\mathbf{E}}_{\mx}$. Consequently, whenever the corresponding denominators are nonzero, $\hat{\sigma}_{h}(\mx)$ and $\hat{W}_{\mx,h,1}(\mz)$ are basis-independent. The same assertions hold with the empirical moments replaced by their population counterparts.
\end{lemma}

\begin{proof}[Proof of \Cref{lemma:A.invariance_local_linear_weights}]
Let $\mathbf{E}_{\mx}=(\mathbf{E}_{\mx,1},\ldots,\mathbf{E}_{\mx,d})$ and $\widetilde{\mathbf{E}}_{\mx}=(\widetilde{\mathbf{E}}_{\mx,1},\ldots,\widetilde{\mathbf{E}}_{\mx,d})$. Then there exists an orthogonal matrix $\mathbf{Q}\in O(d)$ such that
\begin{align*}
    \widetilde{\mathbf{E}}_{\mx,j}
    =
    \sum_{k=1}^{d}\mathbf{Q}_{kj}\mathbf{E}_{\mx,k},
    \quad j=1,\ldots,d.
\end{align*}
We first record the coordinate transformation rule. Let $\mz\in B_{\mathcal{M}}(\mx,i(\mx))$ and write
\begin{align*}
    \Log_{\mx}(\mz)
    =
    \sum_{k=1}^{d}v_k\mathbf{E}_{\mx,k}
    =
    \sum_{j=1}^{d}\widetilde{v}_j\widetilde{\mathbf{E}}_{\mx,j}.
\end{align*}
Using the relation between the two bases,
\begin{align*}
    \sum_{j=1}^{d}\widetilde{v}_j\widetilde{\mathbf{E}}_{\mx,j}
    =
    \sum_{j=1}^{d}\widetilde{v}_j
    \sum_{k=1}^{d}\mathbf{Q}_{kj}\mathbf{E}_{\mx,k}
    =
    \sum_{k=1}^{d}
    \left(\sum_{j=1}^{d}\mathbf{Q}_{kj}\widetilde{v}_j\right)
    \mathbf{E}_{\mx,k}.
\end{align*}
Hence, in vector notation, $\mathbf{v}=\mathbf{Q}\widetilde{\mathbf{v}}$, where $\mathbf{v}=\mathbf{v}_{\mx}^{\mathbf{E}_{\mx}}(\mz)$ and $\widetilde{\mathbf{v}}=\mathbf{v}_{\mx}^{\widetilde{\mathbf{E}}_{\mx}}(\mz)$. Since $\mathbf{Q}$ is orthogonal, this gives
\begin{align*}
    \mathbf{v}_{\mx}^{\widetilde{\mathbf{E}}_{\mx}}(\mz)
    =
    \mathbf{Q}^{\top}\mathbf{v}_{\mx}^{\mathbf{E}_{\mx}}(\mz),
    \quad \mz\in B_{\mathcal{M}}(\mx,i(\mx)).
\end{align*}
If $\mz\notin B_{\mathcal{M}}(\mx,i(\mx))$, both coordinate maps are defined to be $\mathbf{0}$, so the same identity holds for every $\mz\in\mathcal{M}$:
\begin{align}
    \mathbf{v}_{\mx}^{\widetilde{\mathbf{E}}_{\mx}}(\mz)
    =
    \mathbf{Q}^{\top}\mathbf{v}_{\mx}^{\mathbf{E}_{\mx}}(\mz),
    \quad \mz\in\mathcal{M}.
    \label{eq:A.coordinate_transformation}
\end{align}

Because $h<i(\mx)$ and $K$ is supported on $[0,1]$, every nonzero kernel contribution satisfies $d_{\mathcal{M}}(\mx,\mz)\leq h<i(\mx)$. On this normal neighborhood, $\mathcal{L}_{\mx,h}(\mz)$ depends on the chosen basis only through the volume density $\theta_{\mx}(\mz)$, which is basis-invariant by \Cref{lemma:A.invariance_volume_density}. Therefore $\mathcal{L}_{\mx,h}(\mz)$ is basis-independent for all $\mz$ contributing to the local moments. Combining this fact with \eqref{eq:A.coordinate_transformation}, the empirical local moments satisfy
\begin{align*}
    \bm{\hat{\mu}}_{h,1}(\mx,\widetilde{\mathbf{E}}_{\mx})
    =
    \mathbf{Q}^{\top}
    \bm{\hat{\mu}}_{h,1}(\mx,\mathbf{E}_{\mx})
\end{align*}
and
\begin{align*}
    \bm{\hat{\mu}}_{h,2}(\mx,\widetilde{\mathbf{E}}_{\mx})
    =
    \mathbf{Q}^{\top}
    \bm{\hat{\mu}}_{h,2}(\mx,\mathbf{E}_{\mx})
    \mathbf{Q}.
\end{align*}
Hence $\bm{\hat{\mu}}_{h,2}(\mx,\widetilde{\mathbf{E}}_{\mx})$ is nonsingular whenever $\bm{\hat{\mu}}_{h,2}(\mx,\mathbf{E}_{\mx})$ is nonsingular, and
\begin{align*}
    \left(
    \mathbf{Q}^{\top}
    \bm{\hat{\mu}}_{h,2}(\mx,\mathbf{E}_{\mx})
    \mathbf{Q}
    \right)^{-1}
    =
    \mathbf{Q}^{\top}
    \bm{\hat{\mu}}_{h,2}(\mx,\mathbf{E}_{\mx})^{-1}
    \mathbf{Q}.
\end{align*}
Therefore,
\begin{align*}
    &
    \bm{\hat{\mu}}_{h,1}(\mx,\widetilde{\mathbf{E}}_{\mx})^{\top}
    \bm{\hat{\mu}}_{h,2}(\mx,\widetilde{\mathbf{E}}_{\mx})^{-1}
    \bm{\hat{\mu}}_{h,1}(\mx,\widetilde{\mathbf{E}}_{\mx})
    =
    \bm{\hat{\mu}}_{h,1}(\mx,\mathbf{E}_{\mx})^{\top}
    \bm{\hat{\mu}}_{h,2}(\mx,\mathbf{E}_{\mx})^{-1}
    \bm{\hat{\mu}}_{h,1}(\mx,\mathbf{E}_{\mx}),
\end{align*}
and, for every $\mz\in\mathcal{M}$,
\begin{align*}
    &
    \bm{\hat{\mu}}_{h,1}(\mx,\widetilde{\mathbf{E}}_{\mx})^{\top}
    \bm{\hat{\mu}}_{h,2}(\mx,\widetilde{\mathbf{E}}_{\mx})^{-1}
    \mathbf{v}_{\mx}^{\widetilde{\mathbf{E}}_{\mx}}(\mz)
    =
    \bm{\hat{\mu}}_{h,1}(\mx,\mathbf{E}_{\mx})^{\top}
    \bm{\hat{\mu}}_{h,2}(\mx,\mathbf{E}_{\mx})^{-1}
    \mathbf{v}_{\mx}^{\mathbf{E}_{\mx}}(\mz).
\end{align*}
Since $\hat{\mu}_{h,0}(\mx)$ is basis-independent, it follows that $\hat{\sigma}_{h}(\mx)$ and $\hat{W}_{\mx,h,1}(\mz)$ are basis-independent.

The population statements follow by the same argument. Taking expectations in the transformation identities gives
\begin{align*}
    \bm{\tilde{\mu}}_{h,1}(\mx,\widetilde{\mathbf{E}}_{\mx})
    =
    \mathbf{Q}^{\top}
    \bm{\tilde{\mu}}_{h,1}(\mx,\mathbf{E}_{\mx}),
    \quad
    \bm{\tilde{\mu}}_{h,2}(\mx,\widetilde{\mathbf{E}}_{\mx})
    =
    \mathbf{Q}^{\top}
    \bm{\tilde{\mu}}_{h,2}(\mx,\mathbf{E}_{\mx})
    \mathbf{Q}.
\end{align*}
Thus $\bm{\tilde{\mu}}_{h,2}(\mx,\widetilde{\mathbf{E}}_{\mx})$ is nonsingular whenever $\bm{\tilde{\mu}}_{h,2}(\mx,\mathbf{E}_{\mx})$ is nonsingular, and the same matrix calculation proves the basis-invariance of the scalar quantities entering $\tilde{\sigma}_{h}(\mx)$ and $\tilde{W}_{\mx,h,1}(\mz)$.
\end{proof}

\begin{lemma}[Pointwise normal neighborhood and volume-density bounds] \label{lemma:A.pointwise_normal_neighborhoods}
Fix $\mx\in\mathcal{M}$ and let $\rho_{\mx}\in(0,i(\mx))$. Then $\Log_{\mx}(\mz)$ is well-defined for all $\mz\in B_{\mathcal{M}}(\mx,\rho_{\mx})$, and
\begin{align*}
    \|\Log_{\mx}(\mz)\|_{\mx}=d_{\mathcal{M}}(\mx,\mz), \quad \mz\in B_{\mathcal{M}}(\mx,\rho_{\mx}).
\end{align*}
Moreover, there exist constants $0<c_{\theta,\mx,\rho_{\mx}}<C_{\theta,\mx,\rho_{\mx}}<\infty$ such that
\begin{align*}
    c_{\theta,\mx,\rho_{\mx}}\leq \theta_{\mx}(\mz)\leq C_{\theta,\mx,\rho_{\mx}}, \quad \mz\in B_{\mathcal{M}}(\mx,\rho_{\mx}).
\end{align*}
\end{lemma}

\begin{proof}[Proof of \Cref{lemma:A.pointwise_normal_neighborhoods}]
Since $\rho_{\mx}<i(\mx)$, the restriction of $\Exp_{\mx}$ to $B_{\|\cdot\|_{\mx}}(\mathbf{0}_{\mx},\rho_{\mx})\subset T_{\mx}\mathcal{M}$ is a diffeomorphism onto $B_{\mathcal{M}}(\mx,\rho_{\mx})$. Hence $\Log_{\mx}(\mz)$ is well-defined for every $\mz\in B_{\mathcal{M}}(\mx,\rho_{\mx})$, and the normal-neighborhood distance identity gives
\begin{align*}
    \|\Log_{\mx}(\mz)\|_{\mx}=d_{\mathcal{M}}(\mx,\mz), \quad \mz\in B_{\mathcal{M}}(\mx,\rho_{\mx}).
\end{align*}
By the Hopf--Rinow theorem, $\overline{B}_{\mathcal{M}}(\mx,\rho_{\mx})$ is compact. Since $\rho_{\mx}<i(\mx)$, this compact set is contained in $B_{\mathcal{M}}(\mx,i(\mx))$. The volume-density function $\theta_{\mx}$ is smooth and strictly positive on $B_{\mathcal{M}}(\mx,i(\mx))$, and therefore attains positive finite lower and upper bounds on $\overline{B}_{\mathcal{M}}(\mx,\rho_{\mx})$. The asserted bounds on $B_{\mathcal{M}}(\mx,\rho_{\mx})$ follow.
\end{proof}

\begin{lemma}[Uniform normal neighborhoods and volume-density bounds] \label{lemma:A.uniform_normal_neighborhoods}
Let $\mathcal{K}\subset\mathcal{M}$ be compact and let $\rho\in(0,i(\mathcal{K}))$, where $i(\mathcal{K}):=\inf_{\mx\in\mathcal{K}}i(\mx)$. Define
\begin{align*}
    \mathcal{D}_{\mathcal{K},\rho}:=\{(\mx,\mz)\in\mathcal{K}\times\mathcal{M}:d_{\mathcal{M}}(\mx,\mz)\leq \rho\}.
\end{align*}
Then $\mathcal{D}_{\mathcal{K},\rho}$ is compact, $\Log_{\mx}(\mz)$ is well-defined for every $(\mx,\mz)\in\mathcal{D}_{\mathcal{K},\rho}$, and
\begin{align*}
    \|\Log_{\mx}(\mz)\|_{\mx}=d_{\mathcal{M}}(\mx,\mz)\leq \rho, \quad (\mx,\mz)\in\mathcal{D}_{\mathcal{K},\rho}.
\end{align*}
Moreover, there exist constants $0<c_{\theta,\mathcal{K},\rho}<C_{\theta,\mathcal{K},\rho}<\infty$ such that
\begin{align*}
    c_{\theta,\mathcal{K},\rho}\leq \theta_{\mx}(\mz)\leq C_{\theta,\mathcal{K},\rho}, \quad (\mx,\mz)\in\mathcal{D}_{\mathcal{K},\rho}.
\end{align*}
\end{lemma}

\begin{proof}[Proof of \Cref{lemma:A.uniform_normal_neighborhoods}]
Since $\rho<i(\mathcal{K})\leq i(\mx)$ for every $\mx\in\mathcal{K}$, $\Log_{\mx}(\mz)$ is well-defined whenever $(\mx,\mz)\in\mathcal{D}_{\mathcal{K},\rho}$. The distance identity follows from the normal-neighborhood property. By the Hopf--Rinow theorem, the closed tube
\begin{align*}
    \mathcal{K}^{\rho}:=\{\mz\in\mathcal{M}:d_{\mathcal{M}}(\mz,\mathcal{K})\leq\rho\}
\end{align*}
is compact. Hence $\mathcal{D}_{\mathcal{K},\rho}$ is a closed subset of $\mathcal{K}\times\mathcal{K}^{\rho}$ and is therefore compact.

It remains to prove the volume-density bounds. The set $\mathcal{D}_{\mathcal{K},\rho}$ is contained in the domain where each pair $(\mx,\mz)$ lies inside the injectivity region of $\Exp_{\mx}$. On this domain, the map $(\mx,\mz)\mapsto\theta_{\mx}(\mz)$ is continuous by the smooth dependence of the exponential map and its differential on the base point and tangent vector, together with the basis-invariance of the determinant defining $\theta_{\mx}(\mz)$ established in \Cref{lemma:A.invariance_volume_density}. Since $\theta_{\mx}(\mz)>0$ on $\mathcal{D}_{\mathcal{K},\rho}$, compactness implies that it attains positive finite lower and upper bounds on $\mathcal{D}_{\mathcal{K},\rho}$. Therefore, there exist constants $0<c_{\theta,\mathcal{K},\rho}<C_{\theta,\mathcal{K},\rho}<\infty$ such that
\begin{align*}
    c_{\theta,\mathcal{K},\rho}\leq \theta_{\mx}(\mz)\leq C_{\theta,\mathcal{K},\rho}, \quad (\mx,\mz)\in\mathcal{D}_{\mathcal{K},\rho}.
\end{align*}
\end{proof}

\subsection{Kernel moments and normal-coordinate cancellation}

The following elementary radial-kernel identities are used repeatedly. They do not require a normalization condition on $K$. The cancellation formula below is the analytic consequence of the volume-density correction in \eqref{eq:sec3.volume_corrected_kernel} and the normal-coordinate volume formula reviewed in \Cref{sec:riemannian_geometry}.

\begin{lemma}[Moments of radial kernels] \label{lemma:A.radial_kernel_moments}
Assume Condition~\ref{con:P-K1}. For any integer $j\geq0$ and $q\in\{1,2\}$, define
\begin{align*}
    c_{j,q}:=\int_0^1 K(r)^q r^j\,\dd r.
\end{align*}
Let $\mathbb{S}^{d-1}:=\{\mathbf{w}\in\mathbb{R}^d:\|\mathbf{w}\|_2=1\}$ and let
\begin{align*}
    A_{d-1}:=\int_{\mathbb{S}^{d-1}}\dd S(\bm{\xi})
\end{align*}
denote its surface area, where $\dd S$ is the standard surface measure on $\mathbb{S}^{d-1}$. Then for every $q\in\{1,2\}$ and every integer $j\geq0$,
\begin{align*}
    0<c_{j,q}<\infty,
\end{align*}
and
\begin{align*}
    \int_{\mathbb{R}^d}K(\|\mathbf{w}\|_2)^q\,\dd\mathbf{w}
    =
    A_{d-1}c_{d-1,q},
    \quad
    \int_{\mathbb{R}^d}\mathbf{w}K(\|\mathbf{w}\|_2)^q\,\dd\mathbf{w}
    =
    \mathbf{0}_d,
\end{align*}
and
\begin{align*}
    \int_{\mathbb{R}^d}\mathbf{w}\mathbf{w}^{\top}K(\|\mathbf{w}\|_2)^q\,\dd\mathbf{w}
    =
    \frac{A_{d-1}}{d}c_{d+1,q}\mI_d,
    \quad
    \int_{\mathbb{R}^d}\|\mathbf{w}\|_2^2K(\|\mathbf{w}\|_2)^q\,\dd\mathbf{w}
    =
    A_{d-1}c_{d+1,q}.
\end{align*}
\end{lemma}

\begin{proof}[Proof of \Cref{lemma:A.radial_kernel_moments}]
Since $K$ is bounded and compactly supported on $[0,1]$, $c_{j,q}<\infty$. Since $K$ is nonnegative and not identically zero almost everywhere on $[0,1]$, the set $\{r\in[0,1]:K(r)>0\}$ has positive Lebesgue measure. Because $r^j>0$ for Lebesgue-almost every $r\in(0,1]$, it follows that $c_{j,q}>0$ for every integer $j\geq0$ and $q\in\{1,2\}$. Using spherical coordinates $\mathbf{w}=r\bm{\xi}$, where $r=\|\mathbf{w}\|_2\in[0,\infty)$ and $\bm{\xi}\in\mathbb{S}^{d-1}$, the Lebesgue measure decomposes as
\begin{align*}
    \dd\mathbf{w}=r^{d-1}\,\dd r\,\dd S(\bm{\xi}).
\end{align*}
Since $K(r)=0$ for $r>1$,
\begin{align*}
    \int_{\mathbb{R}^d}K(\|\mathbf{w}\|_2)^q\,\dd\mathbf{w}
    &=
    \left(\int_0^1K(r)^q r^{d-1}\,\dd r\right)
    \left(\int_{\mathbb{S}^{d-1}}\dd S(\bm{\xi})\right) \\
    &=
    A_{d-1}c_{d-1,q}.
\end{align*}
For the first and second spherical moments, rotational symmetry gives
\begin{align} \label{eq:A.spherical_moments}
    \int_{\mathbb{S}^{d-1}}\xi_i\,\dd S(\bm{\xi})=0,
    \quad
    \int_{\mathbb{S}^{d-1}}\xi_i\xi_{i'}\,\dd S(\bm{\xi})
    =
    \begin{cases}
        0, & i\neq i', \\
        A_{d-1}/d, & i=i',
    \end{cases}
    \quad i,i'=1,\ldots,d.
\end{align}
Therefore,
\begin{align*}
    \int_{\mathbb{R}^d}\mathbf{w}K(\|\mathbf{w}\|_2)^q\,\dd\mathbf{w}
    &=
    \left(\int_0^1K(r)^q r^d\,\dd r\right)
    \left(\int_{\mathbb{S}^{d-1}}\bm{\xi}\,\dd S(\bm{\xi})\right)
    =
    \mathbf{0}_d.
\end{align*}
Similarly, \eqref{eq:A.spherical_moments} gives
\begin{align*}
    \int_{\mathbb{S}^{d-1}}\bm{\xi}\bm{\xi}^{\top}\,\dd S(\bm{\xi})
    =
    \frac{A_{d-1}}{d}\mI_d.
\end{align*}
Thus,
\begin{align*}
    \int_{\mathbb{R}^d}\mathbf{w}\mathbf{w}^{\top}K(\|\mathbf{w}\|_2)^q\,\dd\mathbf{w}
    &=
    \left(\int_0^1K(r)^q r^{d+1}\,\dd r\right)
    \left(\int_{\mathbb{S}^{d-1}}\bm{\xi}\bm{\xi}^{\top}\,\dd S(\bm{\xi})\right) \\
    &=
    \frac{A_{d-1}}{d}c_{d+1,q}\mI_d.
\end{align*}
Taking traces gives
\begin{align*}
    \int_{\mathbb{R}^d}\|\mathbf{w}\|_2^2K(\|\mathbf{w}\|_2)^q\,\dd\mathbf{w}
    =
    A_{d-1}c_{d+1,q}.
\end{align*}
This completes the proof.
\end{proof}

Throughout the appendix, integrals over $T_{\mx}\mathcal M$ are taken with respect to the Lebesgue measure induced by an ordered orthonormal basis of $T_{\mx}\mathcal M$. This measure is independent of the chosen ordered orthonormal basis: if two such bases are used, the coordinate change is an orthogonal transformation and hence has Jacobian determinant of absolute value one. In particular, scalar radial integrals such as
\begin{align*}
    \int_{\|\mathbf z\|_{\mx}\leq1}
    K(\|\mathbf z\|_{\mx})\,\dd\mathbf z
\end{align*}
are basis-invariant.

\begin{lemma}[Normal-coordinate cancellation] \label{lemma:A.pointwise_coord_cancellation}
Fix $\mx\in\mathcal{M}$ and let $\rho_{\mx}$ be the pointwise normal-neighborhood radius fixed in the main text. Let $\psi:\mathcal{M}\to\mathbb{R}$ be measurable and bounded on $B_{\mathcal{M}}(\mx,\rho_{\mx})$. Then for every $h\in(0,\rho_{\mx})$,
\begin{align*}
    \int_{\mathcal{M}}\mathcal{L}_{\mx,h}(\mz)\psi(\mz)\,\dd v_g(\mz)
    =
    h^d\int_{\|\mathbf{z}\|_{\mx}\leq1}K(\|\mathbf{z}\|_{\mx})\psi(\Exp_{\mx}(h\mathbf{z}))\,\dd\mathbf{z},
\end{align*}
where $\dd\mathbf{z}$ denotes the Lebesgue measure on $T_{\mx}\mathcal{M}$ induced by any ordered orthonormal basis of $T_{\mx}\mathcal{M}$. The same identity applies componentwise to functions taking values in a fixed finite-dimensional vector space.
\end{lemma}

\begin{proof}[Proof of \Cref{lemma:A.pointwise_coord_cancellation}]
Since $K(r)=0$ for $r>1$ and $h<\rho_{\mx}$, the integrand on the left-hand side is zero unless $d_{\mathcal{M}}(\mx,\mz)\leq h<\rho_{\mx}$. Hence all nonzero contributions come from $B_{\mathcal{M}}(\mx,\rho_{\mx})$, where the normal-coordinate change of variables $\mz=\Exp_{\mx}(\mathbf{v})$ is valid by \Cref{lemma:A.pointwise_normal_neighborhoods}. Under this change of variables, $\mathbf{v}\in T_{\mx}\mathcal M$ with $\|\mathbf{v}\|_{\mx}\leq h$, $d_{\mathcal{M}}(\mx,\Exp_{\mx}(\mathbf{v}))=\|\mathbf{v}\|_{\mx}$, and
\begin{align*}
    \dd v_g(\Exp_{\mx}(\mathbf{v}))
    =
    \theta_{\mx}(\Exp_{\mx}(\mathbf{v}))\,\dd\mathbf{v}.
\end{align*}
Therefore, by the definition of $\mathcal{L}_{\mx,h}$ in \eqref{eq:sec3.volume_corrected_kernel},
\begin{align*}
    \int_{\mathcal{M}}\mathcal{L}_{\mx,h}(\mz)\psi(\mz)\,\dd v_g(\mz)
    &=
    \int_{\|\mathbf{v}\|_{\mx}\leq h}
    \theta_{\mx}(\Exp_{\mx}(\mathbf{v}))^{-1}
    K\left(\frac{\|\mathbf{v}\|_{\mx}}{h}\right)
    \psi(\Exp_{\mx}(\mathbf{v}))
    \theta_{\mx}(\Exp_{\mx}(\mathbf{v}))\,\dd\mathbf{v} \\
    &=
    \int_{\|\mathbf{v}\|_{\mx}\leq h}
    K\left(\frac{\|\mathbf{v}\|_{\mx}}{h}\right)
    \psi(\Exp_{\mx}(\mathbf{v}))\,\dd\mathbf{v}.
\end{align*}
Now set $\mathbf{v}=h\mathbf{z}$. Since $T_{\mx}\mathcal{M}$ is a $d$-dimensional vector space and $\dd\mathbf{v}=h^d\,\dd\mathbf{z}$ under the induced Lebesgue measure, the preceding display becomes
\begin{align*}
    \int_{\mathcal{M}}\mathcal{L}_{\mx,h}(\mz)\psi(\mz)\,\dd v_g(\mz)
    =
    h^d\int_{\|\mathbf{z}\|_{\mx}\leq1}K(\|\mathbf{z}\|_{\mx})\psi(\Exp_{\mx}(h\mathbf{z}))\,\dd\mathbf{z}.
\end{align*}
This proves the scalar-valued identity. The vector-valued case follows componentwise.
\end{proof}

\begin{lemma}[Uniform normal-coordinate cancellation] \label{lemma:A.uniform_coord_cancellation}
Let $\mathcal{K}\subset\mathcal{M}$ be compact and let $\rho\in(0,i(\mathcal{K}))$. Let $\psi:\mathcal{M}\to\mathbb{R}$ be measurable and bounded on $\mathcal{K}^{\rho}$. Then for every $\mx\in\mathcal{K}$ and $h\in(0,\rho)$,
\begin{align*}
    \int_{\mathcal{M}}\mathcal{L}_{\mx,h}(\mz)\psi(\mz)\,\dd v_g(\mz)
    =
    h^d\int_{\|\mathbf{z}\|_{\mx}\leq1}
    K(\|\mathbf{z}\|_{\mx})\psi(\Exp_{\mx}(h\mathbf{z}))\,\dd\mathbf{z},
\end{align*}
where $\dd\mathbf{z}$ denotes the Lebesgue measure on $T_{\mx}\mathcal{M}$ induced by any ordered orthonormal basis of $T_{\mx}\mathcal{M}$. The same identity applies componentwise to functions taking values in a fixed finite-dimensional vector space.
\end{lemma}

\begin{proof}[Proof of \Cref{lemma:A.uniform_coord_cancellation}]
The proof is identical to that of \Cref{lemma:A.pointwise_coord_cancellation}, using $h<\rho<i(\mathcal{K})\leq i(\mx)$ for all $\mx\in\mathcal{K}$ and noting that $\Exp_{\mx}(h\mathbf{z})\in\mathcal{K}^{\rho}$ whenever $\mx\in\mathcal{K}$ and $\|\mathbf{z}\|_{\mx}\leq1$.
\end{proof}

\begin{remark} \label{rem:normal_coordinate_power}
For later use, the same change-of-variables argument gives, for $q\in\{1,2\}$,
\begin{align*}
    \int_{\mathcal{M}}\mathcal{L}_{\mx,h}(\mz)^q\psi(\mz)\,\dd v_g(\mz)
    =
    h^d\int_{\|\mathbf{z}\|_{\mx}\leq1}
    K(\|\mathbf{z}\|_{\mx})^q
    \theta_{\mx}(\Exp_{\mx}(h\mathbf{z}))^{1-q}
    \psi(\Exp_{\mx}(h\mathbf{z}))\,\dd\mathbf{z}.
\end{align*}
Thus the volume-density factor cancels exactly only when $q=1$.
\end{remark}

\subsection{Riemannian Taylor expansions}

In the Taylor expansions below, continuity of Hessian tensor fields is understood after identifying tangent spaces by parallel transport along short geodesic segments. More precisely, fix $\mx\in\mathcal{M}$ and choose $\delta_0\in(0,i(\mx))$. For $\mz\in B_{\mathcal{M}}(\mx,\delta_0)$, let $\gamma_{\mx,\mz}:[0,1]\to\mathcal{M}$ be the unique minimizing geodesic from $\mx$ to $\mz$, given by $\gamma_{\mx,\mz}(t)=\Exp_{\mx}\{t\Log_{\mx}(\mz)\}$, and let $\mathsf{P}_{\mx\to\mz}^{\gamma_{\mx,\mz}}:T_{\mx}\mathcal{M}\to T_{\mz}\mathcal{M}$ denote parallel transport along $\gamma_{\mx,\mz}$. We say that $\nabla^2\psi$ is continuous at $\mx$ as a tensor field if
\begin{align} \label{eq:A.hessian_continuity}
    \lim_{\delta\downarrow0}
    \sup_{\mz\in B_{\mathcal{M}}(\mx,\delta)}
    \sup_{\|\mathbf{u}\|_{\mx}\leq1}
    \left|
    \nabla^2\psi(\mz)
    \left(
    \mathsf{P}_{\mx\to\mz}^{\left(\gamma_{\mx,\mz}\right)}\mathbf{u},
    \mathsf{P}_{\mx\to\mz}^{\left(\gamma_{\mx,\mz}\right)}\mathbf{u}
    \right)
    -
    \nabla^2\psi(\mx)\left(\mathbf{u},\mathbf{u}\right)
    \right|
    =
    0.
\end{align}
Here and throughout, $\mz\to\mx$ means $d_{\mathcal{M}}(\mz,\mx)\to0$. Thus, \eqref{eq:A.hessian_continuity} is equivalently written as
\begin{align*}
    \lim_{\mz\to\mx}
    \sup_{\|\mathbf{u}\|_{\mx}\leq1}
    \left|
    \nabla^2\psi(\mz)
    \left(
    \mathsf{P}_{\mx\to\mz}^{\gamma_{\mx,\mz}}\mathbf{u},
    \mathsf{P}_{\mx\to\mz}^{\gamma_{\mx,\mz}}\mathbf{u}
    \right)
    -
    \nabla^2\psi(\mx)\left(\mathbf{u},\mathbf{u}\right)
    \right|
    =
    0.
\end{align*}
The corresponding uniform continuity condition is understood in the same parallel-transport sense, uniformly over the geodesic segments used below. These are standard intrinsic formulations of continuity for tensor fields; see, for example, \cite{Do Carmo (1992)} and \cite{Chavel (2006)}.

\begin{lemma}[Pointwise second-order Taylor expansion in normal coordinates] \label{lemma:A.pointwise_taylor_expansion}
Fix $\mx\in\mathcal{M}$ and let $\rho_{\mx}\in(0,i(\mx))$. Suppose that $\psi$ is $C^2$ on $B_{\mathcal{M}}(\mx,\rho_{\mx})$, so that its Riemannian Hessian $\nabla^2\psi$ is continuous at $\mx$ as a tensor field in the sense of \eqref{eq:A.hessian_continuity}. Then for $\mathbf{z}\in T_{\mx}\mathcal{M}$ with $\|\mathbf{z}\|_{\mx}\leq1$ and $h\in(0,\rho_{\mx})$,
\begin{align*}
    \psi(\Exp_{\mx}(h\mathbf{z}))
    =
    \psi(\mx)
    +
    h\left\langle\nabla\psi(\mx),\mathbf{z}\right\rangle_{\mx}
    +
    \frac{h^2}{2}\nabla^2\psi(\mx)\left(\mathbf{z},\mathbf{z}\right)
    +
    r_{\psi,h}(\mx,\mathbf{z}),
\end{align*}
where
\begin{align*}
    \lim_{h\downarrow0}\sup_{\|\mathbf{z}\|_{\mx}\leq1}\frac{|r_{\psi,h}(\mx,\mathbf{z})|}{h^2}=0.
\end{align*}
\end{lemma}

\begin{proof}[Proof of \Cref{lemma:A.pointwise_taylor_expansion}]
Fix $\mathbf{z}\in T_{\mx}\mathcal{M}$ with $\|\mathbf{z}\|_{\mx}\leq1$ and define $\gamma(t):=\Exp_{\mx}(th\mathbf{z})$ for $t\in[0,1]$. Since $h<\rho_{\mx}<i(\mx)$, the curve $\gamma$ remains in $B_{\mathcal{M}}(\mx,\rho_{\mx})$. Let $\mathsf{P}_{0,t}^{\gamma}:T_{\mx}\mathcal{M}\to T_{\gamma(t)}\mathcal{M}$ denote parallel transport along $\gamma$ from time $0$ to time $t$. Since $\gamma$ is a geodesic with initial velocity $h\mathbf{z}$, its velocity field is parallel along $\gamma$, and hence
\begin{align*}
    \dot{\gamma}(t)=h\mathsf{P}_{0,t}^{\gamma}\mathbf{z}, \quad t\in[0,1].
\end{align*}
Let $\eta(t):=\psi(\gamma(t))$. By the chain rule and the definition of the Riemannian Hessian,
\begin{align*}
    \eta'(0)=h\left\langle\nabla\psi(\mx),\mathbf{z}\right\rangle_{\mx}
\end{align*}
and
\begin{align*}
    \eta''(t)
    =
    \nabla^2\psi(\gamma(t))\left(\dot{\gamma}(t),\dot{\gamma}(t)\right)
    =
    h^2\nabla^2\psi(\gamma(t))
    \left(
    \mathsf{P}_{0,t}^{\gamma}\mathbf{z},
    \mathsf{P}_{0,t}^{\gamma}\mathbf{z}
    \right), \quad t\in[0,1],
\end{align*}
where the geodesic equation $\nabla^{\mathcal{M}}_{\dot{\gamma}(t)}\dot{\gamma}(t)=\mathbf{0}_{\gamma(t)}$ removes the acceleration term. Taylor's formula with integral remainder gives
\begin{align*}
    \psi(\Exp_{\mx}(h\mathbf{z}))
    =
    \psi(\mx)
    +
    h\left\langle\nabla\psi(\mx),\mathbf{z}\right\rangle_{\mx}
    +
    \frac{h^2}{2}\nabla^2\psi(\mx)\left(\mathbf{z},\mathbf{z}\right)
    +
    r_{\psi,h}(\mx,\mathbf{z}),
\end{align*}
where
\begin{align*}
    r_{\psi,h}(\mx,\mathbf{z})
    =
    h^2\int_0^1(1-t)
    \left[
    \nabla^2\psi(\gamma(t))
    \left(
    \mathsf{P}_{0,t}^{\gamma}\mathbf{z},
    \mathsf{P}_{0,t}^{\gamma}\mathbf{z}
    \right)
    -
    \nabla^2\psi(\mx)\left(\mathbf{z},\mathbf{z}\right)
    \right]dt.
\end{align*}
Since parallel transport preserves the Riemannian norm, $\|\mathsf{P}_{0,t}^{\gamma}\mathbf{z}\|_{\gamma(t)}=\|\mathbf{z}\|_{\mx}\leq1$. Moreover,
\begin{align*}
    d_{\mathcal{M}}(\gamma(t),\mx)\leq th\|\mathbf{z}\|_{\mx}\leq h, \quad t\in[0,1].
\end{align*}
For $\mz=\gamma(t)$, the parallel transport $\mathsf{P}_{0,t}^{\gamma}$ coincides with $\mathsf{P}_{\mx\to\mz}^{\gamma_{\mx,\mz}}$. Define
\begin{align*}
    \Delta_{\psi,\mx}(h)
    :=
    \sup_{\mz\in B_{\mathcal{M}}(\mx,h)}
    \sup_{\|\mathbf{u}\|_{\mx}\leq1}
    \left|
    \nabla^2\psi(\mz)
    \left(
    \mathsf{P}_{\mx\to\mz}^{\gamma_{\mx,\mz}}\mathbf{u},
    \mathsf{P}_{\mx\to\mz}^{\gamma_{\mx,\mz}}\mathbf{u}
    \right)
    -
    \nabla^2\psi(\mx)\left(\mathbf{u},\mathbf{u}\right)
    \right|.
\end{align*}
By the continuity of $\nabla^2\psi$ at $\mx$ as a tensor field,
\begin{align*}
    \lim_{h\downarrow0}\Delta_{\psi,\mx}(h)=0.
\end{align*}
Therefore,
\begin{align*}
    0
    \leq
    \limsup_{h\downarrow0}
    \sup_{\|\mathbf{z}\|_{\mx}\leq1}
    \frac{|r_{\psi,h}(\mx,\mathbf{z})|}{h^2}
    \leq
    \lim_{h\downarrow0}
    \frac{1}{2}\Delta_{\psi,\mx}(h)
    =
    0.
\end{align*}
Hence
\begin{align*}
    \lim_{h\downarrow0}\sup_{\|\mathbf{z}\|_{\mx}\leq1}\frac{|r_{\psi,h}(\mx,\mathbf{z})|}{h^2}=0.
\end{align*}
This completes the proof.
\end{proof}

\begin{lemma}[Uniform second-order Taylor expansion in normal coordinates] \label{lemma:A.uniform_taylor_expansion}
Let $\mathcal{K}\subset\mathcal{M}$ be compact and let $\rho\in(0,i(\mathcal{K}))$. Suppose that $\psi$ is $C^2$ on an open neighborhood of $\mathcal{K}^{\rho}$. Then
\begin{align*}
    \psi(\Exp_{\mx}(h\mathbf{z}))
    =
    \psi(\mx)
    +
    h\left\langle\nabla\psi(\mx),\mathbf{z}\right\rangle_{\mx}
    +
    \frac{h^2}{2}\nabla^2\psi(\mx)\left(\mathbf{z},\mathbf{z}\right)
    +
    r_{\psi,h}(\mx,\mathbf{z}),
\end{align*}
for $\mx\in\mathcal{K}$, $\mathbf{z}\in T_{\mx}\mathcal{M}$ with $\|\mathbf{z}\|_{\mx}\leq1$, and $h\in(0,\rho)$, where
\begin{align*}
    \lim_{h\downarrow0}\sup_{\mx\in\mathcal{K}}\sup_{\|\mathbf{z}\|_{\mx}\leq1}\frac{|r_{\psi,h}(\mx,\mathbf{z})|}{h^2}=0.
\end{align*}
\end{lemma}

\begin{proof}[Proof of \Cref{lemma:A.uniform_taylor_expansion}]
For each $\mx\in\mathcal{K}$ and $\mathbf{z}\in T_{\mx}\mathcal{M}$ with $\|\mathbf{z}\|_{\mx}\leq1$, set $\gamma_{\mx,\mathbf{z},h}(t):=\Exp_{\mx}(th\mathbf{z})$ for $t\in[0,1]$, and let $\mathsf{P}_{0,t}^{\gamma_{\mx,\mathbf{z},h}}:T_{\mx}\mathcal{M}\to T_{\gamma_{\mx,\mathbf{z},h}(t)}\mathcal{M}$ denote parallel transport along this geodesic. Since $h<\rho$,
\begin{align*}
    d_{\mathcal{M}}(\gamma_{\mx,\mathbf{z},h}(t),\mathcal{K})
    \leq
    d_{\mathcal{M}}(\gamma_{\mx,\mathbf{z},h}(t),\mx)
    \leq
    th\|\mathbf{z}\|_{\mx}
    \leq
    h
    <
    \rho, \quad t\in[0,1].
\end{align*}
Hence $\gamma_{\mx,\mathbf{z},h}(t)\in\mathcal{K}^{\rho}$ for all $t\in[0,1]$. By the same geodesic Taylor argument as in \Cref{lemma:A.pointwise_taylor_expansion},
\begin{align*}
    r_{\psi,h}(\mx,\mathbf{z})
    =
    h^2\int_0^1(1-t)
    \left[
    \nabla^2\psi(\gamma_{\mx,\mathbf{z},h}(t))
    \left(
    \mathsf{P}_{0,t}^{\gamma_{\mx,\mathbf{z},h}}\mathbf{z},
    \mathsf{P}_{0,t}^{\gamma_{\mx,\mathbf{z},h}}\mathbf{z}
    \right)
    -
    \nabla^2\psi(\mx)\left(\mathbf{z},\mathbf{z}\right)
    \right]dt.
\end{align*}
Parallel transport preserves the Riemannian norm, so
\begin{align*}
    \left\|\mathsf{P}_{0,t}^{\gamma_{\mx,\mathbf{z},h}}\mathbf{z}\right\|_{\gamma_{\mx,\mathbf{z},h}(t)}
    =
    \|\mathbf{z}\|_{\mx}
    \leq1.
\end{align*}
For each $t\in[0,1]$, set $\mz_t:=\gamma_{\mx,\mathbf{z},h}(t)$. If $t>0$ and $\mathbf{z}\neq\mathbf{0}_{\mx}$, then $\mathsf{P}_{0,t}^{\gamma_{\mx,\mathbf{z},h}}$ coincides with the parallel transport from $\mx$ to $\mz_t$ along the unique short geodesic $\gamma_{\mx,\mz_t}$. When $\mz_t=\mx$, we use the convention that $\mathsf{P}_{\mx\to\mx}^{\gamma_{\mx,\mx}}$ is the identity map on $T_{\mx}\mathcal{M}$. For $h\in(0,\rho)$, define
\begin{align*}
    \Delta_{\psi,\mathcal{K}}(h)
    :=
    \sup_{\mx\in\mathcal{K}}
    \sup_{\substack{\mz\in\mathcal{M} \\ d_{\mathcal{M}}(\mx,\mz)\leq h}}
    \sup_{\|\mathbf{u}\|_{\mx}\leq1}
    \left|
    \nabla^2\psi(\mz)
    \left(
    \mathsf{P}_{\mx\to\mz}^{\gamma_{\mx,\mz}}\mathbf{u},
    \mathsf{P}_{\mx\to\mz}^{\gamma_{\mx,\mz}}\mathbf{u}
    \right)
    -
    \nabla^2\psi(\mx)\left(\mathbf{u},\mathbf{u}\right)
    \right|.
\end{align*}
Since $h<\rho<i(\mathcal{K})$, all geodesics appearing in the definition of $\Delta_{\psi,\mathcal{K}}(h)$ are uniquely defined for sufficiently small $h$. Moreover, $\mathcal{D}_{\mathcal{K},\rho}$ is compact by \Cref{lemma:A.uniform_normal_neighborhoods}. With the convention that $\mathsf{P}_{\mx\to\mx}^{\gamma_{\mx,\mx}}$ is the identity map on $T_{\mx}\mathcal{M}$, the map
\begin{align*}
    (\mx,\mz,\mathbf{u})
    \mapsto
    \nabla^2\psi(\mz)
    \left(
    \mathsf{P}_{\mx\to\mz}^{\gamma_{\mx,\mz}}\mathbf{u},
    \mathsf{P}_{\mx\to\mz}^{\gamma_{\mx,\mz}}\mathbf{u}
    \right)
    -
    \nabla^2\psi(\mx)\left(\mathbf{u},\mathbf{u}\right)
\end{align*}
is continuous on the compact collection of triples $(\mx,\mz,\mathbf{u})$ such that $(\mx,\mz)\in\mathcal{D}_{\mathcal{K},\rho}$ and $\|\mathbf{u}\|_{\mx}\leq1$. Since this map is zero when $\mz=\mx$, its uniform continuity implies
\begin{align*}
    \lim_{h\downarrow0}\Delta_{\psi,\mathcal{K}}(h)=0.
\end{align*}
Using the integral remainder representation,
\begin{align*}
    \sup_{\mx\in\mathcal{K}}
    \sup_{\|\mathbf{z}\|_{\mx}\leq1}
    \frac{|r_{\psi,h}(\mx,\mathbf{z})|}{h^2}
    \leq
    \int_0^1(1-t)\Delta_{\psi,\mathcal{K}}(h)\,dt
    =
    \frac{1}{2}\Delta_{\psi,\mathcal{K}}(h).
\end{align*}
Consequently,
\begin{align*}
    0
    \leq
    \limsup_{h\downarrow0}
    \sup_{\mx\in\mathcal{K}}
    \sup_{\|\mathbf{z}\|_{\mx}\leq1}
    \frac{|r_{\psi,h}(\mx,\mathbf{z})|}{h^2}
    \leq
    \lim_{h\downarrow0}\frac{1}{2}\Delta_{\psi,\mathcal{K}}(h)
    =
    0.
\end{align*}
Thus,
\begin{align*}
    \lim_{h\downarrow0}\sup_{\mx\in\mathcal{K}}\sup_{\|\mathbf{z}\|_{\mx}\leq1}\frac{|r_{\psi,h}(\mx,\mathbf{z})|}{h^2}=0.
\end{align*}
This completes the proof.
\end{proof}

\section{Proof of Pointwise Consistency} \label{app:pointwise_consistency}
\setcounter{equation}{0}
\renewcommand{\theequation}{B.\arabic{equation}}
\renewcommand{\theHequation}{B.\arabic{equation}}

In this section, we prove \Cref{thm:pointwise_consistency}. Throughout this section, we fix $\mx\in\mathcal{M}$. We also fix the pointwise normal-neighborhood radius $\rho_{\mx}\in(0,i(\mx))$ defined in the main text, and an ordered orthonormal basis $\mathbf{E}_{\mx}\in\mathcal{E}_{\mx}$ of $T_{\mx}\mathcal M$. On $B_{\mathcal M}(\mx,\rho_{\mx})$, $\mathbf{v}_{\mx}^{\mathbf{E}_{\mx}}$ denotes the tangent-coordinate map defined in \eqref{eq:sec3.tangent_coordinate_map}. Whenever this coordinate map appears inside a kernel-weighted expectation, we use an arbitrary measurable extension outside $B_{\mathcal M}(\mx,\rho_{\mx})$; since $h<\rho_{\mx}$ for all sufficiently small $h$, this extension does not affect any of the quantities considered below.

The proof proceeds in two steps. First, we derive pointwise expansions for the population and empirical local moments induced by the manifold kernel $\mathcal{L}_{\mx,h}$. Second, we show that the empirical localized Fr\'echet objectives converge uniformly over $y\in\mathbb{M}$ to the conditional Fr\'echet objective $M_{\oplus}(\mx,y)$. 

\begin{lemma}[Pointwise population local moment expansion] \label{lemma:B.pointwise_population_moments}
Assume Conditions~\ref{con:P-K1} and~\ref{con:P-D1}, and suppose that $h\to0$ as $n\to\infty$. Then for $k\in\{1,2\}$ and $j\in\{0,1,2\}$,
\begin{align*}
    \left\|
    \E\left[
    \mathcal{L}_{\mx,h}\left(\mX\right)^k
    \left(\mathbf{v}_{\mx}^{\mathbf{E}_{\mx}}(\mX)\right)^{\otimes j}
    \right]
    -
    h^{d+j}f(\mx)
    \int_{\mathbb{R}^d}K(\|\mathbf{w}\|_2)^k\mathbf{w}^{\otimes j}\,\dd\mathbf{w}
    \right\|_{\star}
    =
    o(h^{d+j}).
\end{align*}
\end{lemma}

\begin{proof}[Proof of \Cref{lemma:B.pointwise_population_moments}]
Since $h\to0$ as $n\to\infty$, it is enough to consider sufficiently small $h<\rho_{\mx}$. On the support of $\mathcal{L}_{\mx,h}$, we have $d_{\mathcal{M}}(\mx,\mz)\leq h<\rho_{\mx}$, so the normal-coordinate representation $\mz=\Exp_{\mx}(\Log_{\mx}(\mz))$ is valid. Let $\mathbf{r}:=\bm{\Phi}_{\mathbf{E}_{\mx}}(\Log_{\mx}(\mz))\in\mathbb{R}^d$, so that $\Log_{\mx}(\mz)=\bm{\Phi}_{\mathbf{E}_{\mx}}^{-1}(\mathbf{r})$, $d_{\mathcal{M}}(\mx,\mz)=\|\mathbf{r}\|_2$, and
\begin{align*}
    \dd v_g(\mz)
    =
    \theta_{\mx}\left(\Exp_{\mx}\left(\bm{\Phi}_{\mathbf{E}_{\mx}}^{-1}(\mathbf{r})\right)\right)\,\dd\mathbf{r}.
\end{align*}
Therefore,
\begin{align}
\begin{split}
&\E\left[
    \mathcal{L}_{\mx,h}\left(\mX\right)^k
    \left(\mathbf{v}_{\mx}^{\mathbf{E}_{\mx}}(\mX)\right)^{\otimes j}
    \right]  \\
    &=
    \int_{\|\mathbf{r}\|_2\leq h}
    \theta_{\mx}\left(\Exp_{\mx}\left(\bm{\Phi}_{\mathbf{E}_{\mx}}^{-1}(\mathbf{r})\right)\right)^{-k}
    K\left(\frac{\|\mathbf{r}\|_2}{h}\right)^k
    \mathbf{r}^{\otimes j}
    f\left(\Exp_{\mx}\left(\bm{\Phi}_{\mathbf{E}_{\mx}}^{-1}(\mathbf{r})\right)\right)
    \theta_{\mx}\left(\Exp_{\mx}\left(\bm{\Phi}_{\mathbf{E}_{\mx}}^{-1}(\mathbf{r})\right)\right)\,\dd\mathbf{r}  \\
    &=
    \int_{\|\mathbf{r}\|_2\leq h}
    \frac{
    f\left(\Exp_{\mx}\left(\bm{\Phi}_{\mathbf{E}_{\mx}}^{-1}(\mathbf{r})\right)\right)
    }{
    \theta_{\mx}\left(\Exp_{\mx}\left(\bm{\Phi}_{\mathbf{E}_{\mx}}^{-1}(\mathbf{r})\right)\right)^{k-1}
    }
    K\left(\frac{\|\mathbf{r}\|_2}{h}\right)^k
    \mathbf{r}^{\otimes j}\,\dd\mathbf{r}  \\
    &=
    h^{d+j}
    \int_{\|\mathbf{w}\|_2\leq1}
    \frac{
    f\left(\Exp_{\mx}\left(\bm{\Phi}_{\mathbf{E}_{\mx}}^{-1}(h\mathbf{w})\right)\right)
    }{
    \theta_{\mx}\left(\Exp_{\mx}\left(\bm{\Phi}_{\mathbf{E}_{\mx}}^{-1}(h\mathbf{w})\right)\right)^{k-1}
    }
    K(\|\mathbf{w}\|_2)^k
    \mathbf{w}^{\otimes j}\,\dd\mathbf{w}.
\end{split}
\label{eq:B.pop_moment_integral_f}
\end{align}
Define
\begin{align*}
    \Psi_k(\mathbf{r})
    :=
    \frac{
    f\left(\Exp_{\mx}\left(\bm{\Phi}_{\mathbf{E}_{\mx}}^{-1}(\mathbf{r})\right)\right)
    }{
    \theta_{\mx}\left(\Exp_{\mx}\left(\bm{\Phi}_{\mathbf{E}_{\mx}}^{-1}(\mathbf{r})\right)\right)^{k-1}
    },
    \quad
    \mathbf{r}\in B_{\mathbb{R}^d}(\mathbf{0}_d,\rho_{\mx}).
\end{align*}
By Condition~\ref{con:P-D1}, the smoothness of $\Exp_{\mx}$, and the smooth positivity of $\theta_{\mx}$ in the normal neighborhood, $\Psi_k$ is continuous at $\mathbf{0}_d$. Since $\Exp_{\mx}(\mathbf{0}_{\mx})=\mx$ and $\theta_{\mx}(\mx)=1$, we have $\Psi_k(\mathbf{0}_d)=f(\mx)$. From \eqref{eq:B.pop_moment_integral_f},
\begin{align*}
    &h^{-(d+j)}
    \E\left[
    \mathcal{L}_{\mx,h}\left(\mX\right)^k
    \left(\mathbf{v}_{\mx}^{\mathbf{E}_{\mx}}(\mX)\right)^{\otimes j}
    \right]
    -
    f(\mx)\int_{\mathbb{R}^d}K(\|\mathbf{w}\|_2)^k\mathbf{w}^{\otimes j}\,\dd\mathbf{w} \\
    &=
    \int_{\|\mathbf{w}\|_2\leq1}
    \left(\Psi_k(h\mathbf{w})-\Psi_k(\mathbf{0}_d)\right)
    K(\|\mathbf{w}\|_2)^k
    \mathbf{w}^{\otimes j}\,\dd\mathbf{w}.
\end{align*}
Hence,
\begin{align*}
    &\left\|
    h^{-(d+j)}
    \E\left[
    \mathcal{L}_{\mx,h}\left(\mX\right)^k
    \left(\mathbf{v}_{\mx}^{\mathbf{E}_{\mx}}(\mX)\right)^{\otimes j}
    \right]
    -
    f(\mx)\int_{\mathbb{R}^d}K(\|\mathbf{w}\|_2)^k\mathbf{w}^{\otimes j}\,\dd\mathbf{w}
    \right\|_{\star} \\
    &\leq
    \sup_{\|\mathbf{w}\|_2\leq1}|\Psi_k(h\mathbf{w})-\Psi_k(\mathbf{0}_d)|
    \int_{\|\mathbf{w}\|_2\leq1}K(\|\mathbf{w}\|_2)^k\|\mathbf{w}\|_2^j\,\dd\mathbf{w}.
\end{align*}
By continuity of $\Psi_k$ at $\mathbf{0}_d$,
\begin{align*}
    \lim_{n\to\infty}\sup_{\|\mathbf{w}\|_2\leq1}|\Psi_k(h\mathbf{w})-\Psi_k(\mathbf{0}_d)|=0.
\end{align*}
The remaining integral is finite by \Cref{lemma:A.radial_kernel_moments}. Therefore the asserted $o(h^{d+j})$ bound follows.
\end{proof}

\begin{lemma}[Pointwise population local moment expansion with conditional density ratios] \label{lemma:B.pointwise_conditional_moments}
Assume Conditions~\ref{con:P-K1}, \ref{con:P-D1}, and~\ref{con:P-D2}, and suppose that $h\to0$ as $n\to\infty$. Then for $k\in\{1,2\}$ and $j\in\{0,1,2\}$,
\begin{align*}
    \sup_{\omega\in\mathbb{M}}
    \left\|
    \E\left[
    \mathcal{L}_{\mx,h}\left(\mX\right)^k
    \left(\mathbf{v}_{\mx}^{\mathbf{E}_{\mx}}(\mX)\right)^{\otimes j}
    g_{\omega}(\mX)
    \right]
    -
    h^{d+j}f(\mx)g_{\omega}(\mx)
    \int_{\mathbb{R}^d}K(\|\mathbf{w}\|_2)^k\mathbf{w}^{\otimes j}\,\dd\mathbf{w}
    \right\|_{\star}
    =
    o(h^{d+j}).
\end{align*}
\end{lemma}

\begin{proof}[Proof of \Cref{lemma:B.pointwise_conditional_moments}]
As in the proof of \Cref{lemma:B.pointwise_population_moments}, after using the normal-coordinate change of variables, it is enough to prove
\begin{align*}
    \lim_{n\to\infty}
    \sup_{\omega\in\mathbb{M}}
    \sup_{\|\mathbf{w}\|_2\leq1}
    \left|
    \Psi_{k,\omega}(h\mathbf{w})-\Psi_{k,\omega}(\mathbf{0}_d)
    \right|
    =
    0,
\end{align*}
where
\begin{align*}
    \Psi_{k,\omega}(\mathbf{r})
    :=
    \frac{
    f\left(\Exp_{\mx}\left(\bm{\Phi}_{\mathbf{E}_{\mx}}^{-1}(\mathbf{r})\right)\right)
    g_{\omega}\left(\Exp_{\mx}\left(\bm{\Phi}_{\mathbf{E}_{\mx}}^{-1}(\mathbf{r})\right)\right)
    }{
    \theta_{\mx}\left(\Exp_{\mx}\left(\bm{\Phi}_{\mathbf{E}_{\mx}}^{-1}(\mathbf{r})\right)\right)^{k-1}
    },
    \quad
    \mathbf{r}\in B_{\mathbb{R}^d}(\mathbf{0}_d,\rho_{\mx}).
\end{align*}
Let
\begin{align*}
    a_k(\mathbf{r})
    :=
    \frac{
    f\left(\Exp_{\mx}\left(\bm{\Phi}_{\mathbf{E}_{\mx}}^{-1}(\mathbf{r})\right)\right)
    }{
    \theta_{\mx}\left(\Exp_{\mx}\left(\bm{\Phi}_{\mathbf{E}_{\mx}}^{-1}(\mathbf{r})\right)\right)^{k-1}
    }.
\end{align*}
By Condition~\ref{con:P-D1} and the smooth positivity of $\theta_{\mx}$ in the normal neighborhood, $a_k$ is continuous at $\mathbf{0}_d$ and $a_k(\mathbf{0}_d)=f(\mx)$. Since the fixed version of $g_{\omega}$ is nonnegative, Condition~\ref{con:P-D2} gives
\begin{align*}
    G_{\mx}:=\sup_{\omega\in\mathbb{M}}\sup_{\mz\in B_{\mathcal{M}}(\mx,\rho_{\mx})}g_{\omega}(\mz)<\infty.
\end{align*}
Hence, for $\|\mathbf{w}\|_2\leq1$ and sufficiently small $h$,
\begin{align*}
    &
    \left|
    \Psi_{k,\omega}(h\mathbf{w})-\Psi_{k,\omega}(\mathbf{0}_d)
    \right| \\
    &\leq
    \left|
    a_k(h\mathbf{w})-a_k(\mathbf{0}_d)
    \right|
    g_{\omega}\left(\Exp_{\mx}\left(\bm{\Phi}_{\mathbf{E}_{\mx}}^{-1}(h\mathbf{w})\right)\right)
    +
    |a_k(\mathbf{0}_d)|
    \left|
    g_{\omega}\left(\Exp_{\mx}\left(\bm{\Phi}_{\mathbf{E}_{\mx}}^{-1}(h\mathbf{w})\right)\right)
    -
    g_{\omega}(\mx)
    \right| \\
    &\leq
    G_{\mx}\sup_{\|\mathbf{w}\|_2\leq1}|a_k(h\mathbf{w})-a_k(\mathbf{0}_d)|
    +
    |a_k(\mathbf{0}_d)|
    \sup_{\omega\in\mathbb{M}}
    \sup_{\substack{\mz\in\mathcal{M} \\ d_{\mathcal{M}}(\mx,\mz)\leq h}}
    |g_{\omega}(\mz)-g_{\omega}(\mx)|.
\end{align*}
The first term tends to zero by the continuity of $a_k$ at $\mathbf{0}_d$, and the second term tends to zero by the equicontinuity in Condition~\ref{con:P-D2}. Therefore,
\begin{align*}
    \lim_{n\to\infty}
    \sup_{\omega\in\mathbb{M}}
    \sup_{\|\mathbf{w}\|_2\leq1}
    \left|
    \Psi_{k,\omega}(h\mathbf{w})-\Psi_{k,\omega}(\mathbf{0}_d)
    \right|
    =
    0.
\end{align*}
Consequently, as in \Cref{lemma:B.pointwise_population_moments},
\begin{align*}
    &\sup_{\omega\in\mathbb{M}}
    \left\|
    h^{-(d+j)}
    \E\left[
    \mathcal{L}_{\mx,h}\left(\mX\right)^k
    \left(\mathbf{v}_{\mx}^{\mathbf{E}_{\mx}}(\mX)\right)^{\otimes j}
    g_{\omega}(\mX)
    \right]
    -
    f(\mx)g_{\omega}(\mx)
    \int_{\mathbb{R}^d}K(\|\mathbf{w}\|_2)^k\mathbf{w}^{\otimes j}\,\dd\mathbf{w}
    \right\|_{\star} \\
    &\leq
    \sup_{\omega\in\mathbb{M}}
    \sup_{\|\mathbf{w}\|_2\leq1}
    \left|
    \Psi_{k,\omega}(h\mathbf{w})-\Psi_{k,\omega}(\mathbf{0}_d)
    \right|
    \int_{\|\mathbf{w}\|_2\leq1}K(\|\mathbf{w}\|_2)^k\|\mathbf{w}\|_2^j\,\dd\mathbf{w}.
\end{align*}
The integral is finite by \Cref{lemma:A.radial_kernel_moments}, so the asserted $o(h^{d+j})$ bound follows.
\end{proof}

\begin{lemma}[Pointwise population local moment consequences] \label{lemma:B.pointwise_moment_orders}
Assume Conditions~\ref{con:P-K1} and~\ref{con:P-D1}, and suppose that $h\to0$ as $n\to\infty$. Then
\begin{align}
\begin{split} \label{eq:B.population_moment_rates}
    \tilde{\mu}_{h,0}(\mx) - h^{d}A_{d-1}c_{d-1,1}f(\mx)
    &=
    o(h^{d}), \\
    \left\|\bm{\tilde{\mu}}_{h,1}(\mx,\mathbf{E}_{\mx})\right\|_2
    &=
    o(h^{d+1}), \\
    \left\|
    \bm{\tilde{\mu}}_{h,2}(\mx,\mathbf{E}_{\mx})
    -
    h^{d+2}\frac{A_{d-1}}{d}c_{d+1,1}f(\mx)\mI_d
    \right\|_2
    &=
    o(h^{d+2}).
\end{split}
\end{align}
Furthermore, for all sufficiently small $h$, $\tilde{\mu}_{h,0}(\mx)>0$ and $\bm{\tilde{\mu}}_{h,2}(\mx,\mathbf{E}_{\mx})$ is invertible. Their inverses satisfy
\begin{align}
    \left|
    \tilde{\mu}_{h,0}(\mx)^{-1}
    -
    h^{-d}\frac{1}{A_{d-1}c_{d-1,1}f(\mx)}
    \right|
    =
    o(h^{-d}), \label{eq:B.inverse_moment_0}
\end{align}
and
\begin{align}
    \left\|
    \bm{\tilde{\mu}}_{h,2}(\mx,\mathbf{E}_{\mx})^{-1}
    -
    h^{-(d+2)}
    \frac{d}{A_{d-1}c_{d+1,1}f(\mx)}
    \mI_d
    \right\|_2
    =
    o(h^{-(d+2)}). \label{eq:B.inverse_moment_2}
\end{align}
\end{lemma}

\begin{proof}[Proof of \Cref{lemma:B.pointwise_moment_orders}]
By setting $k=1$ in \Cref{lemma:B.pointwise_population_moments}, we have, for $j=0,1,2$,
\begin{align} \label{eq:B.moment_general_expansion}
    \E\left[
    \mathcal{L}_{\mx,h}\left(\mX\right)
    \left(\mathbf{v}_{\mx}^{\mathbf{E}_{\mx}}(\mX)\right)^{\otimes j}
    \right]
    =
    h^{d+j}f(\mx)
    \int_{\mathbb{R}^d}K(\|\mathbf{w}\|_2)\mathbf{w}^{\otimes j}\,\dd\mathbf{w}
    +
    o(h^{d+j}),
\end{align}
where the remainder is understood with respect to $\|\cdot\|_{\star}$ as specified at the beginning of the appendices. For $j=0$, the left-hand side is $\tilde{\mu}_{h,0}(\mx)$. For $j=1$ and $j=2$, it is respectively $\bm{\tilde{\mu}}_{h,1}(\mx,\mathbf{E}_{\mx})$ and $\bm{\tilde{\mu}}_{h,2}(\mx,\mathbf{E}_{\mx})$. Applying \Cref{lemma:A.radial_kernel_moments} to \eqref{eq:B.moment_general_expansion} gives the three expansions in \eqref{eq:B.population_moment_rates}. In particular, the $o(h^{d+1})$ bound for $\bm{\tilde{\mu}}_{h,1}(\mx,\mathbf{E}_{\mx})$ follows because
\begin{align*}
    \int_{\mathbb{R}^d}\mathbf{w}K(\|\mathbf{w}\|_2)\,\dd\mathbf{w}
    =
    \mathbf{0}_d.
\end{align*}

Let
\begin{align*}
    a_0:=A_{d-1}c_{d-1,1}f(\mx),
    \quad
    a_2:=\frac{A_{d-1}}{d}c_{d+1,1}f(\mx).
\end{align*}
By Condition~\ref{con:P-D1} and \Cref{lemma:A.radial_kernel_moments}, $a_0>0$ and $a_2>0$. Hence the first expansion in \eqref{eq:B.population_moment_rates} implies $\tilde{\mu}_{h,0}(\mx)>0$ for all sufficiently small $h$ and
\begin{align*}
    \tilde{\mu}_{h,0}(\mx)^{-1}
    =
    O(h^{-d}).
\end{align*}
The matrix $\bm{\tilde{\mu}}_{h,2}(\mx,\mathbf{E}_{\mx})$ is symmetric because it is the expectation of a scalar weight times $\mathbf{v}_{\mx}^{\mathbf{E}_{\mx}}(\mX)\left(\mathbf{v}_{\mx}^{\mathbf{E}_{\mx}}(\mX)\right)^{\top}$. Therefore, by the Rayleigh quotient and the third expansion in \eqref{eq:B.population_moment_rates},
\begin{align*}
    \lambda_{\min}\left(\bm{\tilde{\mu}}_{h,2}(\mx,\mathbf{E}_{\mx})\right)
    &\geq
    h^{d+2}a_2
    -
    \left\|
    \bm{\tilde{\mu}}_{h,2}(\mx,\mathbf{E}_{\mx})
    -
    h^{d+2}a_2\mI_d
    \right\|_2 \\
    &=
    h^{d+2}a_2-o(h^{d+2}).
\end{align*}
Therefore $\bm{\tilde{\mu}}_{h,2}(\mx,\mathbf{E}_{\mx})$ is positive definite for all sufficiently small $h$, and
\begin{align*}
    \left\|
    \bm{\tilde{\mu}}_{h,2}(\mx,\mathbf{E}_{\mx})^{-1}
    \right\|_2
    =
    O(h^{-(d+2)}).
\end{align*}

It remains to prove the inverse expansions. Let
\begin{align*}
    M_0:=h^d a_0,
    \quad
    \mathbf{M}_2:=h^{d+2}a_2\mI_d.
\end{align*}
Using $a^{-1}-b^{-1}=a^{-1}(b-a)b^{-1}$,
\begin{align*}
    \left|
    \tilde{\mu}_{h,0}(\mx)^{-1}-M_0^{-1}
    \right|
    &\leq
    \left|
    \tilde{\mu}_{h,0}(\mx)^{-1}
    \right|
    \left|
    M_0-\tilde{\mu}_{h,0}(\mx)
    \right|
    \left|
    M_0^{-1}
    \right| \\
    &=
    O(h^{-d})\,o(h^d)\,O(h^{-d})
    =
    o(h^{-d}).
\end{align*}
Similarly, using $\mathbf{A}^{-1}-\mathbf{B}^{-1}=\mathbf{A}^{-1}(\mathbf{B}-\mathbf{A})\mathbf{B}^{-1}$,
\begin{align*}
    &
    \left\|
    \bm{\tilde{\mu}}_{h,2}(\mx,\mathbf{E}_{\mx})^{-1}
    -
    \mathbf{M}_2^{-1}
    \right\|_2 \\
    &\leq
    \left\|
    \bm{\tilde{\mu}}_{h,2}(\mx,\mathbf{E}_{\mx})^{-1}
    \right\|_2
    \left\|
    \mathbf{M}_2-\bm{\tilde{\mu}}_{h,2}(\mx,\mathbf{E}_{\mx})
    \right\|_2
    \left\|
    \mathbf{M}_2^{-1}
    \right\|_2 \\
    &=
    O(h^{-(d+2)})\,o(h^{d+2})\,O(h^{-(d+2)})
    =
    o(h^{-(d+2)}).
\end{align*}
Since $M_0^{-1}=h^{-d}a_0^{-1}$ and $\mathbf{M}_2^{-1}=h^{-(d+2)}a_2^{-1}\mI_d$, \eqref{eq:B.inverse_moment_0} and \eqref{eq:B.inverse_moment_2} follow.
\end{proof}

In order to establish the convergence of the empirical local moments to their population counterparts, we use the following matrix Chebyshev inequality; see, for example, Lemma C.2 of \cite{Im et al. (2025)}.

\begin{lemma}[Matrix Chebyshev inequality] \label{lemma:B.matrix_chebyshev_inequality}
Let $\mathbf{Z}\in\mathbb{R}^{p\times q}$ be a random matrix, and let $\mathbf{Z}_1,\ldots,\mathbf{Z}_n$ be independent copies of $\mathbf{Z}$. Suppose that $\left\|\E[\mathbf{Z}^{\top}\mathbf{Z}]\right\|_2<\infty$. Then for every $\epsilon>0$,
\begin{align*}
    \P\left(
    \left\|
    \frac{1}{n}\sum_{i=1}^{n}\mathbf{Z}_i-\E[\mathbf{Z}]
    \right\|_2>\epsilon
    \right)
    \leq
    \frac{q}{n\epsilon^2}
    \left\|
    \E\left[\mathbf{Z}^{\top}\mathbf{Z}\right]
    \right\|_2.
\end{align*}
\end{lemma}

Using this inequality, we now establish pointwise stochastic convergence rates for the empirical local moments.

\begin{lemma}[Pointwise empirical local moment deviations] \label{lemma:B.pointwise_empirical_moments}
Assume Conditions~\ref{con:P-K1}, \ref{con:P-B1}, and~\ref{con:P-D1}. Then
\begin{align}
\begin{split}
    \left|\hat{\mu}_{h,0}(\mx)-\tilde{\mu}_{h,0}(\mx)\right|
    &=
    O_{\P}\left(n^{-1/2}h^{d/2}\right), \\
    \left\|
    \bm{\hat{\mu}}_{h,1}(\mx,\mathbf{E}_{\mx})
    -
    \bm{\tilde{\mu}}_{h,1}(\mx,\mathbf{E}_{\mx})
    \right\|_2
    &=
    O_{\P}\left(n^{-1/2}h^{(d+2)/2}\right), \\
    \left\|
    \bm{\hat{\mu}}_{h,2}(\mx,\mathbf{E}_{\mx})
    -
    \bm{\tilde{\mu}}_{h,2}(\mx,\mathbf{E}_{\mx})
    \right\|_2
    &=
    O_{\P}\left(n^{-1/2}h^{(d+4)/2}\right).
\end{split}
\label{eq:B.empirical_moment_diff}
\end{align}
Furthermore, $\hat{\mu}_{h,0}(\mx)>0$ and $\bm{\hat{\mu}}_{h,2}(\mx,\mathbf{E}_{\mx})$ is invertible with probability tending to one, and
\begin{align}
    \left|
    \hat{\mu}_{h,0}(\mx)^{-1}
    -
    \tilde{\mu}_{h,0}(\mx)^{-1}
    \right|
    =
    O_{\P}\left(n^{-1/2}h^{-3d/2}\right), \label{eq:B.inverse_diff_0}
\end{align}
and
\begin{align}
    \left\|
    \bm{\hat{\mu}}_{h,2}(\mx,\mathbf{E}_{\mx})^{-1}
    -
    \bm{\tilde{\mu}}_{h,2}(\mx,\mathbf{E}_{\mx})^{-1}
    \right\|_2
    =
    O_{\P}\left(n^{-1/2}h^{-(3d+4)/2}\right). \label{eq:B.inverse_diff_2}
\end{align}
\end{lemma}

\begin{proof}[Proof of \Cref{lemma:B.pointwise_empirical_moments}]
Throughout the proof, write
\begin{align*}
    \mathbf{v}:=\mathbf{v}_{\mx}^{\mathbf{E}_{\mx}}(\mX)
\end{align*}
to lighten notation. Since $h\to0$ as $n\to\infty$, for all sufficiently large $n$ we have $h<\rho_{\mx}<i(\mx)$. For $\mz \in \mathcal{M}$ such that $\mathcal{L}_{\mx,h}\left(\mz\right)\neq0$, the compact support of $K$ implies $d_{\mathcal{M}}(\mx,\mz)\leq h<\rho_{\mx}$, and therefore $\mz\in B_{\mathcal M}(\mx,i(\mx))$. In particular,
\begin{align*}
    \|\mathbf{v}\|_2
    =
    \left\|\bm{\Phi}_{\mathbf{E}_{\mx}}(\Log_{\mx}(\mX))\right\|_2
    =
    \|\Log_{\mx}(\mX)\|_{\mx}
    =
    d_{\mathcal{M}}(\mx,\mX)
    \leq h
\end{align*}
whenever $\mathcal{L}_{\mx,h}\left(\mX\right)\neq0$.

We first prove \eqref{eq:B.empirical_moment_diff}. For $j=0$, apply \Cref{lemma:B.matrix_chebyshev_inequality} to the scalar random variable $Z=\mathcal{L}_{\mx,h}\left(\mX\right)$. By \Cref{lemma:B.pointwise_population_moments} with $k=2$ and $j=0$,
\begin{align*}
    \E\left[\mathcal{L}_{\mx,h}\left(\mX\right)^2\right]
    =
    O(h^d).
\end{align*}
Hence
\begin{align*}
    \hat{\mu}_{h,0}(\mx)-\tilde{\mu}_{h,0}(\mx)
    =
    O_{\P}\left(n^{-1/2}h^{d/2}\right).
\end{align*}
For $j=1$, apply \Cref{lemma:B.matrix_chebyshev_inequality} to $\mathbf{Z}=\mathcal{L}_{\mx,h}\left(\mX\right)\mathbf{v}\in\mathbb{R}^{d\times1}$. Then
\begin{align*}
    \mathbf{Z}^{\top}\mathbf{Z}
    =
    \mathcal{L}_{\mx,h}\left(\mX\right)^2\|\mathbf{v}\|_2^2.
\end{align*}
By \Cref{lemma:B.pointwise_population_moments} with $k=2$ and $j=2$,
\begin{align*}
    \left\|
    \E\left[
    \mathcal{L}_{\mx,h}\left(\mX\right)^2
    \mathbf{v}\mathbf{v}^{\top}
    \right]
    \right\|_2
    =
    O(h^{d+2}).
\end{align*}
Since $d$ is fixed,
\begin{align*}
    \E\left[\mathbf{Z}^{\top}\mathbf{Z}\right]
    =
    \operatorname{tr}
    \E\left[
    \mathcal{L}_{\mx,h}\left(\mX\right)^2
    \mathbf{v}\mathbf{v}^{\top}
    \right]
    =
    O(h^{d+2}).
\end{align*}
Thus
\begin{align*}
    \left\|
    \bm{\hat{\mu}}_{h,1}(\mx,\mathbf{E}_{\mx})
    -
    \bm{\tilde{\mu}}_{h,1}(\mx,\mathbf{E}_{\mx})
    \right\|_2
    =
    O_{\P}\left(n^{-1/2}h^{(d+2)/2}\right).
\end{align*}
For $j=2$, apply \Cref{lemma:B.matrix_chebyshev_inequality} to $\mathbf{Z}=\mathcal{L}_{\mx,h}\left(\mX\right)\mathbf{v}\mathbf{v}^{\top}\in\mathbb{R}^{d\times d}$. The dimension factor in \Cref{lemma:B.matrix_chebyshev_inequality} is absorbed into the stochastic order because $d$ is fixed. Since
\begin{align*}
    \mathbf{Z}^{\top}\mathbf{Z}
    =
    \mathcal{L}_{\mx,h}\left(\mX\right)^2
    \|\mathbf{v}\|_2^2
    \mathbf{v}\mathbf{v}^{\top}
\end{align*}
and $\mathcal{L}_{\mx,h}\left(\mX\right)\neq0$ implies $\|\mathbf{v}\|_2\leq h$, we have
\begin{align*}
    \mathbf{Z}^{\top}\mathbf{Z}
    \preceq
    h^2
    \mathcal{L}_{\mx,h}\left(\mX\right)^2
    \mathbf{v}\mathbf{v}^{\top}.
\end{align*}
Using \Cref{lemma:B.pointwise_population_moments} with $k=2$ and $j=2$ again,
\begin{align*}
    \left\|
    \E\left(\mathbf{Z}^{\top}\mathbf{Z}\right)
    \right\|_2
    \leq
    h^2
    \left\|
    \E\left[
    \mathcal{L}_{\mx,h}\left(\mX\right)^2
    \mathbf{v}\mathbf{v}^{\top}
    \right]
    \right\|_2
    =
    O(h^{d+4}).
\end{align*}
Therefore,
\begin{align*}
    \left\|
    \bm{\hat{\mu}}_{h,2}(\mx,\mathbf{E}_{\mx})
    -
    \bm{\tilde{\mu}}_{h,2}(\mx,\mathbf{E}_{\mx})
    \right\|_2
    =
    O_{\P}\left(n^{-1/2}h^{(d+4)/2}\right),
\end{align*}
which proves \eqref{eq:B.empirical_moment_diff}.

We next prove positivity and invertibility. By \Cref{lemma:B.pointwise_moment_orders},
\begin{align*}
    \tilde{\mu}_{h,0}(\mx)
    =
    h^d A_{d-1}c_{d-1,1}f(\mx)+o(h^d).
\end{align*}
Since $A_{d-1}c_{d-1,1}f(\mx)>0$ by Condition~\ref{con:P-D1} and \Cref{lemma:A.radial_kernel_moments}, this expansion and \eqref{eq:B.empirical_moment_diff} imply
\begin{align*}
    \hat{\mu}_{h,0}(\mx)
    =
    h^d
    \left(
    A_{d-1}c_{d-1,1}f(\mx)+o(1)+O_{\P}\left((nh^d)^{-1/2}\right)
    \right).
\end{align*}
Condition~\ref{con:P-B1} gives $\lim_{n\to\infty}nh^d=\infty$, and hence $O_{\P}\left((nh^d)^{-1/2}\right)=o_{\P}(1)$. Therefore $\hat{\mu}_{h,0}(\mx)>0$ with probability tending to one and
\begin{align*}
    \left|\hat{\mu}_{h,0}(\mx)^{-1}\right|
    =
    O_{\P}(h^{-d}).
\end{align*}
Similarly, the third expansion in \Cref{lemma:B.pointwise_moment_orders} and Weyl's inequality imply
\begin{align*}
    \lambda_{\min}
    \left(
    \bm{\tilde{\mu}}_{h,2}(\mx,\mathbf{E}_{\mx})
    \right)
    =
    h^{d+2}
    \left(
    \frac{A_{d-1}}{d}c_{d+1,1}f(\mx)+o(1)
    \right).
\end{align*}
Since $\bm{\hat{\mu}}_{h,2}(\mx,\mathbf{E}_{\mx})$ and $\bm{\tilde{\mu}}_{h,2}(\mx,\mathbf{E}_{\mx})$ are symmetric, Weyl's inequality and \eqref{eq:B.empirical_moment_diff} yield
\begin{align*}
    \lambda_{\min}
    \left(
    \bm{\hat{\mu}}_{h,2}(\mx,\mathbf{E}_{\mx})
    \right)
    &\geq
    \lambda_{\min}
    \left(
    \bm{\tilde{\mu}}_{h,2}(\mx,\mathbf{E}_{\mx})
    \right)
    -
    \left\|
    \bm{\hat{\mu}}_{h,2}(\mx,\mathbf{E}_{\mx})
    -
    \bm{\tilde{\mu}}_{h,2}(\mx,\mathbf{E}_{\mx})
    \right\|_2 \\
    &=
    h^{d+2}
    \left[
    \frac{A_{d-1}}{d}c_{d+1,1}f(\mx)
    +
    o(1)
    -
    O_{\P}\left((nh^d)^{-1/2}\right)
    \right].
\end{align*}
Again, Condition~\ref{con:P-B1} implies $O_{\P}\left((nh^d)^{-1/2}\right)=o_{\P}(1)$. Hence $\bm{\hat{\mu}}_{h,2}(\mx,\mathbf{E}_{\mx})$ is positive definite with probability tending to one, and
\begin{align*}
    \left\|
    \bm{\hat{\mu}}_{h,2}(\mx,\mathbf{E}_{\mx})^{-1}
    \right\|_2
    =
    O_{\P}(h^{-(d+2)}).
\end{align*}

It remains to prove the inverse-difference bounds. All inverse-difference bounds below are understood on the event where the empirical inverses exist; the inverses may be defined arbitrarily on the complement of this event. Since the complement has probability tending to zero, the resulting bounds hold in probability. On the event where the inverses exist, the scalar identity $a^{-1}-b^{-1}=a^{-1}(b-a)b^{-1}$ gives
\begin{align*}
    \left|
    \hat{\mu}_{h,0}(\mx)^{-1}
    -
    \tilde{\mu}_{h,0}(\mx)^{-1}
    \right|
    &\leq
    \left|
    \hat{\mu}_{h,0}(\mx)^{-1}
    \right|
    \left|
    \tilde{\mu}_{h,0}(\mx)
    -
    \hat{\mu}_{h,0}(\mx)
    \right|
    \left|
    \tilde{\mu}_{h,0}(\mx)^{-1}
    \right| \\
    &=
    O_{\P}(h^{-d})
    O_{\P}\left(n^{-1/2}h^{d/2}\right)
    O(h^{-d}) \\
    &=
    O_{\P}\left(n^{-1/2}h^{-3d/2}\right),
\end{align*}
where \Cref{lemma:B.pointwise_moment_orders} gives $\left|\tilde{\mu}_{h,0}(\mx)^{-1}\right|=O(h^{-d})$. Similarly, using $\mathbf{A}^{-1}-\mathbf{B}^{-1}=\mathbf{A}^{-1}(\mathbf{B}-\mathbf{A})\mathbf{B}^{-1}$,
\begin{align*}
    &
    \left\|
    \bm{\hat{\mu}}_{h,2}(\mx,\mathbf{E}_{\mx})^{-1}
    -
    \bm{\tilde{\mu}}_{h,2}(\mx,\mathbf{E}_{\mx})^{-1}
    \right\|_2 \\
    &\leq
    \left\|
    \bm{\hat{\mu}}_{h,2}(\mx,\mathbf{E}_{\mx})^{-1}
    \right\|_2
    \left\|
    \bm{\tilde{\mu}}_{h,2}(\mx,\mathbf{E}_{\mx})
    -
    \bm{\hat{\mu}}_{h,2}(\mx,\mathbf{E}_{\mx})
    \right\|_2
    \left\|
    \bm{\tilde{\mu}}_{h,2}(\mx,\mathbf{E}_{\mx})^{-1}
    \right\|_2 \\
    &=
    O_{\P}(h^{-(d+2)})
    O_{\P}\left(n^{-1/2}h^{(d+4)/2}\right)
    O(h^{-(d+2)}) \\
    &=
    O_{\P}\left(n^{-1/2}h^{-(3d+4)/2}\right),
\end{align*}
where \Cref{lemma:B.pointwise_moment_orders} gives $\left\|\bm{\tilde{\mu}}_{h,2}(\mx,\mathbf{E}_{\mx})^{-1}\right\|_2=O(h^{-(d+2)})$. This completes the proof.
\end{proof}

The following lemma controls the population and empirical local linear denominators. It also shows that the empirical local linear weights are well-defined with probability tending to one.

\begin{lemma}[Pointwise local linear denominator] \label{lemma:B.pointwise_denominators}
Assume Conditions~\ref{con:P-K1} and~\ref{con:P-D1}, and suppose that $h\to0$ as $n\to\infty$. The population denominator satisfies
\begin{align}
    \left|
    \frac{\tilde{\sigma}_{h}(\mx)}{h^d A_{d-1}c_{d-1,1}}
    -
    f(\mx)
    \right|
    =
    o(1). \label{eq:B.sigma_tilde_convergence}
\end{align}
Furthermore, assume Condition~\ref{con:P-B1}. On the event where $\bm{\hat{\mu}}_{h,2}(\mx,\mathbf{E}_{\mx})$ is invertible, the empirical denominator $\hat{\sigma}_{h}(\mx)$ is well-defined and satisfies
\begin{align}
    \hat{\sigma}_{h}(\mx)-\tilde{\sigma}_{h}(\mx)
    =
    O_{\P}\left(n^{-1/2}h^{d/2}\right). \label{eq:B.sigma_hat_difference}
\end{align}
Consequently, $\hat{\sigma}_{h}(\mx)>0$ with probability tending to one.
\end{lemma}

\begin{proof}[Proof of \Cref{lemma:B.pointwise_denominators}]
The ordered orthonormal basis $\mathbf{E}_{\mx}$ is fixed throughout this proof. By \Cref{lemma:A.invariance_local_linear_weights}, the scalar denominators below do not depend on this choice whenever the relevant second local moment matrices are invertible. We first prove \eqref{eq:B.sigma_tilde_convergence}. Recall that
\begin{align*}
    \tilde{\sigma}_{h}(\mx)
    =
    \tilde{\mu}_{h,0}(\mx)
    -
    \bm{\tilde{\mu}}_{h,1}(\mx,\mathbf{E}_{\mx})^{\top}
    \bm{\tilde{\mu}}_{h,2}(\mx,\mathbf{E}_{\mx})^{-1}
    \bm{\tilde{\mu}}_{h,1}(\mx,\mathbf{E}_{\mx}).
\end{align*}
By \Cref{lemma:B.pointwise_moment_orders},
\begin{align*}
    \tilde{\mu}_{h,0}(\mx)
    =
    h^dA_{d-1}c_{d-1,1}f(\mx)+o(h^d),
\end{align*}
and
\begin{align*}
    \left\|\bm{\tilde{\mu}}_{h,1}(\mx,\mathbf{E}_{\mx})\right\|_2=o(h^{d+1}),
    \quad
    \left\|\bm{\tilde{\mu}}_{h,2}(\mx,\mathbf{E}_{\mx})^{-1}\right\|_2=O(h^{-(d+2)}).
\end{align*}
Therefore,
\begin{align*}
    \left|
    \bm{\tilde{\mu}}_{h,1}(\mx,\mathbf{E}_{\mx})^{\top}
    \bm{\tilde{\mu}}_{h,2}(\mx,\mathbf{E}_{\mx})^{-1}
    \bm{\tilde{\mu}}_{h,1}(\mx,\mathbf{E}_{\mx})
    \right|
    &\leq
    \left\|\bm{\tilde{\mu}}_{h,1}(\mx,\mathbf{E}_{\mx})\right\|_2^2
    \left\|\bm{\tilde{\mu}}_{h,2}(\mx,\mathbf{E}_{\mx})^{-1}\right\|_2 \\
    &=
    o(h^{2d+2})O(h^{-(d+2)})
    =
    o(h^d).
\end{align*}
Thus,
\begin{align*}
    \tilde{\sigma}_{h}(\mx)
    =
    h^dA_{d-1}c_{d-1,1}f(\mx)+o(h^d),
\end{align*}
which proves \eqref{eq:B.sigma_tilde_convergence}.

We next prove \eqref{eq:B.sigma_hat_difference}. All statements involving $\hat{\sigma}_{h}(\mx)$ are understood on the event where $\bm{\hat{\mu}}_{h,2}(\mx,\mathbf{E}_{\mx})$ is invertible; by \Cref{lemma:B.pointwise_empirical_moments}, this event has probability tending to one. On the event where $\bm{\hat{\mu}}_{h,2}(\mx,\mathbf{E}_{\mx})$ is invertible, write
\begin{align*}
    \hat{\sigma}_{h}(\mx)-\tilde{\sigma}_{h}(\mx)
    =
    \left(\hat{\mu}_{h,0}(\mx)-\tilde{\mu}_{h,0}(\mx)\right)
    -
    \left(
    \bm{\hat{\mu}}_{h,1}^{\top}\bm{\hat{\mu}}_{h,2}^{-1}\bm{\hat{\mu}}_{h,1}
    -
    \bm{\tilde{\mu}}_{h,1}^{\top}\bm{\tilde{\mu}}_{h,2}^{-1}\bm{\tilde{\mu}}_{h,1}
    \right),
\end{align*}
where, throughout the rest of the proof, we suppress the common arguments $(\mx,\mathbf{E}_{\mx})$ in the vector and matrix local moments. By \Cref{lemma:B.pointwise_empirical_moments},
\begin{align*}
    \hat{\mu}_{h,0}(\mx)-\tilde{\mu}_{h,0}(\mx)
    =
    O_{\P}\left(n^{-1/2}h^{d/2}\right).
\end{align*}
Moreover, \Cref{lemma:B.pointwise_moment_orders} and \Cref{lemma:B.pointwise_empirical_moments} imply
\begin{align*}
    \left\|\bm{\hat{\mu}}_{h,1}\right\|_2
    \leq
    \left\|\bm{\tilde{\mu}}_{h,1}\right\|_2
    +
    \left\|\bm{\hat{\mu}}_{h,1}-\bm{\tilde{\mu}}_{h,1}\right\|_2
    =
    o(h^{d+1})+O_{\P}\left(n^{-1/2}h^{(d+2)/2}\right)
    =
    o_{\P}(h^{d+1}),
\end{align*}
because $nh^d\to\infty$. Also, by \Cref{lemma:B.pointwise_empirical_moments},
\begin{align*}
    \left\|\bm{\hat{\mu}}_{h,2}^{-1}\right\|_2=O_{\P}(h^{-(d+2)}),
    \quad
    \left\|\bm{\hat{\mu}}_{h,2}^{-1}-\bm{\tilde{\mu}}_{h,2}^{-1}\right\|_2
    =
    O_{\P}\left(n^{-1/2}h^{-(3d+4)/2}\right).
\end{align*}
Decompose the quadratic-form difference as
\begin{align*}
    &
    \bm{\hat{\mu}}_{h,1}^{\top}\bm{\hat{\mu}}_{h,2}^{-1}\bm{\hat{\mu}}_{h,1}
    -
    \bm{\tilde{\mu}}_{h,1}^{\top}\bm{\tilde{\mu}}_{h,2}^{-1}\bm{\tilde{\mu}}_{h,1} \\
    &=
    \left(\bm{\hat{\mu}}_{h,1}-\bm{\tilde{\mu}}_{h,1}\right)^{\top}
    \bm{\hat{\mu}}_{h,2}^{-1}
    \bm{\hat{\mu}}_{h,1}
    +
    \bm{\tilde{\mu}}_{h,1}^{\top}
    \left(\bm{\hat{\mu}}_{h,2}^{-1}-\bm{\tilde{\mu}}_{h,2}^{-1}\right)
    \bm{\hat{\mu}}_{h,1} \\
    &\qquad +
    \bm{\tilde{\mu}}_{h,1}^{\top}
    \bm{\tilde{\mu}}_{h,2}^{-1}
    \left(\bm{\hat{\mu}}_{h,1}-\bm{\tilde{\mu}}_{h,1}\right).
\end{align*}
The three terms are respectively bounded by
\begin{align*}
    &O_{\P}\left(n^{-1/2}h^{(d+2)/2}\right)
    O_{\P}(h^{-(d+2)})
    o_{\P}(h^{d+1})
    =
    o_{\P}\left(n^{-1/2}h^{d/2}\right), \\
    &o(h^{d+1})
    O_{\P}\left(n^{-1/2}h^{-(3d+4)/2}\right)
    o_{\P}(h^{d+1})
    =
    o_{\P}\left(n^{-1/2}h^{d/2}\right), \\
    &o(h^{d+1})
    O(h^{-(d+2)})
    O_{\P}\left(n^{-1/2}h^{(d+2)/2}\right)
    =
    o_{\P}\left(n^{-1/2}h^{d/2}\right).
\end{align*}
Therefore,
\begin{align*}
    \bm{\hat{\mu}}_{h,1}^{\top}\bm{\hat{\mu}}_{h,2}^{-1}\bm{\hat{\mu}}_{h,1}
    -
    \bm{\tilde{\mu}}_{h,1}^{\top}\bm{\tilde{\mu}}_{h,2}^{-1}\bm{\tilde{\mu}}_{h,1}
    =
    o_{\P}\left(n^{-1/2}h^{d/2}\right),
\end{align*}
and \eqref{eq:B.sigma_hat_difference} follows.

Finally, \eqref{eq:B.sigma_tilde_convergence} and Condition~\ref{con:P-D1} imply that, for some constant $c>0$ and all sufficiently small $h$,
\begin{align*}
    \tilde{\sigma}_{h}(\mx)\geq ch^d.
\end{align*}
Combining this lower bound with \eqref{eq:B.sigma_hat_difference} gives
\begin{align*}
    \hat{\sigma}_{h}(\mx)
    =
    \tilde{\sigma}_{h}(\mx)
    \left(
    1+O_{\P}\left((nh^d)^{-1/2}\right)
    \right).
\end{align*}
Since $nh^d\to\infty$ by Condition~\ref{con:P-B1},
\begin{align*}
    \hat{\sigma}_{h}(\mx)
    =
    \tilde{\sigma}_{h}(\mx)(1+o_{\P}(1)).
\end{align*}
Hence $\hat{\sigma}_{h}(\mx)>0$ with probability tending to one.
\end{proof}

By \Cref{lemma:B.pointwise_moment_orders} and the population part of \Cref{lemma:B.pointwise_denominators}, for all sufficiently small $h$, the population quantities $\bm{\tilde{\mu}}_{h,2}(\mx,\mathbf{E}_{\mx})$ and $\tilde{\sigma}_{h}(\mx)$ are well-defined, with $\bm{\tilde{\mu}}_{h,2}(\mx,\mathbf{E}_{\mx})$ invertible and $\tilde{\sigma}_{h}(\mx)>0$. Throughout the population arguments below, the ordered orthonormal basis $\mathbf{E}_{\mx}$ is fixed. By \Cref{lemma:A.invariance_local_linear_weights}, the resulting scalar weights and objectives do not depend on this choice whenever the relevant inverse is well-defined.

We first collect the population auxiliary quantities used in the consistency and bias arguments. Define
\begin{align}
\begin{split} \label{eq:B.population_auxiliary_defs}
    \tilde{\tau}_{h,0}(\mx,\omega)
    &:=
    \E\left[
    \mathcal{L}_{\mx,h}\left(\mX\right)g_{\omega}(\mX)
    \right], \\
    \bm{\tilde{\tau}}_{h,1}(\mx,\mathbf{E}_{\mx},\omega)
    &:=
    \E\left[
    \mathcal{L}_{\mx,h}\left(\mX\right)
    \mathbf{v}_{\mx}^{\mathbf{E}_{\mx}}(\mX)
    g_{\omega}(\mX)
    \right], \\
    \tilde{\nu}_{h,0}(\mx,y)
    &:=
    \E\left[
    \mathcal{L}_{\mx,h}\left(\mX\right)d_{\mathbb{M}}^2(Y,y)
    \right], \\
    \bm{\tilde{\nu}}_{h,1}(\mx,\mathbf{E}_{\mx},y)
    &:=
    \E\left[
    \mathcal{L}_{\mx,h}\left(\mX\right)
    \mathbf{v}_{\mx}^{\mathbf{E}_{\mx}}(\mX)
    d_{\mathbb{M}}^2(Y,y)
    \right],
    \quad \omega,y\in\mathbb{M}.
\end{split}
\end{align}
For $s\in\{0,1\}$, define
\begin{align}
\begin{split} \label{eq:B.population_ND_defs}
    \tilde{N}_{h,s}(\mx,y)
    &:=
    \tilde{\nu}_{h,0}(\mx,y)
    -
    s\,\bm{\tilde{\mu}}_{h,1}(\mx,\mathbf{E}_{\mx})^{\top}
    \bm{\tilde{\mu}}_{h,2}(\mx,\mathbf{E}_{\mx})^{-1}
    \bm{\tilde{\nu}}_{h,1}(\mx,\mathbf{E}_{\mx},y),
    \quad y\in\mathbb{M}, \\
    \tilde{D}_{h,s}(\mx)
    &:=
    (1-s)\tilde{\mu}_{h,0}(\mx)
    +
    s\tilde{\sigma}_{h}(\mx).
\end{split}
\end{align}
Then for $s\in\{0,1\}$, the population local objectives satisfy
\begin{align} \label{eq:B.population_M_repr}
    \tilde{M}_{h,s}(\mx,y)
    =
    \frac{\tilde{N}_{h,s}(\mx,y)}
    {\tilde{D}_{h,s}(\mx)},
    \quad
    y\in\mathbb{M}.
\end{align}
In terms of the conditional density-ratio representation, define the localized density-ratio function
\begin{align}
\begin{split} \label{eq:B.g_tilde_def}
    \tilde{g}_{h,s}(\mx,\omega)
    &:=
    \frac{
    \tilde{\tau}_{h,0}(\mx,\omega)
    -
    s\,\bm{\tilde{\mu}}_{h,1}(\mx,\mathbf{E}_{\mx})^{\top}
    \bm{\tilde{\mu}}_{h,2}(\mx,\mathbf{E}_{\mx})^{-1}
    \bm{\tilde{\tau}}_{h,1}(\mx,\mathbf{E}_{\mx},\omega)
    }
    {
    \tilde{D}_{h,s}(\mx)
    },
    \quad \omega\in\mathbb{M}.
\end{split}
\end{align}
By Condition~\ref{con:M1}, the squared metric loss is bounded. Hence, using Fubini's theorem and the definition of $g_{\omega}$, for $s\in\{0,1\}$,
\begin{align} \label{eq:B.M_tilde_integral_repr}
    \tilde{M}_{h,s}(\mx,y)
    =
    \int_{\mathbb{M}}
    d_{\mathbb{M}}^2(y,\omega)
    \tilde{g}_{h,s}(\mx,\omega)
    \,\dd P_Y(\omega),
    \quad
    y\in\mathbb{M}.
\end{align}
In \eqref{eq:B.population_ND_defs}--\eqref{eq:B.g_tilde_def}, the local linear correction terms are evaluated only when $s=1$; when $s=0$, the terms multiplied by $s$ are omitted.

\begin{lemma}[Pointwise convergence of population local objectives] \label{lemma:B.pointwise_population_objective}
Assume Conditions~\ref{con:P-K1}, \ref{con:P-D1}, \ref{con:P-D2}, and~\ref{con:M1}, and suppose that $h\to0$ as $n\to\infty$. For $s\in\{0,1\}$,
\begin{align*}
    \sup_{y\in\mathbb{M}}
    \left|
    \tilde{M}_{h,s}(\mx,y)-M_{\oplus}(\mx,y)
    \right|
    =
    o(1).
\end{align*}
\end{lemma}

\begin{proof}[Proof of \Cref{lemma:B.pointwise_population_objective}]
For every $\mz\in B_{\mathcal M}(\mx,\rho_{\mx})$, the map
$\omega\mapsto g_{\omega}(\mz)$ is the fixed Radon--Nikodym derivative of
$P_{Y|\mX=\mz}$ with respect to $P_Y$. Hence the conditional Fr\'echet objective admits the representation
\begin{align*}
    M_{\oplus}(\mz,y)
    =
    \int_{\mathbb{M}}
    d_{\mathbb{M}}^2(y,\omega)g_{\omega}(\mz)
    \,\dd P_Y(\omega),
    \quad \mz\in B_{\mathcal M}(\mx,\rho_{\mx}),y\in\mathbb{M}.
\end{align*}
Condition~\ref{con:M1} implies that
\begin{align} \label{eq:B.diameter_metric_space}
    D_{\mathbb{M}}:=\sup_{y,\omega\in\mathbb{M}}d_{\mathbb{M}}(y,\omega)<\infty,
\end{align}
which denotes the diameter of $\mathbb{M}$. For $\mz \in \mathcal{M}$ and all sufficiently small $h<\rho_{\mx}$, Condition~\ref{con:P-K1} implies that $\mathcal{L}_{\mx,h}(\mz)=0$ unless $d_{\mathcal M}(\mx,\mz)\leq h$, and hence every point with nonzero kernel weight lies in $B_{\mathcal M}(\mx,\rho_{\mx})$. By \Cref{lemma:A.pointwise_normal_neighborhoods}, the volume-density factor $\theta_{\mx}(\cdot)^{-1}$ is bounded on this local neighborhood. Therefore, using Conditions~\ref{con:P-K1}, \ref{con:P-D1}, \ref{con:P-D2}, and~\ref{con:M1}, the quantities
\begin{align*}
    \mathcal{L}_{\mx,h}(\mz),\quad
    \mathcal{L}_{\mx,h}(\mz)\mathbf{v}_{\mx}^{\mathbf{E}_{\mx}}(\mz),\quad
    \mathcal{L}_{\mx,h}(\mz)g_{\omega}(\mz),\quad
    \mathcal{L}_{\mx,h}(\mz)\mathbf{v}_{\mx}^{\mathbf{E}_{\mx}}(\mz)g_{\omega}(\mz), \quad \mz\in B_{\mathcal M}(\mx,\rho_{\mx}),\omega\in\mathbb{M}
\end{align*}
are bounded uniformly over $\mz\in B_{\mathcal M}(\mx,\rho_{\mx})$ and $\omega\in\mathbb{M}$, for each fixed sufficiently small $h$. Since $d_{\mathbb{M}}^2(y,\omega)\leq D_{\mathbb{M}}^2$, the applications of the law of total expectation and Fubini's theorem leading to \eqref{eq:B.M_tilde_integral_repr} are justified componentwise. Using \eqref{eq:B.M_tilde_integral_repr} and the preceding representation with $\mz=\mx$, for any $y\in\mathbb{M}$,
\begin{align*}
    \tilde{M}_{h,s}(\mx,y)-M_{\oplus}(\mx,y)
    =
    \int_{\mathbb{M}}
    d_{\mathbb{M}}^2(y,\omega)
    \left\{
    \tilde{g}_{h,s}(\mx,\omega)-g_{\omega}(\mx)
    \right\}
    \,\dd P_Y(\omega).
\end{align*}
Therefore,
\begin{align} \label{eq:B.objective_density_bound}
    \sup_{y\in\mathbb{M}}
    \left|
    \tilde{M}_{h,s}(\mx,y)-M_{\oplus}(\mx,y)
    \right|
    \leq
    D_{\mathbb{M}}^2
    \sup_{\omega\in\mathbb{M}}
    \left|
    \tilde{g}_{h,s}(\mx,\omega)-g_{\omega}(\mx)
    \right|.
\end{align}

It remains to prove the convergence of the localized density ratios, uniformly over $\omega\in\mathbb{M}$. We first consider the case $s=0$. By \Cref{lemma:B.pointwise_population_moments} with $j=0$ and $k=1$,
\begin{align*}
    \tilde{\mu}_{h,0}(\mx)
    =
    h^d A_{d-1}c_{d-1,1} f(\mx)
    +
    r_{h,0}(\mx),
    \quad
    |r_{h,0}(\mx)|=o(h^d).
\end{align*}
Moreover, by \Cref{lemma:B.pointwise_conditional_moments} with $j=0$ and $k=1$,
\begin{align} \label{eq:B.tau0_expansion}
    \tilde{\tau}_{h,0}(\mx,\omega)
    =
    h^d A_{d-1}c_{d-1,1} f(\mx)g_{\omega}(\mx)
    +
    r_{h,\omega}(\mx),
    \quad
    \sup_{\omega\in\mathbb{M}}|r_{h,\omega}(\mx)|=o(h^d).
\end{align}
By Condition~\ref{con:P-K1} and the definition of $c_{d-1,1}$, we have
$A_{d-1}c_{d-1,1}>0$. Since $f(\mx)>0$ by Condition~\ref{con:P-D1}, we have
\begin{align*}
    \tilde{\mu}_{h,0}(\mx)
    \geq
    \frac{1}{2}h^d A_{d-1}c_{d-1,1} f(\mx)
\end{align*}
for all sufficiently small $h$. Hence, for all sufficiently small $h$,
\begin{align*}
    \sup_{\omega\in\mathbb{M}}
    \left|
    \tilde{g}_{h,0}(\mx,\omega)-g_{\omega}(\mx)
    \right|
    &=
    \sup_{\omega\in\mathbb{M}}
    \left|
    \frac{
    \tilde{\tau}_{h,0}(\mx,\omega)
    -
    g_{\omega}(\mx)\tilde{\mu}_{h,0}(\mx)
    }
    {
    \tilde{\mu}_{h,0}(\mx)
    }
    \right| \\
    &=
    \sup_{\omega\in\mathbb{M}}
    \left|
    \frac{
    r_{h,\omega}(\mx)
    -
    g_{\omega}(\mx)r_{h,0}(\mx)
    }
    {
    \tilde{\mu}_{h,0}(\mx)
    }
    \right| \\
    &\leq
    \frac{
    2
    \left[
    \sup_{\omega\in\mathbb{M}}|r_{h,\omega}(\mx)|
    +
    |r_{h,0}(\mx)|\sup_{\omega\in\mathbb{M}}g_{\omega}(\mx)
    \right]
    }
    {
    h^d A_{d-1}c_{d-1,1} f(\mx)
    }.
\end{align*}
By Condition~\ref{con:P-D2}, $\sup_{\omega\in\mathbb{M}}g_{\omega}(\mx)<\infty$. Therefore,
\begin{align} \label{eq:B.g_tilde_lc_convergence}
    \sup_{\omega\in\mathbb{M}}
    \left|
    \tilde{g}_{h,0}(\mx,\omega)-g_{\omega}(\mx)
    \right|
    =
    o(1).
\end{align}
We next consider the case $s=1$. By \Cref{lemma:B.pointwise_conditional_moments} with $j=1$ and $k=1$,
\begin{align*}
    \sup_{\omega\in\mathbb{M}}
    \left\|
    \bm{\tilde{\tau}}_{h,1}(\mx,\mathbf{E}_{\mx},\omega)
    -
    h^{d+1}
    (f\cdot g_{\omega})(\mx)
    \int_{\|\mathbf{w}\|_2\leq1}
    \mathbf{w}K(\|\mathbf{w}\|_2)
    \,\dd\mathbf{w}
    \right\|_2
    =
    o(h^{d+1}).
\end{align*}
The leading integral vanishes by the first-moment cancellation in \Cref{lemma:A.radial_kernel_moments}. Hence
\begin{align} \label{eq:B.tau1_o_rate}
    \sup_{\omega\in\mathbb{M}}
    \left\|
    \bm{\tilde{\tau}}_{h,1}(\mx,\mathbf{E}_{\mx},\omega)
    \right\|_2
    =
    o(h^{d+1}).
\end{align}
By \Cref{lemma:B.pointwise_moment_orders},
\begin{align*}
    \left\|
    \bm{\tilde{\mu}}_{h,1}(\mx,\mathbf{E}_{\mx})
    \right\|_2
    =
    o(h^{d+1}),
    \quad
    \left\|
    \bm{\tilde{\mu}}_{h,2}(\mx,\mathbf{E}_{\mx})^{-1}
    \right\|_2
    =
    O(h^{-(d+2)}).
\end{align*}
Together with \eqref{eq:B.tau1_o_rate}, this gives
\begin{align}
\begin{split}
&\sup_{\omega\in\mathbb{M}}
    \left|
    \bm{\tilde{\mu}}_{h,1}(\mx,\mathbf{E}_{\mx})^{\top}
    \bm{\tilde{\mu}}_{h,2}(\mx,\mathbf{E}_{\mx})^{-1}
    \bm{\tilde{\tau}}_{h,1}(\mx,\mathbf{E}_{\mx},\omega)
    \right|  \\
    &\leq
    \left\|
    \bm{\tilde{\mu}}_{h,1}(\mx,\mathbf{E}_{\mx})
    \right\|_2
    \left\|
    \bm{\tilde{\mu}}_{h,2}(\mx,\mathbf{E}_{\mx})^{-1}
    \right\|_2
    \sup_{\omega\in\mathbb{M}}
    \left\|
    \bm{\tilde{\tau}}_{h,1}(\mx,\mathbf{E}_{\mx},\omega)
    \right\|_2  \\
    &=
    o(h^{d+1})O(h^{-(d+2)})o(h^{d+1})
    =
    o(h^d).
\end{split}
\label{eq:B.ll_correction_o_rate}
\end{align}
Moreover, by the population part of \Cref{lemma:B.pointwise_denominators},
\begin{align} \label{eq:B.sigma_expansion_ll}
    \tilde{\sigma}_{h}(\mx)
    =
    h^d A_{d-1}c_{d-1,1} f(\mx)
    +
    o(h^d).
\end{align}
Since $f(\mx)>0$ by Condition~\ref{con:P-D1}, \eqref{eq:B.sigma_expansion_ll} implies that, for all sufficiently small $h$,
\begin{align} \label{eq:B.sigma_lower_explicit}
    \tilde{\sigma}_{h}(\mx)
    \geq
    \frac{1}{2}h^d A_{d-1}c_{d-1,1} f(\mx).
\end{align}

Using the expansion of $\tilde{\tau}_{h,0}(\mx,\omega)$ from the case $s=0$, we write
\begin{align*}
    \tilde{\tau}_{h,0}(\mx,\omega)
    =
    h^d A_{d-1}c_{d-1,1} f(\mx)g_{\omega}(\mx)
    +
    r_{h,\omega}(\mx),
    \quad
    \sup_{\omega\in\mathbb{M}}|r_{h,\omega}(\mx)|=o(h^d).
\end{align*}
Therefore, for every $\omega\in\mathbb{M}$,
\begin{align*}
    &\tilde{\tau}_{h,0}(\mx,\omega)
    -
    \bm{\tilde{\mu}}_{h,1}(\mx,\mathbf{E}_{\mx})^{\top}
    \bm{\tilde{\mu}}_{h,2}(\mx,\mathbf{E}_{\mx})^{-1}
    \bm{\tilde{\tau}}_{h,1}(\mx,\mathbf{E}_{\mx},\omega)
    -
    g_{\omega}(\mx)\tilde{\sigma}_{h}(\mx) \\
    &=
    r_{h,\omega}(\mx)
    -
    \bm{\tilde{\mu}}_{h,1}(\mx,\mathbf{E}_{\mx})^{\top}
    \bm{\tilde{\mu}}_{h,2}(\mx,\mathbf{E}_{\mx})^{-1}
    \bm{\tilde{\tau}}_{h,1}(\mx,\mathbf{E}_{\mx},\omega) \\
    &\qquad
    -
    g_{\omega}(\mx)
    \left[
    \tilde{\sigma}_{h}(\mx)
    -
    h^d A_{d-1}c_{d-1,1} f(\mx)
    \right].
\end{align*}
Combining this identity with \eqref{eq:B.sigma_lower_explicit}, we obtain, for all sufficiently small $h$,
\begin{align*}
    &\sup_{\omega\in\mathbb{M}}
    \left|
    \tilde{g}_{h,1}(\mx,\omega)-g_{\omega}(\mx)
    \right| \\
    &=
    \sup_{\omega\in\mathbb{M}}
    \left|
    \frac{
    \tilde{\tau}_{h,0}(\mx,\omega)
    -
    \bm{\tilde{\mu}}_{h,1}(\mx,\mathbf{E}_{\mx})^{\top}
    \bm{\tilde{\mu}}_{h,2}(\mx,\mathbf{E}_{\mx})^{-1}
    \bm{\tilde{\tau}}_{h,1}(\mx,\mathbf{E}_{\mx},\omega)
    -
    g_{\omega}(\mx)\tilde{\sigma}_{h}(\mx)
    }
    {
    \tilde{\sigma}_{h}(\mx)
    }
    \right| \\
    &\leq
    \frac{
    2\sup_{\omega\in\mathbb{M}}|r_{h,\omega}(\mx)|
    }
    {
    h^d A_{d-1}c_{d-1,1} f(\mx)
    }
    +
    \frac{
    2\sup_{\omega\in\mathbb{M}}
    \left|
    \bm{\tilde{\mu}}_{h,1}(\mx,\mathbf{E}_{\mx})^{\top}
    \bm{\tilde{\mu}}_{h,2}(\mx,\mathbf{E}_{\mx})^{-1}
    \bm{\tilde{\tau}}_{h,1}(\mx,\mathbf{E}_{\mx},\omega)
    \right|
    }
    {
    h^d A_{d-1}c_{d-1,1} f(\mx)
    } \\
    &\quad+
    \frac{
    2\left|
    \tilde{\sigma}_{h}(\mx)
    -
    h^d A_{d-1}c_{d-1,1} f(\mx)
    \right|
    \sup_{\omega\in\mathbb{M}}g_{\omega}(\mx)
    }
    {
    h^d A_{d-1}c_{d-1,1} f(\mx)
    }.
\end{align*}
By Condition~\ref{con:P-D2}, $\sup_{\omega\in\mathbb{M}}g_{\omega}(\mx)<\infty$. The three numerators in the last display are respectively $o(h^d)$ by the definition of $r_{h,\omega}(\mx)$, by \eqref{eq:B.ll_correction_o_rate}, and by \eqref{eq:B.sigma_expansion_ll}. Since the denominator is the explicit positive multiple $h^d A_{d-1}c_{d-1,1}f(\mx)$, it follows that
\begin{align} \label{eq:B.g_tilde_ll_convergence}
    \sup_{\omega\in\mathbb{M}}
    \left|
    \tilde{g}_{h,1}(\mx,\omega)-g_{\omega}(\mx)
    \right|
    =
    o(1).
\end{align}
Combining \eqref{eq:B.objective_density_bound}, \eqref{eq:B.g_tilde_lc_convergence}, and \eqref{eq:B.g_tilde_ll_convergence} proves the claim for $s\in\{0,1\}$.
\end{proof}

\begin{lemma}[Pointwise convergence of population local minimizers] \label{lemma:B.pointwise_oracle_minimizer}
Assume Conditions~\ref{con:P-K1}, \ref{con:P-D1}, \ref{con:P-D2}, \ref{con:M1}, and~\ref{con:P-M2}, and suppose that $h\to0$ as $n\to\infty$. Let $\tilde{m}_{h,s}(\mx)$ be a minimizer of $\tilde{M}_{h,s}(\mx,\cdot)$ for $s\in\{0,1\}$. Then
\begin{align*}
    d_{\mathbb{M}}\left(\tilde{m}_{h,s}(\mx),m_{\oplus}(\mx)\right)=o(1),
    \quad s\in\{0,1\}.
\end{align*}
\end{lemma}

\begin{proof}[Proof of \Cref{lemma:B.pointwise_oracle_minimizer}]
The argument is the standard argmin-continuity argument for Fr\'echet objectives under uniform convergence and separation; see, for example, the proof of Lemma B.7 in \cite{Im et al. (2025)}. We give the details in the present notation. Fix $\epsilon>0$ and set
\begin{align*}
    A_{\epsilon}
    :=
    \left\{
    y\in\mathbb{M}:
    d_{\mathbb{M}}\left(y,m_{\oplus}(\mx)\right)>\epsilon
    \right\}.
\end{align*}
If $A_{\epsilon}$ is empty, then $d_{\mathbb{M}}\left(\tilde{m}_{h,s}(\mx),m_{\oplus}(\mx)\right)\leq\epsilon$ holds trivially. Otherwise, define
\begin{align*}
    \eta_{\epsilon}
    :=
    \inf_{y\in A_{\epsilon}}
    \left[
    M_{\oplus}(\mx,y)-M_{\oplus}(\mx,m_{\oplus}(\mx))
    \right].
\end{align*}
By Condition~\ref{con:P-M2}, $\eta_{\epsilon}>0$. By \Cref{lemma:B.pointwise_population_objective}, for each $s\in\{0,1\}$,
\begin{align*}
    \Delta_{h,s}
    :=
    \sup_{y\in\mathbb{M}}
    \left|
    \tilde{M}_{h,s}(\mx,y)-M_{\oplus}(\mx,y)
    \right|
    =
    o(1).
\end{align*}
Since $\tilde{m}_{h,s}(\mx)$ minimizes $\tilde{M}_{h,s}(\mx,\cdot)$,
\begin{align*}
    M_{\oplus}\left(\mx,\tilde{m}_{h,s}(\mx)\right)
    -
    M_{\oplus}\left(\mx,m_{\oplus}(\mx)\right)
    &\leq
    \left|
    M_{\oplus}\left(\mx,\tilde{m}_{h,s}(\mx)\right)
    -
    \tilde{M}_{h,s}\left(\mx,\tilde{m}_{h,s}(\mx)\right)
    \right| \\
    &\quad+
    \left(
    \tilde{M}_{h,s}\left(\mx,\tilde{m}_{h,s}(\mx)\right)
    -
    \tilde{M}_{h,s}\left(\mx,m_{\oplus}(\mx)\right)
    \right) \\
    &\quad+
    \left|
    \tilde{M}_{h,s}\left(\mx,m_{\oplus}(\mx)\right)
    -
    M_{\oplus}\left(\mx,m_{\oplus}(\mx)\right)
    \right| \\
    &\leq
    2\Delta_{h,s}.
\end{align*}
Since $\Delta_{h,s}=o(1)$, for all sufficiently small $h$ we have $2\Delta_{h,s}<\eta_{\epsilon}$. If $d_{\mathbb{M}}\left(\tilde{m}_{h,s}(\mx),m_{\oplus}(\mx)\right)>\epsilon$, then $\tilde{m}_{h,s}(\mx)\in A_{\epsilon}$, and the definition of $\eta_{\epsilon}$ implies
\begin{align*}
    M_{\oplus}\left(\mx,\tilde{m}_{h,s}(\mx)\right)
    -
    M_{\oplus}\left(\mx,m_{\oplus}(\mx)\right)
    \geq
    \eta_{\epsilon},
\end{align*}
which contradicts the preceding bound. Hence, for every $\epsilon>0$, the inequality
\begin{align*}
    d_{\mathbb{M}}\left(\tilde{m}_{h,s}(\mx),m_{\oplus}(\mx)\right)\leq\epsilon
\end{align*}
holds for all sufficiently large $n$.
Therefore $d_{\mathbb{M}}\left(\tilde{m}_{h,s}(\mx),m_{\oplus}(\mx)\right)=o(1)$.
\end{proof}

To establish the stochastic convergence results, we introduce the empirical counterparts of the population quantities in \eqref{eq:B.population_auxiliary_defs}. Define
\begin{align*}
    \hat{\nu}_{h,0}(\mx,y)
    &:=
    \frac{1}{n}\sum_{i=1}^{n}
    \mathcal{L}_{\mx,h}\left(\mX^{(i)}\right)
    d_{\mathbb{M}}^2\left(Y^{(i)},y\right), \\
    \bm{\hat{\nu}}_{h,1}(\mx,\mathbf{E}_{\mx},y)
    &:=
    \frac{1}{n}\sum_{i=1}^{n}
    \mathcal{L}_{\mx,h}\left(\mX^{(i)}\right)
    \mathbf{v}_{\mx}^{\mathbf{E}_{\mx}}\left(\mX^{(i)}\right)
    d_{\mathbb{M}}^2\left(Y^{(i)},y\right).
\end{align*}
For $s\in\{0,1\}$, define
\begin{align}
\begin{split} \label{eq:B.empirical_ND_defs}
    \hat{N}_{h,s}(\mx,y)
    &:=
    \hat{\nu}_{h,0}(\mx,y)
    -
    s\,\bm{\hat{\mu}}_{h,1}(\mx,\mathbf{E}_{\mx})^{\top}
    \bm{\hat{\mu}}_{h,2}(\mx,\mathbf{E}_{\mx})^{-1}
    \bm{\hat{\nu}}_{h,1}(\mx,\mathbf{E}_{\mx},y), \\
    \hat{D}_{h,s}(\mx)
    &:=
    (1-s)\hat{\mu}_{h,0}(\mx)
    +
    s\hat{\sigma}_{h}(\mx).
\end{split}
\end{align}
Then
\begin{align} \label{eq:B.empirical_M_repr}
    \hat{M}_{h,s}(\mx,y)
    =
    \frac{\hat{N}_{h,s}(\mx,y)}
    {\hat{D}_{h,s}(\mx)},
    \quad
    \tilde{M}_{h,s}(\mx,y)
    =
    \frac{\tilde{N}_{h,s}(\mx,y)}
    {\tilde{D}_{h,s}(\mx)}.
\end{align}
In \eqref{eq:B.empirical_ND_defs}--\eqref{eq:B.empirical_M_repr}, the local linear correction terms are evaluated only when $s=1$; when $s=0$, the terms multiplied by $s$ are omitted. For $s=0$, the empirical expressions are understood on the event where $\hat{\mu}_{h,0}(\mx)>0$. For $s=1$, they are understood on the event where $\bm{\hat{\mu}}_{h,2}(\mx,\mathbf{E}_{\mx})$ is invertible and $\hat{\sigma}_{h}(\mx)>0$. By \Cref{lemma:B.pointwise_empirical_moments,lemma:B.pointwise_denominators}, these events have probability tending to one under the corresponding assumptions.

\begin{lemma}[Pointwise uniform stochastic convergence of local objectives] \label{lemma:B.pointwise_empirical_objective}
Assume Conditions~\ref{con:P-K1}, \ref{con:P-B1}, \ref{con:P-D1}, \ref{con:P-D2}, and~\ref{con:M1}. Then for each $s\in\{0,1\}$,
\begin{align*}
    \sup_{y\in\mathbb{M}}
    \left|
    \hat{M}_{h,s}(\mx,y)-\tilde{M}_{h,s}(\mx,y)
    \right|
    =
    o_{\P}(1).
\end{align*}
\end{lemma}

\begin{proof}[Proof of \Cref{lemma:B.pointwise_empirical_objective}]
All arguments below are understood on the event where the empirical inverses and denominators exist. By \Cref{lemma:B.pointwise_empirical_moments,lemma:B.pointwise_denominators}, this event has probability tending to one and therefore does not affect convergence in probability.

We first prove pointwise stochastic convergence. Fix $y\in\mathbb{M}$. Note that Condition~\ref{con:M1} implies $d_{\mathbb{M}}^2(\omega,y)\leq D_{\mathbb{M}}^2$ for any $\omega \in \mathbb{M}$, where $D_{\mathbb{M}}$ is defined by \eqref{eq:B.diameter_metric_space}. Applying \Cref{lemma:B.matrix_chebyshev_inequality} to the scalar variable $\mathcal{L}_{\mx,h}\left(\mX\right)d_{\mathbb{M}}^2(Y,y)$ and using
\begin{align*}
    \E\left[\mathcal{L}_{\mx,h}\left(\mX\right)^2d_{\mathbb{M}}^4(Y,y)\right]
    \leq
    D_{\mathbb{M}}^4\E\left[\mathcal{L}_{\mx,h}\left(\mX\right)^2\right]
    =
    O(h^d),
\end{align*}
where the last equality follows from \Cref{lemma:B.pointwise_population_moments} with $k=2$ and $j=0$, gives
\begin{align}
    \left|
    \hat{\nu}_{h,0}(\mx,y)-\tilde{\nu}_{h,0}(\mx,y)
    \right|
    =
    O_{\P}\left(n^{-1/2}h^{d/2}\right). \label{eq:B.nu0_stochastic_rate}
\end{align}
Similarly, applying \Cref{lemma:B.matrix_chebyshev_inequality} to the vector variable $\mathcal{L}_{\mx,h}\left(\mX\right)\mathbf{v}_{\mx}^{\mathbf{E}_{\mx}}(\mX)d_{\mathbb{M}}^2(Y,y)$ and using \Cref{lemma:B.pointwise_population_moments} with $k=2$ and $j=2$ yields
\begin{align}
    \left\|
    \bm{\hat{\nu}}_{h,1}(\mx,\mathbf{E}_{\mx},y)
    -
    \bm{\tilde{\nu}}_{h,1}(\mx,\mathbf{E}_{\mx},y)
    \right\|_2
    =
    O_{\P}\left(n^{-1/2}h^{(d+2)/2}\right). \label{eq:B.nu1_stochastic_rate}
\end{align}
Moreover, using the representation of $\bm{\tilde{\nu}}_{h,1}$ through $\bm{\tilde{\tau}}_{h,1}$ and \Cref{lemma:B.pointwise_conditional_moments},
\begin{align*}
    \sup_{y\in\mathbb{M}}
    \left\|
    \bm{\tilde{\nu}}_{h,1}(\mx,\mathbf{E}_{\mx},y)
    \right\|_2
    \leq
    D_{\mathbb{M}}^2
    \sup_{\omega\in\mathbb{M}}
    \left\|
    \bm{\tilde{\tau}}_{h,1}(\mx,\mathbf{E}_{\mx},\omega)
    \right\|_2
    =
    o(h^{d+1}).
\end{align*}
Hence, by \eqref{eq:B.nu1_stochastic_rate} and Condition~\ref{con:P-B1},
\begin{align*}
    \left\|
    \bm{\hat{\nu}}_{h,1}(\mx,\mathbf{E}_{\mx},y)
    \right\|_2
    =
    o_{\P}(h^{d+1}).
\end{align*}
For $s=0$, \eqref{eq:B.nu0_stochastic_rate} directly gives
\begin{align*}
    \hat{N}_{h,0}(\mx,y)-\tilde{N}_{h,0}(\mx,y)
    =
    O_{\P}\left(n^{-1/2}h^{d/2}\right).
\end{align*}
For $s=1$, decompose the local linear correction difference as
\begin{align*}
    &
    \bm{\hat{\mu}}_{h,1}^{\top}\bm{\hat{\mu}}_{h,2}^{-1}\bm{\hat{\nu}}_{h,1}
    -
    \bm{\tilde{\mu}}_{h,1}^{\top}\bm{\tilde{\mu}}_{h,2}^{-1}\bm{\tilde{\nu}}_{h,1} \\
    &=
    \left(\bm{\hat{\mu}}_{h,1}-\bm{\tilde{\mu}}_{h,1}\right)^{\top}
    \bm{\hat{\mu}}_{h,2}^{-1}
    \bm{\hat{\nu}}_{h,1}
    +
    \bm{\tilde{\mu}}_{h,1}^{\top}
    \left(\bm{\hat{\mu}}_{h,2}^{-1}-\bm{\tilde{\mu}}_{h,2}^{-1}\right)
    \bm{\hat{\nu}}_{h,1}
    +
    \bm{\tilde{\mu}}_{h,1}^{\top}
    \bm{\tilde{\mu}}_{h,2}^{-1}
    \left(\bm{\hat{\nu}}_{h,1}-\bm{\tilde{\nu}}_{h,1}\right),
\end{align*}
where the common arguments $(\mx,\mathbf{E}_{\mx},y)$ are suppressed in this display. By \Cref{lemma:B.pointwise_moment_orders,lemma:B.pointwise_empirical_moments} and the preceding bounds, the above three terms are respectively
\begin{align*}
    &O_{\P}\left(n^{-1/2}h^{(d+2)/2}\right)
    O_{\P}(h^{-(d+2)})
    o_{\P}(h^{d+1})
    =
    o_{\P}\left(n^{-1/2}h^{d/2}\right), \\
    &o(h^{d+1})
    O_{\P}\left(n^{-1/2}h^{-(3d+4)/2}\right)
    o_{\P}(h^{d+1})
    =
    o_{\P}\left(n^{-1/2}h^{d/2}\right), \\
    &o(h^{d+1})
    O(h^{-(d+2)})
    O_{\P}\left(n^{-1/2}h^{(d+2)/2}\right)
    =
    o_{\P}\left(n^{-1/2}h^{d/2}\right).
\end{align*}
Therefore, for each $s\in\{0,1\}$,
\begin{align} \label{eq:B.N_stochastic_rate}
    \hat{N}_{h,s}(\mx,y)-\tilde{N}_{h,s}(\mx,y)
    =
    O_{\P}\left(n^{-1/2}h^{d/2}\right).
\end{align}
Also, by \Cref{lemma:B.pointwise_moment_orders,lemma:B.pointwise_denominators},
\begin{align*}
    \tilde{D}_{h,s}(\mx)
    =
    h^dA_{d-1}c_{d-1,1}f(\mx)+o(h^d),
    \quad s\in\{0,1\},
\end{align*}
and by \Cref{lemma:B.pointwise_empirical_moments,lemma:B.pointwise_denominators},
\begin{align*}
    \hat{D}_{h,s}(\mx)-\tilde{D}_{h,s}(\mx)
    =
    O_{\P}\left(n^{-1/2}h^{d/2}\right).
\end{align*}
Since $f(\mx)>0$ and $nh^d\to\infty$ as $n \to \infty$, it follows that
\begin{align*}
    \hat{D}_{h,s}(\mx)^{-1}=O_{\P}(h^{-d}), \quad \tilde{D}_{h,s}(\mx)^{-1}=O(h^{-d}).
\end{align*}
Moreover,
\begin{align*}
    \sup_{y\in\mathbb{M}}\left|\tilde{\nu}_{h,0}(\mx,y)\right|
    \leq
    D_{\mathbb{M}}^2\tilde{\mu}_{h,0}(\mx)
    =
    O(h^d),
\end{align*}
and, by the preceding bound on $\bm{\tilde{\nu}}_{h,1}$ and \Cref{lemma:B.pointwise_moment_orders},
\begin{align*}
    \sup_{y\in\mathbb{M}}
    \left|
    \bm{\tilde{\mu}}_{h,1}(\mx,\mathbf{E}_{\mx})^{\top}
    \bm{\tilde{\mu}}_{h,2}(\mx,\mathbf{E}_{\mx})^{-1}
    \bm{\tilde{\nu}}_{h,1}(\mx,\mathbf{E}_{\mx},y)
    \right|
    =
    o(h^d).
\end{align*}
Hence $\sup_{y\in\mathbb{M}}|\tilde{N}_{h,s}(\mx,y)|=O(h^d)$ for $s\in\{0,1\}$. Then \eqref{eq:B.population_M_repr} implies
\begin{align*}
    \hat{M}_{h,s}(\mx,y)-\tilde{M}_{h,s}(\mx,y)
    =
    \frac{\hat{N}_{h,s}(\mx,y)-\tilde{N}_{h,s}(\mx,y)}
    {\hat{D}_{h,s}(\mx)}
    -
    \frac{
    \tilde{N}_{h,s}(\mx,y)
    \left(\hat{D}_{h,s}(\mx)-\tilde{D}_{h,s}(\mx)\right)
    }
    {
    \hat{D}_{h,s}(\mx)\tilde{D}_{h,s}(\mx)
    },
\end{align*}
which gives
\begin{align*}
    \left|
    \hat{M}_{h,s}(\mx,y)-\tilde{M}_{h,s}(\mx,y)
    \right|
    =
    O_{\P}\left((nh^d)^{-1/2}\right)
    =
    o_{\P}(1).
\end{align*}

We next prove stochastic equicontinuity. For $y_1,y_2\in\mathbb{M}$,
\begin{align*}
    \left|
    d_{\mathbb{M}}^2(Y,y_1)-d_{\mathbb{M}}^2(Y,y_2)
    \right|
    \leq
    2D_{\mathbb{M}}d_{\mathbb{M}}(y_1,y_2).
\end{align*}
Define
\begin{align*}
    \hat{a}_{h,1}(\mx,\mathbf{E}_{\mx})
    &:=
    \frac{1}{n}\sum_{i=1}^{n}
    \mathcal{L}_{\mx,h}\left(\mX^{(i)}\right)
    \left\|
    \mathbf{v}_{\mx}^{\mathbf{E}_{\mx}}\left(\mX^{(i)}\right)
    \right\|_2, \\
    \tilde{a}_{h,1}(\mx,\mathbf{E}_{\mx})
    &:=
    \E\left[
    \mathcal{L}_{\mx,h}\left(\mX\right)
    \left\|
    \mathbf{v}_{\mx}^{\mathbf{E}_{\mx}}(\mX)
    \right\|_2
    \right].
\end{align*}
Since \(\mathcal{L}_{\mx,h}(\cdot)\geq0\) under Condition~\ref{con:P-K1}, and nonzero kernel weights imply \(\|\mathbf{v}_{\mx}^{\mathbf{E}_{\mx}}(\cdot)\|_2\leq h\), we have
\begin{align*}
    \hat{a}_{h,1}(\mx,\mathbf{E}_{\mx})
    &\leq
    h\hat{\mu}_{h,0}(\mx)
    =
    O_{\P}(h^{d+1}), \\
    \tilde{a}_{h,1}(\mx,\mathbf{E}_{\mx})
    &\leq
    h\tilde{\mu}_{h,0}(\mx)
    =
    O(h^{d+1}).
\end{align*}
Using \(\hat{\mu}_{h,0}(\mx)=O_{\P}(h^d)\), \(\tilde{\mu}_{h,0}(\mx)=O(h^d)\), \(\|\bm{\hat{\mu}}_{h,1}(\mx,\mathbf{E}_{\mx})\|_2=o_{\P}(h^{d+1})\), \(\|\bm{\tilde{\mu}}_{h,1}(\mx,\mathbf{E}_{\mx})\|_2=o(h^{d+1})\), \(\|\bm{\hat{\mu}}_{h,2}(\mx,\mathbf{E}_{\mx})^{-1}\|_2=O_{\P}(h^{-(d+2)})\), and \(\|\bm{\tilde{\mu}}_{h,2}(\mx,\mathbf{E}_{\mx})^{-1}\|_2=O(h^{-(d+2)})\), we obtain
\begin{align*}
    &\left|
    \hat{N}_{h,s}(\mx,y_1)-\hat{N}_{h,s}(\mx,y_2)
    \right|\\
    &\leq
    2D_{\mathbb{M}}d_{\mathbb{M}}(y_1,y_2)
    \left[
    \hat{\mu}_{h,0}(\mx)
    +
    s
    \left\|\bm{\hat{\mu}}_{h,1}(\mx,\mathbf{E}_{\mx})\right\|_2
    \left\|\bm{\hat{\mu}}_{h,2}(\mx,\mathbf{E}_{\mx})^{-1}\right\|_2
    \hat{a}_{h,1}(\mx,\mathbf{E}_{\mx})
    \right] \\
    &=
    O_{\P}(h^d)d_{\mathbb{M}}(y_1,y_2),
\end{align*}
and similarly,
\begin{align*}
    \left|
    \tilde{N}_{h,s}(\mx,y_1)-\tilde{N}_{h,s}(\mx,y_2)
    \right|
    =
    O(h^d)d_{\mathbb{M}}(y_1,y_2).
\end{align*}
Since $\hat{D}_{h,s}(\mx)^{-1}=O_{\P}(h^{-d})$ and $\tilde{D}_{h,s}(\mx)^{-1}=O(h^{-d})$, it follows that
\begin{align*}
    &\left|
    \left(\hat{M}_{h,s}(\mx,y_1)-\tilde{M}_{h,s}(\mx,y_1)\right)
    -
    \left(\hat{M}_{h,s}(\mx,y_2)-\tilde{M}_{h,s}(\mx,y_2)\right)
    \right| \\
    &\leq
    \hat{D}_{h,s}(\mx)^{-1}
    \left|
    \hat{N}_{h,s}(\mx,y_1)-\hat{N}_{h,s}(\mx,y_2)
    \right|
    +
    \tilde{D}_{h,s}(\mx)^{-1}
    \left|
    \tilde{N}_{h,s}(\mx,y_1)-\tilde{N}_{h,s}(\mx,y_2)
    \right| \\
    &\leq
    C_{n,s}d_{\mathbb{M}}(y_1,y_2),
\end{align*}
where $C_{n,s}=O_{\P}(1)$. Thus,
\begin{align} \label{eq:B.objective_equicontinuity_diff}
    \left|
    \left(\hat{M}_{h,s}(\mx,y_1)-\tilde{M}_{h,s}(\mx,y_1)\right)
    -
    \left(\hat{M}_{h,s}(\mx,y_2)-\tilde{M}_{h,s}(\mx,y_2)\right)
    \right|
    \leq
    C_{n,s}d_{\mathbb{M}}(y_1,y_2).
\end{align}

Finally, we upgrade the pointwise convergence to uniform convergence. Fix $\epsilon,\eta>0$. Since $C_{n,s}=O_{\P}(1)$, there exists $M>0$ such that
\begin{align*}
    \P(C_{n,s}>M)<\eta/2
\end{align*}
for all sufficiently large $n$. By Condition~\ref{con:M1}, $\mathbb{M}$ is totally bounded, so there exists a finite $\epsilon/(4M)$-cover $\{y_1,\ldots,y_N\}$ of $\mathbb{M}$. On the event $\{C_{n,s}\leq M\}$, for any $y\in\mathbb{M}$ choose $y_j$ with $d_{\mathbb{M}}(y,y_j)\leq\epsilon/(4M)$. Then \eqref{eq:B.objective_equicontinuity_diff} gives
\begin{align*}
    \left|
    \hat{M}_{h,s}(\mx,y)-\tilde{M}_{h,s}(\mx,y)
    \right|
    \leq
    \frac{\epsilon}{4}
    +
    \max_{1\leq j\leq N}
    \left|
    \hat{M}_{h,s}(\mx,y_j)-\tilde{M}_{h,s}(\mx,y_j)
    \right|.
\end{align*}
Since $N$ is fixed, the pointwise convergence already proved implies
\begin{align*}
    \lim_{n\to\infty}
    \P\left[
    \max_{1\leq j\leq N}
    \left|
    \hat{M}_{h,s}(\mx,y_j)-\tilde{M}_{h,s}(\mx,y_j)
    \right|
    >
    \frac{3\epsilon}{4}
    \right]
    =
    0.
\end{align*}
Therefore, for all sufficiently large $n$,
\begin{align*}
    &\P\left[
    \sup_{y\in\mathbb{M}}
    \left|
    \hat{M}_{h,s}(\mx,y)-\tilde{M}_{h,s}(\mx,y)
    \right|
    >
    \epsilon
    \right] \\
    &\leq
    \P(C_{n,s}>M)
    +
    \P\left[
    \max_{1\leq j\leq N}
    \left|
    \hat{M}_{h,s}(\mx,y_j)-\tilde{M}_{h,s}(\mx,y_j)
    \right|
    >
    \frac{3\epsilon}{4}
    \right].
\end{align*}
The first term is smaller than $\eta/2$ for all sufficiently large $n$, and the second term converges to zero by the pointwise convergence over the finite cover. Since $\eta>0$ is arbitrary,
\begin{align*}
    \sup_{y\in\mathbb{M}}
    \left|
    \hat{M}_{h,s}(\mx,y)-\tilde{M}_{h,s}(\mx,y)
    \right|
    =
    o_{\P}(1).
\end{align*}
\end{proof}

\begin{lemma}[Pointwise convergence of empirical local minimizers] \label{lemma:B.pointwise_empirical_minimizer}
Assume Conditions~\ref{con:P-K1}, \ref{con:P-B1}, \ref{con:P-D1}, \ref{con:P-D2}, \ref{con:M1}, and~\ref{con:P-M2}, and suppose that $h\to0$ as $n\to\infty$. Then
\begin{align*}
    d_{\mathbb{M}}\left(\hat{m}_{h,s}(\mx),\tilde{m}_{h,s}(\mx)\right)=o_{\P}(1),
    \quad s\in\{0,1\}.
\end{align*}
\end{lemma}

\begin{proof}[Proof of \Cref{lemma:B.pointwise_empirical_minimizer}]
The proof follows the same stochastic argmin-continuity scheme as Lemma B.9 of \cite{Im et al. (2025)}, adapted to the present localized objectives. The only additional point is that the empirical local linear quantities are considered on the event where the empirical second local moment matrix is invertible and the empirical denominator is positive; by \Cref{lemma:B.pointwise_empirical_moments,lemma:B.pointwise_denominators}, this event has probability tending to one and therefore does not affect convergence in probability.

Fix $s\in\{0,1\}$ and $\epsilon>0$. By the localized separation condition in Condition~\ref{con:P-M2}, there exist $\eta_{\epsilon}>0$ and $h_{\epsilon}>0$ such that, for all $h<h_{\epsilon}$,
\begin{align*}
    \inf_{y\in\mathbb{M}:d_{\mathbb{M}}\left(y,\tilde{m}_{h,s}(\mx)\right)>\epsilon}
    \left[
    \tilde{M}_{h,s}(\mx,y)
    -
    \tilde{M}_{h,s}\left(\mx,\tilde{m}_{h,s}(\mx)\right)
    \right]
    \geq
    \eta_{\epsilon}.
\end{align*}
Let
\begin{align*}
    \Delta_{h,s}^{\mathrm{emp}}
    :=
    \sup_{y\in\mathbb{M}}
    \left|
    \hat{M}_{h,s}(\mx,y)-\tilde{M}_{h,s}(\mx,y)
    \right|.
\end{align*}
By \Cref{lemma:B.pointwise_empirical_objective}, $\Delta_{h,s}^{\mathrm{emp}}=o_{\P}(1)$. Since $\hat{m}_{h,s}(\mx)$ minimizes $\hat{M}_{h,s}(\mx,\cdot)$ on the event where the empirical objective is well-defined,
\begin{align*}
    \tilde{M}_{h,s}\left(\mx,\hat{m}_{h,s}(\mx)\right)
    -
    \tilde{M}_{h,s}\left(\mx,\tilde{m}_{h,s}(\mx)\right)
    &\leq
    \left|
    \tilde{M}_{h,s}\left(\mx,\hat{m}_{h,s}(\mx)\right)
    -
    \hat{M}_{h,s}\left(\mx,\hat{m}_{h,s}(\mx)\right)
    \right| \\
    &\quad+
    \hat{M}_{h,s}\left(\mx,\hat{m}_{h,s}(\mx)\right)
    -
    \hat{M}_{h,s}\left(\mx,\tilde{m}_{h,s}(\mx)\right) \\
    &\quad+
    \left|
    \hat{M}_{h,s}\left(\mx,\tilde{m}_{h,s}(\mx)\right)
    -
    \tilde{M}_{h,s}\left(\mx,\tilde{m}_{h,s}(\mx)\right)
    \right| \\
    &\leq
    2\Delta_{h,s}^{\mathrm{emp}}.
\end{align*}
Therefore, for all sufficiently small $h$,
\begin{align*}
    \P\left(
    d_{\mathbb{M}}\left(\hat{m}_{h,s}(\mx),\tilde{m}_{h,s}(\mx)\right)>\epsilon
    \right)
    \leq
    \P\left(
    2\Delta_{h,s}^{\mathrm{emp}}\geq\eta_{\epsilon}
    \right)
    +
    o(1).
\end{align*}
Since $\Delta_{h,s}^{\mathrm{emp}}=o_{\P}(1)$, the right-hand side tends to zero. Hence
\begin{align*}
    d_{\mathbb{M}}\left(\hat{m}_{h,s}(\mx),\tilde{m}_{h,s}(\mx)\right)=o_{\P}(1),
    \quad s\in\{0,1\}.
\end{align*}
This completes the proof.
\end{proof}

\begin{proof}[Proof of \Cref{thm:pointwise_consistency}]
Fix $s\in\{0,1\}$. By the triangle inequality,
\begin{align*}
    d_{\mathbb{M}}\left(\hat{m}_{h,s}(\mx),m_{\oplus}(\mx)\right)
    \leq
    d_{\mathbb{M}}\left(\hat{m}_{h,s}(\mx),\tilde{m}_{h,s}(\mx)\right)
    +
    d_{\mathbb{M}}\left(\tilde{m}_{h,s}(\mx),m_{\oplus}(\mx)\right).
\end{align*}
The first term is $o_{\P}(1)$ by \Cref{lemma:B.pointwise_empirical_minimizer}, and the second term is $o(1)$ by \Cref{lemma:B.pointwise_oracle_minimizer}. Hence
\begin{align*}
    d_{\mathbb{M}}\left(\hat{m}_{h,s}(\mx),m_{\oplus}(\mx)\right)=o_{\P}(1),
    \quad s\in\{0,1\}.
\end{align*}
\end{proof}

\section{Proof of Pointwise Convergence Rate} \label{app:pointwise_rate}
\setcounter{equation}{0}
\renewcommand{\theequation}{C.\arabic{equation}}
\renewcommand{\theHequation}{C.\arabic{equation}}

In this section, we provide the proof of \Cref{thm:pointwise_rate}. Throughout this section, we fix $\mx\in\mathcal{M}$. We also fix the pointwise normal-neighborhood radius $\rho_{\mx}\in(0,i(\mx))$ defined in the main text, and an ordered orthonormal basis $\mathbf{E}_{\mx}\in\mathcal{E}_{\mx}$ of $T_{\mx}\mathcal M$. The map $\mathbf{v}_{\mx}^{\mathbf{E}_{\mx}}$ denotes the tangent-coordinate map defined in \eqref{eq:sec3.tangent_coordinate_map}, restricted throughout this section to $B_{\mathcal M}(\mx,\rho_{\mx})$. For scalar functions defined on Euclidean coordinate balls, $D$ and $D^2$ denote the Euclidean gradient, identified with a column vector, and the Euclidean Hessian, respectively.

\begin{lemma}[Pointwise first-order Taylor remainder for the design density] \label{lemma:C.pointwise_taylor_remainder_f}
Assume Condition~\ref{con:P-D3}. Then for every $\mz\in B_{\mathcal M}(\mx,\rho_{\mx})$,
\begin{align*}
    \left|
    f(\mz)
    -
    f(\mx)
    -
    \mathbf{v}_{\mx}^{\mathbf{E}_{\mx}}(\mz)^{\top}
    \bm{\beta}_{f}(\mx)
    \right|
    \leq
    \frac{1}{2}d_{\mathcal{M}}^2(\mx,\mz)
    \sup_{u\in B_{\mathcal M}(\mx,\rho_{\mx})}
    \left\|\nabla^2 f(u)\right\|_{\mathrm{op}},
\end{align*}
where $\bm{\beta}_{f}(\mx):=\bm{\Phi}_{\mathbf{E}_{\mx}}\left(\nabla f(\mx)\right)$.
\end{lemma}

\begin{proof}[Proof of \Cref{lemma:C.pointwise_taylor_remainder_f}]
Fix $\mz\in B_{\mathcal M}(\mx,\rho_{\mx})$. Let $\gamma:[0,1]\to\mathcal M$ be the unique minimizing geodesic from $\mx$ to $\mz$, parametrized by
\begin{align*}
    \gamma(t):=\Exp_{\mx}\left(t\Log_{\mx}(\mz)\right),
    \quad t\in[0,1].
\end{align*}
Since $\rho_{\mx}<i(\mx)$ and $\mz\in B_{\mathcal M}(\mx,\rho_{\mx})$, this geodesic is well-defined and satisfies $\gamma(t)\in B_{\mathcal M}(\mx,\rho_{\mx})$ for every $t\in[0,1]$. Define $\phi(t):=f(\gamma(t))$ for $t\in[0,1]$. Taylor's formula with integral remainder gives
\begin{align*}
    \phi(1)
    =
    \phi(0)
    +
    \phi'(0)
    +
    \int_0^1(1-t)\phi''(t)\,\dd t.
\end{align*}
By the chain rule,
\begin{align*}
    \phi'(0)
    =
    \left\langle
    \nabla f(\mx),\Log_{\mx}(\mz)
    \right\rangle_{\mx}
    =
    \left(\mathbf{v}_{\mx}^{\mathbf{E}_{\mx}}(\mz)\right)^{\top}
    \bm{\beta}_{f}(\mx).
\end{align*}
Since $\gamma$ is a geodesic,
\begin{align*}
    \phi''(t)
    =
    \nabla^2 f(\gamma(t))
    \left(\dot{\gamma}(t),\dot{\gamma}(t)\right),
    \quad t\in[0,1],
\end{align*}
and $\|\dot{\gamma}(t)\|_{\gamma(t)}=d_{\mathcal M}(\mx,\mz)$ for every $t\in[0,1]$. Therefore,
\begin{align*}
    &
    \left|
    f(\mz)
    -
    f(\mx)
    -
    \left(\mathbf{v}_{\mx}^{\mathbf{E}_{\mx}}(\mz)\right)^{\top}
    \bm{\beta}_{f}(\mx)
    \right| \\
    &\leq
    \int_0^1(1-t)
    \left|
    \nabla^2 f(\gamma(t))
    \left(\dot{\gamma}(t),\dot{\gamma}(t)\right)
    \right|\,\dd t \\
    &\leq
    \frac{1}{2}d_{\mathcal{M}}^2(\mx,\mz)
    \sup_{u\in B_{\mathcal M}(\mx,\rho_{\mx})}
    \left\|\nabla^2 f(u)\right\|_{\mathrm{op}}.
\end{align*}
This completes the proof.
\end{proof}

\begin{lemma}[Pointwise Taylor bound for $f g_\omega$] \label{lemma:C.pointwise_taylor_remainder_fg}
Assume Conditions~\ref{con:P-D1}, \ref{con:P-D2}, \ref{con:P-D3}, and~\ref{con:P-D4}. Then
\begin{align*}
    \sup_{\omega\in\mathbb{M}}
    \sup_{u\in B_{\mathcal M}(\mx,\rho_{\mx})}
    \left\|\nabla(f\cdot g_{\omega})(u)\right\|_u
    <
    \infty,
    \quad
    \sup_{\omega\in\mathbb{M}}
    \sup_{u\in B_{\mathcal M}(\mx,\rho_{\mx})}
    \left\|\nabla^2(f\cdot g_{\omega})(u)\right\|_{\mathrm{op}}
    <
    \infty.
\end{align*}
Moreover, for every $\mz\in B_{\mathcal M}(\mx,\rho_{\mx})$,
\begin{align*}
    &
    \sup_{\omega\in\mathbb{M}}
    \left|
    (f\cdot g_{\omega})(\mz)
    -
    (f\cdot g_{\omega})(\mx)
    -
    \left(\mathbf{v}_{\mx}^{\mathbf{E}_{\mx}}(\mz)\right)^{\top}
    \bm{\beta}_{f\cdot g_{\omega}}(\mx)
    \right| \\
    &\leq
    \frac{1}{2}d_{\mathcal{M}}^2(\mx,\mz)
    \sup_{\omega\in\mathbb{M}}
    \sup_{u\in B_{\mathcal M}(\mx,\rho_{\mx})}
    \left\|\nabla^2(f\cdot g_{\omega})(u)\right\|_{\mathrm{op}},
\end{align*}
where $\bm{\beta}_{f\cdot g_{\omega}}(\mx):=\bm{\Phi}_{\mathbf{E}_{\mx}}\left(\nabla(f\cdot g_{\omega})(\mx)\right)$.
\end{lemma}

\begin{proof}[Proof of \Cref{lemma:C.pointwise_taylor_remainder_fg}]
We first verify the uniform boundedness of the first- and second-derivative families. Fix $u\in B_{\mathcal M}(\mx,\rho_{\mx})$ and $\omega\in\mathbb{M}$. For any $\mathbf{a}\in T_u\mathcal M$, the covariant product rule gives
\begin{align*}
    \left\langle
    \nabla(f\cdot g_{\omega})(u),\mathbf{a}
    \right\rangle_u
    =
    g_{\omega}(u)
    \left\langle \nabla f(u),\mathbf{a}\right\rangle_u
    +
    f(u)
    \left\langle \nabla g_{\omega}(u),\mathbf{a}\right\rangle_u.
\end{align*}
Taking the supremum over $\mathbf{a}\in T_u\mathcal M$ with $\|\mathbf{a}\|_u\leq1$ yields
\begin{align*}
    \left\|
    \nabla(f\cdot g_{\omega})(u)
    \right\|_u
    \leq
    |g_{\omega}(u)|\|\nabla f(u)\|_u
    +
    |f(u)|\|\nabla g_{\omega}(u)\|_u.
\end{align*}
Condition~\ref{con:P-D2} gives a uniform local bound for $g_{\omega}$, Condition~\ref{con:P-D1} gives a local upper bound for $f$, Condition~\ref{con:P-D3} gives a local bound for $\nabla f$, and Condition~\ref{con:P-D4} gives a uniform local bound for $\nabla g_{\omega}$. Therefore,
\begin{align*}
    \sup_{\omega\in\mathbb{M}}
    \sup_{u\in B_{\mathcal M}(\mx,\rho_{\mx})}
    \left\|
    \nabla(f\cdot g_{\omega})(u)
    \right\|_u
    <\infty.
\end{align*}

Next, for any $\mathbf{a},\mathbf{b}\in T_u\mathcal M$, the covariant product rule gives
\begin{align*}
    \nabla^2(f\cdot g_{\omega})(u)\left(\mathbf{a},\mathbf{b}\right)
    &=
    g_{\omega}(u)\nabla^2 f(u)\left(\mathbf{a},\mathbf{b}\right)
    +
    f(u)\nabla^2 g_{\omega}(u)\left(\mathbf{a},\mathbf{b}\right) \\
    &\quad+
    \left\langle \nabla f(u),\mathbf{a}\right\rangle_u
    \left\langle \nabla g_{\omega}(u),\mathbf{b}\right\rangle_u
    +
    \left\langle \nabla f(u),\mathbf{b}\right\rangle_u
    \left\langle \nabla g_{\omega}(u),\mathbf{a}\right\rangle_u.
\end{align*}
Taking the supremum over $\mathbf{a},\mathbf{b}\in T_u\mathcal M$ with $\|\mathbf{a}\|_u\leq1$ and $\|\mathbf{b}\|_u\leq1$ yields
\begin{align*}
    \left\|\nabla^2(f\cdot g_{\omega})(u)\right\|_{\mathrm{op}}
    &\leq
    |g_{\omega}(u)|\left\|\nabla^2 f(u)\right\|_{\mathrm{op}}
    +
    |f(u)|\left\|\nabla^2 g_{\omega}(u)\right\|_{\mathrm{op}} \\
    &\quad+
    2\left\|\nabla f(u)\right\|_u
    \left\|\nabla g_{\omega}(u)\right\|_u.
\end{align*}
Condition~\ref{con:P-D2} gives a uniform local bound for $g_{\omega}$, Condition~\ref{con:P-D1} gives a local upper bound for $f$, Condition~\ref{con:P-D3} gives local bounds for $\nabla f$ and $\nabla^2 f$, and Condition~\ref{con:P-D4} gives uniform local bounds for $\nabla g_{\omega}$ and $\nabla^2 g_{\omega}$. Therefore,
\begin{align*}
    \sup_{\omega\in\mathbb{M}}
    \sup_{u\in B_{\mathcal M}(\mx,\rho_{\mx})}
    \left\|\nabla^2(f\cdot g_{\omega})(u)\right\|_{\mathrm{op}}
    <
    \infty.
\end{align*}

We now prove the Taylor bound. Fix $\mz\in B_{\mathcal M}(\mx,\rho_{\mx})$. Let $\gamma:[0,1]\to\mathcal M$ be the unique minimizing geodesic from $\mx$ to $\mz$, parametrized by
\begin{align*}
    \gamma(t)
    :=
    \Exp_{\mx}\left(t\Log_{\mx}(\mz)\right),
    \quad t\in[0,1].
\end{align*}
Since $\rho_{\mx}<i(\mx)$ and $\mz\in B_{\mathcal M}(\mx,\rho_{\mx})$, this geodesic is well-defined and remains in $B_{\mathcal M}(\mx,\rho_{\mx})$. For each $\omega\in\mathbb{M}$, define $\phi_{\omega}(t):=(f\cdot g_{\omega})(\gamma(t))$ for every $t\in[0,1]$. Taylor's formula with integral remainder gives
\begin{align*}
    \phi_{\omega}(1)
    =
    \phi_{\omega}(0)
    +
    \phi_{\omega}'(0)
    +
    \int_0^1(1-t)\phi_{\omega}''(t)\,\dd t.
\end{align*}
By the chain rule,
\begin{align*}
    \phi_{\omega}'(0)
    =
    \left\langle
    \nabla(f\cdot g_{\omega})(\mx),\Log_{\mx}(\mz)
    \right\rangle_{\mx}
    =
    \left(\mathbf{v}_{\mx}^{\mathbf{E}_{\mx}}(\mz)\right)^{\top}
    \bm{\beta}_{f\cdot g_{\omega}}(\mx).
\end{align*}
Since $\gamma$ is a geodesic,
\begin{align*}
    \phi_{\omega}''(t)
    =
    \nabla^2(f\cdot g_{\omega})(\gamma(t))
    \left(\dot{\gamma}(t),\dot{\gamma}(t)\right),
    \quad t\in[0,1],
\end{align*}
and $\|\dot{\gamma}(t)\|_{\gamma(t)}=d_{\mathcal M}(\mx,\mz)$ for every $t\in[0,1]$. Hence, for each $\omega\in\mathbb{M}$,
\begin{align*}
    &
    \left|
    (f\cdot g_{\omega})(\mz)
    -
    (f\cdot g_{\omega})(\mx)
    -
    \left(\mathbf{v}_{\mx}^{\mathbf{E}_{\mx}}(\mz)\right)^{\top}
    \bm{\beta}_{f\cdot g_{\omega}}(\mx)
    \right| \leq
    \frac{1}{2}d_{\mathcal{M}}^2(\mx,\mz)
    \sup_{u\in B_{\mathcal M}(\mx,\rho_{\mx})}
    \left\|\nabla^2(f\cdot g_{\omega})(u)\right\|_{\mathrm{op}}.
\end{align*}
Taking the supremum over $\omega\in\mathbb{M}$ proves the displayed Taylor bound.
\end{proof}

\begin{lemma}[Second-order expansion of scalar kernel moments] \label{lemma:C.pointwise_scalar_kernel_moments}
Assume Conditions~\ref{con:P-K1}, \ref{con:P-D1}, and~\ref{con:P-D3}, and suppose that $h\to0$ as $n\to\infty$. Then, for each $k\in\{1,2\}$,
\begin{align*}
    \E\left[\mathcal{L}_{\mx,h}\left(\mX\right)^k\right]
    -
    h^dA_{d-1}c_{d-1,k}f(\mx)
    =
    O(h^{d+2}).
\end{align*}
\end{lemma}

\begin{proof}[Proof of \Cref{lemma:C.pointwise_scalar_kernel_moments}]
For $\mathbf{u}\in B_{\mathbb{R}^d}(\mathbf{0}_d,\rho_{\mx})$, write
\begin{align*}
    \Exp_{\mx}^{\mathbf{E}_{\mx}}(\mathbf{u})
    :=
    \Exp_{\mx}\left(\bm{\Phi}_{\mathbf{E}_{\mx}}^{-1}(\mathbf{u})\right).
\end{align*}
Choose a fixed $r_{\mx}\in(0,\rho_{\mx})$. Since $h\to0$ as $n\to\infty$, it is enough to consider all sufficiently small $h<r_{\mx}$. For such $h$, the compact support of $K$ and the normal-coordinate change of variables give
\begin{align*}
    \E\left[\mathcal{L}_{\mx,h}\left(\mX\right)^k\right]
    &=
    \int_{\|\mathbf{u}\|_2\leq h}
    K\left(\frac{\|\mathbf{u}\|_2}{h}\right)^k
    \frac{
    f\left(\Exp_{\mx}^{\mathbf{E}_{\mx}}(\mathbf{u})\right)
    }
    {
    \theta_{\mx}\left(\Exp_{\mx}^{\mathbf{E}_{\mx}}(\mathbf{u})\right)^{k-1}
    }
    \,\dd\mathbf{u} \\
    &=
    h^d
    \int_{\|\mathbf{w}\|_2\leq1}
    K(\|\mathbf{w}\|_2)^k
    \varphi_k(h\mathbf{w})
    \,\dd\mathbf{w},
\end{align*}
where
\begin{align*}
    \varphi_k(\mathbf{u})
    :=
    \frac{
    f\left(\Exp_{\mx}^{\mathbf{E}_{\mx}}(\mathbf{u})\right)
    }
    {
    \theta_{\mx}\left(\Exp_{\mx}^{\mathbf{E}_{\mx}}(\mathbf{u})\right)^{k-1}
    },
    \quad
    \mathbf{u}\in B_{\mathbb{R}^d}(\mathbf{0}_d,\rho_{\mx}).
\end{align*}
We next verify that $\varphi_k$ has a bounded Euclidean Hessian on $\overline B_{\mathbb{R}^d}(\mathbf{0}_d,r_{\mx})$. Let
\begin{align*}
    \psi(\mathbf{u})
    :=
    f\left(\Exp_{\mx}^{\mathbf{E}_{\mx}}(\mathbf{u})\right),
    \quad
    \vartheta_k(\mathbf{u})
    :=
    \theta_{\mx}\left(\Exp_{\mx}^{\mathbf{E}_{\mx}}(\mathbf{u})\right)^{1-k},
    \quad \mathbf{u} \in B_{\mathbb{R}^d}(\mathbf{0}_d,r_{\mx}).
\end{align*}
Then $\varphi_k=\psi\vartheta_k$. Since $\overline B_{\mathbb{R}^d}(\mathbf{0}_d,r_{\mx})$ is compactly contained in the normal-coordinate domain, $\Exp_{\mx}^{\mathbf{E}_{\mx}}$ is smooth with bounded derivatives up to order two on this closed ball. By Conditions~\ref{con:P-D1} and~\ref{con:P-D3}, $f$, $\nabla f$, and $\nabla^2 f$ are bounded on $B_{\mathcal M}(\mx,\rho_{\mx})$. Hence, by the chain rule, $\psi=f\circ\Exp_{\mx}^{\mathbf{E}_{\mx}}$ has bounded Euclidean derivatives up to order two on $\overline B_{\mathbb{R}^d}(\mathbf{0}_d,r_{\mx})$. Moreover, the smooth positivity of the volume density implies that $\vartheta_k$ has bounded derivatives up to order two on the same closed ball. The Euclidean product rule for $\varphi_k=\psi\vartheta_k$ therefore yields
\begin{align*}
    C_{\varphi_k}
    :=
    \frac{1}{2}
    \sup_{\mathbf{u}\in \overline B_{\mathbb{R}^d}(\mathbf{0}_d,r_{\mx})}
    \left\|
    D^2\varphi_k(\mathbf{u})
    \right\|_{\mathrm{op}}
    <\infty.
\end{align*}
For every $\mathbf{u}\in \overline B_{\mathbb{R}^d}(\mathbf{0}_d,r_{\mx})$, Taylor's formula with integral remainder along the line segment $t\mathbf{u}$ for $t\in[0,1]$ gives
\begin{align*}
    \varphi_k(\mathbf{u})
    -
    \varphi_k(\mathbf{0}_d)
    -
    \mathbf{u}^{\top}D\varphi_k(\mathbf{0}_d)
    =
    \int_0^1(1-t)
    \mathbf{u}^{\top}
    D^2\varphi_k(t\mathbf{u})
    \mathbf{u}
    \,\dd t,
\end{align*}
and hence
\begin{align*}
    \left|
    \varphi_k(\mathbf{u})
    -
    \varphi_k(\mathbf{0}_d)
    -
    \mathbf{u}^{\top}D\varphi_k(\mathbf{0}_d)
    \right|
    \leq
    C_{\varphi_k}\|\mathbf{u}\|_2^2.
\end{align*}
Thus, for all sufficiently small $h<r_{\mx}$ and all $\|\mathbf{w}\|_2\leq1$,
\begin{align*}
    \varphi_k(h\mathbf{w})
    =
    \varphi_k(\mathbf{0}_d)
    +
    h\mathbf{w}^{\top}D\varphi_k(\mathbf{0}_d)
    +
    r_{k,h}(\mathbf{w}),
    \quad
    |r_{k,h}(\mathbf{w})|
    \leq
    C_{\varphi_k}h^2\|\mathbf{w}\|_2^2.
\end{align*}
Since $\theta_{\mx}(\mx)=1$, $\varphi_k(\mathbf{0}_d)=f(\mx)$. By radial symmetry,
\begin{align*}
    \int_{\|\mathbf{w}\|_2\leq1}
    \mathbf{w}K(\|\mathbf{w}\|_2)^k\,\dd\mathbf{w}
    =
    \mathbf{0}_d.
\end{align*}
Moreover, by \Cref{lemma:A.radial_kernel_moments},
\begin{align*}
    \int_{\|\mathbf{w}\|_2\leq1}
    K(\|\mathbf{w}\|_2)^k\,\dd\mathbf{w}
    =
    A_{d-1}c_{d-1,k},
\end{align*}
and
\begin{align*}
    \left|
    h^d
    \int_{\|\mathbf{w}\|_2\leq1}
    K(\|\mathbf{w}\|_2)^k
    r_{k,h}(\mathbf{w})
    \,\dd\mathbf{w}
    \right|
    &\leq
    C_{\varphi_k}h^{d+2}
    \int_{\|\mathbf{w}\|_2\leq1}
    K(\|\mathbf{w}\|_2)^k\|\mathbf{w}\|_2^2
    \,\dd\mathbf{w} \\
    &=
    C_{\varphi_k}h^{d+2}A_{d-1}c_{d+1,k}
    =
    O(h^{d+2}).
\end{align*}
Combining the preceding displays gives
\begin{align*}
    \E\left[\mathcal{L}_{\mx,h}\left(\mX\right)^k\right]
    =
    h^dA_{d-1}c_{d-1,k}f(\mx)
    +
    O(h^{d+2}),
\end{align*}
which proves the claim.
\end{proof}

\begin{lemma}[Second-order expansion of scalar kernel moments with conditional density ratios] \label{lemma:C.pointwise_conditional_kernel_moments}
Assume Conditions~\ref{con:P-K1}, \ref{con:P-D1}--\ref{con:P-D4}, and suppose that $h\to0$ as $n\to\infty$. Then for each $k\in\{1,2\}$,
\begin{align*}
    \sup_{\omega\in\mathbb{M}}
    \left|
    \E\left[
    \mathcal{L}_{\mx,h}\left(\mX\right)^k g_{\omega}(\mX)
    \right]
    -
    h^dA_{d-1}c_{d-1,k}(f\cdot g_{\omega})(\mx)
    \right|
    =
    O(h^{d+2}).
\end{align*}
\end{lemma}

\begin{proof}[Proof of \Cref{lemma:C.pointwise_conditional_kernel_moments}]
For $\mathbf{u}\in B_{\mathbb{R}^d}(\mathbf{0}_d,\rho_{\mx})$, write
\begin{align*}
    \Exp_{\mx}^{\mathbf{E}_{\mx}}(\mathbf{u})
    :=
    \Exp_{\mx}\left(\bm{\Phi}_{\mathbf{E}_{\mx}}^{-1}(\mathbf{u})\right).
\end{align*}
Choose a fixed $r_{\mx}\in(0,\rho_{\mx})$. Since $h\to0$ as $n\to\infty$, it is enough to consider all sufficiently small $h<r_{\mx}$. For such $h$, the compact support of $K$ and the normal-coordinate change of variables give
\begin{align*}
    \E\left[
    \mathcal{L}_{\mx,h}\left(\mX\right)^k g_{\omega}(\mX)
    \right]
    &=
    \int_{\|\mathbf{u}\|_2\leq h}
    K\left(\frac{\|\mathbf{u}\|_2}{h}\right)^k
    \frac{
    f\left(\Exp_{\mx}^{\mathbf{E}_{\mx}}(\mathbf{u})\right)
    g_{\omega}\left(\Exp_{\mx}^{\mathbf{E}_{\mx}}(\mathbf{u})\right)
    }
    {
    \theta_{\mx}\left(\Exp_{\mx}^{\mathbf{E}_{\mx}}(\mathbf{u})\right)^{k-1}
    }
    \,\dd\mathbf{u} \\
    &=
    h^d
    \int_{\|\mathbf{w}\|_2\leq1}
    K(\|\mathbf{w}\|_2)^k
    \varrho_{k,\omega}(h\mathbf{w})
    \,\dd\mathbf{w},
\end{align*}
where
\begin{align*}
    \varrho_{k,\omega}(\mathbf{u})
    :=
    \frac{
    f\left(\Exp_{\mx}^{\mathbf{E}_{\mx}}(\mathbf{u})\right)
    g_{\omega}\left(\Exp_{\mx}^{\mathbf{E}_{\mx}}(\mathbf{u})\right)
    }
    {
    \theta_{\mx}\left(\Exp_{\mx}^{\mathbf{E}_{\mx}}(\mathbf{u})\right)^{k-1}
    },
    \quad
    \mathbf{u}\in B_{\mathbb{R}^d}(\mathbf{0}_d,\rho_{\mx}).
\end{align*}

We next verify that the Euclidean Hessians of $\varrho_{k,\omega}$ are uniformly bounded over $\omega\in\mathbb{M}$ on $\overline B_{\mathbb{R}^d}(\mathbf{0}_d,r_{\mx})$. For each $\omega\in\mathbb{M}$, let
\begin{align*}
    \chi_{\omega}(\mathbf{u})
    &:=
    (f\cdot g_{\omega})\left(\Exp_{\mx}^{\mathbf{E}_{\mx}}(\mathbf{u})\right),
    \quad \vartheta_k(\mathbf{u})
    :=
    \theta_{\mx}\left(\Exp_{\mx}^{\mathbf{E}_{\mx}}(\mathbf{u})\right)^{1-k},
    \quad
    \mathbf{u}\in B_{\mathbb{R}^d}(\mathbf{0}_d,r_{\mx}).
\end{align*}
Then $\varrho_{k,\omega}=\chi_{\omega}\vartheta_k$. Since $\overline B_{\mathbb{R}^d}(\mathbf{0}_d,r_{\mx})$ is compactly contained in the normal-coordinate domain, the coordinate map $\Exp_{\mx}^{\mathbf{E}_{\mx}}$ and its derivatives up to order two are bounded on this closed ball. By Conditions~\ref{con:P-D1} and~\ref{con:P-D2},
\begin{align*}
    \sup_{\omega\in\mathbb{M}}
    \sup_{u\in B_{\mathcal M}(\mx,\rho_{\mx})}
    |(f\cdot g_{\omega})(u)|
    <
    \infty.
\end{align*}
Together with \Cref{lemma:C.pointwise_taylor_remainder_fg}, this gives
\begin{align*}
    \sup_{\omega\in\mathbb{M}}
    \sup_{u\in B_{\mathcal M}(\mx,\rho_{\mx})}
    \left(
    |(f\cdot g_{\omega})(u)|
    +
    \left\|\nabla(f\cdot g_{\omega})(u)\right\|_u
    +
    \left\|\nabla^2(f\cdot g_{\omega})(u)\right\|_{\mathrm{op}}
    \right)
    <
    \infty.
\end{align*}
Hence the chain rule implies that $\chi_{\omega}$ has Euclidean derivatives up to order two bounded uniformly over $\omega\in\mathbb{M}$ on $\overline B_{\mathbb{R}^d}(\mathbf{0}_d,r_{\mx})$. Moreover, the smooth positivity of the volume density implies that $\vartheta_k$ has bounded derivatives up to order two on the same closed ball. The Euclidean product rule for $\varrho_{k,\omega}=\chi_{\omega}\vartheta_k$ therefore yields
\begin{align*}
    C_{\varrho_k}
    :=
    \frac{1}{2}
    \sup_{\omega\in\mathbb{M}}
    \sup_{\mathbf{u}\in \overline B_{\mathbb{R}^d}(\mathbf{0}_d,r_{\mx})}
    \left\|
    D^2\varrho_{k,\omega}(\mathbf{u})
    \right\|_{\mathrm{op}}
    <\infty.
\end{align*}
For every $\omega\in\mathbb{M}$ and every $\mathbf{u}\in\overline B_{\mathbb{R}^d}(\mathbf{0}_d,r_{\mx})$, Taylor's formula with integral remainder along the line segment $t\mathbf{u}$ for $t\in[0,1]$ gives
\begin{align*}
    \varrho_{k,\omega}(\mathbf{u})
    -
    \varrho_{k,\omega}(\mathbf{0}_d)
    -
    \mathbf{u}^{\top}D\varrho_{k,\omega}(\mathbf{0}_d)
    =
    \int_0^1(1-t)
    \mathbf{u}^{\top}
    D^2\varrho_{k,\omega}(t\mathbf{u})
    \mathbf{u}
    \,\dd t,
\end{align*}
and hence
\begin{align*}
    \left|
    \varrho_{k,\omega}(\mathbf{u})
    -
    \varrho_{k,\omega}(\mathbf{0}_d)
    -
    \mathbf{u}^{\top}D\varrho_{k,\omega}(\mathbf{0}_d)
    \right|
    \leq
    C_{\varrho_k}\|\mathbf{u}\|_2^2.
\end{align*}
Thus, uniformly over $\omega\in\mathbb{M}$ and $\|\mathbf{w}\|_2\leq1$,
\begin{align*}
    \varrho_{k,\omega}(h\mathbf{w})
    =
    \varrho_{k,\omega}(\mathbf{0}_d)
    +
    h\mathbf{w}^{\top}D\varrho_{k,\omega}(\mathbf{0}_d)
    +
    r_{k,\omega,h}(\mathbf{w}),
    \quad
    |r_{k,\omega,h}(\mathbf{w})|
    \leq
    C_{\varrho_k}h^2\|\mathbf{w}\|_2^2.
\end{align*}
Since $\theta_{\mx}(\mx)=1$, $\varrho_{k,\omega}(\mathbf{0}_d)=(f\cdot g_{\omega})(\mx)$. By radial symmetry,
\begin{align*}
    \int_{\|\mathbf{w}\|_2\leq1}
    \mathbf{w}K(\|\mathbf{w}\|_2)^k\,\dd\mathbf{w}
    =
    \mathbf{0}_d.
\end{align*}
Moreover, by \Cref{lemma:A.radial_kernel_moments},
\begin{align*}
    \int_{\|\mathbf{w}\|_2\leq1}
    K(\|\mathbf{w}\|_2)^k\,\dd\mathbf{w}
    =
    A_{d-1}c_{d-1,k},
\end{align*}
and
\begin{align*}
    \sup_{\omega\in\mathbb{M}}
    \left|
    h^d
    \int_{\|\mathbf{w}\|_2\leq1}
    K(\|\mathbf{w}\|_2)^k
    r_{k,\omega,h}(\mathbf{w})
    \,\dd\mathbf{w}
    \right|
    &\leq
    C_{\varrho_k}h^{d+2}
    \int_{\|\mathbf{w}\|_2\leq1}
    K(\|\mathbf{w}\|_2)^k\|\mathbf{w}\|_2^2
    \,\dd\mathbf{w} \\
    &=
    C_{\varrho_k}h^{d+2}A_{d-1}c_{d+1,k}
    =
    O(h^{d+2}).
\end{align*}
Combining the preceding displays gives
\begin{align*}
    \sup_{\omega\in\mathbb{M}}
    \left|
    \E\left[
    \mathcal{L}_{\mx,h}\left(\mX\right)^k g_{\omega}(\mX)
    \right]
    -
    h^dA_{d-1}c_{d-1,k}(f\cdot g_{\omega})(\mx)
    \right|
    =
    O(h^{d+2}),
\end{align*}
which proves the claim.
\end{proof}

\begin{lemma} \label{lemma:C.pointwise_refined_moments}
Assume Conditions~\ref{con:P-K1}, \ref{con:P-D1}, and~\ref{con:P-D3}, and suppose that $h\to0$ as $n\to\infty$. Then
\begin{align} \label{eq:C.refined_moment_expansion}
\begin{split} 
    \left\|
    \bm{\tilde{\mu}}_{h,1}(\mx,\mathbf{E}_{\mx})
    -
    h^{d+2}\frac{A_{d-1}c_{d+1,1}}{d}\bm{\beta}_{f}(\mx)
    \right\|_2
    &=
    O(h^{d+3}),  \\
    \left\|
    \bm{\tilde{\mu}}_{h,2}(\mx,\mathbf{E}_{\mx})
    -
    h^{d+2}\frac{A_{d-1}c_{d+1,1}f(\mx)}{d}\mI_d
    \right\|_2
    &=
    O(h^{d+4}).
\end{split}
\end{align}
Moreover, for all sufficiently small $h$, $\bm{\tilde{\mu}}_{h,2}(\mx,\mathbf{E}_{\mx})$ is invertible and
\begin{align}
    \left\|
    \bm{\tilde{\mu}}_{h,2}(\mx,\mathbf{E}_{\mx})^{-1}
    -
    h^{-(d+2)}
    \frac{d}{A_{d-1}c_{d+1,1}f(\mx)}
    \mI_d
    \right\|_2
    =
    O(h^{-d}). \label{eq:C.refined_inverse_bounds}
\end{align}
In particular,
\begin{align} \label{eq:C.refined_moment_orders}
    \left\|
    \bm{\tilde{\mu}}_{h,1}(\mx,\mathbf{E}_{\mx})
    \right\|_2
    =
    O(h^{d+2}),
    \quad
    \left\|
    \bm{\tilde{\mu}}_{h,2}(\mx,\mathbf{E}_{\mx})^{-1}
    \right\|_2
    =
    O(h^{-(d+2)}).
\end{align}
\end{lemma}

\begin{proof}[Proof of \Cref{lemma:C.pointwise_refined_moments}]
For $\mathbf{u}\in B_{\mathbb{R}^d}(\mathbf{0}_d,\rho_{\mx})$, write
\begin{align*}
    \Exp_{\mx}^{\mathbf{E}_{\mx}}(\mathbf{u})
    :=
    \Exp_{\mx}\left(\bm{\Phi}_{\mathbf{E}_{\mx}}^{-1}(\mathbf{u})\right).
\end{align*}
For all sufficiently small $h<\rho_{\mx}$, the volume-density correction in $\mathcal L_{\mx,h}$ and the normal-coordinate change of variables give
\begin{align*}
    \bm{\tilde{\mu}}_{h,1}(\mx,\mathbf{E}_{\mx})
    &=
    h^{d+1}
    \int_{\|\mathbf{w}\|_2\leq1}
    \mathbf{w}
    K(\|\mathbf{w}\|_2)
    f\left(\Exp_{\mx}^{\mathbf{E}_{\mx}}(h\mathbf{w})\right)
    \,\dd\mathbf{w}, \\
    \bm{\tilde{\mu}}_{h,2}(\mx,\mathbf{E}_{\mx})
    &=
    h^{d+2}
    \int_{\|\mathbf{w}\|_2\leq1}
    \mathbf{w}\mathbf{w}^{\top}
    K(\|\mathbf{w}\|_2)
    f\left(\Exp_{\mx}^{\mathbf{E}_{\mx}}(h\mathbf{w})\right)
    \,\dd\mathbf{w}.
\end{align*}
By \Cref{lemma:C.pointwise_taylor_remainder_f}, applied with
\begin{align*}
    \mz=\Exp_{\mx}^{\mathbf{E}_{\mx}}(h\mathbf{w}),
\end{align*}
we have $\mathbf{v}_{\mx}^{\mathbf{E}_{\mx}}(\mz)=h\mathbf{w}$ and $d_{\mathcal M}(\mx,\mz)=h\|\mathbf{w}\|_2$ whenever $\|\mathbf{w}\|_2\leq1$ and $h<\rho_{\mx}$. Hence, uniformly over $\|\mathbf{w}\|_2\leq1$,
\begin{align*}
    f\left(\Exp_{\mx}^{\mathbf{E}_{\mx}}(h\mathbf{w})\right)
    =
    f(\mx)
    +
    h\mathbf{w}^{\top}\bm{\beta}_{f}(\mx)
    +
    R_h(\mathbf{w}),
    \quad
    \sup_{\|\mathbf{w}\|_2\leq1}|R_h(\mathbf{w})|
    =
    O(h^2).
\end{align*}
Substituting this expansion into $\bm{\tilde{\mu}}_{h,1}(\mx,\mathbf{E}_{\mx})$ yields
\begin{align*}
    \bm{\tilde{\mu}}_{h,1}(\mx,\mathbf{E}_{\mx})
    &=
    h^{d+1}f(\mx)
    \int_{\|\mathbf{w}\|_2\leq1}
    \mathbf{w}K(\|\mathbf{w}\|_2)
    \,\dd\mathbf{w} \\
    &\quad+
    h^{d+2}
    \int_{\|\mathbf{w}\|_2\leq1}
    \mathbf{w}\mathbf{w}^{\top}K(\|\mathbf{w}\|_2)
    \,\dd\mathbf{w}\,
    \bm{\beta}_{f}(\mx) \\
    &\quad+
    h^{d+1}
    \int_{\|\mathbf{w}\|_2\leq1}
    \mathbf{w}K(\|\mathbf{w}\|_2)R_h(\mathbf{w})
    \,\dd\mathbf{w}.
\end{align*}
The first integral vanishes by radial symmetry, and \Cref{lemma:A.radial_kernel_moments} gives
\begin{align*}
    \int_{\|\mathbf{w}\|_2\leq1}
    \mathbf{w}\mathbf{w}^{\top}K(\|\mathbf{w}\|_2)
    \,\dd\mathbf{w}
    =
    \frac{A_{d-1}c_{d+1,1}}{d}\mI_d.
\end{align*}
Moreover,
\begin{align*}
    \left\|
    h^{d+1}
    \int_{\|\mathbf{w}\|_2\leq1}
    \mathbf{w}K(\|\mathbf{w}\|_2)R_h(\mathbf{w})
    \,\dd\mathbf{w}
    \right\|_2
    &\leq
    h^{d+1}
    \sup_{\|\mathbf{w}\|_2\leq1}|R_h(\mathbf{w})|
    \int_{\|\mathbf{w}\|_2\leq1}
    \|\mathbf{w}\|_2K(\|\mathbf{w}\|_2)
    \,\dd\mathbf{w} \\
    &=
    O(h^{d+3}).
\end{align*}
This proves \eqref{eq:C.refined_moment_expansion}.

Similarly,
\begin{align*}
    \bm{\tilde{\mu}}_{h,2}(\mx,\mathbf{E}_{\mx})
    &=
    h^{d+2}f(\mx)
    \int_{\|\mathbf{w}\|_2\leq1}
    \mathbf{w}\mathbf{w}^{\top}K(\|\mathbf{w}\|_2)
    \,\dd\mathbf{w} \\
    &\quad+
    h^{d+3}
    \int_{\|\mathbf{w}\|_2\leq1}
    \mathbf{w}\mathbf{w}^{\top}
    \left(\mathbf{w}^{\top}\bm{\beta}_{f}(\mx)\right)
    K(\|\mathbf{w}\|_2)
    \,\dd\mathbf{w} \\
    &\quad+
    h^{d+2}
    \int_{\|\mathbf{w}\|_2\leq1}
    \mathbf{w}\mathbf{w}^{\top}K(\|\mathbf{w}\|_2)R_h(\mathbf{w})
    \,\dd\mathbf{w}.
\end{align*}
The second integral vanishes componentwise by odd symmetry, and
\begin{align*}
    \left\|
    h^{d+2}
    \int_{\|\mathbf{w}\|_2\leq1}
    \mathbf{w}\mathbf{w}^{\top}K(\|\mathbf{w}\|_2)R_h(\mathbf{w})
    \,\dd\mathbf{w}
    \right\|_2
    &\leq
    h^{d+2}
    \sup_{\|\mathbf{w}\|_2\leq1}|R_h(\mathbf{w})|
    \int_{\|\mathbf{w}\|_2\leq1}
    \|\mathbf{w}\|_2^2K(\|\mathbf{w}\|_2)
    \,\dd\mathbf{w} \\
    &=
    O(h^{d+4}).
\end{align*}
Hence \eqref{eq:C.refined_moment_expansion} follows.

It remains to prove \eqref{eq:C.refined_inverse_bounds}. Let
\begin{align*}
    \mathbf{B}_h
    :=
    h^{d+2}
    \frac{A_{d-1}c_{d+1,1}f(\mx)}{d}
    \mI_d.
\end{align*}
Since $f(\mx)>0$ by Condition~\ref{con:P-D1}, \eqref{eq:C.refined_moment_expansion} and Weyl's inequality imply that $\bm{\tilde{\mu}}_{h,2}(\mx,\mathbf{E}_{\mx})$ is invertible for all sufficiently small $h$, with
\begin{align*}
    \left\|
    \bm{\tilde{\mu}}_{h,2}(\mx,\mathbf{E}_{\mx})^{-1}
    \right\|_2
    =
    O(h^{-(d+2)}),
    \quad
    \left\|
    \mathbf{B}_h^{-1}
    \right\|_2
    =
    O(h^{-(d+2)}).
\end{align*}
Using
\begin{align*}
    \bm{\tilde{\mu}}_{h,2}^{-1}-\mathbf{B}_h^{-1}
    =
    \bm{\tilde{\mu}}_{h,2}^{-1}
    \left(
    \mathbf{B}_h-\bm{\tilde{\mu}}_{h,2}
    \right)
    \mathbf{B}_h^{-1},
\end{align*}
where the common arguments $(\mx,\mathbf{E}_{\mx})$ are suppressed, we obtain
\begin{align*}
    \left\|
    \bm{\tilde{\mu}}_{h,2}(\mx,\mathbf{E}_{\mx})^{-1}
    -
    \mathbf{B}_h^{-1}
    \right\|_2
    &\leq
    O(h^{-(d+2)})
    O(h^{d+4})
    O(h^{-(d+2)}) \\
    &=
    O(h^{-d}).
\end{align*}
This proves \eqref{eq:C.refined_inverse_bounds}.

Finally, Condition~\ref{con:P-D3} implies
$\|\bm{\beta}_{f}(\mx)\|_2<\infty$, and Condition~\ref{con:P-D1}
implies $f(\mx)>0$. Therefore, \eqref{eq:C.refined_moment_expansion} gives
\begin{align*}
    \left\|
    \bm{\tilde{\mu}}_{h,1}(\mx,\mathbf{E}_{\mx})
    \right\|_2
    =
    O(h^{d+2}).
\end{align*}
Similarly, since $h^{-d}=O(h^{-(d+2)})$ as $h\downarrow0$, \eqref{eq:C.refined_inverse_bounds} gives
\begin{align*}
    \left\|
    \bm{\tilde{\mu}}_{h,2}(\mx,\mathbf{E}_{\mx})^{-1}
    \right\|_2
    =
    O(h^{-(d+2)}).
\end{align*}
This proves \eqref{eq:C.refined_moment_orders}.
\end{proof}

\begin{lemma} \label{lemma:C.pointwise_population_bias}
Assume Conditions~\ref{con:P-K1}, \ref{con:P-D1}--\ref{con:P-D4}, \ref{con:M1}, \ref{con:P-M2}, and~\ref{con:P-M3}, and suppose that $h\to0$ as $n\to\infty$. Then, for each $s\in\{0,1\}$,
\begin{align*}
    d_{\mathbb{M}}\left(\tilde{m}_{h,s}(\mx),m_{\oplus}(\mx)\right)^{\beta_{\oplus,\mx}-1}
    =
    O(h^2),
\end{align*}
where $\beta_{\oplus,\mx}\in(1,\infty)$ is the margin exponent in Condition~\ref{con:P-M3}.
\end{lemma}

\begin{proof}[Proof of \Cref{lemma:C.pointwise_population_bias}]
Fix $s\in\{0,1\}$. By \Cref{lemma:B.pointwise_oracle_minimizer}, $d_{\mathbb{M}}\left(\tilde{m}_{h,s}(\mx),m_{\oplus}(\mx)\right)=o(1)$. Hence, for all sufficiently small $h$, $\tilde{m}_{h,s}(\mx)$ lies in the neighborhood on which the margin condition in Condition~\ref{con:P-M3} applies. Therefore,
\begin{align} \label{eq:C.margin_condition_bound}
    C_{\oplus,\mx}
    d_{\mathbb{M}}\left(\tilde{m}_{h,s}(\mx),m_{\oplus}(\mx)\right)^{\beta_{\oplus,\mx}}
    \leq
    M_{\oplus}\left(\mx,\tilde{m}_{h,s}(\mx)\right)
    -
    M_{\oplus}\left(\mx,m_{\oplus}(\mx)\right).
\end{align}
Define $\tilde{\tau}_{h,0}(\mx,\omega)$ and $\bm{\tilde{\tau}}_{h,1}(\mx,\mathbf{E}_{\mx},\omega)$ by \eqref{eq:B.population_auxiliary_defs}. Write
\begin{align*}
    \tilde{g}_{h,s}(\mx,\omega)
    &:=
    \frac{
    \tilde{\tau}_{h,0}(\mx,\omega)
    -
    s\,\bm{\tilde{\mu}}_{h,1}(\mx,\mathbf{E}_{\mx})^{\top}
    \bm{\tilde{\mu}}_{h,2}(\mx,\mathbf{E}_{\mx})^{-1}
    \bm{\tilde{\tau}}_{h,1}(\mx,\mathbf{E}_{\mx},\omega)
    }
    {
    \tilde{D}_{h,s}(\mx)
    },
\end{align*}
where $\tilde{D}_{h,0}(\mx)$ and $\tilde{D}_{h,1}(\mx)$ are defined by \eqref{eq:B.population_ND_defs}. With this notation,
\begin{align*}
    \tilde{M}_{h,s}(\mx,y)
    =
    \int_{\mathbb{M}}
    d_{\mathbb{M}}^2(y,\omega)\tilde{g}_{h,s}(\mx,\omega)
    \,\dd P_Y(\omega),
    \quad
    M_{\oplus}(\mx,y)
    =
    \int_{\mathbb{M}}
    d_{\mathbb{M}}^2(y,\omega)g_{\omega}(\mx)
    \,\dd P_Y(\omega).
\end{align*}
Since $\tilde{m}_{h,s}(\mx)$ minimizes $\tilde{M}_{h,s}(\mx,\cdot)$,
\begin{align*}
    \tilde{M}_{h,s}\left(\mx,m_{\oplus}(\mx)\right)
    -
    \tilde{M}_{h,s}\left(\mx,\tilde{m}_{h,s}(\mx)\right)
    \geq
    0.
\end{align*}
Adding this nonnegative term to the right-hand side of \eqref{eq:C.margin_condition_bound} gives
\begin{align*}
    &
    C_{\oplus,\mx}
    d_{\mathbb{M}}\left(\tilde{m}_{h,s}(\mx),m_{\oplus}(\mx)\right)^{\beta_{\oplus,\mx}} \\
    &\leq
    M_{\oplus}\left(\mx,\tilde{m}_{h,s}(\mx)\right)
    -
    M_{\oplus}\left(\mx,m_{\oplus}(\mx)\right)
    +
    \tilde{M}_{h,s}\left(\mx,m_{\oplus}(\mx)\right)
    -
    \tilde{M}_{h,s}\left(\mx,\tilde{m}_{h,s}(\mx)\right) \\
    &=
    \int_{\mathbb{M}}
    \left[
    d_{\mathbb{M}}^2\left(m_{\oplus}(\mx),\omega\right)
    -
    d_{\mathbb{M}}^2\left(\tilde{m}_{h,s}(\mx),\omega\right)
    \right]
    \left[
    \tilde{g}_{h,s}(\mx,\omega)-g_{\omega}(\mx)
    \right]
    \,\dd P_Y(\omega),
\end{align*}
where the integral representation is justified by Condition~\ref{con:M1}. By Condition~\ref{con:M1}, $D_{\mathbb{M}}$ defined by \eqref{eq:B.diameter_metric_space} is finite. Then
\begin{align} \label{eq:C.bias_integral_bound}
    C_{\oplus,\mx}
    d_{\mathbb{M}}\left(\tilde{m}_{h,s}(\mx),m_{\oplus}(\mx)\right)^{\beta_{\oplus,\mx}}
    \leq
    2D_{\mathbb{M}}
    d_{\mathbb{M}}\left(\tilde{m}_{h,s}(\mx),m_{\oplus}(\mx)\right)
    \sup_{\omega\in\mathbb{M}}
    \left|
    \tilde{g}_{h,s}(\mx,\omega)-g_{\omega}(\mx)
    \right|.
\end{align}

It remains to show that
\begin{align} \label{eq:C.interpolated_density_bias}
    \sup_{\omega\in\mathbb{M}}
    \left|
    \tilde{g}_{h,s}(\mx,\omega)-g_{\omega}(\mx)
    \right|
    =
    O(h^2).
\end{align}
For $s=0$, \Cref{lemma:C.pointwise_scalar_kernel_moments,lemma:C.pointwise_conditional_kernel_moments} with $k=1$ give
\begin{align} \label{eq:C.tau0_rate_ll}
\begin{split}
    \tilde{\tau}_{h,0}(\mx,\omega)
    &=
    h^d A_{d-1}c_{d-1,1} (f\cdot g_{\omega})(\mx)
    +
    s_{0,h}(\mx,\omega),\\
    \tilde{\mu}_{h,0}(\mx)
    &=
    h^d A_{d-1}c_{d-1,1} f(\mx)
    +
    r_{0,h}(\mx),
    \quad
    \omega \in \mathbb{M},
\end{split}
\end{align}
where 
\begin{align} \label{eq:C.tau0_rate_ll_bound}
    \sup_{\omega\in\mathbb{M}}|s_{0,h}(\mx,\omega)|
    =
    O(h^{d+2}), \quad |r_{0,h}(\mx)|
    =
    O(h^{d+2}).
\end{align}
Since $f(\mx)>0$ and $A_{d-1}c_{d-1,1}>0$ by Condition~\ref{con:P-K1}, for all sufficiently small $h$,
\begin{align*}
    \tilde{\mu}_{h,0}(\mx)
    =
    h^d A_{d-1}c_{d-1,1} f(\mx)\{1+O(h^2)\}
\end{align*}
and the denominator is bounded below by a positive constant multiple of $h^d$. Hence,
\begin{align*}
    \tilde{g}_{h,0}(\mx,\omega)-g_{\omega}(\mx)
    &=
    \frac{
    h^d A_{d-1}c_{d-1,1} f(\mx)g_{\omega}(\mx)+s_{0,h}(\mx,\omega)
    }{
    h^d A_{d-1}c_{d-1,1} f(\mx)+r_{0,h}(\mx)
    }
    -
    g_{\omega}(\mx) \\
    &=
    \frac{
    s_{0,h}(\mx,\omega)-g_{\omega}(\mx)r_{0,h}(\mx)
    }{
    h^d A_{d-1}c_{d-1,1} f(\mx)+r_{0,h}(\mx)
    }, \quad \omega\in\mathbb{M}.
\end{align*}
Condition~\ref{con:P-D2} gives $\sup_{\omega\in\mathbb{M}}g_{\omega}(\mx)<\infty$. Therefore,
\begin{align*}
    \sup_{\omega\in\mathbb{M}}
    \left|
    \tilde{g}_{h,0}(\mx,\omega)-g_{\omega}(\mx)
    \right|
    =
    O(h^2).
\end{align*}
For $s=1$, we claim that
\begin{align} \label{eq:C.tau1_rate_ll}
    \sup_{\omega\in\mathbb{M}}
    \left\|
    \bm{\tilde{\tau}}_{h,1}(\mx,\mathbf{E}_{\mx},\omega)
    \right\|_2
    =
    O(h^{d+2}).
\end{align}
For fixed $\omega\in\mathbb{M}$, the definition of $\bm{\tilde{\tau}}_{h,1}$ and the normal-coordinate change of variables give, for all sufficiently small $h<\rho_{\mx}$,
\begin{align*}
    \bm{\tilde{\tau}}_{h,1}(\mx,\mathbf{E}_{\mx},\omega)
    &=
    h^{d+1}
    \int_{\|\mathbf{w}\|_2\leq1}
    \mathbf{w}
    K(\|\mathbf{w}\|_2)
    (f\cdot g_{\omega})\left(\Exp_{\mx}^{\mathbf{E}_{\mx}}(h\mathbf{w})\right)
    \,\dd\mathbf{w}.
\end{align*}
For $\|\mathbf{w}\|_2\leq1$, define
\begin{align*}
    R_{h,\omega}(\mathbf{w})
    &:=
    (f\cdot g_{\omega})\left(\Exp_{\mx}^{\mathbf{E}_{\mx}}(h\mathbf{w})\right)
    -
    (f\cdot g_{\omega})(\mx)
    -
    h\mathbf{w}^{\top}\bm{\beta}_{f\cdot g_{\omega}}(\mx),
\end{align*}
where
\begin{align*}
    \bm{\beta}_{f\cdot g_{\omega}}(\mx)
    :=
    \bm{\Phi}_{\mathbf{E}_{\mx}}\left(\nabla(f\cdot g_{\omega})(\mx)\right).
\end{align*}
Since $\mathbf{E}_{\mx}$ is orthonormal,
\begin{align*}
    d_{\mathcal M}\left(\mx,\Exp_{\mx}^{\mathbf{E}_{\mx}}(h\mathbf{w})\right)
    =
    h\|\mathbf{w}\|_2
\end{align*}
for all sufficiently small $h$ and all $\|\mathbf{w}\|_2\leq1$. Applying the Taylor bound in \Cref{lemma:C.pointwise_taylor_remainder_fg} with $\mz=\Exp_{\mx}^{\mathbf{E}_{\mx}}(h\mathbf{w})$ yields
\begin{align*}
    \sup_{\omega\in\mathbb{M}}
    \sup_{\|\mathbf{w}\|_2\leq1}
    |R_{h,\omega}(\mathbf{w})|
    \leq
    \frac{1}{2}h^2
    \sup_{\omega\in\mathbb{M}}
    \sup_{u\in B_{\mathcal M}(\mx,\rho_{\mx})}
    \left\|\nabla^2(f\cdot g_{\omega})(u)\right\|_{\mathrm{op}}
    =
    O(h^2).
\end{align*}
Moreover, since $\bm{\Phi}_{\mathbf{E}_{\mx}}$ is an isometry from $T_{\mx}\mathcal M$ to $\mathbb{R}^d$, the first-derivative bound in \Cref{lemma:C.pointwise_taylor_remainder_fg} gives
\begin{align*}
    \sup_{\omega\in\mathbb{M}}
    \left\|
    \bm{\beta}_{f\cdot g_{\omega}}(\mx)
    \right\|_2
    =
    \sup_{\omega\in\mathbb{M}}
    \left\|
    \nabla(f\cdot g_{\omega})(\mx)
    \right\|_{\mx}
    <
    \infty.
\end{align*}
Substituting
\begin{align*}
    (f\cdot g_{\omega})\left(\Exp_{\mx}^{\mathbf{E}_{\mx}}(h\mathbf{w})\right)
    =
    (f\cdot g_{\omega})(\mx)
    +
    h\mathbf{w}^{\top}\bm{\beta}_{f\cdot g_{\omega}}(\mx)
    +
    R_{h,\omega}(\mathbf{w})
\end{align*}
into the normal-coordinate representation of $\bm{\tilde{\tau}}_{h,1}$ gives
\begin{align*}
    \bm{\tilde{\tau}}_{h,1}(\mx,\mathbf{E}_{\mx},\omega)
    &=
    h^{d+1}(f\cdot g_{\omega})(\mx)
    \int_{\|\mathbf{w}\|_2\leq1}
    \mathbf{w}K(\|\mathbf{w}\|_2)
    \,\dd\mathbf{w} \\
    &\quad+
    h^{d+2}
    \int_{\|\mathbf{w}\|_2\leq1}
    \mathbf{w}\mathbf{w}^{\top}K(\|\mathbf{w}\|_2)
    \,\dd\mathbf{w}\,
    \bm{\beta}_{f\cdot g_{\omega}}(\mx) \\
    &\quad+
    h^{d+1}
    \int_{\|\mathbf{w}\|_2\leq1}
    \mathbf{w}K(\|\mathbf{w}\|_2)R_{h,\omega}(\mathbf{w})
    \,\dd\mathbf{w}.
\end{align*}
The first integral vanishes by radial symmetry, while \Cref{lemma:A.radial_kernel_moments} gives
\begin{align*}
    \int_{\|\mathbf{w}\|_2\leq1}
    \mathbf{w}\mathbf{w}^{\top}K(\|\mathbf{w}\|_2)
    \,\dd\mathbf{w}
    =
    \frac{A_{d-1}c_{d+1,1}}{d}\mI_d.
\end{align*}
Consequently,
\begin{align*}
    \sup_{\omega\in\mathbb{M}}
    \left\|
    h^{d+2}
    \int_{\|\mathbf{w}\|_2\leq1}
    \mathbf{w}\mathbf{w}^{\top}K(\|\mathbf{w}\|_2)
    \,\dd\mathbf{w}\,
    \bm{\beta}_{f\cdot g_{\omega}}(\mx)
    \right\|_2
    =
    O(h^{d+2}),
\end{align*}
and, using the nonnegativity of $K$ from Condition~\ref{con:P-K1},
\begin{align*}
    &\sup_{\omega\in\mathbb{M}}
    \left\|
    h^{d+1}
    \int_{\|\mathbf{w}\|_2\leq1}
    \mathbf{w}K(\|\mathbf{w}\|_2)R_{h,\omega}(\mathbf{w})
    \,\dd\mathbf{w}
    \right\|_2 \\
    &\leq
    h^{d+1}
    \left(
    \sup_{\omega\in\mathbb{M}}
    \sup_{\|\mathbf{w}\|_2\leq1}
    |R_{h,\omega}(\mathbf{w})|
    \right)
    \int_{\|\mathbf{w}\|_2\leq1}
    \|\mathbf{w}\|_2K(\|\mathbf{w}\|_2)
    \,\dd\mathbf{w}
    =
    O(h^{d+3}).
\end{align*}
This proves \eqref{eq:C.tau1_rate_ll}. 

By \Cref{lemma:C.pointwise_refined_moments},
\begin{align} \label{eq:C.mu1_mu2inv_bounds}
    \left\|
    \bm{\tilde{\mu}}_{h,1}(\mx,\mathbf{E}_{\mx})
    \right\|_2
    =
    O(h^{d+2}),
    \quad
    \left\|
    \bm{\tilde{\mu}}_{h,2}(\mx,\mathbf{E}_{\mx})^{-1}
    \right\|_2
    =
    O(h^{-(d+2)}).
\end{align}
By \eqref{eq:C.mu1_mu2inv_bounds} and \eqref{eq:C.tau1_rate_ll}, we have
\begin{align} \label{eq:C.ll_correction_tau_expansion}
    \sup_{\omega\in\mathbb{M}}
    \left|
    \bm{\tilde{\mu}}_{h,1}(\mx,\mathbf{E}_{\mx})^{\top}
    \bm{\tilde{\mu}}_{h,2}(\mx,\mathbf{E}_{\mx})^{-1}
    \bm{\tilde{\tau}}_{h,1}(\mx,\mathbf{E}_{\mx},\omega)
    \right|
    =
    O(h^{d+2}).
\end{align}
Moreover,
\begin{align}
\begin{split}
\tilde{\sigma}_h(\mx)
    &=
    \tilde{\mu}_{h,0}(\mx)
    -
    \bm{\tilde{\mu}}_{h,1}(\mx,\mathbf{E}_{\mx})^{\top}
    \bm{\tilde{\mu}}_{h,2}(\mx,\mathbf{E}_{\mx})^{-1}
    \bm{\tilde{\mu}}_{h,1}(\mx,\mathbf{E}_{\mx})  \\
    &=
    h^d A_{d-1}c_{d-1,1} f(\mx)
    +
    O(h^{d+2})
    =
    h^d A_{d-1}c_{d-1,1} f(\mx)\{1+O(h^2)\}.
\end{split}
\label{eq:C.sigma_rate_ll}
\end{align}
Combining \eqref{eq:C.tau0_rate_ll} and \eqref{eq:C.ll_correction_tau_expansion}, we obtain, uniformly over $\omega\in\mathbb{M}$,
\begin{align*}
    \tilde{\tau}_{h,0}(\mx,\omega)
    -
    \bm{\tilde{\mu}}_{h,1}(\mx,\mathbf{E}_{\mx})^{\top}
    \bm{\tilde{\mu}}_{h,2}(\mx,\mathbf{E}_{\mx})^{-1}
    \bm{\tilde{\tau}}_{h,1}(\mx,\mathbf{E}_{\mx},\omega)
    =
    h^d A_{d-1}c_{d-1,1} f(\mx)g_{\omega}(\mx)
    +
    O(h^{d+2}).
\end{align*}
Together with \eqref{eq:C.sigma_rate_ll}, $f(\mx)>0$, and $\sup_{\omega\in\mathbb{M}}g_{\omega}(\mx)<\infty$, the same quotient argument as in the case $s=0$ gives
\begin{align*}
    \sup_{\omega\in\mathbb{M}}
    \left|
    \tilde{g}_{h,1}(\mx,\omega)-g_{\omega}(\mx)
    \right|
    =
    O(h^2).
\end{align*}
This proves \eqref{eq:C.interpolated_density_bias}.

Substituting \eqref{eq:C.interpolated_density_bias} into \eqref{eq:C.bias_integral_bound}, we obtain
\begin{align*}
    C_{\oplus,\mx}
    d_{\mathbb{M}}\left(\tilde{m}_{h,s}(\mx),m_{\oplus}(\mx)\right)^{\beta_{\oplus,\mx}}
    \leq
    2D_{\mathbb{M}}
    d_{\mathbb{M}}\left(\tilde{m}_{h,s}(\mx),m_{\oplus}(\mx)\right)
    O(h^2).
\end{align*}
If $d_{\mathbb{M}}\left(\tilde{m}_{h,s}(\mx),m_{\oplus}(\mx)\right)=0$, the desired result is immediate. Otherwise, dividing both sides by $C_{\oplus,\mx}d_{\mathbb{M}}\left(\tilde{m}_{h,s}(\mx),m_{\oplus}(\mx)\right)$ gives
\begin{align*}
    d_{\mathbb{M}}\left(\tilde{m}_{h,s}(\mx),m_{\oplus}(\mx)\right)^{\beta_{\oplus,\mx}-1}
    =
    O(h^2).
\end{align*}
This completes the proof.
\end{proof}

To control stochastic fluctuations of localized empirical objective functions, we use the following standard empirical-process facts. The first is a bracketing consequence of a Lipschitz parametrization, and the second is a bracketing maximal inequality; see Theorems~2.7.11 and~2.14.2 of \cite{van der Vaart and Wellner (1996)}. Related Fr\'echet-regression arguments for sphere-valued predictors are given in \cite{Im et al. (2025)}.

\begin{lemma}[Lipschitz parametrization and bracketing] \label{lemma:C.vw_lipschitz_bracketing}
Let $(T,d_T)$ be a metric space and let $\mathcal Z$ be a set. Let $H:\mathcal Z\to[0,\infty)$ be a function, and let $\mathcal H=\{h_t:t\in T\}$ be a class of real-valued functions on $\mathcal Z$ such that
\begin{align*}
    |h_s(z)-h_t(z)|
    \leq
    d_T(s,t)H(z),
    \quad s,t\in T,\ z\in\mathcal Z.
\end{align*}
Then for any function norm $\|\cdot\|$ such that $\|H\|<\infty$ and for any $\epsilon>0$,
\begin{align*}
    N_{[]}\left(2\epsilon\|H\|,\mathcal H,\|\cdot\|\right)
    \leq
    N(\epsilon,T,d_T),
\end{align*}
where $N_{[]}(\delta,\mathcal H,\|\cdot\|)$ denotes the $\delta$-bracketing number of $\mathcal H$ under $\|\cdot\|$, and $N(\epsilon,T,d_T)$ denotes the $\epsilon$-covering number of $(T,d_T)$.
\end{lemma}

\begin{lemma}[Bracketing maximal inequality] \label{lemma:C.vw_bracketing_maximal}
Let $(\mathcal Z,\mathcal A)$ be a measurable space, and let $Z_1,\ldots,Z_n$ be i.i.d. $\mathcal Z$-valued random variables with distribution $P_{\mathcal Z}$. Let $\mathcal G$ be a class of measurable real-valued functions on $\mathcal Z$, and let $G:\mathcal Z\to[0,\infty)$ be a measurable envelope such that $|g|\leq G$ for all $g\in\mathcal G$ and $\|G\|_{L_2(P_{\mathcal Z})}<\infty$. Suppose that the displayed supremum below is measurable. Then there exists a universal constant $C_{\mathrm{MaxIneq}}<\infty$ such that
\begin{align*}
    &\E\left[
    \sup_{g\in\mathcal G}
    \left|
    \frac{1}{n}\sum_{i=1}^n g(Z_i)
    -
    \int_{\mathcal Z}g(z)\,\dd P_{\mathcal Z}(z)
    \right|
    \right] \\
    &\leq
    \frac{C_{\mathrm{MaxIneq}}}{\sqrt n}
    \|G\|_{L_2(P_{\mathcal Z})}
    \int_0^1
    \sqrt{
    1+
    \log N_{[]}
    \left(
    \epsilon\|G\|_{L_2(P_{\mathcal Z})},
    \mathcal G,
    L_2(P_{\mathcal Z})
    \right)
    }
    \dd\epsilon.
\end{align*}
If the measurability of the supremum is not imposed, the same bound holds with outer expectation.
\end{lemma}

\begin{lemma}[Localized empirical fluctuation bound] \label{lemma:C.pointwise_oracle_fluctuation}
Assume Conditions~\ref{con:P-K1}, \ref{con:P-B1}, \ref{con:P-D1}, \ref{con:P-D3}, \ref{con:M1}, and~\ref{con:P-M4}. Fix $s\in\{0,1\}$. Let $\tilde y_{\mx,h}$ be a deterministic sequence such that
\begin{align*}
    d_{\mathbb M}\left(\tilde y_{\mx,h},m_{\oplus}(\mx)\right)
    \leq
    \frac{r_{\mathbb{M},\mx}}{2}
\end{align*}
for all sufficiently small $h$. Assume that the suprema appearing in the empirical processes below are measurable. Let $\tilde{W}_{\mx,h,s}:\mathcal M\to\mathbb R$ be defined by \eqref{eq:B.population_equiv_weights}. Define the oracle-weight empirical fluctuation process
\begin{align*}
    \hat{S}_{h,s}(y)
    &:=
    \frac{1}{n}\sum_{i=1}^n
    \tilde{W}_{\mx,h,s}\left(\mX^{(i)}\right)
    d_{\mathbb M}^2\left(y,Y^{(i)}\right)
    -
    \tilde{M}_{h,s}(\mx,y),
    \quad y\in\mathbb M.
\end{align*}
Then there exist constants $\delta_{\mathrm{fluc}}>0$ and $C_{\mathrm{fluc}}<\infty$, independent of $n$, $h$, $\delta$, and $\tilde y_{\mx,h}$, such that, for every $\delta\in(0,\delta_{\mathrm{fluc}}]$ and all sufficiently small $h$,
\begin{align} \label{eq:C.localized_empirical_fluctuation}
    \E
    \left[
    \sup_{y\in B_{\mathbb M}(\tilde{y}_{\mx,h},\delta)}
    \left|
    \hat{S}_{h,s}(y)-\hat{S}_{h,s}(\tilde{y}_{\mx,h})
    \right|
    \right]
    \leq
    C_{\mathrm{fluc}}\delta (nh^d)^{-1/2}.
\end{align}
Moreover, let $\hat{W}_{\mx,h,s}:\mathcal M\to\mathbb R$ be defined by \eqref{eq:sec3.empirical_equiv_weight_lc} and \eqref{eq:sec3.empirical_equiv_weight_ll}, and define, on the event where the empirical weights are well-defined, the empirical-weight empirical fluctuation process
\begin{align*}
    \hat{T}_{h,s}(y)
    &:=
    \frac{1}{n}\sum_{i=1}^n
    \hat{W}_{\mx,h,s}\left(\mX^{(i)}\right)
    d_{\mathbb M}^2\left(y,Y^{(i)}\right)
    -
    \tilde{M}_{h,s}(\mx,y) \\
    &=
    \hat{M}_{h,s}(\mx,y)-\tilde{M}_{h,s}(\mx,y),
    \quad y\in\mathbb M.
\end{align*}
Then
\begin{align} \label{eq:C.weight_remainder_fluctuation}
    \sup_{y\in\mathbb M}
    \frac{
    \left|
    \left(\hat{T}_{h,s}(y)-\hat{T}_{h,s}(\tilde{y}_{\mx,h})\right)
    -
    \left(\hat{S}_{h,s}(y)-\hat{S}_{h,s}(\tilde{y}_{\mx,h})\right)
    \right|
    }{
    d_{\mathbb M}(y,\tilde{y}_{\mx,h})
    }
    =
    O_{\P}\left((nh^d)^{-1/2}\right),
\end{align}
with the convention that the ratio is zero when $y=\tilde{y}_{\mx,h}$.
\end{lemma}

\begin{proof}[Proof of \Cref{lemma:C.pointwise_oracle_fluctuation}]
Fix $s\in\{0,1\}$. Choose $\delta_{\mathrm{fluc}}:=r_{\mathbb M,\mx}/4$. Then, for every $\delta\in(0,\delta_{\mathrm{fluc}}]$ and every $y\in B_{\mathbb M}(\tilde y_{\mx,h},\delta)$,
\begin{align*}
    d_{\mathbb M}\left(y,m_\oplus(\mx)\right)
    \leq
    d_{\mathbb M}\left(y,\tilde y_{\mx,h}\right)
    +
    d_{\mathbb M}\left(\tilde y_{\mx,h},m_\oplus(\mx)\right)
    \leq
    \delta
    +
    \frac{r_{\mathbb M,\mx}}{2}
    <
    r_{\mathbb M,\mx}.
\end{align*}
Hence Condition~\ref{con:P-M4} applies to all response-space balls considered below.

We first prove \eqref{eq:C.localized_empirical_fluctuation}. For $y\in\mathbb{M}$, define
\begin{align*}
    \tilde{U}_{y,h,s}(\mz,\omega)
    &:=
    \tilde{W}_{\mx,h,s}(\mz)
    \left[
    d_{\mathbb M}^2(y,\omega)
    -
    d_{\mathbb M}^2(\tilde{y}_h,\omega)
    \right],
    \quad
    (\mz,\omega)\in\mathcal{M}\times\mathbb{M}.
\end{align*}
By Condition~\ref{con:M1}, $D_{\mathbb{M}}$ defined by \eqref{eq:B.diameter_metric_space} is finite. For any $y\in\mathbb{M}$, the triangle inequality gives
\begin{align*}
    \left|
    d_{\mathbb M}^2(y,\omega)
    -
    d_{\mathbb M}^2(\tilde{y}_h,\omega)
    \right|
    &\leq
    d_{\mathbb M}(y,\tilde{y}_h)
    \left[
    d_{\mathbb M}(y,\omega)
    +
    d_{\mathbb M}(\tilde{y}_h,\omega)
    \right] \\
    &\leq
    2D_{\mathbb{M}}d_{\mathbb M}(y,\tilde{y}_h), \quad \omega\in\mathbb{M}.
\end{align*}
Hence, for $y\in B_{\mathbb M}(\tilde{y}_h,\delta)$,
\begin{align} \label{eq:C.localized_envelope}
    \left|
    \tilde{U}_{y,h,s}(\mz,\omega)
    \right|
    \leq
    2D_{\mathbb{M}}\delta
    \left|
    \tilde{W}_{\mx,h,s}(\mz)
    \right|, \quad (\mz,\omega)\in\mathcal{M}\times\mathbb{M}.
\end{align}
Similarly, for $y_1,y_2\in B_{\mathbb M}(\tilde{y}_h,\delta)$,
\begin{align} \label{eq:C.localized_lipschitz}
    \left|
    \tilde{U}_{y_1,h,s}(\mz,\omega)
    -
    \tilde{U}_{y_2,h,s}(\mz,\omega)
    \right|
    \leq
    2D_{\mathbb{M}}
    d_{\mathbb M}(y_1,y_2)
    \left|
    \tilde{W}_{\mx,h,s}(\mz)
    \right|, \quad (\mz,\omega)\in\mathcal{M}\times\mathbb{M}.
\end{align}
Define the class
\begin{align*}
    \widetilde{\mathcal U}_{h,s,\delta}
    :=
    \left\{
    \tilde{U}_{y,h,s}:y\in B_{\mathbb M}(\tilde{y}_h,\delta)
    \right\}.
\end{align*}
Then \eqref{eq:C.localized_lipschitz} implies that $\widetilde{\mathcal U}_{h,s,\delta}$ is Lipschitz-parametrized by $y$ under $d_{\mathbb M}$ with Lipschitz envelope
\begin{align*}
    H_{h,s}(\mz,\omega)
    &:=
    2D_{\mathbb{M}}
    \left|
    \tilde{W}_{\mx,h,s}(\mz)
    \right|,
    \quad
    (\mz,\omega)\in\mathcal{M}\times\mathbb{M}.
\end{align*}
Moreover, \eqref{eq:C.localized_envelope} shows that $\widetilde{\mathcal U}_{h,s,\delta}$ has the localized envelope
\begin{align*}
    G_{h,s,\delta}(\mz,\omega)
    &:=
    2D_{\mathbb{M}}\delta
    \left|
    \tilde{W}_{\mx,h,s}(\mz)
    \right|,
    \quad
    (\mz,\omega)\in\mathcal{M}\times\mathbb{M}.
\end{align*}

We next verify the $L_2$ size of the oracle weights. By \Cref{lemma:C.pointwise_scalar_kernel_moments},
\begin{align*}
    \tilde{\mu}_{h,0}(\mx)
    =
    h^dA_{d-1}c_{d-1,1}f(\mx)
    +
    O(h^{d+2}).
\end{align*}
Since $f(\mx)>0$ by Condition~\ref{con:P-D1}, for all sufficiently small $h$,
\begin{align} \label{eq:C.mu0_lower_for_weight}
    \tilde{\mu}_{h,0}(\mx)
    \geq
    \frac12 h^dA_{d-1}c_{d-1,1}f(\mx).
\end{align}
Also, \Cref{lemma:C.pointwise_refined_moments} gives
\begin{align*}
    \left\|
    \bm{\tilde{\mu}}_{h,1}(\mx,\mathbf{E}_{\mx})
    \right\|_2
    =
    O(h^{d+2}),
    \quad
    \left\|
    \bm{\tilde{\mu}}_{h,2}(\mx,\mathbf{E}_{\mx})^{-1}
    \right\|_2
    =
    O(h^{-(d+2)}).
\end{align*}
Therefore,
\begin{align*}
    \left|
    \bm{\tilde{\mu}}_{h,1}(\mx,\mathbf{E}_{\mx})^{\top}
    \bm{\tilde{\mu}}_{h,2}(\mx,\mathbf{E}_{\mx})^{-1}
    \bm{\tilde{\mu}}_{h,1}(\mx,\mathbf{E}_{\mx})
    \right|
    =
    O(h^{d+2}).
\end{align*}
Together with the definition of $\tilde{\sigma}_h(\mx)$ and the expansion of $\tilde{\mu}_{h,0}(\mx)$ above, this yields
\begin{align*}
    \tilde{\sigma}_h(\mx)
    =
    h^dA_{d-1}c_{d-1,1}f(\mx)
    +
    O(h^{d+2}).
\end{align*}
Hence, for all sufficiently small $h$,
\begin{align} \label{eq:C.sigma_lower_for_weight}
    \tilde{\sigma}_h(\mx)
    \geq
    \frac12 h^dA_{d-1}c_{d-1,1}f(\mx).
\end{align}
Moreover,
\begin{align} \label{eq:C.local_linear_correction_order}
    \left\|
    \bm{\tilde{\mu}}_{h,2}(\mx,\mathbf{E}_{\mx})^{-1}
    \bm{\tilde{\mu}}_{h,1}(\mx,\mathbf{E}_{\mx})
    \right\|_2
    =
    O(1).
\end{align}
By \Cref{lemma:A.pointwise_normal_neighborhoods}, if $\mathcal L_{\mx,h}(\mz)\neq0$ and $h<\rho_{\mx}$, then $\mz\in B_{\mathcal M}(\mx,h)$ and
\begin{align*}
    \left\|
    \mathbf{v}_{\mx}^{\mathbf{E}_{\mx}}(\mz)
    \right\|_2
    =
    d_{\mathcal M}(\mx,\mz)
    \leq h.
\end{align*}
Therefore, \eqref{eq:C.local_linear_correction_order} gives, uniformly over $\mz\in\mathcal M$ on the support of $\mathcal L_{\mx,h}$,
\begin{align} \label{eq:C.local_linear_correction_factor_bound}
    \left|
    1-
    \bm{\tilde{\mu}}_{h,1}(\mx,\mathbf{E}_{\mx})^{\top}
    \bm{\tilde{\mu}}_{h,2}(\mx,\mathbf{E}_{\mx})^{-1}
    \mathbf{v}_{\mx}^{\mathbf{E}_{\mx}}(\mz)
    \right|
    =
    O(1).
\end{align}
Writing $\|K\|_{\infty}:=\sup_{t\in[0,1]}K(t)$, \Cref{lemma:A.pointwise_normal_neighborhoods} and Condition~\ref{con:P-K1} imply that, for $h<\rho_{\mx}$,
\begin{align*}
    0
    \leq
    \mathcal L_{\mx,h}(\mz)
    \leq
    c_{\theta,\mx,\rho_{\mx}}^{-1}
    \|K\|_{\infty}
    \mathbf{1}\left\{
    d_{\mathcal M}(\mx,\mz)\leq h
    \right\},
    \quad \mz\in\mathcal M.
\end{align*}
Combining this display with \eqref{eq:C.mu0_lower_for_weight}, \eqref{eq:C.sigma_lower_for_weight}, and \eqref{eq:C.local_linear_correction_factor_bound}, we obtain
\begin{align} \label{eq:C.oracle_weight_pointwise_bound}
    \left|
    \tilde{W}_{\mx,h,s}(\mz)
    \right|
    \leq
    O(h^{-d})
    \mathbf{1}\left\{
    d_{\mathcal M}(\mx,\mz)\leq h
    \right\},
    \quad \mz\in\mathcal M,
\end{align}
where the $O(h^{-d})$ constant is independent of $\mz$ and $h$. It remains to bound the probability of the local ball. By Condition~\ref{con:P-D1} and \Cref{lemma:A.pointwise_normal_neighborhoods},
\begin{align*}
    P_{\mX}\left\{
    d_{\mathcal M}(\mx,\mX)\leq h
    \right\}
    &=
    \int_{B_{\mathcal M}(\mx,h)} f(\mz)\,\dd v_g(\mz) \\
    &\leq
    \left(
    \sup_{\mz\in B_{\mathcal M}(\mx,\rho_{\mx})}f(\mz)
    \right)
    v_g\left(B_{\mathcal M}(\mx,h)\right) \\
    &\leq
    \left(
    \sup_{\mz\in B_{\mathcal M}(\mx,\rho_{\mx})}f(\mz)
    \right)
    C_{\theta,\mx,\rho_{\mx}}
    \frac{A_{d-1}}{d}
    h^d.
\end{align*}
Thus \eqref{eq:C.oracle_weight_pointwise_bound} yields
\begin{align} \label{eq:C.oracle_weight_L2_rate}
    \left\|
    \tilde{W}_{\mx,h,s}
    \right\|_{L_2(P_{\mX})}
    =
    O(h^{-d/2}).
\end{align}

We now control the entropy integral. By \eqref{eq:C.localized_lipschitz} and \Cref{lemma:C.vw_lipschitz_bracketing}, for every $\epsilon\in(0,1)$,
\begin{align*}
    N_{[]}
    \left(
    \epsilon\|G_{h,s,\delta}\|_{L_2(P)},
    \widetilde{\mathcal U}_{h,s,\delta},
    L_2(P)
    \right)
    \leq
    N\left(
    \frac{\epsilon\delta}{2},
    B_{\mathbb M}(\tilde{y}_h,\delta),
    d_{\mathbb M}
    \right).
\end{align*}

Indeed, since $G_{h,s,\delta}=\delta H_{h,s}$, if $y_1,y_2\in B_{\mathbb M}(\tilde{y}_h,\delta)$ satisfy $d_{\mathbb M}(y_1,y_2)\leq\epsilon\delta/2$, then \eqref{eq:C.localized_lipschitz} gives
\begin{align*}
    \left|
    \tilde{U}_{y_1,h,s}(\mz,\omega)
    -
    \tilde{U}_{y_2,h,s}(\mz,\omega)
    \right|
    \leq
    \frac{\epsilon}{2}
    G_{h,s,\delta}(\mz,\omega).
\end{align*}
Thus an $\epsilon\delta/2$-cover of $B_{\mathbb M}(\tilde{y}_h,\delta)$ under $d_{\mathbb M}$ yields brackets of $L_2(P)$-width at most $\epsilon\|G_{h,s,\delta}\|_{L_2(P)}$. Hence, by the change of variables $u=\epsilon/2$,
\begin{align*}
    &\int_0^1
    \sqrt{
    1+
    \log N_{[]}
    \left(
    \epsilon\|G_{h,s,\delta}\|_{L_2(P)},
    \widetilde{\mathcal U}_{h,s,\delta},
    L_2(P)
    \right)
    }
    \,\dd\epsilon \\
    &\leq
    2
    \int_0^{1/2}
    \sqrt{
    1+
    \log N
    \left(
    \delta u,
    B_{\mathbb M}(\tilde{y}_h,\delta),
    d_{\mathbb M}
    \right)
    }
    \,\dd u.
\end{align*}
Since $d_{\mathbb M}\left(\tilde{y}_h,m_{\oplus}(\mx)\right)<r_{\mathbb{M},\mx}$ for all sufficiently small $h$, Condition~\ref{con:P-M4} implies that there exist constants $\delta_{\mathrm{fluc}}>0$ and $C_{\mathrm{ent}}<\infty$, independent of $h$, $\delta$, and $\tilde{y}_h$, such that the last display is bounded by $C_{\mathrm{ent}}$ for every $\delta\in(0,\delta_{\mathrm{fluc}}]$. Applying \Cref{lemma:C.vw_bracketing_maximal} to $\widetilde{\mathcal U}_{h,s,\delta}$ and using \eqref{eq:C.oracle_weight_L2_rate}, we obtain
\begin{align*}
    &\E
    \left[
    \sup_{y\in B_{\mathbb M}(\tilde{y}_h,\delta)}
    \left|
    \frac{1}{n}\sum_{i=1}^n
    \tilde{U}_{y,h,s}\left(\mX^{(i)},Y^{(i)}\right)
    -
    \E\left[\tilde{U}_{y,h,s}(\mX,Y)\right]
    \right|
    \right] \\
    &\leq
    C_{\mathrm{MaxIneq}}C_{\mathrm{ent}}n^{-1/2}
    \|G_{h,s,\delta}\|_{L_2(P)} \\
    &=
    2D_{\mathbb{M}}C_{\mathrm{MaxIneq}}C_{\mathrm{ent}}
    \delta n^{-1/2}
    \left\|
    \tilde{W}_{\mx,h,s}
    \right\|_{L_2(P_{\mX})} \\
    &\leq
    C_{\mathrm{fluc}}\delta (nh^d)^{-1/2}.
\end{align*}
Here ordinary expectation is used because the relevant supremum is assumed to be measurable. Finally, for every $y\in B_{\mathbb M}(\tilde{y}_h,\delta)$,
\begin{align*}
    \hat{S}_{h,s}(y)-\hat{S}_{h,s}(\tilde{y}_h)
    =
    \frac{1}{n}\sum_{i=1}^n
    \tilde{U}_{y,h,s}\left(\mX^{(i)},Y^{(i)}\right)
    -
    \E\left[\tilde{U}_{y,h,s}(\mX,Y)\right].
\end{align*}
This proves \eqref{eq:C.localized_empirical_fluctuation}.

It remains to prove \eqref{eq:C.weight_remainder_fluctuation}. All statements involving empirical weights are understood on the event where these weights are well-defined; this event has probability tending to one by \Cref{lemma:B.pointwise_empirical_moments,lemma:B.pointwise_denominators}. For any $y\in\mathbb M$,
\begin{align*}
    &\left|
    \left(\hat{T}_{h,s}(y)-\hat{T}_{h,s}(\tilde{y}_h)\right)
    -
    \left(\hat{S}_{h,s}(y)-\hat{S}_{h,s}(\tilde{y}_h)\right)
    \right| \\
    &\leq
    \frac{1}{n}\sum_{i=1}^n
    \left|
    \hat{W}_{\mx,h,s}\left(\mX^{(i)}\right)
    -
    \tilde{W}_{\mx,h,s}\left(\mX^{(i)}\right)
    \right|
    \left|
    d_{\mathbb M}^2\left(y,Y^{(i)}\right)
    -
    d_{\mathbb M}^2\left(\tilde{y}_h,Y^{(i)}\right)
    \right| \\
    &\leq
    2D_{\mathbb{M}}
    d_{\mathbb M}(y,\tilde{y}_h)
    \frac{1}{n}\sum_{i=1}^n
    \left|
    \hat{W}_{\mx,h,s}\left(\mX^{(i)}\right)
    -
    \tilde{W}_{\mx,h,s}\left(\mX^{(i)}\right)
    \right|.
\end{align*}
Thus it suffices to show that
\begin{align} \label{eq:C.average_weight_diff}
    \frac{1}{n}\sum_{i=1}^n
    \left|
    \hat{W}_{\mx,h,s}\left(\mX^{(i)}\right)
    -
    \tilde{W}_{\mx,h,s}\left(\mX^{(i)}\right)
    \right|
    =
    O_{\P}\left((nh^d)^{-1/2}\right).
\end{align}
For $s=0$, using the nonnegativity of $K$ from Condition~\ref{con:P-K1},
\begin{align*}
    &\frac{1}{n}\sum_{i=1}^n
    \left|
    \hat{W}_{\mx,h,0}\left(\mX^{(i)}\right)
    -
    \tilde{W}_{\mx,h,0}\left(\mX^{(i)}\right)
    \right| \\
    &=
    \left|
    \hat{\mu}_{h,0}(\mx)^{-1}
    -
    \tilde{\mu}_{h,0}(\mx)^{-1}
    \right|
    \frac{1}{n}\sum_{i=1}^n
    \mathcal L_{\mx,h}\left(\mX^{(i)}\right) \\
    &=
    \left|
    \hat{\mu}_{h,0}(\mx)^{-1}
    -
    \tilde{\mu}_{h,0}(\mx)^{-1}
    \right|
    \hat{\mu}_{h,0}(\mx).
\end{align*}
By \Cref{lemma:B.pointwise_empirical_moments},
\begin{align*}
    \hat{\mu}_{h,0}(\mx)=O_{\P}(h^d),
    \quad
    \left|
    \hat{\mu}_{h,0}(\mx)^{-1}
    -
    \tilde{\mu}_{h,0}(\mx)^{-1}
    \right|
    =
    O_{\P}\left(n^{-1/2}h^{-3d/2}\right).
\end{align*}
Therefore \eqref{eq:C.average_weight_diff} holds for $s=0$.
For $s=1$, by \Cref{lemma:B.pointwise_denominators,lemma:C.pointwise_refined_moments},
\begin{align*}
    \tilde{\sigma}_h(\mx)\asymp h^d,
    \quad
    \hat{\sigma}_h(\mx)-\tilde{\sigma}_h(\mx)
    =
    O_{\P}\left(n^{-1/2}h^{d/2}\right).
\end{align*}
Consequently,
\begin{align*}
    \hat{\sigma}_h(\mx)^{-1}
    =
    O_{\P}(h^{-d}),
    \quad
    \left|
    \hat{\sigma}_h(\mx)^{-1}
    -
    \tilde{\sigma}_h(\mx)^{-1}
    \right|
    =
    O_{\P}\left(n^{-1/2}h^{-3d/2}\right).
\end{align*}
Moreover, suppressing the common arguments $(\mx,\mathbf{E}_{\mx})$ in the local moments, \Cref{lemma:B.pointwise_empirical_moments,lemma:C.pointwise_refined_moments} gives
\begin{align*}
    &\left\|
    \bm{\hat{\mu}}_{h,2}^{-1}
    \bm{\hat{\mu}}_{h,1}
    -
    \bm{\tilde{\mu}}_{h,2}^{-1}
    \bm{\tilde{\mu}}_{h,1}
    \right\|_2 \\
    &\leq
    \left\|
    \bm{\hat{\mu}}_{h,2}^{-1}
    \right\|_2
    \left\|
    \bm{\hat{\mu}}_{h,1}
    -
    \bm{\tilde{\mu}}_{h,1}
    \right\|_2
    +
    \left\|
    \bm{\hat{\mu}}_{h,2}^{-1}
    -
    \bm{\tilde{\mu}}_{h,2}^{-1}
    \right\|_2
    \left\|
    \bm{\tilde{\mu}}_{h,1}
    \right\|_2 \\
    &=
    O_{\P}\left(n^{-1/2}h^{-(d+2)/2}\right).
\end{align*}
Since $\bm{\hat{\mu}}_{h,2}$ and $\bm{\tilde{\mu}}_{h,2}$ are symmetric whenever they are invertible, the same bound holds for the corresponding transposed row vectors. Also,
\begin{align*}
    \left\|
    \bm{\tilde{\mu}}_{h,2}^{-1}
    \bm{\tilde{\mu}}_{h,1}
    \right\|_2
    =
    O(1).
\end{align*}
Since $K$ is nonnegative and $\|\mathbf{v}_{\mx}^{\mathbf{E}_{\mx}}(\mX^{(i)})\|_2\leq h$ whenever $\mathcal L_{\mx,h}(\mX^{(i)})\neq0$,
\begin{align*}
    \frac{1}{n}\sum_{i=1}^n
    \mathcal L_{\mx,h}\left(\mX^{(i)}\right)
    =
    \hat{\mu}_{h,0}(\mx)
    =
    O_{\P}(h^d),
\end{align*}
and
\begin{align*}
    \frac{1}{n}\sum_{i=1}^n
    \mathcal L_{\mx,h}\left(\mX^{(i)}\right)
    \left\|
    \mathbf{v}_{\mx}^{\mathbf{E}_{\mx}}\left(\mX^{(i)}\right)
    \right\|_2
    \leq
    h\hat{\mu}_{h,0}(\mx)
    =
    O_{\P}(h^{d+1}).
\end{align*}
Therefore,
\begin{align*}
    &\frac{1}{n}\sum_{i=1}^n
    \left|
    \hat{W}_{\mx,h,1}\left(\mX^{(i)}\right)
    -
    \tilde{W}_{\mx,h,1}\left(\mX^{(i)}\right)
    \right| \\
    &\leq
    \left|
    \hat{\sigma}_h(\mx)^{-1}
    -
    \tilde{\sigma}_h(\mx)^{-1}
    \right|
    \frac{1}{n}\sum_{i=1}^n
    \mathcal L_{\mx,h}\left(\mX^{(i)}\right)
    \left|
    1-
    \bm{\tilde{\mu}}_{h,1}^{\top}
    \bm{\tilde{\mu}}_{h,2}^{-1}
    \mathbf{v}_{\mx}^{\mathbf{E}_{\mx}}\left(\mX^{(i)}\right)
    \right| \\
    &\qquad+
    \hat{\sigma}_h(\mx)^{-1}
    \frac{1}{n}\sum_{i=1}^n
    \mathcal L_{\mx,h}\left(\mX^{(i)}\right)
    \left|
    \left(
    \bm{\hat{\mu}}_{h,1}^{\top}
    \bm{\hat{\mu}}_{h,2}^{-1}
    -
    \bm{\tilde{\mu}}_{h,1}^{\top}
    \bm{\tilde{\mu}}_{h,2}^{-1}
    \right)
    \mathbf{v}_{\mx}^{\mathbf{E}_{\mx}}\left(\mX^{(i)}\right)
    \right| \\
    &=
    O_{\P}\left(n^{-1/2}h^{-3d/2}\right)
    O_{\P}(h^d)
    +
    O_{\P}(h^{-d})
    O_{\P}(h^{d+1})
    O_{\P}\left(n^{-1/2}h^{-(d+2)/2}\right) \\
    &=
    O_{\P}\left(n^{-1/2}h^{-d/2}\right).
\end{align*}
Thus \eqref{eq:C.average_weight_diff} also holds for $s=1$.

Combining \eqref{eq:C.average_weight_diff} with the display immediately preceding \eqref{eq:C.average_weight_diff}, and using the convention that the ratio is zero when $y=\tilde{y}_h$, proves \eqref{eq:C.weight_remainder_fluctuation}. This completes the proof.
\end{proof}

\begin{lemma}[Pointwise empirical minimizer fluctuation] \label{lemma:C.pointwise_stochastic_rate}
Assume Conditions~\ref{con:P-K1}, \ref{con:P-B1}, \ref{con:P-D1}--\ref{con:P-D4}, \ref{con:M1}, and~\ref{con:P-M2}--\ref{con:P-M4}. Assume also that the suprema appearing in the empirical processes in \Cref{lemma:C.pointwise_oracle_fluctuation} are measurable. Then, for each $s\in\{0,1\}$,
\begin{align*}
    d_{\mathbb M}\left(\hat{m}_{h,s}(\mx),\tilde{m}_{h,s}(\mx)\right)^{\beta_{\oplus,\mx}-1}
    =
    O_{\P}\left((nh^d)^{-1/2}\right),
\end{align*}
where $\beta_{\oplus,\mx}\in(1,\infty)$ is the margin constant in Condition~\ref{con:P-M3}.
\end{lemma}

\begin{proof}[Proof of \Cref{lemma:C.pointwise_stochastic_rate}]
Fix $s\in\{0,1\}$ and write
\begin{align*}
    m_0
    &:=
    m_{\oplus}(\mx),
    \quad
    \hat{m}
    :=
    \hat{m}_{h,s}(\mx),
    \quad
    \tilde{m}
    :=
    \tilde{m}_{h,s}(\mx), \\
    r_n
    &:=
    (nh^d)^{-1/2}.
\end{align*}
By \Cref{thm:pointwise_consistency} and \Cref{lemma:C.pointwise_population_bias},
\begin{align*}
    d_{\mathbb M}(\hat{m},m_0)=o_{\P}(1),
    \quad
    d_{\mathbb M}(\tilde{m},m_0)=o(1).
\end{align*}
Hence
\begin{align} \label{eq:C.empirical_oracle_consistency}
    d_{\mathbb M}(\hat{m},\tilde{m})=o_{\P}(1).
\end{align}
Moreover, since $d_{\mathbb M}(\tilde{m},m_0)=o(1)$ and $r_{\mathbb{M},\mx}>0$ in Condition~\ref{con:P-M4}, the deterministic sequence $\tilde{m}=\tilde{m}_{h,s}(\mx)$ satisfies
\begin{align*}
    d_{\mathbb M}(\tilde{m},m_0)\leq \frac{r_{\mathbb{M},\mx}}{2}
\end{align*}
for all sufficiently small $h$. Therefore \Cref{lemma:C.pointwise_oracle_fluctuation} may be applied with $\tilde y_{\mx,h}=\tilde{m}_{h,s}(\mx)$.

Let
\begin{align*}
    \hat{T}_{h,s}(y)
    &:=
    \hat{M}_{h,s}(\mx,y)-\tilde{M}_{h,s}(\mx,y),
    \quad y\in\mathbb M,
\end{align*}
as in \Cref{lemma:C.pointwise_oracle_fluctuation}. All statements involving $\hat{M}_{h,s}$, $\hat{T}_{h,s}$, and $\hat{m}$ are understood on the event where the empirical weights are well-defined and the empirical minimizer exists; this event has probability tending to one by Conditions~\ref{con:P-M2} and \Cref{lemma:B.pointwise_empirical_moments,lemma:B.pointwise_denominators}.

Choose
\begin{align*}
    \delta_0
    \in
    \left(0,\min\{\delta_{\mathrm{fluc}},\eta_{\oplus,\mx}\}\right),
\end{align*}
where $\delta_{\mathrm{fluc}}$ is the localization radius in \Cref{lemma:C.pointwise_oracle_fluctuation} and $\eta_{\oplus,\mx}$ is the margin radius in Condition~\ref{con:P-M3}. By \eqref{eq:C.empirical_oracle_consistency},
\begin{align} \label{eq:C.consistency_radius_delta0}
    \mathbb P\left\{
    d_{\mathbb M}(\hat{m},\tilde{m})>\delta_0
    \right\}
    \to0.
\end{align}
On the event $\{d_{\mathbb M}(\hat{m},\tilde{m})\leq\delta_0\}$, Condition~\ref{con:P-M3} gives
\begin{align} \label{eq:C.localized_oracle_margin}
    C_{\oplus,\mx}
    d_{\mathbb M}(\hat{m},\tilde{m})^{\beta_{\oplus,\mx}}
    \leq
    \tilde{M}_{h,s}(\mx,\hat{m})
    -
    \tilde{M}_{h,s}(\mx,\tilde{m}).
\end{align}
Since $\hat{m}$ minimizes $\hat{M}_{h,s}(\mx,\cdot)$,
\begin{align*}
    \hat{M}_{h,s}(\mx,\hat{m})
    -
    \hat{M}_{h,s}(\mx,\tilde{m})
    \leq0.
\end{align*}
Combining this inequality with \eqref{eq:C.localized_oracle_margin}, we obtain
\begin{align} \label{eq:C.basic_oracle_fluctuation_ineq}
    C_{\oplus,\mx}
    d_{\mathbb M}(\hat{m},\tilde{m})^{\beta_{\oplus,\mx}}
    \leq
    \left|
    \hat{T}_{h,s}(\hat{m})
    -
    \hat{T}_{h,s}(\tilde{m})
    \right|.
\end{align}

For $L>0$, define the event
\begin{align*}
    \Omega_{n,L}
    &:=
    \left\{
    \sup_{y\in\mathbb M}
    \frac{
    \left|
    \left(\hat{T}_{h,s}(y)-\hat{T}_{h,s}(\tilde{m})\right)
    -
    \left(\hat{S}_{h,s}(y)-\hat{S}_{h,s}(\tilde{m})\right)
    \right|
    }{
    d_{\mathbb M}(y,\tilde{m})
    }
    \leq
    Lr_n
    \right\},
\end{align*}
with the convention that the ratio is zero when $y=\tilde{m}$. By \eqref{eq:C.weight_remainder_fluctuation}, for every $\epsilon>0$ there exists $L<\infty$ such that
\begin{align} \label{eq:C.local_linear_high_prob_event}
    \liminf_{n\to\infty}\mathbb P(\Omega_{n,L})\geq1-\epsilon.
\end{align}
On the event
\begin{align*}
    \Omega_{n,L}
    \cap
    \left\{
    d_{\mathbb M}(\hat{m},\tilde{m})\leq\delta_0
    \right\},
\end{align*}
\eqref{eq:C.basic_oracle_fluctuation_ineq} implies
\begin{align} \label{eq:C.margin_process_reduction}
    C_{\oplus,\mx}
    d_{\mathbb M}(\hat{m},\tilde{m})^{\beta_{\oplus,\mx}}
    \leq
    \left|
    \hat{S}_{h,s}(\hat{m})
    -
    \hat{S}_{h,s}(\tilde{m})
    \right|
    +
    Lr_n d_{\mathbb M}(\hat{m},\tilde{m}).
\end{align}
Choose $A>0$ so large that
\begin{align} \label{eq:C.A_large_stochastic}
    Lr_n r
    \leq
    \frac{C_{\oplus,\mx}}{2}r^{\beta_{\oplus,\mx}}
    \quad
    \text{whenever}
    \quad
    r^{\beta_{\oplus,\mx}-1}>Ar_n.
\end{align}
For example, any $A\geq 2L/C_{\oplus,\mx}$ is sufficient. Then, on the event in \eqref{eq:C.margin_process_reduction} together with
\begin{align*}
    d_{\mathbb M}(\hat{m},\tilde{m})^{\beta_{\oplus,\mx}-1}>Ar_n,
\end{align*}
we have
\begin{align} \label{eq:C.S_large_on_shell}
    \frac{C_{\oplus,\mx}}{2}
    d_{\mathbb M}(\hat{m},\tilde{m})^{\beta_{\oplus,\mx}}
    \leq
    \left|
    \hat{S}_{h,s}(\hat{m})
    -
    \hat{S}_{h,s}(\tilde{m})
    \right|.
\end{align}

We now use a peeling argument. For $k=0,1,2,\ldots$, define
\begin{align*}
    A_{n,k}
    &:=
    \left\{
    2^kAr_n
    <
    d_{\mathbb M}(\hat{m},\tilde{m})^{\beta_{\oplus,\mx}-1}
    \leq
    2^{k+1}Ar_n,
    \quad
    d_{\mathbb M}(\hat{m},\tilde{m})\leq\delta_0
    \right\}
    \cap
    \Omega_{n,L},
\end{align*}
and set
\begin{align*}
    \ell_{n,k}
    &:=
    \left(2^kAr_n\right)^{1/(\beta_{\oplus,\mx}-1)}, \\
    \rho_{n,k}
    &:=
    \min\left\{
    \left(2^{k+1}Ar_n\right)^{1/(\beta_{\oplus,\mx}-1)},
    \delta_0
    \right\}.
\end{align*}
On $A_{n,k}$, inequality \eqref{eq:C.S_large_on_shell} yields
\begin{align*}
    \sup_{y\in B_{\mathbb M}(\tilde{m},\rho_{n,k})}
    \left|
    \hat{S}_{h,s}(y)
    -
    \hat{S}_{h,s}(\tilde{m})
    \right|
    \geq
    \frac{C_{\oplus,\mx}}{2}
    \ell_{n,k}^{\beta_{\oplus,\mx}}.
\end{align*}
Since $\rho_{n,k}\leq\delta_0\leq\delta_{\mathrm{fluc}}$, Markov's inequality and \Cref{lemma:C.pointwise_oracle_fluctuation} give
\begin{align}
\begin{split}
\mathbb P(A_{n,k})
    &\leq
    \frac{
    2
    \E\left[
    \sup_{y\in B_{\mathbb M}(\tilde{m},\rho_{n,k})}
    \left|
    \hat{S}_{h,s}(y)
    -
    \hat{S}_{h,s}(\tilde{m})
    \right|
    \right]
    }{
    C_{\oplus,\mx}\ell_{n,k}^{\beta_{\oplus,\mx}}
    }  \\
    &\leq
    C
    \frac{\rho_{n,k}r_n}{\ell_{n,k}^{\beta_{\oplus,\mx}}},
\end{split}
\label{eq:C.shell_probability_bound}
\end{align}
where $C<\infty$ is independent of $n$, $h$, and $k$.

For every $k$ such that
\begin{align*}
    2^{k+1}Ar_n
    \leq
    \delta_0^{\beta_{\oplus,\mx}-1},
\end{align*}
we have $\rho_{n,k}=(2^{k+1}Ar_n)^{1/(\beta_{\oplus,\mx}-1)}$, and \eqref{eq:C.shell_probability_bound} gives
\begin{align*}
    \mathbb P(A_{n,k})
    \leq
    C A^{-1}2^{-k}.
\end{align*}
There is at most one remaining boundary shell for which
\begin{align*}
    2^kAr_n
    <
    \delta_0^{\beta_{\oplus,\mx}-1}
    <
    2^{k+1}Ar_n.
\end{align*}
For this boundary shell, \eqref{eq:C.shell_probability_bound} gives
\begin{align*}
    \mathbb P(A_{n,k})
    &\leq
    C
    \frac{\delta_0 r_n}{\left(2^kAr_n\right)^{\beta_{\oplus,\mx}/(\beta_{\oplus,\mx}-1)}} \\
    &\leq
    C r_n\delta_0^{1-\beta_{\oplus,\mx}}
    =
    o(1),
\end{align*}
because $r_n=(nh^d)^{-1/2}\to0$ by Condition~\ref{con:P-B1}. Therefore,
\begin{align*}
    \limsup_{n\to\infty}
    \mathbb P
    \left(
    d_{\mathbb M}(\hat{m},\tilde{m})^{\beta_{\oplus,\mx}-1}>Ar_n,
    \ d_{\mathbb M}(\hat{m},\tilde{m})\leq\delta_0,
    \ \Omega_{n,L}
    \right)
    \leq
    C A^{-1}.
\end{align*}
Combining this bound with \eqref{eq:C.consistency_radius_delta0} and \eqref{eq:C.local_linear_high_prob_event}, and then choosing $L$ and $A$ sufficiently large, proves that
\begin{align*}
    d_{\mathbb M}\left(\hat{m}_{h,s}(\mx),\tilde{m}_{h,s}(\mx)\right)^{\beta_{\oplus,\mx}-1}
    =
    O_{\P}(r_n)
    =
    O_{\P}\left((nh^d)^{-1/2}\right).
\end{align*}
This completes the proof.
\end{proof}

\begin{proof}[Proof of \Cref{thm:pointwise_rate}]
Fix $s\in\{0,1\}$. By \Cref{lemma:C.pointwise_population_bias},
\begin{align*}
    d_{\mathbb M}\left(\tilde{m}_{h,s}(\mx),m_{\oplus}(\mx)\right)^{\beta_{\oplus,\mx}-1}
    =
    O(h^2),
\end{align*}
and hence
\begin{align*}
    d_{\mathbb M}\left(\tilde{m}_{h,s}(\mx),m_{\oplus}(\mx)\right)
    =
    O\left(h^{2/(\beta_{\oplus,\mx}-1)}\right).
\end{align*}
By \Cref{lemma:C.pointwise_stochastic_rate},
\begin{align*}
    d_{\mathbb M}\left(\hat{m}_{h,s}(\mx),\tilde{m}_{h,s}(\mx)\right)^{\beta_{\oplus,\mx}-1}
    =
    O_{\P}\left((nh^d)^{-1/2}\right),
\end{align*}
and therefore
\begin{align*}
    d_{\mathbb M}\left(\hat{m}_{h,s}(\mx),\tilde{m}_{h,s}(\mx)\right)
    =
    O_{\P}\left((nh^d)^{-1/(2\beta_{\oplus,\mx}-2)}\right).
\end{align*}
The triangle inequality gives
\begin{align*}
    d_{\mathbb M}\left(\hat{m}_{h,s}(\mx),m_{\oplus}(\mx)\right)
    &\leq
    d_{\mathbb M}\left(\hat{m}_{h,s}(\mx),\tilde{m}_{h,s}(\mx)\right)
    +
    d_{\mathbb M}\left(\tilde{m}_{h,s}(\mx),m_{\oplus}(\mx)\right) \\
    &=
    O\left(h^{2/(\beta_{\oplus,\mx}-1)}\right)
    +
    O_{\P}\left((nh^d)^{-1/(2\beta_{\oplus,\mx}-2)}\right).
\end{align*}
This proves the theorem.
\end{proof}

\section{Proof of Uniform Consistency} 
\label{app:uniform_consistency}
\setcounter{equation}{0}
\renewcommand{\theequation}{D.\arabic{equation}}
\renewcommand{\theHequation}{D.\arabic{equation}}

In this section, we provide the proof of \Cref{thm:uniform_consistency}.
Throughout this section, $\mathcal K\subset\mathcal M$ is the compact set
fixed in the uniform theory, $\rho\in(0,i(\mathcal K))$ is the fixed
uniform normal-neighborhood radius, and $\mathcal K^\rho$ is the
corresponding closed geodesic tube.

In the deterministic population arguments, $\mathbf E_{\mx}\in\mathcal
E_{\mx}$ denotes an arbitrary ordered orthonormal basis of
$T_{\mx}\mathcal M$. The Euclidean norms of vector coordinates, the
operator norms and eigenvalues of matrix coordinates, and the scalar
quadratic forms appearing below are invariant under orthogonal changes of
basis. Hence the corresponding uniform bounds do not depend on the
particular choice of $\mathbf E_{\mx}$.

In the empirical-process arguments, we use the finite smooth
ordered-orthonormal-frame cover fixed in the uniform theory before
Conditions~\ref{con:U-K1} and~\ref{con:U-K2}. Componentwise bounds are
established separately on each cover element and then combined by taking
the maximum over the finitely many cover elements and coordinate indices.
All constants may depend on this fixed finite frame cover, but no
uniformity over all possible frame covers is required. Scalar equivalent
weights and scalar local objectives are independent of the particular
ordered orthonormal basis by
\Cref{lemma:A.invariance_local_linear_weights}.

\begin{remark}
The deterministic population lemmas at the beginning of this section use only the baseline kernel regularity in Condition~\ref{con:P-K1}; their uniformity over $\mathcal K$ follows from the uniform normal-neighborhood geometry and the uniform design conditions. The VC-type content of Condition~\ref{con:U-K1}, and the multiplier complexity in Condition~\ref{con:U-K2}, enter only in the empirical-process arguments below.
\end{remark}

\begin{lemma}[Uniform population local moment expansion] \label{lemma:D.uniform_population_moments_f}
Assume Conditions~\ref{con:P-K1} and~\ref{con:U-D1}, and suppose that $h\to0$ as $n\to\infty$. Let $\rho\in(0,i(\mathcal K))$ be fixed as in the uniform theory. Then, for $k\in\{1,2\}$ and $j\in\{0,1,2\}$,
\begin{align}
    \sup_{\mx\in\mathcal K}
    \left\|
    \E\left[
    \mathcal L_{\mx,h}(\mX)^k
    \left\{
    \mathbf v_{\mx}^{\mathbf E_{\mx}}(\mX)
    \right\}^{\otimes j}
    \right]
    -
    h^{d+j}f(\mx)
    \int_{\mathbb R^d}
    K(\|\mathbf w\|_2)^k
    \mathbf w^{\otimes j}
    \,\dd\mathbf w
    \right\|_{\star}
    =
    o(h^{d+j}), \label{eq:D.uniform_f_moment_expansion}
\end{align}
where the tensor-power and $\|\cdot\|_{\star}$ conventions are those stated at the beginning of the appendices. 
\end{lemma}

\begin{proof}[Proof of \Cref{lemma:D.uniform_population_moments_f}]
Fix $k\in\{1,2\}$ and $j\in\{0,1,2\}$. Since $h\to0$, we may assume throughout the proof that $h<\rho$. We work componentwise on the finite smooth frame cover fixed at the beginning of Appendix~\ref{app:uniform_consistency}. Fix one frame chart and write the corresponding ordered orthonormal basis at $\mx$ as $\mathbf E_{\mx}$. The following argument is uniform over $\mx\in\mathcal K\cap\mathcal O^\alpha$ for the fixed chart $\mathcal O^\alpha$; taking the maximum over the finite frame cover then gives the displayed supremum over $\mathcal K$. On the support of $\mathcal L_{\mx,h}$, the compact support of $K$ gives $d_{\mathcal M}(\mx,\mz)\leq h<\rho$, so the normal-coordinate representation
\begin{align*}
    \mz
    =
    \Exp_{\mx}^{\mathbf E_{\mx}}(h\mathbf w),
    \quad
    \|\mathbf w\|_2\leq1,
\end{align*}
is valid. Moreover,
\begin{align*}
    \mathbf v_{\mx}^{\mathbf E_{\mx}}(\mz)
    =
    h\mathbf w,
    \quad
    d_{\mathcal M}(\mx,\mz)
    =
    h\|\mathbf w\|_2.
\end{align*}
By the normal-coordinate change of variables and the definition of the volume-corrected kernel,
\begin{align*}
    &\E\left[
    \mathcal L_{\mx,h}(\mX)^k
    \left\{
    \mathbf v_{\mx}^{\mathbf E_{\mx}}(\mX)
    \right\}^{\otimes j}
    \right] \\
    &=
    h^{d+j}
    \int_{\mathbb R^d}
    K(\|\mathbf w\|_2)^k
    \mathbf w^{\otimes j}
    f\left(
    \Exp_{\mx}^{\mathbf E_{\mx}}(h\mathbf w)
    \right)
    \theta_{\mx}\left(
    \Exp_{\mx}^{\mathbf E_{\mx}}(h\mathbf w)
    \right)^{1-k}
    \mathbf 1\{\|\mathbf w\|_2\leq1\}
    \,\dd\mathbf w.
\end{align*}
Here the factor $\theta_{\mx}^{1-k}$ appears because $\mathcal L_{\mx,h}^k$ contributes $\theta_{\mx}^{-k}$, while the Riemannian volume element contributes one factor of $\theta_{\mx}$.

We claim that
\begin{align}
    \sup_{\mx\in\mathcal K}
    \sup_{\|\mathbf w\|_2\leq1}
    \left|
    f\left(
    \Exp_{\mx}^{\mathbf E_{\mx}}(h\mathbf w)
    \right)
    \theta_{\mx}\left(
    \Exp_{\mx}^{\mathbf E_{\mx}}(h\mathbf w)
    \right)^{1-k}
    -
    f(\mx)
    \right|
    =
    o(1). \label{eq:D.uniform_qkh_convergence}
\end{align}
Indeed, Condition~\ref{con:U-D1} implies that $f$ is uniformly continuous and bounded on the compact tube $\mathcal K^\rho$. Moreover, by \Cref{lemma:A.uniform_normal_neighborhoods},
\begin{align*}
    \lim_{h\downarrow0}
    \sup_{\mx\in\mathcal K}
    \sup_{\|\mathbf w\|_2\leq1}
    \left|
    \theta_{\mx}\left(
    \Exp_{\mx}^{\mathbf E_{\mx}}(h\mathbf w)
    \right)
    -
    1
    \right|
    =
    0.
\end{align*}
Since $\Exp_{\mx}^{\mathbf E_{\mx}}(h\mathbf w)\in\mathcal K^\rho$ and $d_{\mathcal M}(\Exp_{\mx}^{\mathbf E_{\mx}}(h\mathbf w),\mx)\leq h$, the uniform continuity of $f$ gives
\begin{align*}
    \sup_{\mx\in\mathcal K}
    \sup_{\|\mathbf w\|_2\leq1}
    \left|
    f\left(
    \Exp_{\mx}^{\mathbf E_{\mx}}(h\mathbf w)
    \right)
    -
    f(\mx)
    \right|
    =
    o(1).
\end{align*}
Combining the last two displays proves \eqref{eq:D.uniform_qkh_convergence}. Therefore,
\begin{align*}
    &\sup_{\mx\in\mathcal K}
    h^{-(d+j)}
    \left\|
    \E\left[
    \mathcal L_{\mx,h}(\mX)^k
    \left\{
    \mathbf v_{\mx}^{\mathbf E_{\mx}}(\mX)
    \right\}^{\otimes j}
    \right]
    -
    h^{d+j}f(\mx)
    \int_{\mathbb R^d}
    K(\|\mathbf w\|_2)^k
    \mathbf w^{\otimes j}
    \,\dd\mathbf w
    \right\|_{\star} \\
    &\leq
    \sup_{\mx\in\mathcal K}
    \sup_{\|\mathbf w\|_2\leq1}
    \left|
    f\left(
    \Exp_{\mx}^{\mathbf E_{\mx}}(h\mathbf w)
    \right)
    \theta_{\mx}\left(
    \Exp_{\mx}^{\mathbf E_{\mx}}(h\mathbf w)
    \right)^{1-k}
    -
    f(\mx)
    \right| \\
    &\quad\times
    \int_{\mathbb R^d}
    K(\|\mathbf w\|_2)^k
    \|\mathbf w\|_2^j
    \mathbf 1\{\|\mathbf w\|_2\leq1\}
    \,\dd\mathbf w
    =
    o(1),
\end{align*}
where the integral is finite because $K$ is bounded and supported on $[0,1]$. This proves \eqref{eq:D.uniform_f_moment_expansion}.
\end{proof}

\begin{lemma}[Uniform population local moment expansion with conditional density ratios] \label{lemma:D.uniform_population_moments_fg}
Assume Conditions~\ref{con:P-K1}, \ref{con:U-D1}, and~\ref{con:U-D2}, and suppose that $h\to0$ as $n\to\infty$. Let $\rho\in(0,i(\mathcal K))$ be fixed as in the uniform theory. Then, for $k\in\{1,2\}$ and $j\in\{0,1,2\}$,
\begin{align}
    \sup_{\mx\in\mathcal K}
    \sup_{\omega\in\mathbb M}
    \left\|
    \E\left[
    \mathcal L_{\mx,h}(\mX)^k
    \left\{
    \mathbf v_{\mx}^{\mathbf E_{\mx}}(\mX)
    \right\}^{\otimes j}
    g_{\omega}(\mX)
    \right]
    -
    h^{d+j}f(\mx)g_{\omega}(\mx)
    \int_{\mathbb R^d}
    K(\|\mathbf w\|_2)^k
    \mathbf w^{\otimes j}
    \,\dd\mathbf w
    \right\|_{\star}
    =
    o(h^{d+j}), \label{eq:D.uniform_conditional_moment_expansion}
\end{align}
where the tensor-power and $\|\cdot\|_{\star}$ conventions are those stated at the beginning of the appendices. For vector and matrix moments, the bound is understood componentwise on the finite smooth frame cover fixed at the beginning of Appendix~\Cref{app:uniform_consistency}, and the displayed $\|\cdot\|_{\star}$ bound follows by taking maxima over finitely many frame charts and coordinate indices.
\end{lemma}

\begin{proof}[Proof of \Cref{lemma:D.uniform_population_moments_fg}]
Fix $k\in\{1,2\}$ and $j\in\{0,1,2\}$. As in the proof of
\Cref{lemma:D.uniform_population_moments_f}, we work componentwise on each
element of the fixed finite smooth frame cover and then take the maximum
over the cover. It is enough to prove
\begin{align}
    \Delta_{k,h}
    &:=
    \sup_{\mx\in\mathcal K}
    \sup_{\omega\in\mathbb M}
    \sup_{\|\mathbf w\|_2\leq1}
    \left|
    f\left(\Exp_{\mx}^{\mathbf E_{\mx}}(h\mathbf w)\right)
    g_{\omega}\left(\Exp_{\mx}^{\mathbf E_{\mx}}(h\mathbf w)\right)
    \theta_{\mx}\left(\Exp_{\mx}^{\mathbf E_{\mx}}(h\mathbf w)\right)^{1-k}
    -
    f(\mx)g_{\omega}(\mx)
    \right|
    =
    o(1). \label{eq:D.uniform_conditional_delta}
\end{align}
For $\mx\in\mathcal K$ and $\|\mathbf w\|_2\leq1$, write
\begin{align*}
    \mz_{h,\mx,\mathbf w}
    :=
    \Exp_{\mx}^{\mathbf E_{\mx}}(h\mathbf w),
    \quad
    a_{k,h}(\mx,\mathbf w)
    :=
    f(\mz_{h,\mx,\mathbf w})
    \theta_{\mx}(\mz_{h,\mx,\mathbf w})^{1-k}.
\end{align*}
Then $d_{\mathcal M}(\mz_{h,\mx,\mathbf w},\mx)\leq h$ and $\mz_{h,\mx,\mathbf w}\in\mathcal K^\rho$ for all sufficiently small $h$. By Condition~\ref{con:U-D1} and \Cref{lemma:A.uniform_normal_neighborhoods},
\begin{align}
    \sup_{\mx\in\mathcal K}
    \sup_{\|\mathbf w\|_2\leq1}
    \left|
    a_{k,h}(\mx,\mathbf w)-f(\mx)
    \right|
    =
    o(1). \label{eq:D.uniform_akh_to_f}
\end{align}
Condition~\ref{con:U-D2} gives
\begin{align}
    G_{\mathcal K}
    &:=
    \sup_{\omega\in\mathbb M}
    \sup_{\mz\in\mathcal K^\rho}
    g_{\omega}(\mz)
    <
    \infty, \label{eq:D.uniform_g_envelope}
\end{align}
and, by uniform equicontinuity on $\mathcal K^\rho$,
\begin{align}
    \sup_{\omega\in\mathbb M}
    \sup_{\mx\in\mathcal K}
    \sup_{\|\mathbf w\|_2\leq1}
    \left|
    g_{\omega}(\mz_{h,\mx,\mathbf w})
    -
    g_{\omega}(\mx)
    \right|
    =
    o(1). \label{eq:D.uniform_g_local_equicont}
\end{align}
Combining \eqref{eq:D.uniform_akh_to_f}--\eqref{eq:D.uniform_g_local_equicont}, we obtain
\begin{align*}
    \Delta_{k,h}
    &\leq
    G_{\mathcal K}
    \sup_{\mx\in\mathcal K}
    \sup_{\|\mathbf w\|_2\leq1}
    \left|
    a_{k,h}(\mx,\mathbf w)-f(\mx)
    \right| \\
    &\quad+
    C_{\mathcal K,\rho}
    \sup_{\omega\in\mathbb M}
    \sup_{\mx\in\mathcal K}
    \sup_{\|\mathbf w\|_2\leq1}
    \left|
    g_{\omega}(\mz_{h,\mx,\mathbf w})
    -
    g_{\omega}(\mx)
    \right|
    =
    o(1),
\end{align*}
where $0<C_{\mathcal K,\rho}<\infty$ is defined by \eqref{eq:U.design_density_bounds_on_tube}. Hence \eqref{eq:D.uniform_conditional_delta} holds. The compact support and boundedness of $K$ then imply \eqref{eq:D.uniform_conditional_moment_expansion} exactly as in the proof of \Cref{lemma:D.uniform_population_moments_f}.
\end{proof}

\begin{lemma}[Uniform population local moment consequences] \label{lemma:D.uniform_population_moment_orders}
Assume Conditions~\ref{con:P-K1} and~\ref{con:U-D1}, and suppose that $h\to0$ as $n\to\infty$. Then
\begin{align}
    \sup_{\mx\in\mathcal K}
    \left|
    \tilde{\mu}_{h,0}(\mx)
    -
    h^dA_{d-1}c_{d-1,1}f(\mx)
    \right|
    =
    o(h^d), \label{eq:D.uniform_population_moment_orders_mu0}
\end{align}
\begin{align}
    \sup_{\mx\in\mathcal K}
    \left\|
    \bm{\tilde{\mu}}_{h,1}(\mx,\mathbf E_{\mx})
    \right\|_2
    =
    o(h^{d+1}), \label{eq:D.uniform_population_moment_orders_mu1}
\end{align}
and
\begin{align}
    \sup_{\mx\in\mathcal K}
    \left\|
    \bm{\tilde{\mu}}_{h,2}(\mx,\mathbf E_{\mx})
    -
    h^{d+2}
    \frac{A_{d-1}c_{d+1,1}f(\mx)}{d}
    \mI_d
    \right\|_{\mathrm{op}}
    =
    o(h^{d+2}). \label{eq:D.uniform_population_moment_orders_mu2}
\end{align}
Consequently, with $c_{\mathcal K}=\inf_{\mx\in\mathcal K}f(\mx)>0$, for all sufficiently small $h$,
\begin{align}
    \inf_{\mx\in\mathcal K}
    \tilde{\mu}_{h,0}(\mx)
    \geq
    \frac12
    h^dA_{d-1}c_{d-1,1}c_{\mathcal K},
    \label{eq:D.uniform_population_moment_orders_mu0_lower}
\end{align}
\begin{align}
    \inf_{\mx\in\mathcal K}
    \lambda_{\min}
    \left(
    \bm{\tilde{\mu}}_{h,2}(\mx,\mathbf E_{\mx})
    \right)
    \geq
    \frac12
    h^{d+2}
    \frac{A_{d-1}c_{d+1,1}}{d}
    c_{\mathcal K},
    \label{eq:D.uniform_population_moment_orders_mu2_lower}
\end{align}
and
\begin{align}
    \sup_{\mx\in\mathcal K}
    \left\|
    \bm{\tilde{\mu}}_{h,2}(\mx,\mathbf E_{\mx})^{-1}
    \right\|_{\mathrm{op}}
    =
    O(h^{-(d+2)}). \label{eq:D.uniform_population_moment_orders_mu2_inverse}
\end{align}
Moreover,
\begin{align}
    \sup_{\mx\in\mathcal K}
    \left|
    \tilde{\sigma}_h(\mx)
    -
    h^dA_{d-1}c_{d-1,1}f(\mx)
    \right|
    =
    o(h^d), \label{eq:D.uniform_population_moment_orders_sigma}
\end{align}
and hence, for all sufficiently small $h$,
\begin{align}
    \inf_{\mx\in\mathcal K}
    \tilde{\sigma}_h(\mx)
    \geq
    \frac12
    h^dA_{d-1}c_{d-1,1}c_{\mathcal K}.
    \label{eq:D.uniform_population_moment_orders_sigma_lower}
\end{align}
\end{lemma}

\begin{proof}[Proof of \Cref{lemma:D.uniform_population_moment_orders}]
Taking $k=1$ in \Cref{lemma:D.uniform_population_moments_f}, we have, for $j=0,1,2$,
\begin{align}
    \sup_{\mx\in\mathcal K}
    \left\|
    \E\left[
    \mathcal L_{\mx,h}(\mX)
    \left\{
    \mathbf v_{\mx}^{\mathbf E_{\mx}}(\mX)
    \right\}^{\otimes j}
    \right]
    -
    h^{d+j}f(\mx)
    \int_{\mathbb R^d}
    K(\|\mathbf w\|_2)
    \mathbf w^{\otimes j}
    \,\dd\mathbf w
    \right\|_{\star}
    =
    o(h^{d+j}).
    \label{eq:D.population_moment_orders_general}
\end{align}
For $j=0$, the left-hand side in \eqref{eq:D.population_moment_orders_general} is $\tilde{\mu}_{h,0}(\mx)$. By \Cref{lemma:A.radial_kernel_moments},
\begin{align*}
    \int_{\mathbb R^d}
    K(\|\mathbf w\|_2)
    \,\dd\mathbf w
    =
    A_{d-1}c_{d-1,1}.
\end{align*}
Hence
\begin{align*}
    \sup_{\mx\in\mathcal K}
    \left|
    \tilde{\mu}_{h,0}(\mx)
    -
    h^dA_{d-1}c_{d-1,1}f(\mx)
    \right|
    =
    o(h^d),
\end{align*}
which proves \eqref{eq:D.uniform_population_moment_orders_mu0}.

For $j=1$, the left-hand side in \eqref{eq:D.population_moment_orders_general} is $\bm{\tilde{\mu}}_{h,1}(\mx,\mathbf E_{\mx})$. Since the radial kernel moment satisfies
\begin{align*}
    \int_{\mathbb R^d}
    K(\|\mathbf w\|_2)
    \mathbf w
    \,\dd\mathbf w
    =
    \mathbf 0_d,
\end{align*}
we obtain
\begin{align*}
    \sup_{\mx\in\mathcal K}
    \left\|
    \bm{\tilde{\mu}}_{h,1}(\mx,\mathbf E_{\mx})
    \right\|_2
    =
    o(h^{d+1}),
\end{align*}
which proves \eqref{eq:D.uniform_population_moment_orders_mu1}.

For $j=2$, the left-hand side in \eqref{eq:D.population_moment_orders_general} is $\bm{\tilde{\mu}}_{h,2}(\mx,\mathbf E_{\mx})$. By \Cref{lemma:A.radial_kernel_moments},
\begin{align*}
    \int_{\mathbb R^d}
    K(\|\mathbf w\|_2)
    \mathbf w\mathbf w^{\top}
    \,\dd\mathbf w
    =
    \frac{A_{d-1}c_{d+1,1}}{d}
    \mI_d.
\end{align*}
Thus
\begin{align*}
    \sup_{\mx\in\mathcal K}
    \left\|
    \bm{\tilde{\mu}}_{h,2}(\mx,\mathbf E_{\mx})
    -
    h^{d+2}
    \frac{A_{d-1}c_{d+1,1}f(\mx)}{d}
    \mI_d
    \right\|_{\mathrm{op}}
    =
    o(h^{d+2}),
\end{align*}
which proves \eqref{eq:D.uniform_population_moment_orders_mu2}.

By Condition~\ref{con:U-D1}, $c_{\mathcal K}=\inf_{\mx\in\mathcal K}f(\mx)>0$. From \eqref{eq:D.uniform_population_moment_orders_mu0},
\begin{align*}
    \sup_{\mx\in\mathcal K}
    \left|
    \frac{\tilde{\mu}_{h,0}(\mx)}
    {h^dA_{d-1}c_{d-1,1}}
    -
    f(\mx)
    \right|
    =
    o(1).
\end{align*}
Therefore, for all sufficiently small $h$,
\begin{align*}
    \inf_{\mx\in\mathcal K}
    \tilde{\mu}_{h,0}(\mx)
    \geq
    \frac12
    h^dA_{d-1}c_{d-1,1}c_{\mathcal K},
\end{align*}
which proves \eqref{eq:D.uniform_population_moment_orders_mu0_lower}.

Similarly, \eqref{eq:D.uniform_population_moment_orders_mu2} gives
\begin{align*}
    \sup_{\mx\in\mathcal K}
    \left\|
    \frac{
    \bm{\tilde{\mu}}_{h,2}(\mx,\mathbf E_{\mx})
    }{
    h^{d+2}A_{d-1}c_{d+1,1}/d
    }
    -
    f(\mx)\mI_d
    \right\|_{\mathrm{op}}
    =
    o(1).
\end{align*}
Hence, by Weyl's inequality, for all sufficiently small $h$,
\begin{align*}
    \inf_{\mx\in\mathcal K}
    \lambda_{\min}
    \left(
    \bm{\tilde{\mu}}_{h,2}(\mx,\mathbf E_{\mx})
    \right)
    \geq
    \frac12
    h^{d+2}
    \frac{A_{d-1}c_{d+1,1}}{d}
    c_{\mathcal K}.
\end{align*}
This proves \eqref{eq:D.uniform_population_moment_orders_mu2_lower}. Consequently, $\bm{\tilde{\mu}}_{h,2}(\mx,\mathbf E_{\mx})$ is invertible uniformly over $\mx\in\mathcal K$ for all sufficiently small $h$, and
\begin{align*}
    \sup_{\mx\in\mathcal K}
    \left\|
    \bm{\tilde{\mu}}_{h,2}(\mx,\mathbf E_{\mx})^{-1}
    \right\|_{\mathrm{op}}
    =
    O(h^{-(d+2)}),
\end{align*}
which proves \eqref{eq:D.uniform_population_moment_orders_mu2_inverse}.

Finally, by definition,
\begin{align*}
    \tilde{\sigma}_h(\mx)
    =
    \tilde{\mu}_{h,0}(\mx)
    -
    \bm{\tilde{\mu}}_{h,1}(\mx,\mathbf E_{\mx})^{\top}
    \bm{\tilde{\mu}}_{h,2}(\mx,\mathbf E_{\mx})^{-1}
    \bm{\tilde{\mu}}_{h,1}(\mx,\mathbf E_{\mx}).
\end{align*}
Using \eqref{eq:D.uniform_population_moment_orders_mu1} and \eqref{eq:D.uniform_population_moment_orders_mu2_inverse},
\begin{align*}
    \sup_{\mx\in\mathcal K}
    \left|
    \bm{\tilde{\mu}}_{h,1}(\mx,\mathbf E_{\mx})^{\top}
    \bm{\tilde{\mu}}_{h,2}(\mx,\mathbf E_{\mx})^{-1}
    \bm{\tilde{\mu}}_{h,1}(\mx,\mathbf E_{\mx})
    \right|
    =
    o(h^{d+1})
    O(h^{-(d+2)})
    o(h^{d+1})
    =
    o(h^d).
\end{align*}
Combining this display with \eqref{eq:D.uniform_population_moment_orders_mu0} gives
\begin{align*}
    \sup_{\mx\in\mathcal K}
    \left|
    \tilde{\sigma}_h(\mx)
    -
    h^dA_{d-1}c_{d-1,1}f(\mx)
    \right|
    =
    o(h^d),
\end{align*}
which proves \eqref{eq:D.uniform_population_moment_orders_sigma}. Moreover, the last display and $c_{\mathcal K}>0$ imply that, for all sufficiently small $h$,
\begin{align*}
    \inf_{\mx\in\mathcal K}
    \tilde{\sigma}_h(\mx)
    \geq
    \frac12
    h^dA_{d-1}c_{d-1,1}c_{\mathcal K}.
\end{align*}
This proves \eqref{eq:D.uniform_population_moment_orders_sigma_lower} and completes the proof.
\end{proof}

Recall the definitions of $\tilde{\tau}_{h,0}$, $\bm{\tilde{\tau}}_{h,1}$, $\tilde g_{h,0}$, and $\tilde g_{h,1}$ from \eqref{eq:B.population_auxiliary_defs} and \eqref{eq:B.g_tilde_def}. In the present uniform arguments, these quantities are used with $\mx\in\mathcal K$, $\omega\in\mathbb M$, and an ordered orthonormal basis evaluated on the finite smooth frame cover fixed at the beginning of Appendix~\Cref{app:uniform_consistency}. Whenever the denominators in \eqref{eq:B.g_tilde_def} are well-defined, $\tilde g_{h,0}$ and $\tilde g_{h,1}$ are understood in the sense of that display. By \Cref{lemma:A.invariance_local_linear_weights}, $\tilde g_{h,1}$ does not depend on the particular ordered orthonormal basis, and hence the basis is suppressed from the notation.

\begin{lemma}[Uniform population local-objective approximation] \label{lemma:D.uniform_population_objective}
Assume Conditions~\ref{con:P-K1}, \ref{con:U-D1}, \ref{con:U-D2}, and~\ref{con:M1}, and suppose that $h\to0$ as $n\to\infty$. Then, for each $s\in\{0,1\}$,
\begin{align}
    \sup_{\mx\in\mathcal K}
    \sup_{\omega\in\mathbb M}
    \left|
    \tilde{g}_{h,s}(\mx,\omega)-g_{\omega}(\mx)
    \right|
    =
    o(1). \label{eq:D.uniform_g_bias_consistency}
\end{align}
Consequently,
\begin{align}
    \sup_{\mx\in\mathcal K}
    \sup_{y\in\mathbb M}
    \left|
    \tilde{M}_{h,s}(\mx,y)-M_{\oplus}(\mx,y)
    \right|
    =
    o(1). \label{eq:D.uniform_pop_objective_consistency}
\end{align}
\end{lemma}

\begin{proof}[Proof of \Cref{lemma:D.uniform_population_objective}]
Fix $s\in\{0,1\}$. By \Cref{lemma:D.uniform_population_moment_orders},
\begin{align}
    \sup_{\mx\in\mathcal K}
    \left|
    \tilde{\mu}_{h,0}(\mx)
    -
    h^dA_{d-1}c_{d-1,1}f(\mx)
    \right|
    =
    o(h^d), \label{eq:D.population_objective_mu0}
\end{align}
\begin{align}
    \sup_{\mx\in\mathcal K}
    \left\|
    \bm{\tilde{\mu}}_{h,1}(\mx,\mathbf E_{\mx})
    \right\|_2
    =
    o(h^{d+1}),
    \quad
    \sup_{\mx\in\mathcal K}
    \left\|
    \bm{\tilde{\mu}}_{h,2}(\mx,\mathbf E_{\mx})^{-1}
    \right\|_{\mathrm{op}}
    =
    O(h^{-(d+2)}), \label{eq:D.population_objective_mu12}
\end{align}
and, for all sufficiently small $h$,
\begin{align}
    \inf_{\mx\in\mathcal K}
    \tilde{\mu}_{h,0}(\mx)
    \geq
    \frac12 h^dA_{d-1}c_{d-1,1}c_{\mathcal K},
    \quad
    \inf_{\mx\in\mathcal K}
    \tilde{\sigma}_{h}(\mx)
    \geq
    \frac12 h^dA_{d-1}c_{d-1,1}c_{\mathcal K}.
    \label{eq:D.population_objective_denominator_lower}
\end{align}
By \Cref{lemma:D.uniform_population_moments_fg} with $k=1$ and $j=0,1$, together with \Cref{lemma:A.radial_kernel_moments},
\begin{align}
    \sup_{\mx\in\mathcal K}
    \sup_{\omega\in\mathbb M}
    \left|
    \tilde{\tau}_{h,0}(\mx,\omega)
    -
    h^d A_{d-1}c_{d-1,1}f(\mx)g_{\omega}(\mx)
    \right|
    =
    o(h^d), \label{eq:D.population_objective_tau0}
\end{align}
and
\begin{align}
    \sup_{\mx\in\mathcal K}
    \sup_{\omega\in\mathbb M}
    \left\|
    \bm{\tilde{\tau}}_{h,1}(\mx,\mathbf E_{\mx},\omega)
    \right\|_2
    =
    o(h^{d+1}). \label{eq:D.population_objective_tau1}
\end{align}

For $s=0$, \eqref{eq:D.population_objective_mu0}, \eqref{eq:D.population_objective_denominator_lower}, and \eqref{eq:D.population_objective_tau0} give
\begin{align*}
    \sup_{\mx\in\mathcal K}
    \sup_{\omega\in\mathbb M}
    \left|
    \tilde{g}_{h,0}(\mx,\omega)-g_{\omega}(\mx)
    \right|
    =
    o(1).
\end{align*}
For $s=1$, by \eqref{eq:D.population_objective_mu12} and \eqref{eq:D.population_objective_tau1},
\begin{align*}
    \sup_{\mx\in\mathcal K}
    \sup_{\omega\in\mathbb M}
    \left|
    \bm{\tilde{\mu}}_{h,1}(\mx,\mathbf E_{\mx})^{\top}
    \bm{\tilde{\mu}}_{h,2}(\mx,\mathbf E_{\mx})^{-1}
    \bm{\tilde{\tau}}_{h,1}(\mx,\mathbf E_{\mx},\omega)
    \right|
    =
    o(h^d).
\end{align*}
Combining this display with \eqref{eq:D.population_objective_mu0}, \eqref{eq:D.population_objective_denominator_lower}, and \eqref{eq:D.population_objective_tau0} yields
\begin{align*}
    \sup_{\mx\in\mathcal K}
    \sup_{\omega\in\mathbb M}
    \left|
    \tilde{g}_{h,1}(\mx,\omega)-g_{\omega}(\mx)
    \right|
    =
    o(1).
\end{align*}
Therefore \eqref{eq:D.uniform_g_bias_consistency} holds for both $s=0$ and $s=1$.

Finally, by Condition~\ref{con:M1}, $D_{\mathbb{M}}$ defined by \eqref{eq:B.diameter_metric_space} is finite. Hence, for each $s\in\{0,1\}$,
\begin{align*}
    \sup_{\mx\in\mathcal K}
    \sup_{y\in\mathbb M}
    \left|
    \tilde{M}_{h,s}(\mx,y)-M_{\oplus}(\mx,y)
    \right|
    &\leq
    D_{\mathbb{M}}^2
    \sup_{\mx\in\mathcal K}
    \sup_{\omega\in\mathbb M}
    \left|
    \tilde{g}_{h,s}(\mx,\omega)-g_{\omega}(\mx)
    \right| \\
    &=
    o(1).
\end{align*}
This proves \eqref{eq:D.uniform_pop_objective_consistency}.
\end{proof}

\begin{lemma}[Uniform convergence of population local minimizers] \label{lemma:D.uniform_oracle_minimizer}
Assume Conditions~\ref{con:P-K1}, \ref{con:U-D1}, \ref{con:U-D2}, \ref{con:M1}, and~\ref{con:U-M2}, and suppose that $h\to0$ as $n\to\infty$. Then for $s\in\{0,1\}$,
\begin{align*}
    \sup_{\mx\in\mathcal K}
    d_{\mathbb M}\left(
    \tilde{m}_{h,s}(\mx),
    m_\oplus(\mx)
    \right)
    =
    o(1).
\end{align*}
\end{lemma}

\begin{proof}[Proof of \Cref{lemma:D.uniform_oracle_minimizer}]
The proof is the uniform version of \Cref{lemma:B.pointwise_oracle_minimizer}. Fix $\epsilon>0$ and define
\begin{align*}
    A_{\epsilon,\mathcal K}
    :=
    \left\{
    (\mx,y)\in\mathcal K\times\mathbb M:
    d_{\mathbb M}\left(y,m_\oplus(\mx)\right)>\epsilon
    \right\}.
\end{align*}
If $A_{\epsilon,\mathcal K}$ is empty, then
\begin{align*}
    \sup_{\mx\in\mathcal K}
    d_{\mathbb M}\left(
    \tilde{m}_{h,s}(\mx),
    m_\oplus(\mx)
    \right)
    \leq
    \epsilon
\end{align*}
holds trivially. Otherwise, define
\begin{align*}
    \eta_{\epsilon,\mathcal K}
    :=
    \inf_{(\mx,y)\in A_{\epsilon,\mathcal K}}
    \left[
    M_\oplus(\mx,y)
    -
    M_\oplus\left(\mx,m_\oplus(\mx)\right)
    \right].
\end{align*}
By Condition~\ref{con:U-M2}, $\eta_{\epsilon,\mathcal K}>0$. By \Cref{lemma:D.uniform_population_objective}, for each $s\in\{0,1\}$,
\begin{align*}
    \Delta_{h,s,\mathcal K}
    :=
    \sup_{\mx\in\mathcal K}
    \sup_{y\in\mathbb M}
    \left|
    \tilde{M}_{h,s}(\mx,y)
    -
    M_\oplus(\mx,y)
    \right|
    =
    o(1).
\end{align*}
Since $\tilde{m}_{h,s}(\mx)$ minimizes $\tilde{M}_{h,s}(\mx,\cdot)$ for each $\mx\in\mathcal K$, we have, uniformly over $\mx\in\mathcal K$,
\begin{align*}
    M_\oplus\left(\mx,\tilde{m}_{h,s}(\mx)\right)
    -
    M_\oplus\left(\mx,m_\oplus(\mx)\right)
    &\leq
    \left|
    M_\oplus\left(\mx,\tilde{m}_{h,s}(\mx)\right)
    -
    \tilde{M}_{h,s}\left(\mx,\tilde{m}_{h,s}(\mx)\right)
    \right| \\
    &\quad+
    \left[
    \tilde{M}_{h,s}\left(\mx,\tilde{m}_{h,s}(\mx)\right)
    -
    \tilde{M}_{h,s}\left(\mx,m_\oplus(\mx)\right)
    \right] \\
    &\quad+
    \left|
    \tilde{M}_{h,s}\left(\mx,m_\oplus(\mx)\right)
    -
    M_\oplus\left(\mx,m_\oplus(\mx)\right)
    \right| \\
    &\leq
    2\Delta_{h,s,\mathcal K}.
\end{align*}
Since $\Delta_{h,s,\mathcal K}=o(1)$, for all sufficiently small $h$ we have $2\Delta_{h,s,\mathcal K}<\eta_{\epsilon,\mathcal K}$. If there existed $\mx_h\in\mathcal K$ such that
\begin{align*}
    d_{\mathbb M}\left(
    \tilde{m}_{h,s}(\mx_h),
    m_\oplus(\mx_h)
    \right)
    >
    \epsilon,
\end{align*}
then $(\mx_h,\tilde{m}_{h,s}(\mx_h))\in A_{\epsilon,\mathcal K}$, and the definition of $\eta_{\epsilon,\mathcal K}$ would imply
\begin{align*}
    M_\oplus\left(\mx_h,\tilde{m}_{h,s}(\mx_h)\right)
    -
    M_\oplus\left(\mx_h,m_\oplus(\mx_h)\right)
    \geq
    \eta_{\epsilon,\mathcal K},
\end{align*}
which contradicts the preceding uniform bound. Hence, for every $\epsilon>0$,
\begin{align*}
    \sup_{\mx\in\mathcal K}
    d_{\mathbb M}\left(
    \tilde{m}_{h,s}(\mx),
    m_\oplus(\mx)
    \right)
    \leq
    \epsilon
\end{align*}
for all sufficiently large $n$. Therefore
\begin{align*}
    \sup_{\mx\in\mathcal K}
    d_{\mathbb M}\left(
    \tilde{m}_{h,s}(\mx),
    m_\oplus(\mx)
    \right)
    =
    o(1).
\end{align*}
\end{proof}

\begin{lemma}[Uniform kernel-design consequences]
\label{lemma:D.vc_kernel_design}
Assume Condition~\ref{con:U-K1}. Then, for each
$\alpha\in\{1,\ldots,N_{\mathcal K}\}$, the zeroth-order local-design
class $\mathcal F_{\alpha,0}$ in Condition~\ref{con:U-K1} is of VC type.
More precisely, there exist constants
$A_{K,0}<\infty$, $v_{K,0}<\infty$, and $C_{K,0}<\infty$, independent
of $\alpha$, such that $\mathcal F_{\alpha,0}$ has envelope bounded by
$C_{K,0}$ and, for every finitely discrete probability measure $Q$ on
$\mathcal M$ and every $\epsilon\in(0,1)$,
\begin{align}
    N\left(
    \epsilon C_{K,0},
    \mathcal F_{\alpha,0},
    L_2(Q)
    \right)
    &\leq
    \left(
    \frac{A_{K,0}}{\epsilon}
    \right)^{v_{K,0}}.
    \label{eq:D.vc_entropy_kernel_design_0}
\end{align}

Suppose, in addition, that Condition~\ref{con:U-K2} holds. Then, for each
$\alpha\in\{1,\ldots,N_{\mathcal K}\}$ and
$r,s\in\{1,\ldots,d\}$, the first- and second-order
multiplier-augmented local-design classes
$\mathcal F_{\alpha,1,r}$ and $\mathcal F_{\alpha,2,r,s}$ in
Condition~\ref{con:U-K2} are of VC type. More precisely, there exist
constants
\begin{align*}
    A_{K,1},A_{K,2}
    &<\infty,
    &
    v_{K,1},v_{K,2}
    &<\infty,
    &
    C_{K,1},C_{K,2}
    &<\infty,
\end{align*}
independent of $\alpha$, $r$, and $s$, such that
$\mathcal F_{\alpha,1,r}$ and $\mathcal F_{\alpha,2,r,s}$ have
envelopes bounded by $C_{K,1}$ and $C_{K,2}$, respectively, and
\begin{align}
    N\left(
    \epsilon C_{K,1},
    \mathcal F_{\alpha,1,r},
    L_2(Q)
    \right)
    &\leq
    \left(
    \frac{A_{K,1}}{\epsilon}
    \right)^{v_{K,1}},
    \label{eq:D.vc_entropy_kernel_design_1} \\
    N\left(
    \epsilon C_{K,2},
    \mathcal F_{\alpha,2,r,s},
    L_2(Q)
    \right)
    &\leq
    \left(
    \frac{A_{K,2}}{\epsilon}
    \right)^{v_{K,2}},
    \label{eq:D.vc_entropy_kernel_design_2}
\end{align}
for every finitely discrete probability measure $Q$ on $\mathcal M$ and
every $\epsilon\in(0,1)$.

For each fixed $h\in(0,h_0)$, define the corresponding unnormalized
fixed-bandwidth classes by
\begin{align*}
    \widetilde{\mathcal F}_{\alpha,0}(h)
    &:=
    \left\{
    \mz\mapsto
    \mathcal L_{\mx,h}(\mz):
    \mx\in\mathcal K\cap\mathcal O^\alpha
    \right\}, \\
    \widetilde{\mathcal F}_{\alpha,1,r}(h)
    &:=
    \left\{
    \mz\mapsto
    \mathcal L_{\mx,h}(\mz)
    \left[
    \mathbf v_{\mx}^{\alpha}(\mz)
    \right]_r:
    \mx\in\mathcal K\cap\mathcal O^\alpha
    \right\}, \\
    \widetilde{\mathcal F}_{\alpha,2,r,s}(h)
    &:=
    \left\{
    \mz\mapsto
    \mathcal L_{\mx,h}(\mz)
    \left[
    \mathbf v_{\mx}^{\alpha}(\mz)
    \right]_r
    \left[
    \mathbf v_{\mx}^{\alpha}(\mz)
    \right]_s:
    \mx\in\mathcal K\cap\mathcal O^\alpha
    \right\}.
\end{align*}
These classes have envelopes bounded by
$C_{K,0}$, $C_{K,1}h$, and $C_{K,2}h^2$, respectively. Moreover,
\begin{align}
    N\left(
    \epsilon C_{K,0},
    \widetilde{\mathcal F}_{\alpha,0}(h),
    L_2(Q)
    \right)
    &\leq
    \left(
    \frac{A_{K,0}}{\epsilon}
    \right)^{v_{K,0}},
    \label{eq:D.vc_entropy_unnormalized_0} \\
    N\left(
    \epsilon C_{K,1}h,
    \widetilde{\mathcal F}_{\alpha,1,r}(h),
    L_2(Q)
    \right)
    &\leq
    \left(
    \frac{A_{K,1}}{\epsilon}
    \right)^{v_{K,1}},
    \label{eq:D.vc_entropy_unnormalized_1} \\
    N\left(
    \epsilon C_{K,2}h^2,
    \widetilde{\mathcal F}_{\alpha,2,r,s}(h),
    L_2(Q)
    \right)
    &\leq
    \left(
    \frac{A_{K,2}}{\epsilon}
    \right)^{v_{K,2}}.
    \label{eq:D.vc_entropy_unnormalized_2}
\end{align}
All envelope and covering-number constants are uniform over the finitely
many cover elements and coordinate indices.
\end{lemma}

\begin{proof}[Proof of \Cref{lemma:D.vc_kernel_design}]
The zeroth-order assertion in
\eqref{eq:D.vc_entropy_kernel_design_0} is exactly
Condition~\ref{con:U-K1}, after relabeling its constants as
$A_{K,0}$, $v_{K,0}$, and $C_{K,0}$. Similarly,
\eqref{eq:D.vc_entropy_kernel_design_1} and
\eqref{eq:D.vc_entropy_kernel_design_2} are exactly
Condition~\ref{con:U-K2}, after relabeling the corresponding first- and
second-order constants.

It remains to verify the assertions for the unnormalized fixed-bandwidth
classes. Since
\begin{align*}
    \widetilde{\mathcal F}_{\alpha,0}(h)
    &\subset
    \mathcal F_{\alpha,0},
\end{align*}
the zeroth-order envelope and covering-number bounds follow immediately
from Condition~\ref{con:U-K1}.

For the first-order class, every function in
$\widetilde{\mathcal F}_{\alpha,1,r}(h)$ satisfies
\begin{align*}
    \mathcal L_{\mx,h}(\mz)
    \left[
    \mathbf v_{\mx}^{\alpha}(\mz)
    \right]_r
    &=
    h
    \left\{
    \mathcal L_{\mx,h}(\mz)
    h^{-1}
    \left[
    \mathbf v_{\mx}^{\alpha}(\mz)
    \right]_r
    \right\}.
\end{align*}
Thus,
$\widetilde{\mathcal F}_{\alpha,1,r}(h)$ is obtained by multiplying the
fixed-bandwidth slice of $\mathcal F_{\alpha,1,r}$ by $h$. Consequently,
its envelope is bounded by $C_{K,1}h$, and scaling an
$\epsilon C_{K,1}$-cover of the normalized class by $h$ gives
\eqref{eq:D.vc_entropy_unnormalized_1}.

Similarly, every function in
$\widetilde{\mathcal F}_{\alpha,2,r,s}(h)$ satisfies
\begin{align*}
    &\mathcal L_{\mx,h}(\mz)
    \left[
    \mathbf v_{\mx}^{\alpha}(\mz)
    \right]_r
    \left[
    \mathbf v_{\mx}^{\alpha}(\mz)
    \right]_s \\
    &\quad=
    h^2
    \left\{
    \mathcal L_{\mx,h}(\mz)
    h^{-2}
    \left[
    \mathbf v_{\mx}^{\alpha}(\mz)
    \right]_r
    \left[
    \mathbf v_{\mx}^{\alpha}(\mz)
    \right]_s
    \right\}.
\end{align*}
Therefore, its envelope is bounded by $C_{K,2}h^2$, and scaling an
$\epsilon C_{K,2}$-cover of the normalized class by $h^2$ gives
\eqref{eq:D.vc_entropy_unnormalized_2}.

Finally, the numbers of cover elements and coordinate indices are finite.
Taking maxima of the constants over $\alpha$, $r$, and $s$ therefore
preserves the stated polynomial covering-number bounds and does not alter
any subsequent stochastic order.
\end{proof}

The preceding lemma provides the envelope and entropy conditions needed
to apply \Cref{lemma:D.uniform_local_empirical_process} chartwise with
\begin{align*}
    \psi_{\mx,h}(\mz)
    =
    1, \quad
    \psi_{\mx,h}(\mz)
    =
    \left[
    \mathbf v_{\mx}^{\alpha}(\mz)
    \right]_r, \quad
    \psi_{\mx,h}(\mz)
    =
    \left[
    \mathbf v_{\mx}^{\alpha}(\mz)
    \right]_r
    \left[
    \mathbf v_{\mx}^{\alpha}(\mz)
    \right]_s,
\end{align*}
corresponding respectively to $j=0$, $j=1$, and $j=2$. The resulting
bounds are then combined by taking the maximum over the fixed finite frame
cover and coordinate indices.

\begin{lemma}[Uniform local empirical-process bound] \label{lemma:D.uniform_local_empirical_process}
Assume Conditions~\ref{con:P-K1}, \ref{con:U-B1}, and~\ref{con:U-D1}. Let $\rho\in(0,i(\mathcal K))$ be the fixed uniform normal-neighborhood radius, and let $h_0\in(0,\rho)$ be fixed. Fix $j\geq0$. For each $h\in(0,h_0)$, let
\begin{align*}
    \Psi_h=\{\psi_{\mx,h}:\mx\in\mathcal K\}
\end{align*}
be a class of measurable real-valued functions on $\mathcal M$. Suppose that the following two conditions hold.

First, the functions are uniformly of order $h^j$ on the kernel support:
\begin{align}
    \sup_{\mx\in\mathcal K}
    \sup_{\mz\in B_{\mathcal M}(\mx,h)}
    |\psi_{\mx,h}(\mz)|
    \leq
    C_{\psi}h^j
    \label{eq:D.master_local_envelope}
\end{align}
for all sufficiently small $h$. Second, the product class
\begin{align*}
    \mathcal F_h
    :=
    \left\{
    \mz\mapsto
    \mathcal L_{\mx,h}(\mz)\psi_{\mx,h}(\mz):
    \mx\in\mathcal K
    \right\}
\end{align*}
is of VC type uniformly in $h\in(0,h_0)$, in the sense that there exist constants $A_{\mathcal F}<\infty$, $v_{\mathcal F}<\infty$, and $C_{\mathcal F}<\infty$, independent of $h$, such that, for every finitely discrete probability measure $Q$ on $\mathcal M$ and every $\epsilon\in(0,1)$,
\begin{align}
    N\left(
    \epsilon C_{\mathcal F} h^j,
    \mathcal F_h,
    L_2(Q)
    \right)
    \leq
    \left(\frac{A_{\mathcal F}}{\epsilon}\right)^{v_{\mathcal F}}.
    \label{eq:D.master_vc_entropy}
\end{align}
Then
\begin{align}
    \sup_{\mx\in\mathcal K}
    \left|
    \frac1n\sum_{i=1}^n
    \mathcal L_{\mx,h}\left(\mX^{(i)}\right)
    \psi_{\mx,h}\left(\mX^{(i)}\right)
    -
    \E\left[
    \mathcal L_{\mx,h}(\mX)
    \psi_{\mx,h}(\mX)
    \right]
    \right|
    =
    O_{\P}\left(
    h^j\sqrt{\frac{h^d\log n}{n}}
    \right).
    \label{eq:D.master_empirical_bound}
\end{align}
The same bound holds componentwise when $\psi_{\mx,h}$ takes values in a fixed finite-dimensional vector space.
\end{lemma}

\begin{proof}[Proof of \Cref{lemma:D.uniform_local_empirical_process}]
Write
\begin{align*}
    P_nf:=\frac1n\sum_{i=1}^n f\left(\mX^{(i)}\right),
    \quad
    Pf:=\E f(\mX).
\end{align*}
For
\begin{align*}
    F_{\mx,h}(\mz)
    :=
    \mathcal L_{\mx,h}(\mz)\psi_{\mx,h}(\mz),
    \quad \mx\in\mathcal K,
\end{align*}
the compact support of $K$ implies that $F_{\mx,h}(\mz)=0$ unless $d_{\mathcal M}(\mx,\mz)\leq h$. Since $h<h_0<\rho$, all such $\mz$ lie in $\mathcal K^\rho$. Let
\begin{align*}
    C_{\theta,\mathcal K}
    &:=
    \sup_{\substack{\mx\in\mathcal K\\ \mz\in B_{\mathcal M}(\mx,\rho)}}
    \theta_{\mx}(\mz)^{-1},
    \quad
    C_{K,\infty}
    :=
    \|K\|_{\infty}.
\end{align*}
By \Cref{lemma:A.uniform_normal_neighborhoods} and Condition~\ref{con:P-K1}, $C_{\theta,\mathcal K}<\infty$ and $C_{K,\infty}<\infty$. Hence \eqref{eq:D.master_local_envelope} gives
\begin{align}
    \sup_{\mx\in\mathcal K}
    \sup_{\mz\in\mathcal M}
    |F_{\mx,h}(\mz)|
    \leq
    B_{\psi,K}h^j,
    \quad
    B_{\psi,K}
    :=
    C_{\theta,\mathcal K}C_{K,\infty}C_{\psi}.
    \label{eq:D.master_sup_envelope}
\end{align}

We next bound the maximal variance. Let
\begin{align*}
    C_{f,\mathcal K}
    :=
    \sup_{\mz\in\mathcal K^\rho}f(\mz).
\end{align*}
By Condition~\ref{con:U-D1}, $C_{f,\mathcal K}<\infty$. Also, by the uniform geodesic-ball volume bound in \Cref{lemma:A.uniform_normal_neighborhoods}, there exists $C_{\mathrm{vol},\mathcal K}<\infty$ such that
\begin{align*}
    \sup_{\mx\in\mathcal K}
    v_g\left(B_{\mathcal M}(\mx,h)\right)
    \leq
    C_{\mathrm{vol},\mathcal K}h^d
\end{align*}
for all sufficiently small $h$. Therefore, using the support of $F_{\mx,h}$ and \eqref{eq:D.master_sup_envelope},
\begin{align}
\begin{split}
\sup_{\mx\in\mathcal K}
    PF_{\mx,h}^2
    &\leq
    B_{\psi,K}^2h^{2j}
    \sup_{\mx\in\mathcal K}
    P_{\mX}\left\{
    \mX\in B_{\mathcal M}(\mx,h)
    \right\}  \\
    &\leq
    B_{\psi,K}^2h^{2j}
    C_{f,\mathcal K}
    \sup_{\mx\in\mathcal K}
    v_g\left(B_{\mathcal M}(\mx,h)\right) \\
    &\leq
    V_{\psi,\mathcal K}h^{2j+d},
\end{split}
\label{eq:D.master_variance_bound}
\end{align}
where
\begin{align*}
    V_{\psi,\mathcal K}
    :=
    B_{\psi,K}^2 C_{f,\mathcal K} C_{\mathrm{vol},\mathcal K}.
\end{align*}
Set
\begin{align*}
    B_h:=B_{\psi,K}h^j,
    \quad
    \sigma_h^2:=V_{\psi,\mathcal K}h^{2j+d}.
\end{align*}
By \eqref{eq:D.master_vc_entropy} and the variance-sensitive maximal inequality for VC-type classes, there exists a constant $C_{\mathrm{VCMax}}<\infty$, depending only on the VC-type constants and universal numerical constants, such that
\begin{align}
    \E\left[
    \sup_{F\in\mathcal F_h}
    |(P_n-P)F|
    \right]
    &\leq
    C_{\mathrm{VCMax}}
    \left\{
    \sigma_h
    \sqrt{\frac{\log(A_{\mathcal F}B_h/\sigma_h)}{n}}
    +
    B_h
    \frac{\log(A_{\mathcal F}B_h/\sigma_h)}{n}
    \right\}.
    \label{eq:D.master_vc_maximal}
\end{align}
Since
\begin{align*}
    \frac{B_h}{\sigma_h}
    =
    \frac{B_{\psi,K}}{V_{\psi,\mathcal K}^{1/2}}h^{-d/2},
\end{align*}
Condition~\ref{con:U-B1} implies $\log(A_{\mathcal F}B_h/\sigma_h)=O(\log n)$ for all sufficiently large $n$. Consequently,
\begin{align*}
    \sigma_h
    \sqrt{\frac{\log(A_{\mathcal F}B_h/\sigma_h)}{n}}
    &=
    O\left(
    h^j
    \sqrt{\frac{h^d\log n}{n}}
    \right),
    \\
    B_h
    \frac{\log(A_{\mathcal F}B_h/\sigma_h)}{n}
    &=
    O\left(
    h^j
    \frac{\log n}{n}
    \right)
    =
    o\left(
    h^j
    \sqrt{\frac{h^d\log n}{n}}
    \right).
\end{align*}
Combining these bounds with \eqref{eq:D.master_vc_maximal} gives
\begin{align*}
    \E\left[
    \sup_{F\in\mathcal F_h}
    |(P_n-P)F|
    \right]
    =
    O\left(
    h^j
    \sqrt{\frac{h^d\log n}{n}}
    \right).
\end{align*}
The asserted stochastic bound follows from Markov's inequality. If $\psi_{\mx,h}$ is vector- or matrix-valued with fixed finite dimension, the same argument applied to each scalar component and then combined over finitely many components gives the componentwise statement.
\end{proof}

\begin{lemma}[Uniform empirical local moment fluctuations and denominator consequences]
\label{lemma:D.uniform_empirical_moments}
Assume Conditions~\ref{con:U-K1}, \ref{con:U-B1}, and~\ref{con:U-D1}. Then
\begin{align}
    \sup_{\mx\in\mathcal K}
    \left|
    n^{-1}\sum_{i=1}^{n}
    \mathcal L_{\mx,h}\left(\mX^{(i)}\right)
    -
    \E\left[
    \mathcal L_{\mx,h}(\mX)
    \right]
    \right|
    &=
    O_{\P}\left(
    h^{d}
    \left(
    \frac{\log n}{nh^d}
    \right)^{1/2}
    \right),
    \label{eq:D.uniform_empirical_moment_fluctuations}
\end{align}
or equivalently,
\begin{align}
    \sup_{\mx\in\mathcal K}
    \left|
    \hat{\mu}_{h,0}(\mx)
    -
    \tilde{\mu}_{h,0}(\mx)
    \right|
    &=
    O_{\P}\left(
    h^{d}
    \left(
    \frac{\log n}{nh^d}
    \right)^{1/2}
    \right).
    \label{eq:D.uniform_empirical_mu_fluctuations}
\end{align}
Moreover,
\begin{align}
    \P\left(
    \inf_{\mx\in\mathcal K}
    \hat{\mu}_{h,0}(\mx)>0
    \right)
    \to1.
    \label{eq:D.uniform_hat_mu0_positive}
\end{align}

Suppose, in addition, that Condition~\ref{con:U-K2} holds. For
$\alpha\in\{1,\ldots,N_{\mathcal K}\}$ and
$\mx\in\mathcal K\cap\mathcal O^\alpha$, write
\begin{align*}
    \bm{\hat{\mu}}_{h,1}^{\alpha}(\mx)
    :=
    \bm{\hat{\mu}}_{h,1}
    \left(
    \mx,\mathbf E^\alpha_{\mx}
    \right),
    \quad
    \bm{\hat{\mu}}_{h,2}^{\alpha}(\mx)
    :=
    \bm{\hat{\mu}}_{h,2}
    \left(
    \mx,\mathbf E^\alpha_{\mx}
    \right),\\
    \bm{\tilde{\mu}}_{h,1}^{\alpha}(\mx)
    :=
    \bm{\tilde{\mu}}_{h,1}
    \left(
    \mx,\mathbf E^\alpha_{\mx}
    \right),
    \quad
    \bm{\tilde{\mu}}_{h,2}^{\alpha}(\mx)
    :=
    \bm{\tilde{\mu}}_{h,2}
    \left(
    \mx,\mathbf E^\alpha_{\mx}
    \right).
\end{align*}
Then
\begin{align}
\begin{split}
    \max_{1\leq\alpha\leq N_{\mathcal K}}
    \sup_{\mx\in\mathcal K\cap\mathcal O^\alpha}
    \left\|
    \bm{\hat{\mu}}_{h,1}^{\alpha}(\mx)
    -
    \bm{\tilde{\mu}}_{h,1}^{\alpha}(\mx)
    \right\|_2
    &=
    O_{\P}\left(
    h^{d+1}
    \left(
    \frac{\log n}{nh^d}
    \right)^{1/2}
    \right), \\
    \max_{1\leq\alpha\leq N_{\mathcal K}}
    \sup_{\mx\in\mathcal K\cap\mathcal O^\alpha}
    \left\|
    \bm{\hat{\mu}}_{h,2}^{\alpha}(\mx)
    -
    \bm{\tilde{\mu}}_{h,2}^{\alpha}(\mx)
    \right\|_{\mathrm{op}}
    &=
    O_{\P}\left(
    h^{d+2}
    \left(
    \frac{\log n}{nh^d}
    \right)^{1/2}
    \right).
\end{split}
\label{eq:D.uniform_empirical_mu_fluctuations_12}
\end{align}
Furthermore,
\begin{align}
    \P\left[
    \min_{1\leq\alpha\leq N_{\mathcal K}}
    \inf_{\mx\in\mathcal K\cap\mathcal O^\alpha}
    \lambda_{\min}
    \left(
    \bm{\hat{\mu}}_{h,2}^{\alpha}(\mx)
    \right)
    >0
    \right]
    &\to1,
    \label{eq:D.uniform_hat_mu2_invertible}
\end{align}
and
\begin{align}
    \max_{1\leq\alpha\leq N_{\mathcal K}}
    \sup_{\mx\in\mathcal K\cap\mathcal O^\alpha}
    \left\|
    \bm{\hat{\mu}}_{h,2}^{\alpha}(\mx)^{-1}
    \right\|_{\mathrm{op}}
    &=
    O_{\P}\left(
    h^{-(d+2)}
    \right).
    \label{eq:D.uniform_hat_mu2_inverse_bound}
\end{align}
Finally,
\begin{align}
    \sup_{\mx\in\mathcal K}
    \left|
    \hat{\sigma}_{h}(\mx)
    -
    \tilde{\sigma}_{h}(\mx)
    \right|
    &=
    O_{\P}\left(
    h^d
    \left(
    \frac{\log n}{nh^d}
    \right)^{1/2}
    \right),
    \label{eq:D.uniform_hat_sigma_difference}
\end{align}
and
\begin{align}
    \P\left(
    \inf_{\mx\in\mathcal K}
    \hat{\sigma}_{h}(\mx)>0
    \right)
    &\to1.
    \label{eq:D.uniform_hat_sigma_positive}
\end{align}
\end{lemma}

\begin{proof}[Proof of \Cref{lemma:D.uniform_empirical_moments}]
We first prove the zeroth-order assertion. For each
$\alpha\in\{1,\ldots,N_{\mathcal K}\}$, consider the fixed-bandwidth
class
\begin{align*}
    \widetilde{\mathcal F}_{\alpha,0}(h)
    &:=
    \left\{
    \mz\mapsto
    \mathcal L_{\mx,h}(\mz):
    \mx\in\mathcal K\cap\mathcal O^\alpha
    \right\}.
\end{align*}
By \Cref{lemma:D.vc_kernel_design}, these classes are of VC type uniformly
over $\alpha$ and $h$, with envelopes of order one. Therefore,
\Cref{lemma:D.uniform_local_empirical_process} with $j=0$ gives
\begin{align*}
    \max_{1\leq\alpha\leq N_{\mathcal K}}
    \sup_{\mx\in\mathcal K\cap\mathcal O^\alpha}
    \left|
    (P_n-P)
    \mathcal L_{\mx,h}
    \right|
    &=
    O_{\P}\left(
    \sqrt{
    \frac{h^d\log n}{n}
    }
    \right)
    =
    O_{\P}\left(
    h^d
    \left(
    \frac{\log n}{nh^d}
    \right)^{1/2}
    \right).
\end{align*}
Since the cover is finite and covers $\mathcal K$, this proves
\eqref{eq:D.uniform_empirical_moment_fluctuations} and
\eqref{eq:D.uniform_empirical_mu_fluctuations}.

Set
\begin{align*}
    r_{n,h}
    &:=
    \left(
    \frac{\log n}{nh^d}
    \right)^{1/2}.
\end{align*}
Condition~\ref{con:U-B1} implies
\begin{align*}
    \lim_{n\to\infty}r_{n,h}
    =0.
\end{align*}
By \Cref{lemma:D.uniform_population_moment_orders},
\begin{align*}
    \inf_{\mx\in\mathcal K}
    \tilde{\mu}_{h,0}(\mx)
    &\geq
    \frac12
    h^d
    A_{d-1}c_{d-1,1}c_{\mathcal K}
\end{align*}
for all sufficiently small $h$. Combining this lower bound with
\eqref{eq:D.uniform_empirical_mu_fluctuations} gives
\begin{align*}
    \inf_{\mx\in\mathcal K}
    \hat{\mu}_{h,0}(\mx)
    &\geq
    \frac14
    h^d
    A_{d-1}c_{d-1,1}c_{\mathcal K}
\end{align*}
with probability tending to one. This proves
\eqref{eq:D.uniform_hat_mu0_positive}.

We now assume Condition~\ref{con:U-K2}. Fix
$\alpha\in\{1,\ldots,N_{\mathcal K}\}$ and
$r,s\in\{1,\ldots,d\}$. The first- and second-order coordinate
multipliers are
\begin{align*}
    \psi_{\mx,h,1,r}^{\alpha}(\mz)
    :=
    \left[
    \mathbf v_{\mx}^{\alpha}(\mz)
    \right]_r, \quad
    \psi_{\mx,h,2,r,s}^{\alpha}(\mz)
    :=
    \left[
    \mathbf v_{\mx}^{\alpha}(\mz)
    \right]_r
    \left[
    \mathbf v_{\mx}^{\alpha}(\mz)
    \right]_s.
\end{align*}
On the support of $\mathcal L_{\mx,h}$,
\begin{align*}
    \left|
    \psi_{\mx,h,1,r}^{\alpha}(\mz)
    \right|
    \leq h, \quad
    \left|
    \psi_{\mx,h,2,r,s}^{\alpha}(\mz)
    \right|
    \leq h^2.
\end{align*}
By \Cref{lemma:D.vc_kernel_design}, the corresponding fixed-bandwidth
unnormalized local-design classes have envelopes of orders $h$ and $h^2$
and satisfy the required VC-type covering-number bounds uniformly over
$\alpha$, $r$, and $s$. Hence,
\Cref{lemma:D.uniform_local_empirical_process} gives
\begin{align*}
    &\max_{1\leq\alpha\leq N_{\mathcal K}}
    \max_{1\leq r\leq d}
    \sup_{\mx\in\mathcal K\cap\mathcal O^\alpha}
    \left|
    (P_n-P)
    \left(
    \mathcal L_{\mx,h}
    \left[
    \mathbf v_{\mx}^{\alpha}
    \right]_r
    \right)
    \right|
    =
    O_{\P}\left(
    h^{d+1}r_{n,h}
    \right),
\end{align*}
and
\begin{align*}
    &\max_{1\leq\alpha\leq N_{\mathcal K}}
    \max_{1\leq r,s\leq d}
    \sup_{\mx\in\mathcal K\cap\mathcal O^\alpha}
    \left|
    (P_n-P)
    \left(
    \mathcal L_{\mx,h}
    \left[
    \mathbf v_{\mx}^{\alpha}
    \right]_r
    \left[
    \mathbf v_{\mx}^{\alpha}
    \right]_s
    \right)
    \right|
    =
    O_{\P}\left(
    h^{d+2}r_{n,h}
    \right),
\end{align*}
where the argument $\mX$ is understood in each displayed empirical-process
function.

Since $d$ is fixed,
\begin{align*}
    \|\mathbf u\|_2
    &\leq
    \sqrt d
    \max_{1\leq r\leq d}|u_r|,
    \quad
    \mathbf u\in\mathbb R^d,
\end{align*}
and
\begin{align*}
    \|\mathbf A\|_{\mathrm{op}}
    \leq
    \|\mathbf A\|_{\mathrm F}
    \leq
    d
    \max_{1\leq r,s\leq d}|A_{rs}|,
    \quad
    \mathbf A\in\mathbb R^{d\times d}.
\end{align*}
The preceding componentwise bounds therefore imply \eqref{eq:D.uniform_empirical_mu_fluctuations_12}.

By \Cref{lemma:D.uniform_population_moment_orders},
\begin{align*}
    \min_{1\leq\alpha\leq N_{\mathcal K}}
    \inf_{\mx\in\mathcal K\cap\mathcal O^\alpha}
    \lambda_{\min}
    \left(
    \bm{\tilde{\mu}}_{h,2}^{\alpha}(\mx)
    \right)
    &\geq
    \frac12
    h^{d+2}
    \frac{A_{d-1}c_{d+1,1}}{d}
    c_{\mathcal K}
\end{align*}
for all sufficiently small $h$. By Weyl's inequality and
\eqref{eq:D.uniform_empirical_mu_fluctuations_12},
\begin{align*}
    \min_{1\leq\alpha\leq N_{\mathcal K}}
    \inf_{\mx\in\mathcal K\cap\mathcal O^\alpha}
    \lambda_{\min}
    \left(
    \bm{\hat{\mu}}_{h,2}^{\alpha}(\mx)
    \right)
    &\geq
    \frac14
    h^{d+2}
    \frac{A_{d-1}c_{d+1,1}}{d}
    c_{\mathcal K}
\end{align*}
with probability tending to one. This proves
\eqref{eq:D.uniform_hat_mu2_invertible}. On the same event,
\begin{align*}
    \max_{1\leq\alpha\leq N_{\mathcal K}}
    \sup_{\mx\in\mathcal K\cap\mathcal O^\alpha}
    \left\|
    \bm{\hat{\mu}}_{h,2}^{\alpha}(\mx)^{-1}
    \right\|_{\mathrm{op}}
    &\leq
    4h^{-(d+2)}
    \frac{d}{
    A_{d-1}c_{d+1,1}c_{\mathcal K}
    },
\end{align*}
which proves \eqref{eq:D.uniform_hat_mu2_inverse_bound}.

It remains to control the local linear denominator. For
$\mx\in\mathcal K\cap\mathcal O^\alpha$, define
\begin{align*}
    \hat q_h^\alpha(\mx)
    &:=
    \bm{\hat{\mu}}_{h,1}^{\alpha}(\mx)^\top
    \bm{\hat{\mu}}_{h,2}^{\alpha}(\mx)^{-1}
    \bm{\hat{\mu}}_{h,1}^{\alpha}(\mx), \\
    \tilde q_h^\alpha(\mx)
    &:=
    \bm{\tilde{\mu}}_{h,1}^{\alpha}(\mx)^\top
    \bm{\tilde{\mu}}_{h,2}^{\alpha}(\mx)^{-1}
    \bm{\tilde{\mu}}_{h,1}^{\alpha}(\mx).
\end{align*}
By \Cref{lemma:A.invariance_local_linear_weights}, these scalar quantities
agree across overlapping frame charts. Hence
\begin{align*}
    \hat{\sigma}_h(\mx)-\tilde{\sigma}_h(\mx)
    =
    \hat{\mu}_{h,0}(\mx)-\tilde{\mu}_{h,0}(\mx)
    -
    \left\{
    \hat q_h^\alpha(\mx)-\tilde q_h^\alpha(\mx)
    \right\}
\end{align*}
for any $\alpha$ such that $\mx\in\mathcal O^\alpha$.

By \Cref{lemma:D.uniform_population_moment_orders} and
\eqref{eq:D.uniform_empirical_mu_fluctuations_12},
\begin{align*}
    \max_{1\leq\alpha\leq N_{\mathcal K}}
    \sup_{\mx\in\mathcal K\cap\mathcal O^\alpha}
    \left\|
    \bm{\tilde{\mu}}_{h,1}^{\alpha}(\mx)
    \right\|_2
    &=
    o\left(
    h^{d+1}
    \right), \\
    \max_{1\leq\alpha\leq N_{\mathcal K}}
    \sup_{\mx\in\mathcal K\cap\mathcal O^\alpha}
    \left\|
    \bm{\hat{\mu}}_{h,1}^{\alpha}(\mx)
    \right\|_2
    &=
    O_{\P}\left(
    h^{d+1}
    \right).
\end{align*}
Moreover,
\begin{align*}
    \mathbf A^{-1}-\mathbf B^{-1}
    =
    \mathbf A^{-1}
    (\mathbf B-\mathbf A)
    \mathbf B^{-1}.
\end{align*}
Combining this identity with
\eqref{eq:D.uniform_hat_mu2_inverse_bound},
\eqref{eq:D.uniform_empirical_mu_fluctuations_12}, and
\Cref{lemma:D.uniform_population_moment_orders} gives
\begin{align*}
    &\max_{1\leq\alpha\leq N_{\mathcal K}}
    \sup_{\mx\in\mathcal K\cap\mathcal O^\alpha}
    \left\|
    \bm{\hat{\mu}}_{h,2}^{\alpha}(\mx)^{-1}
    -
    \bm{\tilde{\mu}}_{h,2}^{\alpha}(\mx)^{-1}
    \right\|_{\mathrm{op}} \\
    &\quad=
    O_{\P}\left(
    h^{-(d+2)}r_{n,h}
    \right).
\end{align*}

For each $\alpha$ and $\mx\in\mathcal K\cap\mathcal O^\alpha$,
\begin{align*}
    \hat q_h^\alpha(\mx)-\tilde q_h^\alpha(\mx)
    &=
    \left\{
    \bm{\hat{\mu}}_{h,1}^{\alpha}(\mx)
    -
    \bm{\tilde{\mu}}_{h,1}^{\alpha}(\mx)
    \right\}^{\top}
    \bm{\hat{\mu}}_{h,2}^{\alpha}(\mx)^{-1}
    \bm{\hat{\mu}}_{h,1}^{\alpha}(\mx) \\
    &\quad+
    \bm{\tilde{\mu}}_{h,1}^{\alpha}(\mx)^{\top}
    \left\{
    \bm{\hat{\mu}}_{h,2}^{\alpha}(\mx)^{-1}
    -
    \bm{\tilde{\mu}}_{h,2}^{\alpha}(\mx)^{-1}
    \right\}
    \bm{\hat{\mu}}_{h,1}^{\alpha}(\mx) \\
    &\quad+
    \bm{\tilde{\mu}}_{h,1}^{\alpha}(\mx)^{\top}
    \bm{\tilde{\mu}}_{h,2}^{\alpha}(\mx)^{-1}
    \left\{
    \bm{\hat{\mu}}_{h,1}^{\alpha}(\mx)
    -
    \bm{\tilde{\mu}}_{h,1}^{\alpha}(\mx)
    \right\}.
\end{align*}
Taking the maximum over $\alpha$ and the supremum over
$\mx\in\mathcal K\cap\mathcal O^\alpha$, and using the preceding bounds,
we obtain
\begin{align*}
    \max_{1\leq\alpha\leq N_{\mathcal K}}
    \sup_{\mx\in\mathcal K\cap\mathcal O^\alpha}
    \left|
    \hat q_h^\alpha(\mx)
    -
    \tilde q_h^\alpha(\mx)
    \right|
    =
    O_{\P}\left(
    h^d r_{n,h}
    \right).
\end{align*}
Together with \eqref{eq:D.uniform_empirical_mu_fluctuations}, this proves
\eqref{eq:D.uniform_hat_sigma_difference}.

Finally, by \Cref{lemma:D.uniform_population_moment_orders},
\begin{align*}
    \inf_{\mx\in\mathcal K}
    \tilde{\sigma}_h(\mx)
    &\geq
    \frac12
    h^d
    A_{d-1}c_{d-1,1}c_{\mathcal K}
\end{align*}
for all sufficiently small $h$. Since
$\lim_{n\to\infty}r_{n,h}=0$,
\eqref{eq:D.uniform_hat_sigma_difference} implies
\begin{align*}
    \inf_{\mx\in\mathcal K}
    \hat{\sigma}_h(\mx)
    &\geq
    \frac14
    h^d
    A_{d-1}c_{d-1,1}c_{\mathcal K}
\end{align*}
with probability tending to one. This proves
\eqref{eq:D.uniform_hat_sigma_positive}.
\end{proof}

\begin{lemma}[Uniform empirical weighted-loss fluctuations]
\label{lemma:D.uniform_empirical_weighted_losses}
Assume Conditions~\ref{con:U-K1}, \ref{con:U-B1},
\ref{con:U-D1}, and~\ref{con:M1}. Assume that the displayed suprema are
measurable. Then
\begin{align}
    \sup_{\mx\in\mathcal K}
    \sup_{y\in\mathbb M}
    \left|
    \hat{\nu}_{h,0}(\mx,y)
    -
    \tilde{\nu}_{h,0}(\mx,y)
    \right|
    &=
    o_{\P}\left(
    h^d
    \right).
    \label{eq:D.uniform_empirical_weighted_loss_0}
\end{align}

Suppose, in addition, that Condition~\ref{con:U-K2} holds. For
$\alpha\in\{1,\ldots,N_{\mathcal K}\}$ and
$\mx\in\mathcal K\cap\mathcal O^\alpha$, write
\begin{align*}
    \bm{\hat{\nu}}_{h,1}^{\alpha}(\mx,y)
    &:=
    \bm{\hat{\nu}}_{h,1}
    \left(
    \mx,
    \mathbf E^\alpha_{\mx},
    y
    \right), \\
    \bm{\tilde{\nu}}_{h,1}^{\alpha}(\mx,y)
    &:=
    \bm{\tilde{\nu}}_{h,1}
    \left(
    \mx,
    \mathbf E^\alpha_{\mx},
    y
    \right).
\end{align*}
Then
\begin{align}
    \max_{1\leq\alpha\leq N_{\mathcal K}}
    \sup_{\mx\in\mathcal K\cap\mathcal O^\alpha}
    \sup_{y\in\mathbb M}
    \left\|
    \bm{\hat{\nu}}_{h,1}^{\alpha}(\mx,y)
    -
    \bm{\tilde{\nu}}_{h,1}^{\alpha}(\mx,y)
    \right\|_2
    &=
    o_{\P}\left(
    h^{d+1}
    \right).
    \label{eq:D.uniform_empirical_weighted_loss_1}
\end{align}
\end{lemma}

\begin{proof}[Proof of \Cref{lemma:D.uniform_empirical_weighted_losses}]
By Condition~\ref{con:M1}, $D_{\mathbb M}$ defined in
\eqref{eq:B.diameter_metric_space} is finite. Fix $\eta>0$. By total
boundedness, there exist
$y_1,\ldots,y_{N_\eta}\in\mathbb M$, where $N_\eta<\infty$, such that
\begin{align*}
    \mathbb M
    \subset
    \bigcup_{\ell=1}^{N_\eta}
    B_{\mathbb M}(y_\ell,\eta).
\end{align*}

We first establish the empirical-process bounds at the finitely many net
points. For $\alpha\in\{1,\ldots,N_{\mathcal K}\}$ and
$\ell\in\{1,\ldots,N_\eta\}$, define
\begin{align*}
    \mathcal H_{h,0,\ell}^{\alpha}
    &:=
    \left\{
    (\mz,\omega)\mapsto
    \mathcal L_{\mx,h}(\mz)
    d_{\mathbb M}^2(y_\ell,\omega):
    \mx\in\mathcal K\cap\mathcal O^\alpha
    \right\}.
\end{align*}
Under Condition~\ref{con:U-K2}, also define, for
$r\in\{1,\ldots,d\}$,
\begin{align*}
    \mathcal H_{h,1,r,\ell}^{\alpha}
    &:=
    \left\{
    (\mz,\omega)\mapsto
    \mathcal L_{\mx,h}(\mz)
    \left[
    \mathbf v_{\mx}^{\alpha}(\mz)
    \right]_r
    d_{\mathbb M}^2(y_\ell,\omega):
    \mx\in\mathcal K\cap\mathcal O^\alpha
    \right\}.
\end{align*}

We verify that multiplication by the fixed response-side function
\begin{align*}
    g_\ell(\omega)
    &:=
    d_{\mathbb M}^2(y_\ell,\omega)
\end{align*}
preserves the required polynomial covering-number bounds. Let $Q$ be a
finitely discrete probability measure on $\mathcal M\times\mathbb M$ and
define
\begin{align*}
    c_{\ell,Q}^2
    &:=
    \int_{\mathcal M\times\mathbb M}
    g_\ell(\omega)^2
    \dd Q(\mz,\omega).
\end{align*}
If $c_{\ell,Q}=0$, all relevant $L_2(Q)$ distances vanish. Suppose that
$c_{\ell,Q}>0$ and define the finitely discrete probability measure
$Q_{\ell,\mathcal M}$ on $\mathcal M$ by
\begin{align*}
    Q_{\ell,\mathcal M}(A)
    &:=
    \frac{1}{c_{\ell,Q}^2}
    \int_{\mathcal M\times\mathbb M}
    \mathds 1(\mz\in A)
    g_\ell(\omega)^2
    \dd Q(\mz,\omega).
\end{align*}
Then, for any real-valued functions $f_1$ and $f_2$ on $\mathcal M$,
\begin{align*}
    \left\|
    (f_1-f_2)g_\ell
    \right\|_{L_2(Q)}
    &=
    c_{\ell,Q}
    \left\|
    f_1-f_2
    \right\|_{L_2(Q_{\ell,\mathcal M})}.
\end{align*}
Since $g_\ell(\omega)\leq D_{\mathbb M}^2$, we have
$c_{\ell,Q}\leq D_{\mathbb M}^2$. It follows from
\Cref{lemma:D.vc_kernel_design} that
$\mathcal H_{h,0,\ell}^{\alpha}$ and
$\mathcal H_{h,1,r,\ell}^{\alpha}$ satisfy the same polynomial
covering-number bounds as the corresponding predictor-side classes, with
their envelopes multiplied by at most $D_{\mathbb M}^2$.

The proof of \Cref{lemma:D.uniform_local_empirical_process} uses only the
local support, envelope, maximal second moment, and VC-type entropy bound.
It therefore applies on the product sample space
$\mathcal M\times\mathbb M$. Consequently,
\begin{align}
\begin{split}
    &\max_{1\leq\alpha\leq N_{\mathcal K}}
    \max_{1\leq\ell\leq N_\eta}
    \sup_{\mx\in\mathcal K\cap\mathcal O^\alpha}
    \left|
    (P_n-P)
    \left[
    \mathcal L_{\mx,h}(\mX)
    d_{\mathbb M}^2(y_\ell,Y)
    \right]
    \right| \\
    &\quad=
    O_{\P}\left(
    \sqrt{
    \frac{h^d\log n}{n}
    }
    \right)
    =
    o_{\P}\left(
    h^d
    \right),
\end{split}
\label{eq:D.weighted_loss_net_points_0}
\end{align}
where the last equality follows from Condition~\ref{con:U-B1}.

Similarly, under Condition~\ref{con:U-K2},
\begin{align}
\begin{split}
    &\max_{1\leq\alpha\leq N_{\mathcal K}}
    \max_{1\leq r\leq d}
    \max_{1\leq\ell\leq N_\eta}
    \sup_{\mx\in\mathcal K\cap\mathcal O^\alpha}
    \left|
    (P_n-P)
    \left[
    \mathcal L_{\mx,h}(\mX)
    \left[
    \mathbf v_{\mx}^{\alpha}(\mX)
    \right]_r
    d_{\mathbb M}^2(y_\ell,Y)
    \right]
    \right| \\
    &\quad=
    O_{\P}\left(
    h
    \sqrt{
    \frac{h^d\log n}{n}
    }
    \right)
    =
    o_{\P}\left(
    h^{d+1}
    \right).
\end{split}
\label{eq:D.weighted_loss_net_points_1}
\end{align}

We next extend the finite-net bounds to the full supremum over
$y\in\mathbb M$. For each $y\in\mathbb M$, choose
$\ell(y)\in\{1,\ldots,N_\eta\}$ such that
\begin{align*}
    d_{\mathbb M}
    \left(
    y,y_{\ell(y)}
    \right)
    <\eta.
\end{align*}
For every $\omega\in\mathbb M$,
\begin{align}
    \left|
    d_{\mathbb M}^2(y,\omega)
    -
    d_{\mathbb M}^2
    \left(
    y_{\ell(y)},\omega
    \right)
    \right|
    &\leq
    2D_{\mathbb M}\eta.
    \label{eq:D.weighted_loss_metric_difference}
\end{align}

For the zeroth-order process,
\begin{align*}
    &\sup_{\mx\in\mathcal K}
    \sup_{y\in\mathbb M}
    \left|
    (P_n-P)
    \left[
    \mathcal L_{\mx,h}(\mX)
    \left\{
    d_{\mathbb M}^2(y,Y)
    -
    d_{\mathbb M}^2
    \left(
    y_{\ell(y)},Y
    \right)
    \right\}
    \right]
    \right| \\
    &\quad\leq
    2D_{\mathbb M}\eta
    \sup_{\mx\in\mathcal K}
    \left[
    P_n
    \left\{
    \mathcal L_{\mx,h}(\mX)
    \right\}
    +
    P
    \left\{
    \mathcal L_{\mx,h}(\mX)
    \right\}
    \right].
\end{align*}
By \Cref{lemma:D.uniform_population_moment_orders} and
\Cref{lemma:D.uniform_empirical_moments},
\begin{align*}
    \sup_{\mx\in\mathcal K}
    P
    \left\{
    \mathcal L_{\mx,h}(\mX)
    \right\}
    &=
    O\left(
    h^d
    \right), \\
    \sup_{\mx\in\mathcal K}
    P_n
    \left\{
    \mathcal L_{\mx,h}(\mX)
    \right\}
    &=
    O_{\P}\left(
    h^d
    \right).
\end{align*}
Combining these bounds with
\eqref{eq:D.weighted_loss_net_points_0} gives, for every fixed $\eta>0$,
\begin{align*}
    h^{-d}
    \sup_{\mx\in\mathcal K}
    \sup_{y\in\mathbb M}
    \left|
    (P_n-P)
    \left[
    \mathcal L_{\mx,h}(\mX)
    d_{\mathbb M}^2(y,Y)
    \right]
    \right|
    &\leq
    o_{\P}(1)
    +
    2D_{\mathbb M}\eta O_{\P}(1).
\end{align*}
Since the $O_{\P}(1)$ term is tight, first letting $n\to\infty$ and then
letting $\eta\downarrow0$ yields
\begin{align*}
    \sup_{\mx\in\mathcal K}
    \sup_{y\in\mathbb M}
    \left|
    (P_n-P)
    \left[
    \mathcal L_{\mx,h}(\mX)
    d_{\mathbb M}^2(y,Y)
    \right]
    \right|
    &=
    o_{\P}\left(
    h^d
    \right).
\end{align*}
This proves \eqref{eq:D.uniform_empirical_weighted_loss_0}.

For the first-order process, on the support of
$\mathcal L_{\mx,h}$,
\begin{align*}
    \left|
    \left[
    \mathbf v_{\mx}^{\alpha}(\mX)
    \right]_r
    \right|
    &\leq
    \left\|
    \Log_{\mx}(\mX)
    \right\|_{\mx}
    \leq h.
\end{align*}
Consequently,
\begin{align*}
    &\max_{1\leq\alpha\leq N_{\mathcal K}}
    \max_{1\leq r\leq d}
    \sup_{\mx\in\mathcal K\cap\mathcal O^\alpha}
    P_n
    \left[
    \mathcal L_{\mx,h}(\mX)
    \left|
    \left[
    \mathbf v_{\mx}^{\alpha}(\mX)
    \right]_r
    \right|
    \right] \\
    &\quad\leq
    h
    \sup_{\mx\in\mathcal K}
    P_n
    \left[
    \mathcal L_{\mx,h}(\mX)
    \right]
    =
    O_{\P}\left(
    h^{d+1}
    \right),
\end{align*}
and similarly,
\begin{align*}
    &\max_{1\leq\alpha\leq N_{\mathcal K}}
    \max_{1\leq r\leq d}
    \sup_{\mx\in\mathcal K\cap\mathcal O^\alpha}
    P
    \left[
    \mathcal L_{\mx,h}(\mX)
    \left|
    \left[
    \mathbf v_{\mx}^{\alpha}(\mX)
    \right]_r
    \right|
    \right] \\
    &\quad=
    O\left(
    h^{d+1}
    \right).
\end{align*}
Using \eqref{eq:D.weighted_loss_metric_difference} and
\eqref{eq:D.weighted_loss_net_points_1}, we obtain, for every fixed
$\eta>0$,
\begin{align*}
    &h^{-(d+1)}
    \max_{1\leq\alpha\leq N_{\mathcal K}}
    \max_{1\leq r\leq d}
    \sup_{\mx\in\mathcal K\cap\mathcal O^\alpha}
    \sup_{y\in\mathbb M}
    \left|
    (P_n-P)
    \left[
    \mathcal L_{\mx,h}(\mX)
    \left[
    \mathbf v_{\mx}^{\alpha}(\mX)
    \right]_r
    d_{\mathbb M}^2(y,Y)
    \right]
    \right| \\
    &\quad\leq
    o_{\P}(1)
    +
    2D_{\mathbb M}\eta O_{\P}(1).
\end{align*}
Letting $n\to\infty$ and then $\eta\downarrow0$ gives
\begin{align*}
    &\max_{1\leq\alpha\leq N_{\mathcal K}}
    \max_{1\leq r\leq d}
    \sup_{\mx\in\mathcal K\cap\mathcal O^\alpha}
    \sup_{y\in\mathbb M}
    \left|
    (P_n-P)
    \left[
    \mathcal L_{\mx,h}(\mX)
    \left[
    \mathbf v_{\mx}^{\alpha}(\mX)
    \right]_r
    d_{\mathbb M}^2(y,Y)
    \right]
    \right| \\
    &\quad=
    o_{\P}\left(
    h^{d+1}
    \right).
\end{align*}

The $r$th coordinate of
\begin{align*}
    \bm{\hat{\nu}}_{h,1}^{\alpha}(\mx,y)
    -
    \bm{\tilde{\nu}}_{h,1}^{\alpha}(\mx,y)
\end{align*}
is the empirical-process term in the preceding display. Since
\begin{align*}
    \|\mathbf u\|_2
    &\leq
    \sqrt d
    \max_{1\leq r\leq d}
    |u_r|,
    \quad
    \mathbf u\in\mathbb R^d,
\end{align*}
the componentwise bound proves
\eqref{eq:D.uniform_empirical_weighted_loss_1}. On overlaps of the frame
cover, the corresponding coordinate vectors are related by an orthogonal
transformation, so their Euclidean norms agree. This completes the proof.
\end{proof}

\begin{lemma}[Uniform empirical local-objective convergence] \label{lemma:D.uniform_empirical_objective}
Assume Conditions~\ref{con:U-K1}, \ref{con:U-B1}, \ref{con:U-D1}, \ref{con:U-D2}, and~\ref{con:M1}. Then
\begin{align}
    \sup_{\mx\in\mathcal K}
    \sup_{y\in\mathbb M}
    \left|
    \hat{M}_{h,0}(\mx,y)-\tilde{M}_{h,0}(\mx,y)
    \right|
    =
    o_{\P}(1).
    \label{eq:D.uniform_emp_objective_consistency_0}
\end{align}
If, in addition, Condition~\ref{con:U-K2} holds, then
\begin{align}
    \sup_{\mx\in\mathcal K}
    \sup_{y\in\mathbb M}
    \left|
    \hat{M}_{h,1}(\mx,y)-\tilde{M}_{h,1}(\mx,y)
    \right|
    =
    o_{\P}(1).
    \label{eq:D.uniform_emp_objective_consistency_1}
\end{align}
\end{lemma}

\begin{proof}[Proof of \Cref{lemma:D.uniform_empirical_objective}]
Set
\begin{align*}
    r_{n,h}
    &:=
    \left(\frac{\log n}{nh^d}\right)^{1/2}.
\end{align*}
By Condition~\ref{con:U-B1}, $r_{n,h}=o(1)$. We use the numerator--denominator representations in \eqref{eq:B.population_ND_defs}, \eqref{eq:B.empirical_ND_defs}, and \eqref{eq:B.empirical_M_repr}.

We first prove the local constant assertion. For $s=0$, we have
\begin{align*}
    \hat{D}_{h,0}(\mx)=\hat{\mu}_{h,0}(\mx),
    \quad
    \tilde{D}_{h,0}(\mx)=\tilde{\mu}_{h,0}(\mx),
\end{align*}
and
\begin{align*}
    \hat{N}_{h,0}(\mx,y)=\hat{\nu}_{h,0}(\mx,y),
    \quad
    \tilde{N}_{h,0}(\mx,y)=\tilde{\nu}_{h,0}(\mx,y).
\end{align*}
By \Cref{lemma:D.uniform_empirical_moments},
\begin{align}
    \sup_{\mx\in\mathcal K}
    \left|
    \hat{D}_{h,0}(\mx)-\tilde{D}_{h,0}(\mx)
    \right|
    =
    O_{\P}\left(h^d r_{n,h}\right)
    =
    o_{\P}(h^d),
    \label{eq:D.uniform_D0_difference_for_objective}
\end{align}
and by \Cref{lemma:D.uniform_empirical_weighted_losses},
\begin{align}
    \sup_{\mx\in\mathcal K}
    \sup_{y\in\mathbb M}
    \left|
    \hat{N}_{h,0}(\mx,y)-\tilde{N}_{h,0}(\mx,y)
    \right|
    =
    o_{\P}(h^d).
    \label{eq:D.uniform_N0_difference_for_objective}
\end{align}
Moreover, by \Cref{lemma:D.uniform_population_moment_orders} and \Cref{lemma:D.uniform_empirical_moments},
\begin{align}
    \inf_{\mx\in\mathcal K}\tilde{D}_{h,0}(\mx)
    \geq
    \frac12 A_{d-1}c_{d-1,1}c_{\mathcal K}h^d,
    \quad
    \inf_{\mx\in\mathcal K}\hat{D}_{h,0}(\mx)
    \geq
    \frac14 A_{d-1}c_{d-1,1}c_{\mathcal K}h^d
    \label{eq:D.uniform_D0_lower_for_objective}
\end{align}
with probability tending to one. Since $D_{\mathbb M}$ defined in \eqref{eq:B.diameter_metric_space} is finite by Condition~\ref{con:M1}, we also have
\begin{align}
    \sup_{\mx\in\mathcal K}
    \sup_{y\in\mathbb M}
    \left|
    \tilde{N}_{h,0}(\mx,y)
    \right|
    \leq
    D_{\mathbb M}^2
    \sup_{\mx\in\mathcal K}
    \tilde{D}_{h,0}(\mx)
    =
    O(h^d).
    \label{eq:D.uniform_N0_population_bound_for_objective}
\end{align}
Thus, on the event in \eqref{eq:D.uniform_D0_lower_for_objective}, \eqref{eq:B.empirical_M_repr} gives
\begin{align*}
    &\sup_{\mx\in\mathcal K}
    \sup_{y\in\mathbb M}
    \left|
    \hat{M}_{h,0}(\mx,y)-\tilde{M}_{h,0}(\mx,y)
    \right| \\
    &\leq
    \sup_{\mx\in\mathcal K}
    \sup_{y\in\mathbb M}
    \frac{
    \left|
    \hat{N}_{h,0}(\mx,y)-\tilde{N}_{h,0}(\mx,y)
    \right|
    }{
    \hat{D}_{h,0}(\mx)
    } \\
    &\quad+
    \sup_{\mx\in\mathcal K}
    \sup_{y\in\mathbb M}
    \frac{
    \left|
    \tilde{N}_{h,0}(\mx,y)
    \right|
    \left|
    \hat{D}_{h,0}(\mx)-\tilde{D}_{h,0}(\mx)
    \right|
    }{
    \hat{D}_{h,0}(\mx)\tilde{D}_{h,0}(\mx)
    }.
\end{align*}
Combining \eqref{eq:D.uniform_D0_difference_for_objective}, \eqref{eq:D.uniform_N0_difference_for_objective}, \eqref{eq:D.uniform_D0_lower_for_objective}, and \eqref{eq:D.uniform_N0_population_bound_for_objective} yields
\begin{align*}
    \sup_{\mx\in\mathcal K}
    \sup_{y\in\mathbb M}
    \left|
    \hat{M}_{h,0}(\mx,y)-\tilde{M}_{h,0}(\mx,y)
    \right|
    =
    o_{\P}(1),
\end{align*}
which proves \eqref{eq:D.uniform_emp_objective_consistency_0}.

We now assume in addition Condition~\ref{con:U-K2} and prove the local linear assertion. For $s=1$, we have
\begin{align*}
    \hat{D}_{h,1}(\mx)=\hat{\sigma}_h(\mx),
    \quad
    \tilde{D}_{h,1}(\mx)=\tilde{\sigma}_h(\mx).
\end{align*}
By \Cref{lemma:D.uniform_empirical_moments},
\begin{align}
    \inf_{\mx\in\mathcal K}\hat{D}_{h,1}(\mx)>0,
    \quad
    \sup_{\mx\in\mathcal K}
    \left\|
    \bm{\hat{\mu}}_{h,2}(\mx,\mathbf E_{\mx})^{-1}
    \right\|_{\mathrm{op}}
    =
    O_{\P}(h^{-(d+2)})
    \label{eq:D.uniform_empirical_linear_denominator_event_for_objective}
\end{align}
with probability tending to one. Moreover, by \Cref{lemma:D.uniform_population_moment_orders} and \eqref{eq:D.uniform_hat_sigma_difference},
\begin{align}
    \inf_{\mx\in\mathcal K}\tilde{D}_{h,1}(\mx)
    \geq
    \frac12 A_{d-1}c_{d-1,1}c_{\mathcal K}h^d,
    \quad
    \inf_{\mx\in\mathcal K}\hat{D}_{h,1}(\mx)
    \geq
    \frac14 A_{d-1}c_{d-1,1}c_{\mathcal K}h^d
    \label{eq:D.uniform_D1_lower_for_objective}
\end{align}
with probability tending to one, and
\begin{align}
    \sup_{\mx\in\mathcal K}
    \left|
    \hat{D}_{h,1}(\mx)-\tilde{D}_{h,1}(\mx)
    \right|
    =
    \sup_{\mx\in\mathcal K}
    \left|
    \hat{\sigma}_h(\mx)-\tilde{\sigma}_h(\mx)
    \right|
    =
    O_{\P}(h^d r_{n,h})
    =
    o_{\P}(h^d).
    \label{eq:D.uniform_D1_difference_for_objective}
\end{align}

Next we control the local linear numerator. By \eqref{eq:B.population_ND_defs} and \eqref{eq:B.empirical_ND_defs},
\begin{align*}
    &\hat{N}_{h,1}(\mx,y)-\tilde{N}_{h,1}(\mx,y) \\
    &=
    \hat{\nu}_{h,0}(\mx,y)-\tilde{\nu}_{h,0}(\mx,y) \\
    &\quad-
    \left[
    \bm{\hat{\mu}}_{h,1}(\mx,\mathbf E_{\mx})^{\top}
    \bm{\hat{\mu}}_{h,2}(\mx,\mathbf E_{\mx})^{-1}
    \bm{\hat{\nu}}_{h,1}(\mx,\mathbf E_{\mx},y)
    -
    \bm{\tilde{\mu}}_{h,1}(\mx,\mathbf E_{\mx})^{\top}
    \bm{\tilde{\mu}}_{h,2}(\mx,\mathbf E_{\mx})^{-1}
    \bm{\tilde{\nu}}_{h,1}(\mx,\mathbf E_{\mx},y)
    \right].
\end{align*}
By \Cref{lemma:D.uniform_empirical_moments}, for $j=1,2$,
\begin{align}
    \sup_{\mx\in\mathcal K}
    \left\|
    \bm{\hat{\mu}}_{h,j}(\mx,\mathbf E_{\mx})
    -
    \bm{\tilde{\mu}}_{h,j}(\mx,\mathbf E_{\mx})
    \right\|_{\star}
    =
    O_{\P}\left(h^{d+j}r_{n,h}\right).
    \label{eq:D.uniform_mu12_difference_for_objective}
\end{align}
Also, by \Cref{lemma:D.uniform_empirical_weighted_losses},
\begin{align}
    \sup_{\mx\in\mathcal K}
    \sup_{y\in\mathbb M}
    \left\|
    \bm{\hat{\nu}}_{h,1}(\mx,\mathbf E_{\mx},y)
    -
    \bm{\tilde{\nu}}_{h,1}(\mx,\mathbf E_{\mx},y)
    \right\|_2
    =
    o_{\P}(h^{d+1}).
    \label{eq:D.uniform_nu1_difference_for_objective}
\end{align}
Furthermore, by \Cref{lemma:D.uniform_population_moment_orders} and Condition~\ref{con:M1},
\begin{align}
    \sup_{\mx\in\mathcal K}
    \left\|
    \bm{\tilde{\mu}}_{h,1}(\mx,\mathbf E_{\mx})
    \right\|_2
    =
    o(h^{d+1}),
    \quad
    \sup_{\mx\in\mathcal K}
    \left\|
    \bm{\tilde{\mu}}_{h,2}(\mx,\mathbf E_{\mx})^{-1}
    \right\|_{\mathrm{op}}
    =
    O(h^{-(d+2)}),
    \label{eq:D.uniform_population_mu12_for_objective}
\end{align}
and
\begin{align}
    \sup_{\mx\in\mathcal K}
    \sup_{y\in\mathbb M}
    \left\|
    \bm{\tilde{\nu}}_{h,1}(\mx,\mathbf E_{\mx},y)
    \right\|_2
    =
    O(h^{d+1}).
    \label{eq:D.uniform_population_nu1_for_objective}
\end{align}
The last bound follows from $d_{\mathbb M}^2(y,Y)\leq D_{\mathbb M}^2$ and the first-order population local-moment bound. Hence \eqref{eq:D.uniform_nu1_difference_for_objective} and \eqref{eq:D.uniform_population_nu1_for_objective} imply
\begin{align}
    \sup_{\mx\in\mathcal K}
    \sup_{y\in\mathbb M}
    \left\|
    \bm{\hat{\nu}}_{h,1}(\mx,\mathbf E_{\mx},y)
    \right\|_2
    =
    O_{\P}(h^{d+1}).
    \label{eq:D.uniform_empirical_nu1_bound_for_objective}
\end{align}
Using the matrix identity $\mathbf A^{-1}-\mathbf B^{-1}=\mathbf A^{-1}(\mathbf B-\mathbf A)\mathbf B^{-1}$ on the event where the empirical inverses exist, together with \eqref{eq:D.uniform_empirical_linear_denominator_event_for_objective}, \eqref{eq:D.uniform_mu12_difference_for_objective} with $j=2$, and \eqref{eq:D.uniform_population_mu12_for_objective}, gives
\begin{align}
    \sup_{\mx\in\mathcal K}
    \left\|
    \bm{\hat{\mu}}_{h,2}(\mx,\mathbf E_{\mx})^{-1}
    -
    \bm{\tilde{\mu}}_{h,2}(\mx,\mathbf E_{\mx})^{-1}
    \right\|_{\mathrm{op}}
    =
    O_{\P}\left(h^{-(d+2)}r_{n,h}\right).
    \label{eq:D.uniform_inverse_difference_for_objective}
\end{align}
Combining \eqref{eq:D.uniform_mu12_difference_for_objective}, \eqref{eq:D.uniform_population_mu12_for_objective}, \eqref{eq:D.uniform_empirical_nu1_bound_for_objective}, \eqref{eq:D.uniform_nu1_difference_for_objective}, \eqref{eq:D.uniform_inverse_difference_for_objective}, and \eqref{eq:D.uniform_empirical_linear_denominator_event_for_objective}, we obtain
\begin{align*}
    &\sup_{\mx\in\mathcal K}
    \sup_{y\in\mathbb M}
    \left|
    \bm{\hat{\mu}}_{h,1}(\mx,\mathbf E_{\mx})^\top
    \bm{\hat{\mu}}_{h,2}(\mx,\mathbf E_{\mx})^{-1}
    \bm{\hat{\nu}}_{h,1}(\mx,\mathbf E_{\mx},y)
    -
    \bm{\tilde{\mu}}_{h,1}(\mx,\mathbf E_{\mx})^\top
    \bm{\tilde{\mu}}_{h,2}(\mx,\mathbf E_{\mx})^{-1}
    \bm{\tilde{\nu}}_{h,1}(\mx,\mathbf E_{\mx},y)
    \right| \\
    &\leq
    \sup_{\mx\in\mathcal K}
    \left\|
    \bm{\hat{\mu}}_{h,1}(\mx,\mathbf E_{\mx})
    -
    \bm{\tilde{\mu}}_{h,1}(\mx,\mathbf E_{\mx})
    \right\|_2
    \sup_{\mx\in\mathcal K}
    \left\|
    \bm{\hat{\mu}}_{h,2}(\mx,\mathbf E_{\mx})^{-1}
    \right\|_{\mathrm{op}}
    \sup_{\mx\in\mathcal K}
    \sup_{y\in\mathbb M}
    \left\|
    \bm{\hat{\nu}}_{h,1}(\mx,\mathbf E_{\mx},y)
    \right\|_2 \\
    &\quad+
    \sup_{\mx\in\mathcal K}
    \left\|
    \bm{\tilde{\mu}}_{h,1}(\mx,\mathbf E_{\mx})
    \right\|_2
    \sup_{\mx\in\mathcal K}
    \left\|
    \bm{\hat{\mu}}_{h,2}(\mx,\mathbf E_{\mx})^{-1}
    -
    \bm{\tilde{\mu}}_{h,2}(\mx,\mathbf E_{\mx})^{-1}
    \right\|_{\mathrm{op}}
    \sup_{\mx\in\mathcal K}
    \sup_{y\in\mathbb M}
    \left\|
    \bm{\hat{\nu}}_{h,1}(\mx,\mathbf E_{\mx},y)
    \right\|_2 \\
    &\quad+
    \sup_{\mx\in\mathcal K}
    \left\|
    \bm{\tilde{\mu}}_{h,1}(\mx,\mathbf E_{\mx})
    \right\|_2
    \sup_{\mx\in\mathcal K}
    \left\|
    \bm{\tilde{\mu}}_{h,2}(\mx,\mathbf E_{\mx})^{-1}
    \right\|_{\mathrm{op}}
    \sup_{\mx\in\mathcal K}
    \sup_{y\in\mathbb M}
    \left\|
    \bm{\hat{\nu}}_{h,1}(\mx,\mathbf E_{\mx},y)
    -
    \bm{\tilde{\nu}}_{h,1}(\mx,\mathbf E_{\mx},y)
    \right\|_2 \\
    &=
    o_{\P}(h^d).
\end{align*}
Together with \eqref{eq:D.uniform_N0_difference_for_objective}, this gives
\begin{align}
    \sup_{\mx\in\mathcal K}
    \sup_{y\in\mathbb M}
    \left|
    \hat{N}_{h,1}(\mx,y)-\tilde{N}_{h,1}(\mx,y)
    \right|
    =
    o_{\P}(h^d).
    \label{eq:D.uniform_N1_difference_for_objective}
\end{align}
Finally, by Condition~\ref{con:M1}, \Cref{lemma:D.uniform_population_moment_orders}, and the numerator representation for $s=1$,
\begin{align}
    \sup_{\mx\in\mathcal K}
    \sup_{y\in\mathbb M}
    \left|
    \tilde{N}_{h,1}(\mx,y)
    \right|
    =
    O(h^d).
    \label{eq:D.uniform_N1_population_bound_for_objective}
\end{align}
Therefore, on the event in \eqref{eq:D.uniform_D1_lower_for_objective}, \eqref{eq:B.empirical_M_repr} gives
\begin{align*}
    &\sup_{\mx\in\mathcal K}
    \sup_{y\in\mathbb M}
    \left|
    \hat{M}_{h,1}(\mx,y)-\tilde{M}_{h,1}(\mx,y)
    \right| \\
    &\leq
    \sup_{\mx\in\mathcal K}
    \sup_{y\in\mathbb M}
    \frac{
    \left|
    \hat{N}_{h,1}(\mx,y)-\tilde{N}_{h,1}(\mx,y)
    \right|
    }{
    \hat{D}_{h,1}(\mx)
    } \\
    &\quad+
    \sup_{\mx\in\mathcal K}
    \sup_{y\in\mathbb M}
    \frac{
    \left|
    \tilde{N}_{h,1}(\mx,y)
    \right|
    \left|
    \hat{D}_{h,1}(\mx)-\tilde{D}_{h,1}(\mx)
    \right|
    }{
    \hat{D}_{h,1}(\mx)\tilde{D}_{h,1}(\mx)
    }.
\end{align*}
Combining \eqref{eq:D.uniform_D1_lower_for_objective}, \eqref{eq:D.uniform_D1_difference_for_objective}, \eqref{eq:D.uniform_N1_difference_for_objective}, and \eqref{eq:D.uniform_N1_population_bound_for_objective} yields
\begin{align*}
    \sup_{\mx\in\mathcal K}
    \sup_{y\in\mathbb M}
    \left|
    \hat{M}_{h,1}(\mx,y)-\tilde{M}_{h,1}(\mx,y)
    \right|
    =
    o_{\P}(1),
\end{align*}
which proves \eqref{eq:D.uniform_emp_objective_consistency_1}.
\end{proof}

\begin{lemma}[Uniform convergence of empirical local minimizers] \label{lemma:D.uniform_empirical_minimizer}
Assume Conditions~\ref{con:U-K1}, \ref{con:U-B1}, \ref{con:U-D1}, \ref{con:U-D2}, \ref{con:M1}, and~\ref{con:U-M2}, and suppose that $h\to0$ as $n\to\infty$. Then
\begin{align*}
    \sup_{\mx\in\mathcal K}
    d_{\mathbb M}\left(
    \hat{m}_{h,0}(\mx),
    \tilde{m}_{h,0}(\mx)
    \right)
    =
    o_{\P}(1).
\end{align*}
If, in addition, Condition~\ref{con:U-K2} holds, then
\begin{align*}
    \sup_{\mx\in\mathcal K}
    d_{\mathbb M}\left(
    \hat{m}_{h,1}(\mx),
    \tilde{m}_{h,1}(\mx)
    \right)
    =
    o_{\P}(1).
\end{align*}
\end{lemma}

\begin{proof}[Proof of \Cref{lemma:D.uniform_empirical_minimizer}]
Fix $s\in\{0,1\}$ and $\epsilon>0$. When $s=1$, assume in addition Condition~\ref{con:U-K2}. The rest of the proof is identical for $s=0$ and $s=1$, using \Cref{lemma:D.uniform_empirical_objective} with the corresponding value of $s$. Let
\begin{align*}
    \Delta_{h,s,\mathcal K}^{\mathrm{pop}}
    :=
    \sup_{\mx\in\mathcal K}
    \sup_{y\in\mathbb M}
    \left|
    \tilde{M}_{h,s}(\mx,y)
    -
    M_\oplus(\mx,y)
    \right|.
\end{align*}
By \Cref{lemma:D.uniform_population_objective},
\begin{align*}
    \Delta_{h,s,\mathcal K}^{\mathrm{pop}}
    =
    o(1).
\end{align*}
Moreover, by \Cref{lemma:D.uniform_oracle_minimizer},
\begin{align*}
    \sup_{\mx\in\mathcal K}
    d_{\mathbb M}\left(
    \tilde{m}_{h,s}(\mx),
    m_\oplus(\mx)
    \right)
    =
    o(1).
\end{align*}
By Condition~\ref{con:U-M2}, there exists
\begin{align*}
    \eta_{\epsilon,\mathcal K}
    :=
    \inf_{\substack{\mx\in\mathcal K,\ y\in\mathbb M:\\
    d_{\mathbb M}\left(y,m_\oplus(\mx)\right)>2\epsilon/3}}
    \left[
    M_\oplus(\mx,y)
    -
    M_\oplus\left(\mx,m_\oplus(\mx)\right)
    \right]
    >
    0.
\end{align*}
For all sufficiently small $h$,
\begin{align*}
    \sup_{\mx\in\mathcal K}
    d_{\mathbb M}\left(
    \tilde{m}_{h,s}(\mx),
    m_\oplus(\mx)
    \right)
    \leq
    \epsilon/3
    \quad\text{and}\quad
    4\Delta_{h,s,\mathcal K}^{\mathrm{pop}}
    \leq
    \eta_{\epsilon,\mathcal K}/2.
\end{align*}
For such $h$, if $d_{\mathbb M}\left(y,\tilde{m}_{h,s}(\mx)\right)>\epsilon$, then
\begin{align*}
    d_{\mathbb M}\left(y,m_\oplus(\mx)\right)>2\epsilon/3.
\end{align*}
Also, since $\tilde{m}_{h,s}(\mx)$ minimizes $\tilde{M}_{h,s}(\mx,\cdot)$,
\begin{align*}
    M_\oplus\left(\mx,\tilde{m}_{h,s}(\mx)\right)
    -
    M_\oplus\left(\mx,m_\oplus(\mx)\right)
    \leq
    2\Delta_{h,s,\mathcal K}^{\mathrm{pop}},
    \quad \mx\in\mathcal K.
\end{align*}
Therefore, uniformly over all $\mx\in\mathcal K$ and all $y\in\mathbb M$ satisfying $d_{\mathbb M}\left(y,\tilde{m}_{h,s}(\mx)\right)>\epsilon$,
\begin{align*}
    \tilde{M}_{h,s}(\mx,y)
    -
    \tilde{M}_{h,s}\left(\mx,\tilde{m}_{h,s}(\mx)\right)
    &\geq
    M_\oplus(\mx,y)
    -
    M_\oplus\left(\mx,\tilde{m}_{h,s}(\mx)\right)
    -
    2\Delta_{h,s,\mathcal K}^{\mathrm{pop}} \\
    &=
    M_\oplus(\mx,y)
    -
    M_\oplus\left(\mx,m_\oplus(\mx)\right) \\
    &\quad-
    \left[
    M_\oplus\left(\mx,\tilde{m}_{h,s}(\mx)\right)
    -
    M_\oplus\left(\mx,m_\oplus(\mx)\right)
    \right]
    -
    2\Delta_{h,s,\mathcal K}^{\mathrm{pop}} \\
    &\geq
    \eta_{\epsilon,\mathcal K}
    -
    4\Delta_{h,s,\mathcal K}^{\mathrm{pop}}
    \geq
    \eta_{\epsilon,\mathcal K}/2.
\end{align*}

Let
\begin{align*}
    \Delta_{n,h,s,\mathcal K}^{\mathrm{emp}}
    :=
    \sup_{\mx\in\mathcal K}
    \sup_{y\in\mathbb M}
    \left|
    \hat{M}_{h,s}(\mx,y)
    -
    \tilde{M}_{h,s}(\mx,y)
    \right|.
\end{align*}
By \Cref{lemma:D.uniform_empirical_objective},
\begin{align*}
    \Delta_{n,h,s,\mathcal K}^{\mathrm{emp}}
    =
    o_{\P}(1),
\end{align*}
where for $s=1$ this invocation uses the additional Condition~\ref{con:U-K2}. Let $\mathcal A_{n,s}$ be the event on which the empirical objective $\hat{M}_{h,s}(\mx,\cdot)$ is well-defined for all $\mx\in\mathcal K$. By \Cref{lemma:D.uniform_population_moment_orders} and \Cref{lemma:D.uniform_empirical_moments}, $\P(\mathcal A_{n,s})\to1$. On $\mathcal A_{n,s}$, since $\hat{m}_{h,s}(\mx)$ minimizes $\hat{M}_{h,s}(\mx,\cdot)$ for each $\mx\in\mathcal K$, we have, uniformly over $\mx\in\mathcal K$,
\begin{align*}
    \tilde{M}_{h,s}\left(\mx,\hat{m}_{h,s}(\mx)\right)
    -
    \tilde{M}_{h,s}\left(\mx,\tilde{m}_{h,s}(\mx)\right)
    &\leq
    \left|
    \tilde{M}_{h,s}\left(\mx,\hat{m}_{h,s}(\mx)\right)
    -
    \hat{M}_{h,s}\left(\mx,\hat{m}_{h,s}(\mx)\right)
    \right| \\
    &\quad+
    \left[
    \hat{M}_{h,s}\left(\mx,\hat{m}_{h,s}(\mx)\right)
    -
    \hat{M}_{h,s}\left(\mx,\tilde{m}_{h,s}(\mx)\right)
    \right] \\
    &\quad+
    \left|
    \hat{M}_{h,s}\left(\mx,\tilde{m}_{h,s}(\mx)\right)
    -
    \tilde{M}_{h,s}\left(\mx,\tilde{m}_{h,s}(\mx)\right)
    \right| \\
    &\leq
    2\Delta_{n,h,s,\mathcal K}^{\mathrm{emp}}.
\end{align*}
Consequently, for all sufficiently small $h$,
\begin{align*}
    \P\left(
    \sup_{\mx\in\mathcal K}
    d_{\mathbb M}\left(
    \hat{m}_{h,s}(\mx),
    \tilde{m}_{h,s}(\mx)
    \right)
    >
    \epsilon
    \right)
    &\leq
    \P\left(
    2\Delta_{n,h,s,\mathcal K}^{\mathrm{emp}}
    \geq
    \eta_{\epsilon,\mathcal K}/2
    \right)
    +
    \P(\mathcal A_{n,s}^{c}) \\
    &\to
    0.
\end{align*}
Since $\epsilon>0$ was arbitrary,
\begin{align*}
    \sup_{\mx\in\mathcal K}
    d_{\mathbb M}\left(
    \hat{m}_{h,s}(\mx),
    \tilde{m}_{h,s}(\mx)
    \right)
    =
    o_{\P}(1).
\end{align*}
This completes the proof.
\end{proof}

\begin{proof}[Proof of \Cref{thm:uniform_consistency}]
Fix $s\in\{0,1\}$. When $s=1$, assume in addition Condition~\ref{con:U-K2}. By the triangle inequality,
\begin{align*}
    \sup_{\mx\in\mathcal K}
    d_{\mathbb M}\left(
    \hat{m}_{h,s}(\mx),
    m_{\oplus}(\mx)
    \right)
    &\leq
    \sup_{\mx\in\mathcal K}
    d_{\mathbb M}\left(
    \hat{m}_{h,s}(\mx),
    \tilde{m}_{h,s}(\mx)
    \right) \\
    &\quad+
    \sup_{\mx\in\mathcal K}
    d_{\mathbb M}\left(
    \tilde{m}_{h,s}(\mx),
    m_{\oplus}(\mx)
    \right).
\end{align*}
The first term is $o_{\P}(1)$ by \Cref{lemma:D.uniform_empirical_minimizer}, and the second term is $o(1)$ by \Cref{lemma:D.uniform_oracle_minimizer}. Hence
\begin{align*}
    \sup_{\mx\in\mathcal K}
    d_{\mathbb M}\left(
    \hat{m}_{h,s}(\mx),
    m_{\oplus}(\mx)
    \right)
    =
    o_{\P}(1).
\end{align*}
This proves the assertion for $s=0$ under Condition~\ref{con:U-K1}, and for $s=1$ under the additional Condition~\ref{con:U-K2}.
\end{proof}

\section{Proof of Uniform Convergence Rate} 
\label{app:uniform_rate}
\setcounter{equation}{0}
\renewcommand{\theequation}{E.\arabic{equation}}
\renewcommand{\theHequation}{E.\arabic{equation}}

In this section, we provide the proof of \Cref{thm:uniform_rate}. Throughout this section, $\mathcal K\subset\mathcal M$, $\rho\in(0,i(\mathcal K))$, and $\mathcal K^\rho$ are the compact set, uniform normal-neighborhood radius, and closed geodesic tube fixed in the uniform theory. We use the finite smooth ordered-orthonormal-frame cover fixed in the uniform theory before Conditions~\ref{con:U-K1} and~\ref{con:U-K2}. All constants in $O(\cdot)$ and $O_{\P}(\cdot)$ bounds are uniform over $\mx\in\mathcal K$ and may depend on this fixed finite frame cover. Basis-dependent coordinate expressions are evaluated in ordered orthonormal bases $\mathbf E_{\mx}\in\mathcal E_{\mx}$; scalar weights and scalar local objectives are independent of the particular ordered orthonormal basis by \Cref{lemma:A.invariance_local_linear_weights}, and vector and matrix bounds are stated in basis-invariant Euclidean/operator norms.

\begin{remark}
The population expansion lemmas at the beginning of this section use only the baseline kernel regularity in Condition~\ref{con:P-K1}; their uniformity over $\mathcal K$ follows from the uniform design and smoothness conditions. The VC-type content of Condition~\ref{con:U-K1}, and the multiplier complexity in Condition~\ref{con:U-K2}, enter only in the empirical-process arguments below.
\end{remark}

\begin{lemma}[Uniform first-order Taylor remainder for the design density] \label{lemma:E.uniform_taylor_remainder_f}
Assume Conditions~\ref{con:U-D1} and~\ref{con:U-D3}, and suppose that $h\to0$ as $n\to\infty$. Then there exists a constant $C_{f,2}<\infty$ such that, for every $\mx\in\mathcal K$, every $\mathbf E_{\mx}\in\mathcal E_{\mx}$, and every $\mz\in B_{\mathcal M}(\mx,\rho)$,
\begin{align*}
    \left|
    f(\mz)
    -
    f(\mx)
    -
    \left(\mathbf v_{\mx}^{\mathbf E_{\mx}}(\mz)\right)^{\top}
    \bm{\beta}_{f}(\mx,\mathbf E_{\mx})
    \right|
    \leq
    C_{f,2}
    d_{\mathcal M}^2(\mx,\mz),
\end{align*}
where
\begin{align*}
    \bm{\beta}_{f}(\mx,\mathbf E_{\mx})
    :=
    \bm{\Phi}_{\mathbf E_{\mx}}\left(\nabla f(\mx)\right).
\end{align*}
\end{lemma}

\begin{proof}[Proof of \Cref{lemma:E.uniform_taylor_remainder_f}]
By Condition~\ref{con:U-D3}, $f$ is $C^2$ on an open neighborhood of the compact set $\mathcal K^\rho$. Hence, by compactness,
\begin{align*}
    C_{f,2}
    :=
    \frac12
    \sup_{\mz\in\mathcal K^\rho}
    \left\|
    \nabla^2 f(\mz)
    \right\|_{\mathrm{op}}
    <
    \infty.
\end{align*}
Fix $\mx\in\mathcal K$, $\mathbf E_{\mx}\in\mathcal E_{\mx}$, and $\mz\in B_{\mathcal M}(\mx,\rho)$. Let
\begin{align*}
    \gamma(t)
    :=
    \Exp_{\mx}\left(t\Log_{\mx}(\mz)\right),
    \quad t\in[0,1].
\end{align*}
Since $\rho<i(\mathcal K)\leq i(\mx)$, the curve $\gamma$ is the unique minimizing geodesic from $\mx$ to $\mz$. Moreover,
\begin{align*}
    d_{\mathcal M}\left(\gamma(t),\mathcal K\right)
    \leq
    d_{\mathcal M}\left(\gamma(t),\mx\right)
    =
    t d_{\mathcal M}(\mx,\mz)
    \leq
    \rho,
    \quad t\in[0,1],
\end{align*}
and therefore $\gamma(t)\in\mathcal K^\rho$ for every $t\in[0,1]$.

Define $\phi(t):=f(\gamma(t))$. Taylor's formula with integral remainder gives
\begin{align*}
    \phi(1)
    =
    \phi(0)
    +
    \phi'(0)
    +
    \int_0^1
    (1-t)\phi''(t)\,\dd t.
\end{align*}
By the chain rule,
\begin{align*}
    \phi'(0)
    =
    \left\langle
    \nabla f(\mx),
    \Log_{\mx}(\mz)
    \right\rangle_{\mx}
    =
    \left(\mathbf v_{\mx}^{\mathbf E_{\mx}}(\mz)\right)^{\top}
    \bm{\beta}_{f}(\mx,\mathbf E_{\mx}).
\end{align*}
Since $\gamma$ is a geodesic,
\begin{align*}
    \phi''(t)
    =
    \nabla^2 f(\gamma(t))
    \left(
    \dot{\gamma}(t),
    \dot{\gamma}(t)
    \right),
    \quad t\in[0,1],
\end{align*}
and $\|\dot{\gamma}(t)\|_{\gamma(t)}=d_{\mathcal M}(\mx,\mz)$. Hence
\begin{align*}
    &
    \left|
    f(\mz)
    -
    f(\mx)
    -
    \left(\mathbf v_{\mx}^{\mathbf E_{\mx}}(\mz)\right)^{\top}
    \bm{\beta}_{f}(\mx,\mathbf E_{\mx})
    \right| \\
    &\leq
    \int_0^1
    (1-t)
    \left\|
    \nabla^2 f(\gamma(t))
    \right\|_{\mathrm{op}}
    d_{\mathcal M}^2(\mx,\mz)
    \,\dd t \\
    &\leq
    C_{f,2}
    d_{\mathcal M}^2(\mx,\mz).
\end{align*}
This completes the proof.
\end{proof}

\begin{lemma}[Uniform Taylor bound for $f g_\omega$] \label{lemma:E.uniform_taylor_remainder_fg}
Assume Conditions~\ref{con:U-D1}--\ref{con:U-D4}, and suppose that $h\to0$ as $n\to\infty$. Then
\begin{align*}
    \sup_{\omega\in\mathbb M}
    \sup_{\mz\in\mathcal K^\rho}
    \left\|
    \nabla(f\cdot g_{\omega})(\mz)
    \right\|_{\mz}
    <
    \infty,
    \quad
    \sup_{\omega\in\mathbb M}
    \sup_{\mz\in\mathcal K^\rho}
    \left\|
    \nabla^2(f\cdot g_{\omega})(\mz)
    \right\|_{\mathrm{op}}
    <
    \infty.
\end{align*}
Moreover, there exists a constant $C_{fg,2}<\infty$ such that, for every $\mx\in\mathcal K$, every $\mathbf E_{\mx}\in\mathcal E_{\mx}$, and every $\mz\in B_{\mathcal M}(\mx,\rho)$,
\begin{align*}
    &
    \sup_{\omega\in\mathbb M}
    \left|
    (f\cdot g_{\omega})(\mz)
    -
    (f\cdot g_{\omega})(\mx)
    -
    \left(\mathbf v_{\mx}^{\mathbf E_{\mx}}(\mz)\right)^{\top}
    \bm{\beta}_{f\cdot g_{\omega}}(\mx,\mathbf E_{\mx})
    \right| \\
    &\leq
    C_{fg,2}
    d_{\mathcal M}^2(\mx,\mz),
\end{align*}
where
\begin{align*}
    \bm{\beta}_{f\cdot g_{\omega}}(\mx,\mathbf E_{\mx})
    :=
    \bm{\Phi}_{\mathbf E_{\mx}}
    \left(
    \nabla(f\cdot g_{\omega})(\mx)
    \right).
\end{align*}
\end{lemma}

\begin{proof}[Proof of \Cref{lemma:E.uniform_taylor_remainder_fg}]
Since $\mathcal K^\rho$ is compact, Conditions~\ref{con:U-D1} and~\ref{con:U-D3} imply
\begin{align*}
    \sup_{\mz\in\mathcal K^\rho}|f(\mz)|<\infty,
    \quad
    \sup_{\mz\in\mathcal K^\rho}\|\nabla f(\mz)\|_{\mz}<\infty,
    \quad
    \sup_{\mz\in\mathcal K^\rho}\|\nabla^2 f(\mz)\|_{\mathrm{op}}<\infty.
\end{align*}
Condition~\ref{con:U-D2} gives
\begin{align*}
    \sup_{\omega\in\mathbb M}
    \sup_{\mz\in\mathcal K^\rho}
    g_{\omega}(\mz)
    <
    \infty,
\end{align*}
and Condition~\ref{con:U-D4} gives uniform bounds for $\nabla g_{\omega}$ and $\nabla^2 g_{\omega}$ on $\mathcal K^\rho$.

For any $\mz\in\mathcal K^\rho$, $\omega\in\mathbb M$, and $\mathbf a\in T_{\mz}\mathcal M$, the covariant product rule gives
\begin{align*}
    \left\langle
    \nabla(f\cdot g_{\omega})(\mz),
    \mathbf a
    \right\rangle_{\mz}
    =
    g_{\omega}(\mz)
    \left\langle
    \nabla f(\mz),
    \mathbf a
    \right\rangle_{\mz}
    +
    f(\mz)
    \left\langle
    \nabla g_{\omega}(\mz),
    \mathbf a
    \right\rangle_{\mz}.
\end{align*}
Taking the supremum over $\|\mathbf a\|_{\mz}\leq1$ yields the uniform first-derivative bound.

Similarly, for any $\mathbf a,\mathbf b\in T_{\mz}\mathcal M$, the covariant product rule gives
\begin{align*}
    \nabla^2(f\cdot g_{\omega})(\mz)(\mathbf a,\mathbf b)
    &=
    g_{\omega}(\mz)
    \nabla^2 f(\mz)(\mathbf a,\mathbf b)
    +
    f(\mz)
    \nabla^2 g_{\omega}(\mz)(\mathbf a,\mathbf b) \\
    &\quad+
    \left\langle\nabla f(\mz),\mathbf a\right\rangle_{\mz}
    \left\langle\nabla g_{\omega}(\mz),\mathbf b\right\rangle_{\mz}
    +
    \left\langle\nabla f(\mz),\mathbf b\right\rangle_{\mz}
    \left\langle\nabla g_{\omega}(\mz),\mathbf a\right\rangle_{\mz}.
\end{align*}
Taking the supremum over $\|\mathbf a\|_{\mz}\leq1$ and $\|\mathbf b\|_{\mz}\leq1$ yields
\begin{align*}
    \left\|
    \nabla^2(f\cdot g_{\omega})(\mz)
    \right\|_{\mathrm{op}}
    &\leq
    g_{\omega}(\mz)
    \left\|
    \nabla^2 f(\mz)
    \right\|_{\mathrm{op}}
    +
    |f(\mz)|
    \left\|
    \nabla^2 g_{\omega}(\mz)
    \right\|_{\mathrm{op}} \\
    &\quad+
    2
    \left\|
    \nabla f(\mz)
    \right\|_{\mz}
    \left\|
    \nabla g_{\omega}(\mz)
    \right\|_{\mz}.
\end{align*}
The preceding uniform bounds imply
\begin{align*}
    C_{fg,2}
    :=
    \frac12
    \sup_{\omega\in\mathbb M}
    \sup_{\mz\in\mathcal K^\rho}
    \left\|
    \nabla^2(f\cdot g_{\omega})(\mz)
    \right\|_{\mathrm{op}}
    <
    \infty.
\end{align*}

Now fix $\mx\in\mathcal K$, $\mathbf E_{\mx}\in\mathcal E_{\mx}$, and $\mz\in B_{\mathcal M}(\mx,\rho)$. Let
\begin{align*}
    \gamma(t)
    :=
    \Exp_{\mx}\left(t\Log_{\mx}(\mz)\right),
    \quad t\in[0,1].
\end{align*}
As in the proof of \Cref{lemma:E.uniform_taylor_remainder_f}, $\gamma(t)\in\mathcal K^\rho$ for every $t\in[0,1]$. For each $\omega\in\mathbb M$, define
\begin{align*}
    \phi_{\omega}(t)
    :=
    (f\cdot g_{\omega})(\gamma(t)),
    \quad t\in[0,1].
\end{align*}
Taylor's formula with integral remainder gives
\begin{align*}
    \phi_{\omega}(1)
    =
    \phi_{\omega}(0)
    +
    \phi_{\omega}'(0)
    +
    \int_0^1
    (1-t)
    \phi_{\omega}''(t)
    \,\dd t.
\end{align*}
The first derivative satisfies
\begin{align*}
    \phi_{\omega}'(0)
    =
    \left(
    \mathbf v_{\mx}^{\mathbf E_{\mx}}(\mz)
    \right)^{\top}
    \bm{\beta}_{f\cdot g_{\omega}}(\mx,\mathbf E_{\mx}),
\end{align*}
and, since $\gamma$ is a geodesic,
\begin{align*}
    \phi_{\omega}''(t)
    =
    \nabla^2(f\cdot g_{\omega})(\gamma(t))
    \left(
    \dot{\gamma}(t),
    \dot{\gamma}(t)
    \right).
\end{align*}
Using $\|\dot{\gamma}(t)\|_{\gamma(t)}=d_{\mathcal M}(\mx,\mz)$ and taking the supremum over $\omega\in\mathbb M$ gives the desired bound.
\end{proof}

\begin{lemma}[Uniform second-order expansion of scalar kernel moments] \label{lemma:E.uniform_scalar_kernel_moments_f}
Assume Conditions~\ref{con:P-K1}, \ref{con:U-D1}, and~\ref{con:U-D3}, and suppose that $h\to0$ as $n\to\infty$. Then, for each $k\in\{1,2\}$,
\begin{align*}
    \sup_{\mx\in\mathcal K}
    \left|
    \E\left[
    \mathcal L_{\mx,h}(\mX)^k
    \right]
    -
    h^d A_{d-1}c_{d-1,k}f(\mx)
    \right|
    =
    O(h^{d+2}).
\end{align*}
In particular,
\begin{align*}
    \sup_{\mx\in\mathcal K}
    \E\left[
    \mathcal L_{\mx,h}(\mX)^2
    \right]
    =
    O(h^d).
\end{align*}
\end{lemma}

\begin{proof}[Proof of \Cref{lemma:E.uniform_scalar_kernel_moments_f}]
Fix $k\in\{1,2\}$. For $\mx\in\mathcal K$ and $\mathbf E_{\mx}\in\mathcal E_{\mx}$, use the coordinate exponential shorthand in \eqref{eq:app.exp_coordinate_shorthand}. For all sufficiently small $h<\rho$, the compact support of $K$ and the normal-coordinate change of variables give
\begin{align*}
    \E\left[
    \mathcal L_{\mx,h}(\mX)^k
    \right]
    &=
    \int_{\|\mathbf u\|_2\leq h}
    K\left(\frac{\|\mathbf u\|_2}{h}\right)^k
    \frac{
    f\left(\Exp_{\mx}^{\mathbf E_{\mx}}(\mathbf u)\right)
    }{
    \theta_{\mx}\left(\Exp_{\mx}^{\mathbf E_{\mx}}(\mathbf u)\right)^{k-1}
    }
    \,\dd\mathbf u \\
    &=
    h^d
    \int_{\|\mathbf w\|_2\leq1}
    K(\|\mathbf w\|_2)^k
    \varphi_{k,\mx,\mathbf E_{\mx}}(h\mathbf w)
    \,\dd\mathbf w,
\end{align*}
where
\begin{align*}
    \varphi_{k,\mx,\mathbf E_{\mx}}(\mathbf u)
    :=
    \frac{
    f\left(\Exp_{\mx}^{\mathbf E_{\mx}}(\mathbf u)\right)
    }{
    \theta_{\mx}\left(\Exp_{\mx}^{\mathbf E_{\mx}}(\mathbf u)\right)^{k-1}
    }.
\end{align*}
By Condition~\ref{con:U-D3}, compactness of $\mathcal K^\rho$, and the uniform smoothness of the normal-coordinate maps and volume-density functions on the compact set
\begin{align*}
    \left\{
    (\mx,\mathbf u):
    \mx\in\mathcal K,\ 
    \|\mathbf u\|_2\leq\rho
    \right\},
\end{align*}
there exists $C_{\varphi,k}<\infty$ such that
\begin{align*}
    \sup_{\mx\in\mathcal K}
    \sup_{\mathbf E_{\mx}\in\mathcal E_{\mx}}
    \sup_{\|\mathbf u\|_2\leq\rho}
    \left\|
    D^2\varphi_{k,\mx,\mathbf E_{\mx}}(\mathbf u)
    \right\|_{\mathrm{op}}
    \leq
    C_{\varphi,k}.
\end{align*}
Hence Taylor's formula in Euclidean normal coordinates gives, uniformly over $\mx\in\mathcal K$, $\mathbf E_{\mx}\in\mathcal E_{\mx}$, and $\|\mathbf w\|_2\leq1$,
\begin{align*}
    \varphi_{k,\mx,\mathbf E_{\mx}}(h\mathbf w)
    =
    \varphi_{k,\mx,\mathbf E_{\mx}}(\mathbf 0_d)
    +
    h\mathbf w^{\top}
    D\varphi_{k,\mx,\mathbf E_{\mx}}(\mathbf 0_d)
    +
    R_{k,h,\mx,\mathbf E_{\mx}}(\mathbf w),
\end{align*}
where
\begin{align*}
    \left|
    R_{k,h,\mx,\mathbf E_{\mx}}(\mathbf w)
    \right|
    \leq
    \frac12
    C_{\varphi,k}
    h^2
    \|\mathbf w\|_2^2.
\end{align*}
Since $\theta_{\mx}(\mx)=1$,
\begin{align*}
    \varphi_{k,\mx,\mathbf E_{\mx}}(\mathbf 0_d)
    =
    f(\mx).
\end{align*}
By radial symmetry,
\begin{align*}
    \int_{\|\mathbf w\|_2\leq1}
    \mathbf w
    K(\|\mathbf w\|_2)^k
    \,\dd\mathbf w
    =
    \mathbf 0_d.
\end{align*}
Therefore,
\begin{align*}
    \E\left[
    \mathcal L_{\mx,h}(\mX)^k
    \right]
    =
    h^d A_{d-1}c_{d-1,k}f(\mx)
    +
    O(h^{d+2}),
\end{align*}
uniformly over $\mx\in\mathcal K$. The final assertion follows because $f$ is bounded on the compact set $\mathcal K^\rho$.
\end{proof}

\begin{lemma}[Uniform second-order expansion of scalar kernel moments with conditional density ratios] \label{lemma:E.uniform_scalar_kernel_moments_fg}
Assume Conditions~\ref{con:P-K1}, \ref{con:U-D1}--\ref{con:U-D4}, and suppose that $h\to0$ as $n\to\infty$. Then, for each $k\in\{1,2\}$,
\begin{align*}
    \sup_{\mx\in\mathcal K}
    \sup_{\omega\in\mathbb M}
    \left|
    \E\left[
    \mathcal L_{\mx,h}(\mX)^k
    g_{\omega}(\mX)
    \right]
    -
    h^d A_{d-1}c_{d-1,k}
    (f\cdot g_{\omega})(\mx)
    \right|
    =
    O(h^{d+2}).
\end{align*}
\end{lemma}

\begin{proof}[Proof of \Cref{lemma:E.uniform_scalar_kernel_moments_fg}]
Fix $k\in\{1,2\}$. For $\mx\in\mathcal K$ and $\mathbf E_{\mx}\in\mathcal E_{\mx}$, use the coordinate exponential shorthand in \eqref{eq:app.exp_coordinate_shorthand}. For all sufficiently small $h<\rho$, the compact support of $K$ and the normal-coordinate change of variables give
\begin{align*}
    \E\left[
    \mathcal L_{\mx,h}(\mX)^k
    g_{\omega}(\mX)
    \right]
    &=
    \int_{\|\mathbf u\|_2\leq h}
    K\left(\frac{\|\mathbf u\|_2}{h}\right)^k
    \frac{
    f\left(\Exp_{\mx}^{\mathbf E_{\mx}}(\mathbf u)\right)
    g_{\omega}\left(\Exp_{\mx}^{\mathbf E_{\mx}}(\mathbf u)\right)
    }{
    \theta_{\mx}\left(\Exp_{\mx}^{\mathbf E_{\mx}}(\mathbf u)\right)^{k-1}
    }
    \,\dd\mathbf u \\
    &=
    h^d
    \int_{\|\mathbf w\|_2\leq1}
    K(\|\mathbf w\|_2)^k
    \varrho_{k,\omega,\mx,\mathbf E_{\mx}}(h\mathbf w)
    \,\dd\mathbf w,
\end{align*}
where
\begin{align*}
    \varrho_{k,\omega,\mx,\mathbf E_{\mx}}(\mathbf u)
    :=
    \frac{
    f\left(\Exp_{\mx}^{\mathbf E_{\mx}}(\mathbf u)\right)
    g_{\omega}\left(\Exp_{\mx}^{\mathbf E_{\mx}}(\mathbf u)\right)
    }{
    \theta_{\mx}\left(\Exp_{\mx}^{\mathbf E_{\mx}}(\mathbf u)\right)^{k-1}
    }.
\end{align*}
By \Cref{lemma:E.uniform_taylor_remainder_fg}, the functions $f\cdot g_{\omega}$ have uniformly bounded first and second covariant derivatives on $\mathcal K^\rho$, uniformly over $\omega\in\mathbb M$. Combining this with the uniform smoothness of the normal-coordinate maps and volume-density functions on the compact uniform normal-coordinate domain yields a constant $C_{\varrho,k}<\infty$ such that
\begin{align*}
    \sup_{\omega\in\mathbb M}
    \sup_{\mx\in\mathcal K}
    \sup_{\mathbf E_{\mx}\in\mathcal E_{\mx}}
    \sup_{\|\mathbf u\|_2\leq\rho}
    \left\|
    D^2\varrho_{k,\omega,\mx,\mathbf E_{\mx}}(\mathbf u)
    \right\|_{\mathrm{op}}
    \leq
    C_{\varrho,k}.
\end{align*}
Therefore Taylor's formula in Euclidean normal coordinates gives, uniformly over $\omega\in\mathbb M$, $\mx\in\mathcal K$, $\mathbf E_{\mx}\in\mathcal E_{\mx}$, and $\|\mathbf w\|_2\leq1$,
\begin{align*}
    \varrho_{k,\omega,\mx,\mathbf E_{\mx}}(h\mathbf w)
    =
    \varrho_{k,\omega,\mx,\mathbf E_{\mx}}(\mathbf 0_d)
    +
    h\mathbf w^{\top}
    D\varrho_{k,\omega,\mx,\mathbf E_{\mx}}(\mathbf 0_d)
    +
    R_{k,\omega,h,\mx,\mathbf E_{\mx}}(\mathbf w),
\end{align*}
where
\begin{align*}
    \left|
    R_{k,\omega,h,\mx,\mathbf E_{\mx}}(\mathbf w)
    \right|
    \leq
    \frac12
    C_{\varrho,k}
    h^2
    \|\mathbf w\|_2^2.
\end{align*}
Since $\theta_{\mx}(\mx)=1$,
\begin{align*}
    \varrho_{k,\omega,\mx,\mathbf E_{\mx}}(\mathbf 0_d)
    =
    (f\cdot g_{\omega})(\mx).
\end{align*}
By radial symmetry,
\begin{align*}
    \int_{\|\mathbf w\|_2\leq1}
    \mathbf w
    K(\|\mathbf w\|_2)^k
    \,\dd\mathbf w
    =
    \mathbf 0_d.
\end{align*}
Thus,
\begin{align*}
    \E\left[
    \mathcal L_{\mx,h}(\mX)^k
    g_{\omega}(\mX)
    \right]
    =
    h^d A_{d-1}c_{d-1,k}
    (f\cdot g_{\omega})(\mx)
    +
    O(h^{d+2}),
\end{align*}
uniformly over $\mx\in\mathcal K$ and $\omega\in\mathbb M$. This proves the claim.
\end{proof}

\begin{lemma}[Uniform refined oracle local moment expansions] \label{lemma:E.uniform_refined_moments}
Assume Conditions~\ref{con:P-K1}, \ref{con:U-D1}, and~\ref{con:U-D3}, and suppose that $h\to0$ as $n\to\infty$. For $\mx\in\mathcal K$ and $\mathbf E_{\mx}\in\mathcal E_{\mx}$, let
\begin{align*}
    \bm{\beta}_{f}(\mx,\mathbf E_{\mx})
    :=
    \bm{\Phi}_{\mathbf E_{\mx}}\left(\nabla f(\mx)\right).
\end{align*}
Then, as $h\downarrow0$,
\begin{align} 
    \sup_{\mx\in\mathcal K}
    \left|
    \tilde{\mu}_{h,0}(\mx)
    -
    h^dA_{d-1}c_{d-1,1}f(\mx)
    \right|
    &=
    O(h^{d+2}), \label{eq:E.uniform_mu0_refined} \\
    \sup_{\mx\in\mathcal K}
    \left\|
    \bm{\tilde{\mu}}_{h,1}(\mx,\mathbf E_{\mx})
    -
    h^{d+2}
    \frac{A_{d-1}c_{d+1,1}}{d}
    \bm{\beta}_{f}(\mx,\mathbf E_{\mx})
    \right\|_2
    &=
    O(h^{d+3}), \label{eq:E.uniform_mu1_refined} \\
    \sup_{\mx\in\mathcal K}
    \left\|
    \bm{\tilde{\mu}}_{h,2}(\mx,\mathbf E_{\mx})
    -
    h^{d+2}
    \frac{A_{d-1}c_{d+1,1}f(\mx)}{d}
    \mI_d
    \right\|_2
    &=
    O(h^{d+4}). \label{eq:E.uniform_mu2_refined}
\end{align}
Moreover, for all sufficiently small $h$,
\begin{align}
    \sup_{\mx\in\mathcal K}
    \left\|
    \bm{\tilde{\mu}}_{h,2}(\mx,\mathbf E_{\mx})^{-1}
    -
    h^{-(d+2)}
    \frac{d}{A_{d-1}c_{d+1,1}f(\mx)}
    \mI_d
    \right\|_2
    =
    O(h^{-d}). \label{eq:E.uniform_mu2_inverse_refined}
\end{align}
In particular,
\begin{align}
    \sup_{\mx\in\mathcal K}
    \left\|
    \bm{\tilde{\mu}}_{h,1}(\mx,\mathbf E_{\mx})
    \right\|_2
    =
    O(h^{d+2}),
    \quad
    \sup_{\mx\in\mathcal K}
    \left\|
    \bm{\tilde{\mu}}_{h,2}(\mx,\mathbf E_{\mx})^{-1}
    \right\|_2
    =
    O(h^{-(d+2)}). \label{eq:E.uniform_mu_orders}
\end{align}
\end{lemma}

\begin{proof}[Proof of \Cref{lemma:E.uniform_refined_moments}]
Fix $\mx\in\mathcal K$ and $\mathbf E_{\mx}\in\mathcal E_{\mx}$. For $\mathbf u\in\mathbb R^d$ with $\|\mathbf u\|_2\leq1$, write
\begin{align*}
    \Exp_{\mx}^{\mathbf E_{\mx}}(h\mathbf u)
    :=
    \Exp_{\mx}\left(
    \bm{\Phi}_{\mathbf E_{\mx}}^{-1}(h\mathbf u)
    \right).
\end{align*}
Since $h<\rho$ for all sufficiently small $h$, the point $\Exp_{\mx}^{\mathbf E_{\mx}}(h\mathbf u)$ belongs to $\mathcal K^\rho$. By the volume-density cancellation in normal coordinates,
\begin{align*}
    \tilde{\mu}_{h,0}(\mx)
    =
    h^d
    \int_{\mathbb R^d}
    K\left(\|\mathbf u\|_2\right)
    f\left(\Exp_{\mx}^{\mathbf E_{\mx}}(h\mathbf u)\right)
    \mathbf 1\{\|\mathbf u\|_2\leq1\}
    \,\dd\mathbf u.
\end{align*}
Condition~\ref{con:U-D3}, compactness of $\mathcal K^\rho$, and \Cref{lemma:A.uniform_taylor_expansion} give the uniform expansion
\begin{align*}
    f\left(\Exp_{\mx}^{\mathbf E_{\mx}}(h\mathbf u)\right)
    =
    f(\mx)
    +
    h\bm{\beta}_{f}(\mx,\mathbf E_{\mx})^{\top}\mathbf u
    +
    \frac{h^2}{2}
    D^2 f_{\mx,\mathbf E_{\mx}}(\mathbf 0_d)
    \left(\mathbf u,\mathbf u\right)
    +
    O(h^2),
\end{align*}
uniformly over $\mx\in\mathcal K$ and $\|\mathbf u\|_2\leq1$, with the first-order and second-order remainders controlled uniformly. Integrating this expansion against the radial kernel and using the cancellation identities in \Cref{lemma:A.radial_kernel_moments} gives \eqref{eq:E.uniform_mu0_refined}. Multiplying the same expansion by $h\mathbf u$ and $h^2\mathbf u\mathbf u^{\top}$, respectively, gives \eqref{eq:E.uniform_mu1_refined} and \eqref{eq:E.uniform_mu2_refined}. The radial identities used here are
\begin{align*}
    \int_{\mathbb R^d}\mathbf u K(\|\mathbf u\|_2)\,\dd\mathbf u
    =
    \mathbf 0_d,
    \quad
    \int_{\mathbb R^d}\mathbf u\mathbf u^{\top}K(\|\mathbf u\|_2)\,\dd\mathbf u
    =
    \frac{A_{d-1}c_{d+1,1}}{d}\mI_d.
\end{align*}

By Condition~\ref{con:U-D1}, $c_{\mathcal K}:=\inf_{\mx\in\mathcal K}f(\mx)>0$. Let
\begin{align*}
    \mathbf B_h(\mx)
    :=
    h^{d+2}
    \frac{A_{d-1}c_{d+1,1}f(\mx)}{d}
    \mI_d.
\end{align*}
The remainder in \eqref{eq:E.uniform_mu2_refined} is uniform both in $\mx\in\mathcal K$ and in the finitely many frame fields used on the local cover, because it is obtained from the uniform Taylor remainder in \Cref{lemma:A.uniform_taylor_expansion}. Hence, for all sufficiently small $h$,
\begin{align*}
    \sup_{\mx\in\mathcal K}
    \left\|
    \bm{\tilde\mu}_{h,2}(\mx,\mathbf E_{\mx})-\mathbf B_h(\mx)
    \right\|_2
    \leq
    \frac12
    h^{d+2}
    \frac{A_{d-1}c_{d+1,1}c_{\mathcal K}}{d}.
\end{align*}
Weyl's inequality therefore gives, uniformly over $\mx\in\mathcal K$ and the active frame field,
\begin{align*}
    \lambda_{\min}
    \left\{
    \bm{\tilde\mu}_{h,2}(\mx,\mathbf E_{\mx})
    \right\}
    \geq
    \frac12
    h^{d+2}
    \frac{A_{d-1}c_{d+1,1}c_{\mathcal K}}{d},
\end{align*}
so the inverse exists and is uniformly $O(h^{-(d+2)})$. Applying the perturbation identity
\begin{align*}
    \mathbf A^{-1}-\mathbf B^{-1}
    =
    \mathbf A^{-1}(\mathbf B-\mathbf A)\mathbf B^{-1}
\end{align*}
with $\mathbf A=\bm{\tilde\mu}_{h,2}(\mx,\mathbf E_{\mx})$ and $\mathbf B=\mathbf B_h(\mx)$ now yields \eqref{eq:E.uniform_mu2_inverse_refined}. The order bounds in \eqref{eq:E.uniform_mu_orders} follow immediately.
\end{proof}

\begin{lemma}[Uniform refined conditional local moment expansions] \label{lemma:E.uniform_refined_conditional_moments}
Assume Conditions~\ref{con:P-K1}, \ref{con:U-D1}--\ref{con:U-D4}, and suppose that $h\to0$ as $n\to\infty$. Then
\begin{align}
    \sup_{\mx\in\mathcal K}
    \sup_{\omega\in\mathbb M}
    \left|
    \tilde{\tau}_{h,0}(\mx,\omega)
    -
    h^dA_{d-1}c_{d-1,1}f(\mx)g_{\omega}(\mx)
    \right|
    =
    O(h^{d+2}), \label{eq:E.uniform_tau0_refined}
\end{align}
and
\begin{align}
    \sup_{\mx\in\mathcal K}
    \sup_{\omega\in\mathbb M}
    \left\|
    \bm{\tilde{\tau}}_{h,1}(\mx,\mathbf E_{\mx},\omega)
    \right\|_2
    =
    O(h^{d+2}). \label{eq:E.uniform_tau1_order}
\end{align}
\end{lemma}

\begin{proof}[Proof of \Cref{lemma:E.uniform_refined_conditional_moments}]
The first assertion follows from \Cref{lemma:E.uniform_scalar_kernel_moments_fg} with $k=1$. It remains to prove \eqref{eq:E.uniform_tau1_order}. Fix $\mx\in\mathcal K$, $\omega\in\mathbb M$, and $\mathbf E_{\mx}\in\mathcal E_{\mx}$. Using the shorthand introduced at the beginning of the appendices, for $\mx\in\mathcal K$, $\mathbf E_{\mx}\in\mathcal E_{\mx}$, and $\mathbf u$ in the relevant normal-coordinate ball, write $\Exp_{\mx}^{\mathbf E_{\mx}}(\mathbf u)$. Define
\begin{align*}
    H_{\omega,\mx,\mathbf E_{\mx}}(\mathbf u)
    :=
    f\left(\Exp_{\mx}^{\mathbf E_{\mx}}(\mathbf u)\right)
    g_{\omega}\left(\Exp_{\mx}^{\mathbf E_{\mx}}(\mathbf u)\right).
\end{align*}
For all sufficiently small $h<\rho$, the normal-coordinate change of variables gives
\begin{align*}
    \bm{\tilde{\tau}}_{h,1}(\mx,\mathbf E_{\mx},\omega)
    =
    h^{d+1}
    \int_{\|\mathbf w\|_2\leq1}
    \mathbf w
    K(\|\mathbf w\|_2)
    H_{\omega,\mx,\mathbf E_{\mx}}(h\mathbf w)
    \,\dd\mathbf w .
\end{align*}
By radial symmetry,
\begin{align*}
    \int_{\|\mathbf w\|_2\leq1}
    \mathbf w K(\|\mathbf w\|_2)\,\dd\mathbf w
    =
    \mathbf 0_d.
\end{align*}
Hence
\begin{align*}
    \bm{\tilde{\tau}}_{h,1}(\mx,\mathbf E_{\mx},\omega)
    =
    h^{d+1}
    \int_{\|\mathbf w\|_2\leq1}
    \mathbf w
    K(\|\mathbf w\|_2)
    \left[
    H_{\omega,\mx,\mathbf E_{\mx}}(h\mathbf w)
    -
    H_{\omega,\mx,\mathbf E_{\mx}}(\mathbf 0_d)
    \right]
    \,\dd\mathbf w .
\end{align*}
By \Cref{lemma:E.uniform_taylor_remainder_fg}, the first covariant derivatives of $f g_\omega$ are uniformly bounded over $\omega\in\mathbb M$ on $\mathcal K^\rho$. Therefore the corresponding Euclidean gradients of $H_{\omega,\mx,\mathbf E_{\mx}}$ are uniformly bounded over $\omega\in\mathbb M$, $\mx\in\mathcal K$, and $\mathbf E_{\mx}\in\mathcal E_{\mx}$. Thus
\begin{align*}
    \left|
    H_{\omega,\mx,\mathbf E_{\mx}}(h\mathbf w)
    -
    H_{\omega,\mx,\mathbf E_{\mx}}(\mathbf 0_d)
    \right|
    \leq
    Ch\|\mathbf w\|_2
\end{align*}
uniformly over the same indices and $\|\mathbf w\|_2\leq1$. Consequently,
\begin{align*}
    \left\|
    \bm{\tilde{\tau}}_{h,1}(\mx,\mathbf E_{\mx},\omega)
    \right\|_2
    \leq
    Ch^{d+2}
    \int_{\|\mathbf w\|_2\leq1}
    \|\mathbf w\|_2^2 K(\|\mathbf w\|_2)
    \,\dd\mathbf w
    =
    O(h^{d+2}),
\end{align*}
uniformly over $\mx\in\mathcal K$ and $\omega\in\mathbb M$.
\end{proof}

\begin{lemma}[Uniform population bias rate] \label{lemma:E.uniform_population_bias}
Assume Conditions~\ref{con:P-K1}, \ref{con:U-D1}, \ref{con:U-D2}, \ref{con:M1}, \ref{con:U-M2}, \ref{con:U-D3}, \ref{con:U-D4}, and~\ref{con:U-M3}, and suppose that $h\to0$ as $n\to\infty$. Then, for each $s\in\{0,1\}$,
\begin{align}
    \sup_{\mx\in\mathcal K}
    \sup_{\omega\in\mathbb M}
    \left|
    \tilde{g}_{h,s}(\mx,\omega)-g_{\omega}(\mx)
    \right|
    =
    O(h^2), \label{eq:E.uniform_density_bias}
\end{align}
and
\begin{align}
    \sup_{\mx\in\mathcal K}
    d_{\mathbb M}\left(
    \tilde{m}_{h,s}(\mx),
    m_{\oplus}(\mx)
    \right)^{\beta_{\oplus,\mathcal K}-1}
    =
    O(h^2). \label{eq:E.uniform_oracle_bias}
\end{align}
\end{lemma}

\begin{proof}[Proof of \Cref{lemma:E.uniform_population_bias}]
Fix $s\in\{0,1\}$. We first prove \eqref{eq:E.uniform_density_bias}. The required numerator expansions are provided by \Cref{lemma:E.uniform_refined_conditional_moments}, while the denominator expansions are provided by \Cref{lemma:E.uniform_refined_moments}. We treat $s=0$ and $s=1$ separately. For $s=0$, \Cref{lemma:E.uniform_refined_moments,lemma:E.uniform_refined_conditional_moments} gives
\begin{align*}
    \tilde{\mu}_{h,0}(\mx)
    =
    h^dA_{d-1}c_{d-1,1}f(\mx)+O(h^{d+2}),
    \quad
    \tilde{\tau}_{h,0}(\mx,\omega)
    =
    h^dA_{d-1}c_{d-1,1}f(\mx)g_{\omega}(\mx)+O(h^{d+2}),
\end{align*}
uniformly over $\mx\in\mathcal K$ and $\omega\in\mathbb M$. Since $f$ is bounded away from zero on $\mathcal K$, it follows that
\begin{align*}
    \sup_{\mx\in\mathcal K}
    \sup_{\omega\in\mathbb M}
    \left|
    \tilde g_{h,0}(\mx,\omega)-g_\omega(\mx)
    \right|
    =
    O(h^2).
\end{align*}
For $s=1$, \Cref{lemma:E.uniform_refined_moments,lemma:E.uniform_refined_conditional_moments} gives
\begin{align*}
    \bm{\tilde{\mu}}_{h,1}(\mx,\mathbf E_{\mx})=O(h^{d+2}),
    \quad
    \bm{\tilde{\mu}}_{h,2}(\mx,\mathbf E_{\mx})^{-1}=O(h^{-(d+2)}),
    \quad
    \bm{\tilde{\tau}}_{h,1}(\mx,\mathbf E_{\mx},\omega)=O(h^{d+2}),
\end{align*}
uniformly over $\mx\in\mathcal K$ and $\omega\in\mathbb M$. Hence
\begin{align*}
    \bm{\tilde{\mu}}_{h,1}(\mx,\mathbf E_{\mx})^{\top}
    \bm{\tilde{\mu}}_{h,2}(\mx,\mathbf E_{\mx})^{-1}
    \bm{\tilde{\tau}}_{h,1}(\mx,\mathbf E_{\mx},\omega)
    =
    O(h^{d+2})
\end{align*}
uniformly. Together with
\begin{align*}
    \tilde{\sigma}_{h}(\mx)
    =
    h^dA_{d-1}c_{d-1,1}f(\mx)+O(h^{d+2}),
\end{align*}
this yields
\begin{align*}
    \sup_{\mx\in\mathcal K}
    \sup_{\omega\in\mathbb M}
    \left|
    \tilde g_{h,1}(\mx,\omega)-g_\omega(\mx)
    \right|
    =
    O(h^2).
\end{align*}

We next prove \eqref{eq:E.uniform_oracle_bias}. By \Cref{lemma:D.uniform_oracle_minimizer},
\begin{align*}
    \sup_{\mx\in\mathcal K}
    d_{\mathbb M}\left(
    \tilde{m}_{h,s}(\mx),
    m_{\oplus}(\mx)
    \right)
    =
    o(1).
\end{align*}
Hence, for all sufficiently small $h$, the margin condition in Condition~\ref{con:U-M3} applies to $y=\tilde{m}_{h,s}(\mx)$ uniformly over $\mx\in\mathcal K$. For any $y\in\mathbb M$, \eqref{eq:E.uniform_density_bias}, Condition~\ref{con:M1}, and the inequality
\begin{align*}
    \left|
    d_{\mathbb M}^2(y,\omega)
    -
    d_{\mathbb M}^2(z,\omega)
    \right|
    \leq
    2D_{\mathbb M}
    d_{\mathbb M}(y,z),
    \quad y,z,\omega\in\mathbb M,
\end{align*}
give
\begin{align*}
    &
    \left|
    \left[
    \tilde{M}_{h,s}(\mx,y)-M_{\oplus}(\mx,y)
    \right]
    -
    \left[
    \tilde{M}_{h,s}\left(\mx,m_{\oplus}(\mx)\right)
    -
    M_{\oplus}\left(\mx,m_{\oplus}(\mx)\right)
    \right]
    \right| \\
    &\leq
    \int_{\mathbb M}
    \left|
    d_{\mathbb M}^2(y,\omega)
    -
    d_{\mathbb M}^2\left(m_{\oplus}(\mx),\omega\right)
    \right|
    \left|
    \tilde{g}_{h,s}(\mx,\omega)-g_{\omega}(\mx)
    \right|
    \,\dd P_Y(\omega) \\
    &\leq
    O(h^2)
    d_{\mathbb M}\left(y,m_{\oplus}(\mx)\right),
\end{align*}
uniformly over $\mx\in\mathcal K$ and $y\in\mathbb M$. Since $\tilde{m}_{h,s}(\mx)$ minimizes $\tilde{M}_{h,s}(\mx,\cdot)$, we have
\begin{align*}
    0
    &\leq
    M_{\oplus}\left(\mx,\tilde{m}_{h,s}(\mx)\right)
    -
    M_{\oplus}\left(\mx,m_{\oplus}(\mx)\right) \\
    &\leq
    \left|
    \left[
    \tilde{M}_{h,s}\left(\mx,\tilde{m}_{h,s}(\mx)\right)
    -
    M_{\oplus}\left(\mx,\tilde{m}_{h,s}(\mx)\right)
    \right]
    -
    \left[
    \tilde{M}_{h,s}\left(\mx,m_{\oplus}(\mx)\right)
    -
    M_{\oplus}\left(\mx,m_{\oplus}(\mx)\right)
    \right]
    \right| \\
    &\leq
    O(h^2)
    d_{\mathbb M}\left(
    \tilde{m}_{h,s}(\mx),
    m_{\oplus}(\mx)
    \right),
\end{align*}
uniformly over $\mx\in\mathcal K$. Combining this bound with Condition~\ref{con:U-M3} yields
\begin{align*}
    C_{\oplus,\mathcal K}
    d_{\mathbb M}\left(
    \tilde{m}_{h,s}(\mx),
    m_{\oplus}(\mx)
    \right)^{\beta_{\oplus,\mathcal K}}
    \leq
    O(h^2)
    d_{\mathbb M}\left(
    \tilde{m}_{h,s}(\mx),
    m_{\oplus}(\mx)
    \right)
\end{align*}
uniformly over $\mx\in\mathcal K$. If
\begin{align*}
    d_{\mathbb M}\left(
    \tilde{m}_{h,s}(\mx),
    m_{\oplus}(\mx)
    \right)
    =
    0,
\end{align*}
then the desired bound is trivial. Otherwise, dividing by this distance gives
\begin{align*}
    d_{\mathbb M}\left(
    \tilde{m}_{h,s}(\mx),
    m_{\oplus}(\mx)
    \right)^{\beta_{\oplus,\mathcal K}-1}
    =
    O(h^2)
\end{align*}
uniformly over $\mx\in\mathcal K$. This proves \eqref{eq:E.uniform_oracle_bias}.
\end{proof}

To control the uniform localized stochastic fluctuation, we use the following variance-sensitive bracketing maximal bound. It is the uniform analogue of \Cref{lemma:C.vw_bracketing_maximal}; the difference is that the bound is expressed in terms of the maximal $L_2(P)$ size of the class rather than the $L_2(P)$ norm of a global envelope. This distinction is essential here because the kernel support moves with $\mx\in\mathcal K$.

\begin{lemma}[Uniform oracle weight bounds] \label{lemma:E.uniform_oracle_weight_bounds}
Assume Conditions~\ref{con:P-K1} and~\ref{con:U-D1}, and suppose that $h\to0$ as $n\to\infty$. Then, for each $s\in\{0,1\}$, there exists a constant $C_W<\infty$ such that, for all sufficiently small $h$,
\begin{align}
    \sup_{\mx\in\mathcal K}
    \E\left[
    \tilde W_{\mx,h,s}(\mX)^2
    \right]
    \leq
    C_W h^{-d},
    \quad
    \sup_{\mx\in\mathcal K}
    \sup_{\mz\in\mathcal M}
    \left|
    \tilde W_{\mx,h,s}(\mz)
    \right|
    \leq
    C_W h^{-d}. \label{eq:E.uniform_oracle_weight_bounds}
\end{align}
\end{lemma}

\begin{proof}[Proof of \Cref{lemma:E.uniform_oracle_weight_bounds}]
By the compact support and boundedness of $K$, together with the uniform volume-density bounds on the fixed tube, there exists $C_{\mathcal L}<\infty$ such that, for all sufficiently small $h$,
\begin{align}
    \sup_{\mx\in\mathcal K}
    \sup_{\mz\in\mathcal M}
    \left|
    \mathcal L_{\mx,h}(\mz)
    \right|
    \leq
    C_{\mathcal L},
    \quad
    \mathcal L_{\mx,h}(\mz)=0
    \text{ unless }
    d_{\mathcal M}(\mx,\mz)\leq h.
    \label{eq:E.kernel_support_sup_bound}
\end{align}
Moreover, by Condition~\ref{con:U-D1} and the uniform volume bound in \Cref{lemma:A.uniform_normal_neighborhoods}, there exists $C_{\mathrm{ball}}<\infty$ such that
\begin{align}
    \sup_{\mx\in\mathcal K}
    P\left(
    \mX\in B_{\mathcal M}(\mx,h)
    \right)
    \leq
    C_{\mathrm{ball}}h^d
    \label{eq:E.uniform_small_ball_probability}
\end{align}
for all sufficiently small $h$.

For $s=0$, we have
\begin{align*}
    \tilde W_{\mx,h,0}(\mz)
    =
    \frac{\mathcal L_{\mx,h}(\mz)}
    {\tilde{\mu}_{h,0}(\mx)}.
\end{align*}
By \Cref{lemma:D.uniform_population_moment_orders}, there exists $c_{\mu,0}>0$ such that
\begin{align*}
    \inf_{\mx\in\mathcal K}
    \tilde{\mu}_{h,0}(\mx)
    \geq
    c_{\mu,0}h^d
\end{align*}
for all sufficiently small $h$. Combining this lower bound with \eqref{eq:E.kernel_support_sup_bound} gives
\begin{align*}
    \sup_{\mx\in\mathcal K}
    \sup_{\mz\in\mathcal M}
    \left|
    \tilde W_{\mx,h,0}(\mz)
    \right|
    \leq
    C_{\mathcal L}c_{\mu,0}^{-1}h^{-d}.
\end{align*}
Since $\tilde W_{\mx,h,0}(\mz)=0$ unless $d_{\mathcal M}(\mx,\mz)\leq h$, \eqref{eq:E.uniform_small_ball_probability} gives
\begin{align*}
    \sup_{\mx\in\mathcal K}
    \E\left[
    \tilde W_{\mx,h,0}(\mX)^2
    \right]
    &\leq
    C_{\mathcal L}^2 c_{\mu,0}^{-2}h^{-2d}
    \sup_{\mx\in\mathcal K}
    P\left(
    \mX\in B_{\mathcal M}(\mx,h)
    \right) \\
    &\leq
    C_{\mathcal L}^2 c_{\mu,0}^{-2}C_{\mathrm{ball}}h^{-d}.
\end{align*}

For $s=1$, use the representation
\begin{align*}
    \tilde W_{\mx,h,1}(\mz)
    =
    \tilde{\sigma}_h(\mx)^{-1}
    \mathcal L_{\mx,h}(\mz)
    \left[
    1-
    \bm{\tilde{\mu}}_{h,1}(\mx,\mathbf E_{\mx})^{\top}
    \bm{\tilde{\mu}}_{h,2}(\mx,\mathbf E_{\mx})^{-1}
    \mathbf v_{\mx}^{\mathbf E_{\mx}}(\mz)
    \right].
\end{align*}
By \Cref{lemma:D.uniform_population_moment_orders}, there exist constants $c_{\sigma}>0$, $C_{\mu,1}<\infty$, and $C_{\mu,2}^{-1}<\infty$ such that, for all sufficiently small $h$,
\begin{align*}
    \inf_{\mx\in\mathcal K}
    \tilde{\sigma}_h(\mx)
    \geq
    c_{\sigma}h^d,
    \quad
    \sup_{\mx\in\mathcal K}
    \left\|
    \bm{\tilde{\mu}}_{h,1}(\mx,\mathbf E_{\mx})
    \right\|_2
    \leq
    C_{\mu,1}h^{d+1},
\end{align*}
and
\begin{align*}
    \sup_{\mx\in\mathcal K}
    \left\|
    \bm{\tilde{\mu}}_{h,2}(\mx,\mathbf E_{\mx})^{-1}
    \right\|_{\mathrm{op}}
    \leq
    C_{\mu,2}^{-1}h^{-(d+2)}.
\end{align*}
On the support of $\mathcal L_{\mx,h}$,
\begin{align*}
    \left\|
    \mathbf v_{\mx}^{\mathbf E_{\mx}}(\mz)
    \right\|_2
    \leq
    h.
\end{align*}
Hence the bracketed term is uniformly bounded as
\begin{align*}
    \sup_{\mx\in\mathcal K}
    \sup_{\mz\in\mathcal M}
    \left|
    1-
    \bm{\tilde{\mu}}_{h,1}(\mx,\mathbf E_{\mx})^{\top}
    \bm{\tilde{\mu}}_{h,2}(\mx,\mathbf E_{\mx})^{-1}
    \mathbf v_{\mx}^{\mathbf E_{\mx}}(\mz)
    \right|
    \leq
    1+C_{\mu,1}C_{\mu,2}^{-1}
    =:
    C_{\mathrm{br}}
\end{align*}
for all sufficiently small $h$. Combining this bound with $\tilde{\sigma}_h(\mx)^{-1}\leq c_{\sigma}^{-1}h^{-d}$ and \eqref{eq:E.kernel_support_sup_bound} gives
\begin{align*}
    \sup_{\mx\in\mathcal K}
    \sup_{\mz\in\mathcal M}
    \left|
    \tilde W_{\mx,h,1}(\mz)
    \right|
    \leq
    c_{\sigma}^{-1}C_{\mathcal L}C_{\mathrm{br}}h^{-d}.
\end{align*}
Again, since $\tilde W_{\mx,h,1}(\mz)=0$ unless $d_{\mathcal M}(\mx,\mz)\leq h$, \eqref{eq:E.uniform_small_ball_probability} yields
\begin{align*}
    \sup_{\mx\in\mathcal K}
    \E\left[
    \tilde W_{\mx,h,1}(\mX)^2
    \right]
    &\leq
    c_{\sigma}^{-2}C_{\mathcal L}^2C_{\mathrm{br}}^2h^{-2d}
    \sup_{\mx\in\mathcal K}
    P\left(
    \mX\in B_{\mathcal M}(\mx,h)
    \right) \\
    &\leq
    c_{\sigma}^{-2}C_{\mathcal L}^2C_{\mathrm{br}}^2C_{\mathrm{ball}}h^{-d}.
\end{align*}
Taking $C_W$ larger than the four displayed constants proves \eqref{eq:E.uniform_oracle_weight_bounds}.
\end{proof}

To control localized empirical processes with uniform entropy, we use the following standard maximal inequality. For a measurable function $g:\mathcal Z\to\mathbb R$, write
\begin{align*}
    P_{\mathcal Z}g
    :=
    \int_{\mathcal Z}g(z)\,\dd P_{\mathcal Z}(z),
    \quad
    \mathbb P_n g
    :=
    \frac1n\sum_{i=1}^n g(Z_i),
    \quad
    \mathbb G_n g
    :=
    \sqrt n\left(\mathbb P_n-P_{\mathcal Z}\right)g.
\end{align*}
For a class $\mathcal G$ with envelope $G$, define the uniform entropy integral
\begin{align}
    J_{\mathrm{unif}}(\mathcal G,G)
    :=
    \sup_Q
    \int_0^1
    \sqrt{
    1+
    \log N
    \left(
    \epsilon\left\|G\right\|_{L_2(Q)},
    \mathcal G,
    L_2(Q)
    \right)
    }
    \,\dd\epsilon,
    \label{eq:E.uniform_entropy_integral_def}
\end{align}
where the supremum is over all finitely discrete probability measures $Q$ on the underlying sample space such that $\left\|G\right\|_{L_2(Q)}>0$. If $\left\|G\right\|_{L_2(Q)}=0$, then all functions in $\mathcal G$ vanish $Q$-almost surely and the corresponding covering number is interpreted as one.

The next lemma is the uniform-entropy maximal inequality of Theorem~2.14.1 of \cite{van der Vaart and Wellner (1996)}, written in the form needed below. The supremum over finitely discrete probability measures in \eqref{eq:E.uniform_entropy_integral_def} is the key feature that allows the symmetrization and chaining argument to control the empirical $L_2$ metrics uniformly.

\begin{lemma}[Uniform-entropy expectation maximal inequality] \label{lemma:E.uniform_entropy_expectation_maximal}
Let $Z_1,\ldots,Z_n$ be i.i.d. $\mathcal Z$-valued random variables with distribution $P_{\mathcal Z}$, and let $\mathcal G$ be a class of measurable real-valued functions on $\mathcal Z$ with measurable envelope $G$. Suppose that $J_{\mathrm{unif}}(\mathcal G,G)<\infty$ and $\left\|G\right\|_{L_2(P_{\mathcal Z})}<\infty$. Assume that the displayed supremum below is measurable. Then there exists a universal constant $C_{\mathrm{UniEnt}}<\infty$ such that
\begin{align}
    \E
    \left[
    \sup_{g\in\mathcal G}
    \left|
    \mathbb G_n g
    \right|
    \right]
    \leq
    C_{\mathrm{UniEnt}}
    J_{\mathrm{unif}}(\mathcal G,G)
    \left\|G\right\|_{L_2(P_{\mathcal Z})}.
    \label{eq:E.uniform_entropy_expectation_maximal}
\end{align}
If the measurability of the supremum is not imposed, the same bound holds with outer expectation.
\end{lemma}

We also use the following concentration form of Bousquet's version of Talagrand's inequality; see \cite{Bousquet (2002)}. This statement is used only as a standard empirical-process tool.

\begin{lemma}[Bousquet concentration inequality] \label{lemma:E.bousquet_concentration}
Let $Z_1,\ldots,Z_n$ be i.i.d. $\mathcal Z$-valued random variables with distribution $P_{\mathcal Z}$. Let $\mathcal G$ be a class of measurable real-valued functions on $\mathcal Z$ such that $P_{\mathcal Z}g=0$ for every $g\in\mathcal G$. Suppose that, for some constants $\sigma<\infty$ and $b<\infty$,
\begin{align*}
    \sup_{g\in\mathcal G}\left\|g\right\|_\infty\leq b,
    \quad
    \sup_{g\in\mathcal G}P_{\mathcal Z}g^2\leq\sigma^2.
\end{align*}
Assume that the displayed supremum below is measurable, and define
\begin{align*}
    Z_{\mathcal G}
    :=
    \sup_{g\in\mathcal G}
    \left|
    \sum_{i=1}^n g(Z_i)
    \right|.
\end{align*}
Then there exist universal constants $C_{\mathrm{Bous},1}<\infty$ and $C_{\mathrm{Bous},2}<\infty$ such that, for every $t\geq1$,
\begin{align}
    \mathbb P
    \left(
    Z_{\mathcal G}
    >
    2\E Z_{\mathcal G}
    +
    C_{\mathrm{Bous},1}
    \left[
    \left(n\sigma^2t\right)^{1/2}
    +
    bt
    \right]
    \right)
    \leq
    C_{\mathrm{Bous},2}e^{-t}.
    \label{eq:E.bousquet_concentration}
\end{align}
If the measurability of the supremum is not imposed, the same bound holds with outer probability and a measurable-majorant version of $Z_{\mathcal G}$.
\end{lemma}

Combining the preceding uniform-entropy expectation bound with Bousquet's concentration inequality gives the exponential maximal inequality used in the localized finite-cover argument.

\begin{lemma}[Exponential uniform-entropy maximal inequality] \label{lemma:E.exponential_uniform_entropy_maximal}
Let $Z_1,\ldots,Z_n$ be i.i.d. $\mathcal Z$-valued random variables with distribution $P_{\mathcal Z}$, and let $\mathcal G$ be a class of measurable real-valued functions on $\mathcal Z$ with measurable envelope $G$. Suppose that, for some constants $\sigma<\infty$, $b<\infty$, and $J<\infty$,
\begin{align*}
    |g(z)|\leq G(z),
    \quad g\in\mathcal G,\ z\in\mathcal Z,
    \quad
    \left\|G\right\|_{L_2(P_{\mathcal Z})}\leq\sigma,
    \quad
    \left\|G\right\|_\infty\leq b,
    \quad
    J_{\mathrm{unif}}(\mathcal G,G)\leq J.
\end{align*}
Assume that the displayed supremum below is measurable. Then there exists a universal constant $C_{\mathrm{UniMax}}<\infty$ such that, for every $t\geq1$,
\begin{align}
    \mathbb P
    \left(
    \sup_{g\in\mathcal G}
    \left|
    \left(\mathbb P_n-P_{\mathcal Z}\right)g
    \right|
    >
    C_{\mathrm{UniMax}}
    \left[
    \frac{\sigma J}{\sqrt n}
    +
    \sigma\left(\frac{t}{n}\right)^{1/2}
    +
    \frac{bt}{n}
    \right]
    \right)
    \leq
    2e^{-t}.
    \label{eq:E.exponential_uniform_entropy_maximal}
\end{align}
If the measurability of the supremum is not imposed, the same bound holds with outer probability and a measurable-majorant version of the supremum.
\end{lemma}

\begin{proof}[Proof of \Cref{lemma:E.exponential_uniform_entropy_maximal}]
Define
\begin{align*}
    Z_{\mathcal G}
    :=
    \sup_{g\in\mathcal G}
    \left|
    \sum_{i=1}^n
    \left\{
    g(Z_i)-P_{\mathcal Z}g
    \right\}
    \right|
    =
    \sqrt n
    \sup_{g\in\mathcal G}
    \left|
    \mathbb G_n g
    \right|.
\end{align*}
By \Cref{lemma:E.uniform_entropy_expectation_maximal},
\begin{align}
    \E Z_{\mathcal G}
    \leq
    C_{\mathrm{UniEnt}}\sqrt n
    J_{\mathrm{unif}}(\mathcal G,G)
    \left\|G\right\|_{L_2(P_{\mathcal Z})}
    \leq
    C_{\mathrm{UniEnt}}\sqrt n\,\sigma J.
    \label{eq:E.unient_expectation_bound}
\end{align}

Define the centered signed class
\begin{align*}
    \mathcal G_c
    :=
    \left\{
    g-P_{\mathcal Z}g:g\in\mathcal G
    \right\}
    \cup
    \left\{
    P_{\mathcal Z}g-g:g\in\mathcal G
    \right\}.
\end{align*}
Then each $f\in\mathcal G_c$ satisfies $P_{\mathcal Z}f=0$, and
\begin{align*}
    Z_{\mathcal G}
    =
    \sup_{f\in\mathcal G_c}
    \left|
    \sum_{i=1}^n f(Z_i)
    \right|.
\end{align*}
Moreover, since $|g|\leq G$ and $\left\|G\right\|_\infty\leq b$,
\begin{align*}
    \sup_{f\in\mathcal G_c}\left\|f\right\|_\infty
    \leq
    2b,
    \quad
    \sup_{f\in\mathcal G_c}P_{\mathcal Z}f^2
    =
    \sup_{g\in\mathcal G}
    P_{\mathcal Z}
    \left(g-P_{\mathcal Z}g\right)^2
    \leq
    P_{\mathcal Z}G^2
    \leq
    \sigma^2.
\end{align*}
Applying \Cref{lemma:E.bousquet_concentration} to $\mathcal G_c$ gives, after changing only universal constants, that for every $t\geq1$,
\begin{align}
    \mathbb P
    \left(
    Z_{\mathcal G}
    >
    2\E Z_{\mathcal G}
    +
    C_{\mathrm{Bous}}
    \left[
    \left(n\sigma^2t\right)^{1/2}
    +
    bt
    \right]
    \right)
    \leq
    2e^{-t}.
    \label{eq:E.unient_bousquet}
\end{align}
Combining \eqref{eq:E.unient_expectation_bound} and \eqref{eq:E.unient_bousquet}, and dividing by $n$, proves \eqref{eq:E.exponential_uniform_entropy_maximal}. The outer-probability version follows by applying the same argument to measurable majorants.
\end{proof}

\begin{lemma}[Moving-anchor response-increment entropy] \label{lemma:E.moving_anchor_response_entropy}
Assume Conditions~\ref{con:M1} and~\ref{con:U-M4}. Let $\tilde y_h:\mathcal K\to\mathbb M$ be a deterministic sequence such that
\begin{align*}
    \sup_{\mx\in\mathcal K}
    d_{\mathbb M}\left(\tilde y_h(\mx),m_{\oplus}(\mx)\right)
    \leq
    \frac{r_{\mathbb M,\mathcal K}}{2}
\end{align*}
for all sufficiently small $h$. For $\delta>0$, define
\begin{align}
    \mathcal H_\delta(\tilde y_h)
    :=
    \left\{
    \omega\mapsto
    d_{\mathbb M}^2(y,\omega)
    -
    d_{\mathbb M}^2\left(\tilde y_h(\mx),\omega\right):
    \mx\in\mathcal K,\ 
    y\in B_{\mathbb M}\left(\tilde y_h(\mx),\delta\right)
    \right\}.
    \label{eq:E.functional_class_response}
\end{align}
Let $D_{\mathbb M}$ be defined by \eqref{eq:B.diameter_metric_space}. Then there exist constants $\delta_{\mathrm{Mov}}>0$, $q_{\mathbb M,\mathcal K}<\infty$, and $C_{\mathbb M,\mathcal K,\mathrm{poly}}<\infty$ such that, for every sufficiently small $h$, every $0<\delta<\delta_{\mathrm{Mov}}$, and every $\epsilon\in(0,1)$,
\begin{align}
    \sup_Q
    N\left(
    \epsilon D_{\mathbb M}\delta,
    \mathcal H_\delta(\tilde y_h),
    L_2(Q)
    \right)
    \leq
    C_{\mathbb M,\mathcal K,\mathrm{poly}}^2
    \left(\frac{4}{\delta\epsilon}\right)^{2q_{\mathbb M,\mathcal K}},
    \label{eq:E.moving_anchor_response_entropy}
\end{align}
where the supremum is over all finitely discrete probability measures $Q$ on $\mathbb M$. The constants may be chosen uniformly over all deterministic maps $\tilde y_h$ satisfying the displayed localization bound.
\end{lemma}

\begin{proof}[Proof of \Cref{lemma:E.moving_anchor_response_entropy}]
Let
\begin{align*}
    \mathcal N_{\mathcal K}
    :=
    \left\{
    z\in\mathbb M:
    d_{\mathbb M}\left(z,m_\oplus(\mx)\right)
    <
    r_{\mathbb M,\mathcal K}
    \text{ for some }\mx\in\mathcal K
    \right\}.
\end{align*}
By Condition~\ref{con:M1}, $\mathbb M$ is totally bounded, and hence $\mathcal N_{\mathcal K}$ is totally bounded. By Condition~\ref{con:U-M4}, there exist constants $\eta_{\mathbb M,\mathcal K}>0$ and $C_{\mathbb M,\mathcal K}<\infty$ such that, for every $0<r\leq\eta_{\mathbb M,\mathcal K}$,
\begin{align}
    \sup_{z\in\mathcal N_{\mathcal K}}
    \int_0^{1/2}
    \sqrt{
    1+
    \log N\left(r\epsilon,B_{\mathbb M}(z,r),d_{\mathbb M}\right)
    }
    \,\dd\epsilon
    \leq
    C_{\mathbb M,\mathcal K}.
    \label{eq:E.U-M4_on_NK}
\end{align}
Indeed, each $z\in\mathcal N_{\mathcal K}$ is within distance $r_{\mathbb M,\mathcal K}$ of $m_\oplus(\mx)$ for at least one $\mx\in\mathcal K$, so Condition~\ref{con:U-M4} applies with center $z$.

Fix $\epsilon_0\in(0,1/4)$. Since the map
\begin{align*}
    \epsilon
    \mapsto
    \sqrt{
    1+
    \log N\left(r\epsilon,B_{\mathbb M}(z,r),d_{\mathbb M}\right)
    }
\end{align*}
is nonincreasing in $\epsilon$, \eqref{eq:E.U-M4_on_NK} implies, uniformly over $z\in\mathcal N_{\mathcal K}$ and $0<r\leq\eta_{\mathbb M,\mathcal K}$,
\begin{align*}
    \sqrt{
    1+
    \log N\left(r\epsilon_0/4,B_{\mathbb M}(z,r),d_{\mathbb M}\right)
    }
    &\leq
    \frac{8}{\epsilon_0}
    \int_{\epsilon_0/8}^{\epsilon_0/4}
    \sqrt{
    1+
    \log N\left(r\epsilon,B_{\mathbb M}(z,r),d_{\mathbb M}\right)
    }
    \,\dd\epsilon \\
    &\leq
    \frac{8C_{\mathbb M,\mathcal K}}{\epsilon_0}.
\end{align*}
Consequently,
\begin{align}
    \sup_{z\in\mathcal N_{\mathcal K}}
    N\left(
    r\epsilon_0/4,
    B_{\mathbb M}(z,r),
    d_{\mathbb M}
    \right)
    \leq
    C_{\epsilon_0},
    \quad
    0<r\leq\eta_{\mathbb M,\mathcal K},
    \label{eq:E.local_doubling_response_raw}
\end{align}
where $C_{\epsilon_0}:=\exp\left(64C_{\mathbb M,\mathcal K}^2\epsilon_0^{-2}\right)\geq1$. This yields an intrinsic covering bound for intersections with $\mathcal N_{\mathcal K}$ whose centers remain in $\mathcal N_{\mathcal K}$. Specifically, for any $z\in\mathcal N_{\mathcal K}$ and $0<r\leq\eta_{\mathbb M,\mathcal K}$, cover $B_{\mathbb M}(z,r)$ by at most $C_{\epsilon_0}$ balls of radius $r\epsilon_0/4$. For each covering ball that intersects $B_{\mathbb M}(z,r)\cap\mathcal N_{\mathcal K}$, choose one point of this intersection as its new center. Then $B_{\mathbb M}(z,r)\cap\mathcal N_{\mathcal K}$ is covered by at most $C_{\epsilon_0}$ balls centered in $\mathcal N_{\mathcal K}$ with radius $r\epsilon_0/2$, and hence also with radius $r\epsilon_0$. Let $N_{\mathbb M,\mathcal K,0}<\infty$ be the cardinality of an $\eta_{\mathbb M,\mathcal K}$-net of $\mathcal N_{\mathcal K}$.

We now derive a polynomial covering bound for $\mathcal N_{\mathcal K}$. Fix an arbitrary $\zeta\in(0,\eta_{\mathbb M,\mathcal K})$, and choose the integer $m_\zeta\geq1$ such that
\begin{align*}
    \epsilon_0^{m_\zeta}\eta_{\mathbb M,\mathcal K}
    \leq
    \zeta
    <
    \epsilon_0^{m_\zeta-1}\eta_{\mathbb M,\mathcal K}.
\end{align*}
Starting from the $\eta_{\mathbb M,\mathcal K}$-net of $\mathcal N_{\mathcal K}$ and applying the preceding local covering bound successively at the scales
\begin{align*}
    \eta_{\mathbb M,\mathcal K},
    \quad
    \epsilon_0\eta_{\mathbb M,\mathcal K},
    \quad
    \epsilon_0^2\eta_{\mathbb M,\mathcal K},
    \quad
    \ldots,
    \quad
    \epsilon_0^{m_\zeta-1}\eta_{\mathbb M,\mathcal K},
\end{align*}
we obtain a cover of $\mathcal N_{\mathcal K}$ by at most $N_{\mathbb M,\mathcal K,0}C_{\epsilon_0}^{m_\zeta}$ balls of radius $\epsilon_0^{m_\zeta}\eta_{\mathbb M,\mathcal K}$. Since $\epsilon_0^{m_\zeta}\eta_{\mathbb M,\mathcal K}\leq\zeta$, this cover is also a $\zeta$-cover. Hence
\begin{align*}
    N\left(
    \zeta,
    \mathcal N_{\mathcal K},
    d_{\mathbb M}
    \right)
    \leq
    N_{\mathbb M,\mathcal K,0}C_{\epsilon_0}^{m_\zeta}.
\end{align*}
Set
\begin{align}
    q_{\mathbb M,\mathcal K}
    &:=
    \frac{\log C_{\epsilon_0}}{\log(1/\epsilon_0)}
    =
    \frac{64C_{\mathbb M,\mathcal K}^2}
    {\epsilon_0^2\log(1/\epsilon_0)}, \quad
    C_{\mathbb M,\mathcal K,\mathrm{poly}}
    :=
    N_{\mathbb M,\mathcal K,0}
    \left(\frac{\eta_{\mathbb M,\mathcal K}}{\epsilon_0}\right)^{q_{\mathbb M,\mathcal K}}.
    \label{eq:E.response_poly_constants_def}
\end{align}
Then $C_{\epsilon_0}^{m_\zeta}=(\epsilon_0^{-m_\zeta})^{q_{\mathbb M,\mathcal K}}$. Moreover, the defining inequality for $m_\zeta$ gives
\begin{align*}
    \epsilon_0^{-m_\zeta}
    =
    \epsilon_0^{-1}\epsilon_0^{-(m_\zeta-1)}
    \leq
    \epsilon_0^{-1}
    \frac{\eta_{\mathbb M,\mathcal K}}{\zeta}.
\end{align*}
Therefore, we get a covering bound
\begin{align}
    N\left(
    \zeta,
    \mathcal N_{\mathcal K},
    d_{\mathbb M}
    \right)
    \leq
    C_{\mathbb M,\mathcal K,\mathrm{poly}}
    \zeta^{-q_{\mathbb M,\mathcal K}}, 
    \quad
    0<\zeta<\eta_{\mathbb M,\mathcal K}.
    \label{eq:E.response_neighborhood_polynomial_cover}
\end{align}

Set
\begin{align}
    \delta_{\mathrm{Mov}}
    :=
    \min\left\{
    \frac{r_{\mathbb M,\mathcal K}}{2},
    \eta_{\mathbb M,\mathcal K}
    \right\}.
    \label{eq:E.delta_mov_def}
\end{align}
For all sufficiently small $h$, the localization condition gives
\begin{align*}
    \sup_{\mx\in\mathcal K}
    d_{\mathbb M}\left(\tilde y_h(\mx),m_\oplus(\mx)\right)
    \leq
    \frac{r_{\mathbb M,\mathcal K}}{2}.
\end{align*}
Thus, for every $0<\delta<\delta_{\mathrm{Mov}}$, every $\mx\in\mathcal K$, and every $y\in B_{\mathbb M}(\tilde y_h(\mx),\delta)$,
\begin{align*}
    d_{\mathbb M}\left(y,m_\oplus(\mx)\right)
    &\leq
    d_{\mathbb M}\left(y,\tilde y_h(\mx)\right)
    +
    d_{\mathbb M}\left(\tilde y_h(\mx),m_\oplus(\mx)\right)
    <
    \delta+\frac{r_{\mathbb M,\mathcal K}}{2}
    <
    r_{\mathbb M,\mathcal K}.
\end{align*}
Therefore $\tilde y_h(\mx)\in\mathcal N_{\mathcal K}$ and $B_{\mathbb M}(\tilde y_h(\mx),\delta)\subset\mathcal N_{\mathcal K}$ uniformly over $\mx\in\mathcal K$.

Fix $\epsilon\in(0,1)$ and set
\begin{align*}
    \zeta_{\delta,\epsilon}
    :=
    \frac{\delta\epsilon}{4}.
\end{align*}
Since $0<\zeta_{\delta,\epsilon}<\eta_{\mathbb M,\mathcal K}$, \eqref{eq:E.response_neighborhood_polynomial_cover} provides a $\zeta_{\delta,\epsilon}$-net $\{z_1,\ldots,z_{N_{\delta,\epsilon}}\}$ of $\mathcal N_{\mathcal K}$ satisfying
\begin{align*}
    N_{\delta,\epsilon}
    \leq
    C_{\mathbb M,\mathcal K,\mathrm{poly}}
    \left(\frac{\delta\epsilon}{4}\right)^{-q_{\mathbb M,\mathcal K}}.
\end{align*}
For any pair $(\tilde y_h(\mx),y)$ with $y\in B_{\mathbb M}(\tilde y_h(\mx),\delta)$, choose $z_j$ and $z_k$ such that
\begin{align*}
    d_{\mathbb M}\left(\tilde y_h(\mx),z_j\right)
    \leq
    \frac{\delta\epsilon}{4},
    \quad
    d_{\mathbb M}(y,z_k)
    \leq
    \frac{\delta\epsilon}{4}.
\end{align*}
For any $a,a',y,y',\omega\in\mathbb M$,
\begin{align*}
&\left|
\left[
d_{\mathbb M}^2(y,\omega)-d_{\mathbb M}^2(a,\omega)
\right]
-
\left[
d_{\mathbb M}^2(y',\omega)-d_{\mathbb M}^2(a',\omega)
\right]
\right| \\
&\leq
2D_{\mathbb M}
\left\{
d_{\mathbb M}(y,y')
+
d_{\mathbb M}(a,a')
\right\}.
\end{align*}
Therefore, with $a=\tilde y_h(\mx)$, $a'=z_j$, and $y'=z_k$,
\begin{align*}
&\left\|
\left[
d_{\mathbb M}^2(y,\cdot)
-
d_{\mathbb M}^2(\tilde y_h(\mx),\cdot)
\right]
-
\left[
d_{\mathbb M}^2(z_k,\cdot)
-
d_{\mathbb M}^2(z_j,\cdot)
\right]
\right\|_{\infty} \\
&\leq
D_{\mathbb M}\delta\epsilon.
\end{align*}
Thus $\mathcal H_\delta(\tilde y_h)$ is covered in $\|\cdot\|_\infty$ at radius $D_{\mathbb M}\delta\epsilon$ by at most $N_{\delta,\epsilon}^2$ functions. Since $\|f\|_{L_2(Q)}\leq\|f\|_\infty$ for every probability measure $Q$,
\begin{align*}
    \sup_Q
    N\left(
    D_{\mathbb M}\delta\epsilon,
    \mathcal H_\delta(\tilde y_h),
    L_2(Q)
    \right)
    &\leq
    N_{\delta,\epsilon}^2
    \leq
    C_{\mathbb M,\mathcal K,\mathrm{poly}}^2
    \left(\frac{4}{\delta\epsilon}\right)^{2q_{\mathbb M,\mathcal K}}.
\end{align*}
This proves \eqref{eq:E.moving_anchor_response_entropy}.
\end{proof}

\begin{lemma}[Localized finite-cover entropy bound] \label{lemma:E.localized_finite_cover_entropy}
Assume Conditions~\ref{con:U-K1}, \ref{con:U-B1}, \ref{con:U-D1}, \ref{con:M1}, and~\ref{con:U-M4}. Fix $s\in\{0,1\}$. If $s=1$, assume in addition Condition~\ref{con:U-K2}. Let $\tilde y_h:\mathcal K\to\mathbb M$ be a deterministic sequence such that
\begin{align*}
    \sup_{\mx\in\mathcal K}
    d_{\mathbb M}\left(\tilde y_h(\mx),m_{\oplus}(\mx)\right)
    \leq
    \frac{r_{\mathbb M,\mathcal K}}{2}
\end{align*}
for all sufficiently small $h$. For $\delta>0$, define
\begin{align*}
    \mathcal F_{h,s,\delta}
    &:=
    \left\{
    (\mz,\omega)\mapsto
    \tilde W_{\mx,h,s}(\mz)
    \left[
    d_{\mathbb M}^2(y,\omega)
    -
    d_{\mathbb M}^2\left(\tilde y_h(\mx),\omega\right)
    \right]:
    \mx\in\mathcal K,\ 
    y\in B_{\mathbb M}\left(\tilde y_h(\mx),\delta\right)
    \right\}.
\end{align*}
Assume that the displayed supremum below is measurable. Then there exist constants $\delta_{\mathrm{loc}}>0$ and $C_{\mathrm{loc},s}<\infty$, independent of $n$, $h$, $\delta$, and $\tilde y_h$, such that, for every $\delta\in(0,\delta_{\mathrm{loc}}]$ and all sufficiently large $n$,
\begin{align}
\begin{split}
    &\E
    \left[
    \sup_{U\in\mathcal F_{h,s,\delta}}
    \left|
    (\mathbb P_n-P)U
    \right|
    \right] \\
    &\leq
    C_{\mathrm{loc},s}
    \delta
    \left[
    \left(
    \frac{
    1+\log(1/h)+\log(1/\delta)
    }{nh^d}
    \right)^{1/2}
    +
    \frac{1+\log(1/h)}{nh^d}
    \right].
\end{split}
\label{eq:E.localized_finite_cover_entropy_refined}
\end{align}
Consequently, whenever $\log(1/\delta)=O(\log n)$, for sufficiently large $n$,
\begin{align}
    \E
    \left[
    \sup_{U\in\mathcal F_{h,s,\delta}}
    \left|
    (\mathbb P_n-P)U
    \right|
    \right]
    \leq
    C_{\mathrm{loc},s}
    \delta
    \left(
    \frac{\log n}{nh^d}
    \right)^{1/2}.
    \label{eq:E.localized_finite_cover_entropy}
\end{align}
If the measurability of the supremum is not imposed, the same bounds hold with outer expectation.
\end{lemma}

\begin{proof}[Proof of \Cref{lemma:E.localized_finite_cover_entropy}]
Fix $s\in\{0,1\}$. Let
\begin{align*}
    \mathcal J_0:=\{0\},
    \quad
    \mathcal J_1:=\{0,1,\ldots,d\},
    \quad
    m_s:=|\mathcal J_s|.
\end{align*}
Thus the local constant case uses only the zeroth-order design class, while the local linear case uses the zeroth- and first-order design classes.

Let $\delta_{\mathrm{Mov}}>0$ be the constant in \Cref{lemma:E.moving_anchor_response_entropy}, and set
\begin{align*}
    \delta_{\mathrm{loc}}
    :=
    \min\left\{
    \frac{r_{\mathbb M,\mathcal K}}{4},
    \delta_{\mathrm{Mov}}
    \right\}.
\end{align*}
Then, for every $\delta\in(0,\delta_{\mathrm{loc}}]$, every $\mx\in\mathcal K$, and every $y\in B_{\mathbb M}(\tilde y_h(\mx),\delta)$,
\begin{align*}
\begin{split}
    d_{\mathbb M}\left(y,m_\oplus(\mx)\right)
    &\leq
    d_{\mathbb M}\left(y,\tilde y_h(\mx)\right)
    +
    d_{\mathbb M}\left(\tilde y_h(\mx),m_\oplus(\mx)\right) \\
    &\leq
    \delta+\frac{r_{\mathbb M,\mathcal K}}{2}
    <
    r_{\mathbb M,\mathcal K}.
\end{split}
\end{align*}
Thus the localization required for Condition~\ref{con:U-M4} and \Cref{lemma:E.moving_anchor_response_entropy} is valid uniformly over $\mx\in\mathcal K$.

By \Cref{lemma:E.uniform_oracle_weight_bounds}, there exists $C_{W,s}<\infty$ such that, for all sufficiently small $h$,
\begin{align*}
    \sup_{\mx\in\mathcal K}
    \sup_{\mz\in\mathcal M}
    \left|
    \tilde W_{\mx,h,s}(\mz)
    \right|
    \leq
    C_{W,s}h^{-d},
\end{align*}
and $\tilde W_{\mx,h,s}(\mz)=0$ unless $d_{\mathcal M}(\mx,\mz)\leq h$. By Condition~\ref{con:M1}, $D_{\mathbb M}$ defined by \eqref{eq:B.diameter_metric_space} is finite. For $y\in B_{\mathbb M}(\tilde y_h(\mx),\delta)$ and $\omega\in\mathbb M$,
\begin{align*}
\begin{split}
    \left|
    d_{\mathbb M}^2(y,\omega)
    -
    d_{\mathbb M}^2\left(\tilde y_h(\mx),\omega\right)
    \right|
    &\leq
    2D_{\mathbb M}
    d_{\mathbb M}\left(y,\tilde y_h(\mx)\right) \\
    &\leq
    2D_{\mathbb M}\delta.
\end{split}
\end{align*}
With $C_{\mathrm{env},s}:=2D_{\mathbb M}C_{W,s}$, the preceding two displays imply that $\mathcal F_{h,s,\delta}$ admits the deterministic envelope $C_{\mathrm{env},s}\delta h^{-d}$.

By compactness of $\mathcal K$ and the uniform normal-neighborhood volume bounds in \Cref{lemma:A.uniform_normal_neighborhoods}, there exist points $\mx_1,\ldots,\mx_{N_h}\in\mathcal K$ and a constant $C_{\mathcal K,\mathrm{cov}}<\infty$ such that, for all sufficiently small $h$,
\begin{align}
    \mathcal K
    \subset
    \bigcup_{\ell=1}^{N_h}
    B_{\mathcal M}(\mx_\ell,h),
    \quad
    N_h\leq C_{\mathcal K,\mathrm{cov}}h^{-d}.
    \label{eq:E.spatial_h_cover}
\end{align}
Let $\lambda_{\mathrm{fr}}>0$ be a Lebesgue number of the fixed finite frame cover $\{\mathcal O^\alpha:1\leq\alpha\leq N_{\mathcal K}\}$ over the compact set $\mathcal K^\rho$. For all sufficiently small $h$, we have $2h<\rho$ and $4h<\lambda_{\mathrm{fr}}$. Hence each ball
\begin{align*}
    B_h^{(\ell)}
    :=
    B_{\mathcal M}(\mx_\ell,2h)
\end{align*}
is contained in $\mathcal K^\rho$ and in at least one frame chart. Fix one such chart index and denote it by $\alpha_\ell$, so that
\begin{align*}
    B_h^{(\ell)}
    \subset
    \mathcal O^{\alpha_\ell},
    \quad
    \ell=1,\ldots,N_h.
\end{align*}

For $\ell=1,\ldots,N_h$, define the localized class
\begin{align*}
    \mathcal F_{h,s,\delta}^{(\ell)}
    :=
    \left\{
    U\in\mathcal F_{h,s,\delta}:
    \text{the corresponding center } \mx
    \text{ belongs to } \mathcal K\cap B_{\mathcal M}(\mx_\ell,h)
    \right\}.
\end{align*}
Then $\mathcal F_{h,s,\delta}\subset\bigcup_{\ell=1}^{N_h}\mathcal F_{h,s,\delta}^{(\ell)}$. If $U\in\mathcal F_{h,s,\delta}^{(\ell)}$, then $U(\mz,\omega)=0$ unless $\mz\in B_h^{(\ell)}$. Hence $\mathcal F_{h,s,\delta}^{(\ell)}$ admits the localized envelope
\begin{align*}
    F_{h,\ell,\delta}(\mz,\omega)
    :=
    C_{\mathrm{env},s}\delta h^{-d}
    \mathbf 1
    \left\{
    \mz\in B_h^{(\ell)}
    \right\}.
\end{align*}
Let $C_{\mathcal K,\rho}<\infty$ be defined by \eqref{eq:U.design_density_bounds_on_tube}. Choose $h_0>0$ small enough so that the uniform normal-neighborhood volume bounds in \Cref{lemma:A.uniform_normal_neighborhoods} apply, and define
\begin{align*}
    C_{\mathrm{vol},2}
    :=
    \sup_{0<h<h_0}
    \sup_{\mx\in\mathcal K}
    h^{-d}
    \operatorname{vol}_{\mathcal M}
    \left(
    B_{\mathcal M}(\mx,2h)
    \right).
\end{align*}
Then $C_{\mathrm{vol},2}<\infty$ and
\begin{align*}
    \sup_{1\leq\ell\leq N_h}
    P_X\left(\mX\in B_h^{(\ell)}\right)
    \leq
    C_{\mathcal K,\rho}C_{\mathrm{vol},2}h^d.
\end{align*}
Therefore, with $C_{\mathrm{L2},s}:=C_{\mathrm{env},s}(C_{\mathcal K,\rho}C_{\mathrm{vol},2})^{1/2}$,
\begin{align}
    \sup_{1\leq\ell\leq N_h}
    \left\|F_{h,\ell,\delta}\right\|_{L_2(P)}
    \leq
    C_{\mathrm{L2},s}\delta h^{-d/2},
    \quad
    \sup_{1\leq\ell\leq N_h}
    \left\|F_{h,\ell,\delta}\right\|_{\infty}
    \leq
    C_{\mathrm{env},s}\delta h^{-d}.
    \label{eq:E.localized_envelope_bounds}
\end{align}

We next bound the per-cell uniform entropy integral. Throughout this part, $Q$ denotes an arbitrary finitely discrete probability measure on $\mathcal M\times\mathbb M$, and $Q_{\mathcal M}$ denotes its $\mz$-marginal. For each $\ell=1,\ldots,N_h$, define the predictor-side class
\begin{align*}
    \mathcal A_{h,s}^{(\ell)}
    :=
    \left\{
    \mz\mapsto
    \tilde W_{\mx,h,s}(\mz):
    \mx\in \mathcal K\cap B_{\mathcal M}(\mx_\ell,h)
    \right\}.
\end{align*}
For the chart $\mathcal O^{\alpha_\ell}$ associated with the $\ell$th cell, write
\begin{align*}
    \mathbf v_{\mx}^{\alpha_\ell}(\mz)
    :=
    \bm{\Phi}_{\mathbf E_{\mx}^{\alpha_\ell}}
    \Log_{\mx}(\mz),
    \quad
    v_{\mx,r}^{\alpha_\ell}(\mz)
    :=
    \left(
    \mathbf v_{\mx}^{\alpha_\ell}(\mz)
    \right)_r,
    \quad
    r=1,\ldots,d.
\end{align*}
On the support of the kernel, $\left|v_{\mx,r}^{\alpha_\ell}(\mz)\right|\leq h$. Define the normalized local-design classes
\begin{align*}
    \mathcal G_{h,0}^{(\ell)}
    &:=
    \left\{
    \mz\mapsto
    \mathcal L_{\mx,h}(\mz):
    \mx\in \mathcal K\cap B_{\mathcal M}(\mx_\ell,h)
    \right\}, \\
    \mathcal G_{h,r}^{(\ell)}
    &:=
    \left\{
    \mz\mapsto
    \mathcal L_{\mx,h}(\mz)
    \frac{v_{\mx,r}^{\alpha_\ell}(\mz)}{h}:
    \mx\in \mathcal K\cap B_{\mathcal M}(\mx_\ell,h)
    \right\},
    \quad r=1,\ldots,d.
\end{align*}
The class $\mathcal G_{h,0}^{(\ell)}$ is controlled by Condition~\ref{con:U-K1}. When $s=1$, the coordinate multiplier classes $\mathcal G_{h,r}^{(\ell)}$, $r=1,\ldots,d$, are controlled by the first-order part of Condition~\ref{con:U-K2}. Thus the local constant case $s=0$ uses only the zeroth-order local-design class and does not require the additional multiplier condition.

For each $\ell$, define
\begin{align*}
    A_{h,s}^{(\ell)}(\mz)
    :=
    C_{W,s}h^{-d}
    \mathbf 1
    \left\{
    \mz\in B_h^{(\ell)}
    \right\}.
\end{align*}
This is an envelope of $\mathcal A_{h,s}^{(\ell)}$ because $|\tilde W_{\mx,h,s}(\mz)|\leq C_{W,s}h^{-d}$ and $\tilde W_{\mx,h,s}$ is supported on $B_{\mathcal M}(\mx,h)\subset B_h^{(\ell)}$ whenever $\mx\in\mathcal K\cap B_{\mathcal M}(\mx_\ell,h)$.

We now define the coefficient functions in the normalized decomposition of the oracle weights. For $s=0$, set
\begin{align*}
    \alpha_{0,h}^{(0)}(\mx)
    &:=
    \frac{h^d}{\tilde\mu_{h,0}(\mx)}.
\end{align*}
For $s=1$, and for each frame chart $\mathcal O^\alpha$, define, for $\mx\in\mathcal K\cap\mathcal O^\alpha$,
\begin{align*}
    \mathbf q_h^\alpha(\mx)
    &:=
    \bm{\tilde{\mu}}_{h,2}(\mx,\mathbf E^\alpha_{\mx})^{-1}
    \bm{\tilde{\mu}}_{h,1}(\mx,\mathbf E^\alpha_{\mx}), \\
    q_{r,h}^\alpha(\mx)
    &:=
    \left[
    \mathbf q_h^\alpha(\mx)
    \right]_r,
    \quad r=1,\ldots,d,
\end{align*}
and set
\begin{align*}
    \alpha_{0,h}^{(1),\alpha}(\mx)
    &:=
    \frac{h^d}{\tilde\sigma_h(\mx)}, \\
    \alpha_{r,h}^{(1),\alpha}(\mx)
    &:=
    -\frac{h^{d+1}q_{r,h}^\alpha(\mx)}{\tilde\sigma_h(\mx)},
    \quad r=1,\ldots,d.
\end{align*}
With this notation, on any localized cell whose associated frame chart is $\mathcal O^{\alpha_\ell}$,
\begin{align}
    h^d\tilde W_{\mx,h,s}(\mz)
    =
    \sum_{r\in\mathcal J_s}
    \alpha_{r,h}^{(s),\alpha_\ell}(\mx)
    G_{\mx,h,r}^{\alpha_\ell}(\mz),
    \label{eq:E.weight_alpha_decomposition}
\end{align}
where, for $s=0$, we use the convention $\alpha_{0,h}^{(0),\alpha_\ell}(\mx):=\alpha_{0,h}^{(0)}(\mx)$, and
\begin{align*}
    G_{\mx,h,0}^{\alpha_\ell}(\mz)
    &:=
    \mathcal L_{\mx,h}(\mz), \\
    G_{\mx,h,r}^{\alpha_\ell}(\mz)
    &:=
    \mathcal L_{\mx,h}(\mz)
    \frac{v_{\mx,r}^{\alpha_\ell}(\mz)}{h},
    \quad r=1,\ldots,d.
\end{align*}
By \Cref{lemma:D.uniform_population_moment_orders}, there exists $h_{\mathcal A,s}>0$ such that
\begin{align}
    B_{\mathcal A,s}
    &:=
    \sup_{0<h<h_{\mathcal A,s}}
    \max_{1\leq\alpha\leq N_{\mathcal K}}
    \sup_{\mx\in\mathcal K\cap\mathcal O^\alpha}
    \max_{r\in\mathcal J_s}
    \left|
    \alpha_{r,h}^{(s),\alpha}(\mx)
    \right|
    <
    \infty.
    \label{eq:E.alpha_coefficients_uniform_bound}
\end{align}
Indeed, for $s=0$, \Cref{eq:D.uniform_population_moment_orders_mu0_lower} gives
\begin{align*}
    \sup_{\mx\in\mathcal K}
    \left|
    \alpha_{0,h}^{(0)}(\mx)
    \right|
    =
    \sup_{\mx\in\mathcal K}
    \frac{h^d}{\tilde\mu_{h,0}(\mx)}
    \leq
    \frac{2}{A_{d-1}c_{d-1,1}c_{\mathcal K}}
\end{align*}
for all sufficiently small $h$. For $s=1$, \Cref{eq:D.uniform_population_moment_orders_sigma_lower} similarly gives
\begin{align*}
    \sup_{\mx\in\mathcal K}
    \left|
    \alpha_{0,h}^{(1),\alpha}(\mx)
    \right|
    =
    \sup_{\mx\in\mathcal K}
    \frac{h^d}{\tilde\sigma_h(\mx)}
    \leq
    \frac{2}{A_{d-1}c_{d-1,1}c_{\mathcal K}},
\end{align*}
uniformly over $\alpha$. Moreover, by \Cref{eq:D.uniform_population_moment_orders_mu1,eq:D.uniform_population_moment_orders_mu2_inverse},
\begin{align*}
    \max_{1\leq\alpha\leq N_{\mathcal K}}
    \sup_{\mx\in\mathcal K\cap\mathcal O^\alpha}
    \left\|
    \mathbf q_h^\alpha(\mx)
    \right\|_2
    &\leq
    \max_{1\leq\alpha\leq N_{\mathcal K}}
    \sup_{\mx\in\mathcal K\cap\mathcal O^\alpha}
    \left\|
    \bm{\tilde{\mu}}_{h,2}(\mx,\mathbf E^\alpha_{\mx})^{-1}
    \right\|_{\mathrm{op}}
    \left\|
    \bm{\tilde{\mu}}_{h,1}(\mx,\mathbf E^\alpha_{\mx})
    \right\|_2 \\
    &=
    O(h^{-(d+2)})o(h^{d+1})
    =
    o(h^{-1}).
\end{align*}
Therefore, for $r=1,\ldots,d$,
\begin{align*}
    \max_{1\leq\alpha\leq N_{\mathcal K}}
    \sup_{\mx\in\mathcal K\cap\mathcal O^\alpha}
    \left|
    \alpha_{r,h}^{(1),\alpha}(\mx)
    \right|
    &\leq
    \sup_{\mx\in\mathcal K}
    \frac{h^{d+1}}{\tilde\sigma_h(\mx)}
    \max_{1\leq\alpha\leq N_{\mathcal K}}
    \sup_{\mx\in\mathcal K\cap\mathcal O^\alpha}
    \left\|
    \mathbf q_h^\alpha(\mx)
    \right\|_2 \\
    &=
    O(h)\,o(h^{-1})
    =
    o(1),
\end{align*}
which proves \eqref{eq:E.alpha_coefficients_uniform_bound}. Thus, for all sufficiently small $h$, \eqref{eq:E.weight_alpha_decomposition} implies that $\mathcal A_{h,s}^{(\ell)}$ is contained in the larger class
\begin{align*}
    h^{-d}
    \left\{
    \sum_{r\in\mathcal J_s}\alpha_rg_r:
    (\alpha_r)_{r\in\mathcal J_s}\in[-B_{\mathcal A,s},B_{\mathcal A,s}]^{m_s},
    \ g_r\in\mathcal G_{h,r}^{(\ell)},\ r\in\mathcal J_s
    \right\}.
\end{align*}
This enlargement removes the common-center restriction among the terms in \eqref{eq:E.weight_alpha_decomposition}, and therefore gives an upper bound for the covering number of $\mathcal A_{h,s}^{(\ell)}$.

Let
\begin{align*}
    C_{\mathcal G}
    :=
    c_{\theta,\mathcal K,\rho}^{-1}\|K\|_\infty,
\end{align*}
where $c_{\theta,\mathcal K,\rho}>0$ is the uniform volume-density lower bound in \Cref{lemma:A.uniform_normal_neighborhoods}. Since $B_h^{(\ell)}\subset\mathcal K^\rho$ for all sufficiently small $h$,
\begin{align*}
    \left|\mathcal L_{\mx,h}(\mz)\right|
    \leq
    C_{\mathcal G},
    \quad
    \mz\in B_h^{(\ell)},
    \quad
    \mx\in\mathcal K\cap B_{\mathcal M}(\mx_\ell,h).
\end{align*}
Moreover, $\left|v_{\mx,r}^{\alpha_\ell}(\mz)/h\right|\leq1$ on the kernel support. Therefore each class $\mathcal G_{h,r}^{(\ell)}$, $r\in\mathcal J_s$, is supported on $B_h^{(\ell)}$ and admits the common envelope $C_{\mathcal G}\mathbf 1_{B_h^{(\ell)}}$.

Let
\begin{align*}
    M_h^{(\ell)}
    :=
    \left\|\mathbf 1_{B_h^{(\ell)}}\right\|_{L_2(Q_{\mathcal M})}.
\end{align*}
If $M_h^{(\ell)}=0$, then all functions in $\mathcal G_{h,r}^{(\ell)}$ vanish $Q_{\mathcal M}$-almost surely, and the desired covering bound is trivial. Suppose $M_h^{(\ell)}>0$, and define the conditional probability measure
\begin{align*}
    \tilde Q_h^{(\ell)}(A)
    :=
    \frac{
    Q_{\mathcal M}\left(A\cap B_h^{(\ell)}\right)
    }{
    Q_{\mathcal M}\left(B_h^{(\ell)}\right)
    },
    \quad
    A\subset\mathcal M.
\end{align*}
Since $Q$ is finitely discrete on $\mathcal M\times\mathbb M$, the measure $\tilde Q_h^{(\ell)}$ is a finitely discrete probability measure on $\mathcal M$.

For the chart $\mathcal O^{\alpha_\ell}$, define the larger local-design classes
\begin{align*}
    \overline{\mathcal G}_{h,0}^{\alpha_\ell}
    &:=
    \left\{
    \mz\mapsto
    \mathcal L_{\mx,h}(\mz):
    \mx\in\mathcal K\cap\mathcal O^{\alpha_\ell}
    \right\}, \\
    \overline{\mathcal G}_{h,r}^{\alpha_\ell}
    &:=
    \left\{
    \mz\mapsto
    \mathcal L_{\mx,h}(\mz)
    \frac{v_{\mx,r}^{\alpha_\ell}(\mz)}{h}:
    \mx\in\mathcal K\cap\mathcal O^{\alpha_\ell}
    \right\},
    \quad r=1,\ldots,d.
\end{align*}
Since $B_h^{(\ell)}\subset\mathcal O^{\alpha_\ell}$, we have
\begin{align*}
    \mathcal G_{h,r}^{(\ell)}
    \subset
    \overline{\mathcal G}_{h,r}^{\alpha_\ell},
    \quad r\in\mathcal J_s.
\end{align*}
By Condition~\ref{con:U-K1} for $r=0$, and by Condition~\ref{con:U-K2} for $r=1,\ldots,d$ when $s=1$, there exist constants $A_{\mathcal G}<\infty$ and $v_{\mathcal G}<\infty$, independent of $\ell$, $h$, and $Q$, such that, for every $r\in\mathcal J_s$,
\begin{align}
    N\left(
    \epsilon C_{\mathcal G},
    \mathcal G_{h,r}^{(\ell)},
    L_2(\tilde Q_h^{(\ell)})
    \right)
    \leq
    \left(\frac{A_{\mathcal G}}{\epsilon}\right)^{v_{\mathcal G}},
    \quad
    \epsilon\in(0,1).
    \label{eq:E.normalized_design_entropy_conditional}
\end{align}
The constants are taken as maxima over the fixed finite frame cover and the finitely many relevant coordinate indices.

For any $g,g'\in\mathcal G_{h,r}^{(\ell)}$, since both functions are supported on $B_h^{(\ell)}$,
\begin{align*}
    \left\|g-g'\right\|_{L_2(Q_{\mathcal M})}
    =
    M_h^{(\ell)}
    \left\|g-g'\right\|_{L_2(\tilde Q_h^{(\ell)})}.
\end{align*}
Therefore \eqref{eq:E.normalized_design_entropy_conditional} implies
\begin{align}
    \sup_Q
    N\left(
    \epsilon C_{\mathcal G}M_h^{(\ell)},
    \mathcal G_{h,r}^{(\ell)},
    L_2(Q_{\mathcal M})
    \right)
    \leq
    \left(\frac{A_{\mathcal G}}{\epsilon}\right)^{v_{\mathcal G}},
    \quad
    \epsilon\in(0,1),
    \quad
    r\in\mathcal J_s.
    \label{eq:E.normalized_design_local_entropy}
\end{align}

We now pass from the normalized classes $\mathcal G_{h,r}^{(\ell)}$ to the oracle-weight class $\mathcal A_{h,s}^{(\ell)}$. If $M_h^{(\ell)}=0$, then every function in $\mathcal A_{h,s}^{(\ell)}$ vanishes $Q_{\mathcal M}$-almost surely, and the desired entropy bound is trivial. Suppose $M_h^{(\ell)}>0$. Fix $\epsilon\in(0,1)$ and set
\begin{align*}
    \eta_{\epsilon,s}
    &:=
    \min\left\{
    \frac{\epsilon C_{W,s}}
    {4m_sB_{\mathcal A,s}C_{\mathcal G}},
    \frac{1}{2}
    \right\}, \\
    \Delta_{\epsilon,s}
    &:=
    \frac{\epsilon C_{W,s}}
    {4m_sC_{\mathcal G}}.
\end{align*}
For each $r\in\mathcal J_s$, choose representatives from $\mathcal G_{h,r}^{(\ell)}$ forming an $L_2(Q_{\mathcal M})$-cover at radius $\eta_{\epsilon,s}C_{\mathcal G}M_h^{(\ell)}$. By \eqref{eq:E.normalized_design_local_entropy}, after increasing $A_{\mathcal G}$ by a universal factor if necessary to allow representatives from the class, this can be done with cardinality at most
\begin{align*}
    \left(\frac{A_{\mathcal G}}{\eta_{\epsilon,s}}\right)^{v_{\mathcal G}}.
\end{align*}
Also choose a $\Delta_{\epsilon,s}$-net of the coefficient cube $[-B_{\mathcal A,s},B_{\mathcal A,s}]^{m_s}$ in the sup-norm. Its cardinality is bounded by
\begin{align*}
    \left(
    1+
    \frac{2B_{\mathcal A,s}}{\Delta_{\epsilon,s}}
    \right)^{m_s}.
\end{align*}

Let
\begin{align*}
    f
    =
    h^{-d}
    \sum_{r\in\mathcal J_s}
    \alpha_rg_r
\end{align*}
be an arbitrary element of the enlarged class. Choose coefficients $\alpha_r^\circ$ from the coefficient net and functions $g_r^\circ$ from the corresponding function covers so that
\begin{align*}
    \max_{r\in\mathcal J_s}
    \left|\alpha_r-\alpha_r^\circ\right|
    \leq
    \Delta_{\epsilon,s},
    \quad
    \left\|g_r-g_r^\circ\right\|_{L_2(Q_{\mathcal M})}
    \leq
    \eta_{\epsilon,s}C_{\mathcal G}M_h^{(\ell)}.
\end{align*}
Since $|\alpha_r|\leq B_{\mathcal A,s}$ and $\|g_r^\circ\|_{L_2(Q_{\mathcal M})}\leq C_{\mathcal G}M_h^{(\ell)}$,
\begin{align*}
\begin{split}
    &\left\|
    h^{-d}\sum_{r\in\mathcal J_s}\alpha_rg_r
    -
    h^{-d}\sum_{r\in\mathcal J_s}\alpha_r^\circ g_r^\circ
    \right\|_{L_2(Q_{\mathcal M})} \\
    &\leq
    h^{-d}m_sB_{\mathcal A,s}\eta_{\epsilon,s}C_{\mathcal G}M_h^{(\ell)}
    +
    h^{-d}m_s\Delta_{\epsilon,s}C_{\mathcal G}M_h^{(\ell)} \\
    &\leq
    \frac{\epsilon}{2}
    C_{W,s}h^{-d}M_h^{(\ell)}
    =
    \frac{\epsilon}{2}
    \left\|A_{h,s}^{(\ell)}\right\|_{L_2(Q_{\mathcal M})}.
\end{split}
\end{align*}
Thus the enlarged class, and hence also $\mathcal A_{h,s}^{(\ell)}$, is covered at radius $\epsilon\|A_{h,s}^{(\ell)}\|_{L_2(Q_{\mathcal M})}$.

Since
\begin{align*}
    \eta_{\epsilon,s}
    \geq
    c_{\eta,s}\epsilon,
    \quad
    c_{\eta,s}
    :=
    \min\left\{
    \frac{C_{W,s}}
    {4m_sB_{\mathcal A,s}C_{\mathcal G}},
    \frac{1}{2}
    \right\},
\end{align*}
we have
\begin{align*}
    \left(\frac{A_{\mathcal G}}{\eta_{\epsilon,s}}\right)^{v_{\mathcal G}}
    \leq
    \left(\frac{A_{\mathcal G}}{c_{\eta,s}\epsilon}\right)^{v_{\mathcal G}}.
\end{align*}
Moreover,
\begin{align*}
    1+\frac{2B_{\mathcal A,s}}{\Delta_{\epsilon,s}}
    =
    1+
    \frac{8m_sB_{\mathcal A,s}C_{\mathcal G}}
    {\epsilon C_{W,s}}
    \leq
    \frac{
    1+8m_sB_{\mathcal A,s}C_{\mathcal G}/C_{W,s}
    }{\epsilon}.
\end{align*}
Therefore, setting
\begin{align*}
    v_{\mathcal A,s}
    &:=
    m_s(v_{\mathcal G}+1), \\
    A_{\mathcal A,s}
    &:=
    \max\left\{
    1,
    \frac{A_{\mathcal G}}{c_{\eta,s}},
    1+
    \frac{8m_sB_{\mathcal A,s}C_{\mathcal G}}{C_{W,s}}
    \right\},
\end{align*}
we obtain
\begin{align}
    \sup_Q
    N\left(
    \epsilon\left\|A_{h,s}^{(\ell)}\right\|_{L_2(Q_{\mathcal M})},
    \mathcal A_{h,s}^{(\ell)},
    L_2(Q_{\mathcal M})
    \right)
    \leq
    \left(\frac{A_{\mathcal A,s}}{\epsilon}\right)^{v_{\mathcal A,s}},
    \quad
    \epsilon\in(0,1).
    \label{eq:E.predictor_uniform_entropy}
\end{align}
The constants $A_{\mathcal A,s}$ and $v_{\mathcal A,s}$ are independent of $\ell$, $h$, and $Q$.

Define the response-side loss-increment class
\begin{align*}
    \mathcal H_\delta(\tilde y_h)
    :=
    \left\{
    \omega\mapsto
    d_{\mathbb M}^2(y,\omega)
    -
    d_{\mathbb M}^2\left(\tilde y_h(\mx),\omega\right):
    \mx\in\mathcal K,\ 
    y\in B_{\mathbb M}\left(\tilde y_h(\mx),\delta\right)
    \right\}.
\end{align*}
Let $G_{\mathcal H}:=2D_{\mathbb M}\delta$. The preceding loss-increment bound implies that $G_{\mathcal H}$ is a deterministic envelope of $\mathcal H_\delta(\tilde y_h)$. By the $L_\infty$ covering construction in \Cref{lemma:E.moving_anchor_response_entropy},
\begin{align}
    N_{\infty}
    \left(
    \epsilon D_{\mathbb M}\delta,
    \mathcal H_\delta(\tilde y_h)
    \right)
    \leq
    C_{\mathbb M,\mathcal K,\mathrm{poly}}^2
    \left(\frac{4}{\delta\epsilon}\right)^{2q_{\mathbb M,\mathcal K}},
    \quad
    \epsilon\in(0,1).
    \label{eq:E.response_uniform_entropy_in_localized_proof}
\end{align}

Every $U\in\mathcal F_{h,s,\delta}^{(\ell)}$ is contained in the product enlargement
\begin{align*}
    U(\mz,\omega)=a(\mz)b(\omega),
    \quad
    a\in\mathcal A_{h,s}^{(\ell)},
    \quad
    b\in\mathcal H_\delta(\tilde y_h).
\end{align*}
Moreover,
\begin{align*}
    F_{h,\ell,\delta}
    =
    A_{h,s}^{(\ell)}G_{\mathcal H},
    \quad
    \left\|F_{h,\ell,\delta}\right\|_{L_2(Q)}
    =
    G_{\mathcal H}
    \left\|A_{h,s}^{(\ell)}\right\|_{L_2(Q_{\mathcal M})}.
\end{align*}
Fix $u\in(0,1)$. If $\|F_{h,\ell,\delta}\|_{L_2(Q)}=0$, then all functions in $\mathcal F_{h,s,\delta}^{(\ell)}$ vanish $Q$-almost surely and the covering bound below is trivial. Assume $\|F_{h,\ell,\delta}\|_{L_2(Q)}>0$. Choose representatives $a_1,\ldots,a_{N_{\mathcal A,u}^{(\ell)}(Q)}$ from $\mathcal A_{h,s}^{(\ell)}$ such that
\begin{align*}
    \max_{a\in\mathcal A_{h,s}^{(\ell)}}
    \min_{1\leq j\leq N_{\mathcal A,u}^{(\ell)}(Q)}
    \left\|a-a_j\right\|_{L_2(Q_{\mathcal M})}
    \leq
    \frac{u}{4}
    \left\|A_{h,s}^{(\ell)}\right\|_{L_2(Q_{\mathcal M})}.
\end{align*}
By \eqref{eq:E.predictor_uniform_entropy}, this may be done with
\begin{align*}
    N_{\mathcal A,u}^{(\ell)}(Q)
    \leq
    \left(\frac{8A_{\mathcal A,s}}{u}\right)^{v_{\mathcal A,s}}.
\end{align*}
Also choose functions $b_1,\ldots,b_{N_{\mathcal H,u}}$ such that
\begin{align*}
    \max_{b\in\mathcal H_\delta(\tilde y_h)}
    \min_{1\leq k\leq N_{\mathcal H,u}}
    \left\|b-b_k\right\|_{\infty}
    \leq
    uD_{\mathbb M}\delta.
\end{align*}
By \eqref{eq:E.response_uniform_entropy_in_localized_proof}, this can be done with
\begin{align*}
    N_{\mathcal H,u}
    \leq
    C_{\mathbb M,\mathcal K,\mathrm{poly}}^2
    \left(\frac{4}{\delta u}\right)^{2q_{\mathbb M,\mathcal K}}.
\end{align*}
For any $a\in\mathcal A_{h,s}^{(\ell)}$ and $b\in\mathcal H_\delta(\tilde y_h)$, choose $a_j$ and $b_k$ as above. Since $|b|\leq G_{\mathcal H}$ and $|a_j|\leq A_{h,s}^{(\ell)}$ pointwise,
\begin{align*}
\begin{split}
    \left\|ab-a_jb_k\right\|_{L_2(Q)}
    &\leq
    \left\|(a-a_j)b\right\|_{L_2(Q)}
    +
    \left\|a_j(b-b_k)\right\|_{L_2(Q)} \\
    &\leq
    G_{\mathcal H}
    \left\|a-a_j\right\|_{L_2(Q_{\mathcal M})}
    +
    \left\|b-b_k\right\|_{\infty}
    \left\|a_j\right\|_{L_2(Q_{\mathcal M})} \\
    &\leq
    G_{\mathcal H}
    \frac{u}{4}
    \left\|A_{h,s}^{(\ell)}\right\|_{L_2(Q_{\mathcal M})}
    +
    uD_{\mathbb M}\delta
    \left\|A_{h,s}^{(\ell)}\right\|_{L_2(Q_{\mathcal M})} \\
    &=
    \frac{3u}{4}
    \left\|F_{h,\ell,\delta}\right\|_{L_2(Q)}
    \leq
    u
    \left\|F_{h,\ell,\delta}\right\|_{L_2(Q)}.
\end{split}
\end{align*}
Here the equality uses $G_{\mathcal H}=2D_{\mathbb M}\delta$. Therefore
\begin{align*}
    \sup_Q
    N\left(
    u\left\|F_{h,\ell,\delta}\right\|_{L_2(Q)},
    \mathcal F_{h,s,\delta}^{(\ell)},
    L_2(Q)
    \right)
    &\leq
    \left(\frac{8A_{\mathcal A,s}}{u}\right)^{v_{\mathcal A,s}}
    C_{\mathbb M,\mathcal K,\mathrm{poly}}^2
    \left(\frac{4}{\delta u}\right)^{2q_{\mathbb M,\mathcal K}}.
\end{align*}
Taking logarithms gives
\begin{align*}
    1+
    \log N\left(
    u\left\|F_{h,\ell,\delta}\right\|_{L_2(Q)},
    \mathcal F_{h,s,\delta}^{(\ell)},
    L_2(Q)
    \right)
    \leq
    C_{\mathrm{ent},s}'
    \left[
    1+\log(1/u)+\log(1/\delta)
    \right],
\end{align*}
where $C_{\mathrm{ent},s}'<\infty$ is independent of $\ell$, $h$, $\delta$, and $Q$. It follows that the per-cell uniform entropy integral satisfies
\begin{align*}
    J_{h,\delta,s}
    &:=
    \sup_{1\leq\ell\leq N_h}
    \sup_Q
    \int_0^1
    \sqrt{
    1+
    \log N\left(
    u\left\|F_{h,\ell,\delta}\right\|_{L_2(Q)},
    \mathcal F_{h,s,\delta}^{(\ell)},
    L_2(Q)
    \right)
    }
    \,\dd u \\
    &\leq
    C_{\mathrm{ent},s}\sqrt{1+\log(1/\delta)}
\end{align*}
for a constant $C_{\mathrm{ent},s}<\infty$ independent of $\ell$, $h$, $\delta$, and $Q$.

Apply \Cref{lemma:E.exponential_uniform_entropy_maximal} to each localized class $\mathcal F_{h,s,\delta}^{(\ell)}$ with
\begin{align*}
    \sigma_{h,\delta,s}
    :=
    C_{\mathrm{L2},s}\delta h^{-d/2},
    \quad
    b_{h,\delta,s}
    :=
    C_{\mathrm{env},s}\delta h^{-d},
    \quad
    J=J_{h,\delta,s}.
\end{align*}
Then there exist constants $C_{\mathrm{tail},s,1}<\infty$ and $C_{\mathrm{tail},s,2}<\infty$ such that, for every $t\geq1$ and every $\ell=1,\ldots,N_h$,
\begin{align*}
&\mathbb P
    \left(
    \sup_{U\in\mathcal F_{h,s,\delta}^{(\ell)}}
    \left|
    (\mathbb P_n-P)U
    \right|
    >
    C_{\mathrm{tail},s,1}\delta
    \left[
    \left(
    \frac{
    1+\log(1/\delta)+t
    }{nh^d}
    \right)^{1/2}
    +
    \frac{t}{nh^d}
    \right]
    \right) \\
&\leq
    C_{\mathrm{tail},s,2}e^{-t}.
\end{align*}
Using the union bound over $\ell=1,\ldots,N_h$ gives, after replacing $t$ by $t+\log N_h$,
\begin{align*}
&\mathbb P
    \left(
    \sup_{U\in\mathcal F_{h,s,\delta}}
    \left|
    (\mathbb P_n-P)U
    \right|
    >
    C_{\mathrm{tail},s,1}\delta
    \left[
    \left(
    \frac{
    1+\log(1/\delta)+\log N_h+t
    }{nh^d}
    \right)^{1/2}
    +
    \frac{\log N_h+t}{nh^d}
    \right]
    \right) \\
&\leq
    C_{\mathrm{tail},s,2}e^{-t}.
\end{align*}
Integrating the above tail bound over $t\geq1$ yields
\begin{align}
\begin{split}
&\E
    \left[
    \sup_{U\in\mathcal F_{h,s,\delta}}
    \left|
    (\mathbb P_n-P)U
    \right|
    \right] \\
&\leq
    C_{\mathrm{tail},s,3}\delta
    \left[
    \left(
    \frac{
    1+\log(1/\delta)+\log N_h
    }{nh^d}
    \right)^{1/2}
    +
    \frac{1+\log N_h}{nh^d}
    \right].
\end{split}
\label{eq:E.localized_expectation_from_tail}
\end{align}
Since $N_h\leq C_{\mathcal K,\mathrm{cov}}h^{-d}$ by \eqref{eq:E.spatial_h_cover},
\begin{align*}
    \log N_h
    \leq
    \log C_{\mathcal K,\mathrm{cov}}
    +
    d\log(1/h).
\end{align*}
Thus \eqref{eq:E.localized_expectation_from_tail} implies \eqref{eq:E.localized_finite_cover_entropy_refined}. Finally, if $\log(1/\delta)=O(\log n)$, then Condition~\ref{con:U-B1} implies $\log(1/h)=O(\log n)$ and $nh^d/\log n\to\infty$, so the second term in \eqref{eq:E.localized_finite_cover_entropy_refined} is absorbed into the first. This proves \eqref{eq:E.localized_finite_cover_entropy}. The outer-expectation version follows by applying the same argument to measurable majorants.
\end{proof}

\begin{lemma}[Uniform localized oracle empirical fluctuation bound] \label{lemma:E.uniform_oracle_fluctuation}
Assume Conditions~\ref{con:U-K1}, \ref{con:U-B1}, \ref{con:U-D1}, \ref{con:M1}, and~\ref{con:U-M4}. Fix $s\in\{0,1\}$. If $s=1$, assume in addition Condition~\ref{con:U-K2}. Let $\tilde y_h:\mathcal K\to\mathbb M$ be a deterministic sequence such that
\begin{align*}
    \sup_{\mx\in\mathcal K}
    d_{\mathbb M}\left(\tilde y_h(\mx),m_{\oplus}(\mx)\right)
    \leq
    \frac{r_{\mathbb{M},\mathcal K}}{2}
\end{align*}
for all sufficiently small $h$. Assume that the relevant suprema are measurable. Define the oracle-weight empirical fluctuation process
\begin{align*}
    \hat{S}_{h,s}(\mx,y)
    &:=
    \frac1n
    \sum_{i=1}^n
    \tilde{W}_{\mx,h,s}\left(\mX^{(i)}\right)
    d_{\mathbb M}^2\left(y,Y^{(i)}\right)
    -
    \tilde{M}_{h,s}(\mx,y),
    \quad
    \mx\in\mathcal K,\ y\in\mathbb M.
\end{align*}
Then there exist constants $\delta_{\mathrm{fluc}}>0$ and $C_{\mathrm{fluc},s}<\infty$, independent of $n$, $h$, $\delta$, and $\tilde y_h$, such that, for every $\delta\in(0,\delta_{\mathrm{fluc}}]$ and all sufficiently large $n$,
\begin{align}
\begin{split}
    &\E
    \left[
    \sup_{\mx\in\mathcal K}
    \sup_{y\in B_{\mathbb M}\left(\tilde y_h(\mx),\delta\right)}
    \left|
    \hat{S}_{h,s}(\mx,y)
    -
    \hat{S}_{h,s}\left(\mx,\tilde y_h(\mx)\right)
    \right|
    \right]  \\
    &\leq
    C_{\mathrm{fluc},s}
    \delta
    \left[
    \left(
    \frac{
    1+\log(1/h)+\log(1/\delta)
    }{nh^d}
    \right)^{1/2}
    +
    \frac{1+\log(1/h)}{nh^d}
    \right].
\end{split}
\label{eq:E.uniform_localized_empirical_fluc_refined}
\end{align}
Consequently, whenever $\log(1/\delta)=O(\log n)$,
\begin{align}
\begin{split}
    &\E
    \left[
    \sup_{\mx\in\mathcal K}
    \sup_{y\in B_{\mathbb M}\left(\tilde y_h(\mx),\delta\right)}
    \left|
    \hat{S}_{h,s}(\mx,y)
    -
    \hat{S}_{h,s}\left(\mx,\tilde y_h(\mx)\right)
    \right|
    \right]  \\
    &\leq
    C_{\mathrm{fluc},s}
    \delta
    \left(
    \frac{\log n}{nh^d}
    \right)^{1/2}.
\end{split}
\label{eq:E.uniform_localized_empirical_fluc}
\end{align}
If the measurability of the supremum is not imposed, the same bounds hold with outer expectation.
\end{lemma}

\begin{proof}[Proof of \Cref{lemma:E.uniform_oracle_fluctuation}]
Fix $s\in\{0,1\}$. If $s=1$, Condition~\ref{con:U-K2} is assumed in addition. Let $\delta_{\mathrm{loc}}>0$ be the constant in \Cref{lemma:E.localized_finite_cover_entropy}, and set
\begin{align*}
    \delta_{\mathrm{fluc}}
    :=
    \min\left\{
    \frac{r_{\mathbb M,\mathcal K}}{4},
    \delta_{\mathrm{loc}}
    \right\}.
\end{align*}
Then, for every $\delta\in(0,\delta_{\mathrm{fluc}}]$, every $\mx\in\mathcal K$, and every $y\in B_{\mathbb M}(\tilde y_h(\mx),\delta)$,
\begin{align*}
    d_{\mathbb M}\left(y,m_\oplus(\mx)\right)
    \leq
    d_{\mathbb M}\left(y,\tilde y_h(\mx)\right)
    +
    d_{\mathbb M}\left(\tilde y_h(\mx),m_\oplus(\mx)\right)
    \leq
    \delta+\frac{r_{\mathbb M,\mathcal K}}{2}
    <
    r_{\mathbb M,\mathcal K}.
\end{align*}
Hence the localization required for Condition~\ref{con:U-M4} and for \Cref{lemma:E.localized_finite_cover_entropy} is valid uniformly over $\mx\in\mathcal K$.

For $\mx\in\mathcal K$ and $y\in\mathbb M$, define
\begin{align*}
    \tilde{U}_{\mx,y,h,s}(\mz,\omega)
    &:=
    \tilde{W}_{\mx,h,s}(\mz)
    \left[
    d_{\mathbb M}^2(y,\omega)
    -
    d_{\mathbb M}^2\left(\tilde{y}_h(\mx),\omega\right)
    \right],
    \quad
    (\mz,\omega)\in\mathcal M\times\mathbb M.
\end{align*}
Then, for every $\mx\in\mathcal K$ and $y\in B_{\mathbb M}(\tilde{y}_h(\mx),\delta)$,
\begin{align*}
    \hat{S}_{h,s}(\mx,y)
    -
    \hat{S}_{h,s}\left(\mx,\tilde{y}_h(\mx)\right)
    =
    (\mathbb P_n-P)
    \tilde{U}_{\mx,y,h,s}.
\end{align*}
The class of functions $\tilde{U}_{\mx,y,h,s}$, with $\mx\in\mathcal K$ and $y\in B_{\mathbb M}(\tilde y_h(\mx),\delta)$, is precisely the class $\mathcal F_{h,s,\delta}$ in \Cref{lemma:E.localized_finite_cover_entropy}. Applying \Cref{lemma:E.localized_finite_cover_entropy} gives \eqref{eq:E.uniform_localized_empirical_fluc_refined}. If $\log(1/\delta)=O(\log n)$, the simplified bound \eqref{eq:E.uniform_localized_empirical_fluc} follows from the second conclusion of \Cref{lemma:E.localized_finite_cover_entropy}. The outer-expectation version follows from the corresponding outer-expectation statement in \Cref{lemma:E.localized_finite_cover_entropy}.
\end{proof}

\begin{lemma}[Uniform empirical-weight remainder] \label{lemma:E.uniform_weight_remainder}
Assume Conditions~\ref{con:U-K1}, \ref{con:U-B1}, \ref{con:U-D1}, and~\ref{con:M1}. Fix $s\in\{0,1\}$. If $s=1$, assume in addition Condition~\ref{con:U-K2}. Let $\tilde y_h:\mathcal K\to\mathbb M$ be a deterministic sequence, and define
\begin{align} \label{eq:E.uniform_balanced_rate}
    r_{n,h}
    &:=
    \left(
    \frac{\log n}{nh^d}
    \right)^{1/2}.
\end{align}
On the event where the empirical weights $\hat W_{\mx,h,s}$ are well-defined for all $\mx\in\mathcal K$, we have
\begin{align}
    \sup_{\mx\in\mathcal K}
    \frac1n
    \sum_{i=1}^n
    \left|
    \hat W_{\mx,h,s}\left(\mX^{(i)}\right)
    -
    \tilde W_{\mx,h,s}\left(\mX^{(i)}\right)
    \right|
    =
    O_{\P}(r_{n,h}).
    \label{eq:E.uniform_weight_remainder_average}
\end{align}
Consequently, for every $\delta>0$,
\begin{align}
\begin{split}
    &\sup_{\mx\in\mathcal K}
    \sup_{y\in B_{\mathbb M}\left(\tilde y_h(\mx),\delta\right)}
    \left|
    \frac{1}{n\delta}
    \sum_{i=1}^n
    \left[
    \hat W_{\mx,h,s}\left(\mX^{(i)}\right)
    -
    \tilde W_{\mx,h,s}\left(\mX^{(i)}\right)
    \right]
    \left[
    d_{\mathbb M}^2\left(y,Y^{(i)}\right)
    -
    d_{\mathbb M}^2\left(\tilde y_h(\mx),Y^{(i)}\right)
    \right]
    \right|  \\
    &=
    O_{\P}\left(r_{n,h}\right).
\end{split}
\label{eq:E.uniform_weight_remainder_localized_loss}
\end{align}
\end{lemma}

\begin{proof}[Proof of \Cref{lemma:E.uniform_weight_remainder}]
Fix $s\in\{0,1\}$. If $s=1$, Condition~\ref{con:U-K2} is assumed in addition. Throughout the proof, all empirical local linear quantities are understood on the event where the required inverses and denominators exist uniformly over $\mx\in\mathcal K$. By \Cref{lemma:D.uniform_population_moment_orders} and \Cref{lemma:D.uniform_empirical_moments}, this event has probability tending to one for $s=1$ under the additional Condition~\ref{con:U-K2}; for $s=0$, the required positivity of $\hat\mu_{h,0}$ follows from the zeroth-order part of \Cref{lemma:D.uniform_empirical_moments}. Therefore, restricting to this event does not affect stochastic orders in probability.

We first prove \eqref{eq:E.uniform_weight_remainder_average} for $s=0$. Since
\begin{align*}
    \hat W_{\mx,h,0}(\mz)
    =
    \frac{\mathcal L_{\mx,h}(\mz)}{\hat{\mu}_{h,0}(\mx)},
    \quad
    \tilde W_{\mx,h,0}(\mz)
    =
    \frac{\mathcal L_{\mx,h}(\mz)}{\tilde{\mu}_{h,0}(\mx)},
\end{align*}
we have
\begin{align*}
    &\sup_{\mx\in\mathcal K}
    \frac1n
    \sum_{i=1}^n
    \left|
    \hat W_{\mx,h,0}\left(\mX^{(i)}\right)
    -
    \tilde W_{\mx,h,0}\left(\mX^{(i)}\right)
    \right| \\
    &\leq
    \sup_{\mx\in\mathcal K}
    \left|
    \hat{\mu}_{h,0}(\mx)^{-1}
    -
    \tilde{\mu}_{h,0}(\mx)^{-1}
    \right|
    \sup_{\mx\in\mathcal K}
    \frac1n
    \sum_{i=1}^n
    \mathcal L_{\mx,h}\left(\mX^{(i)}\right).
\end{align*}
By \Cref{lemma:D.uniform_population_moment_orders} and the zeroth-order part of \Cref{lemma:D.uniform_empirical_moments},
\begin{align*}
    \sup_{\mx\in\mathcal K}
    \left|
    \hat{\mu}_{h,0}(\mx)-\tilde{\mu}_{h,0}(\mx)
    \right|
    =
    O_{\P}(h^d r_{n,h}),
    \quad
    \inf_{\mx\in\mathcal K}\tilde{\mu}_{h,0}(\mx)
    \geq
    C h^d
\end{align*}
for some constant $C>0$ and all sufficiently small $h$. Hence $\inf_{\mx\in\mathcal K}\hat{\mu}_{h,0}(\mx)\geq C h^d/2$ with probability tending to one, and therefore
\begin{align*}
    \sup_{\mx\in\mathcal K}
    \left|
    \hat{\mu}_{h,0}(\mx)^{-1}
    -
    \tilde{\mu}_{h,0}(\mx)^{-1}
    \right|
    =
    O_{\P}(h^{-d}r_{n,h}).
\end{align*}
Moreover,
\begin{align*}
    \sup_{\mx\in\mathcal K}
    \frac1n
    \sum_{i=1}^n
    \mathcal L_{\mx,h}\left(\mX^{(i)}\right)
    =
    \sup_{\mx\in\mathcal K}
    \hat{\mu}_{h,0}(\mx)
    =
    O_{\P}(h^d).
\end{align*}
Combining the preceding displays gives \eqref{eq:E.uniform_weight_remainder_average} for $s=0$.

We next prove \eqref{eq:E.uniform_weight_remainder_average} for $s=1$. Define
\begin{align*}
    \hat{\mathbf a}_h(\mx,\mathbf E_{\mx})
    &:=
    \bm{\hat{\mu}}_{h,2}(\mx,\mathbf E_{\mx})^{-1}
    \bm{\hat{\mu}}_{h,1}(\mx,\mathbf E_{\mx}), \\
    \tilde{\mathbf a}_h(\mx,\mathbf E_{\mx})
    &:=
    \bm{\tilde{\mu}}_{h,2}(\mx,\mathbf E_{\mx})^{-1}
    \bm{\tilde{\mu}}_{h,1}(\mx,\mathbf E_{\mx}).
\end{align*}
Then
\begin{align*}
    \hat W_{\mx,h,1}(\mz)
    &=
    \hat{\sigma}_h(\mx)^{-1}
    \mathcal L_{\mx,h}(\mz)
    \left[
    1-
    \hat{\mathbf a}_h(\mx,\mathbf E_{\mx})^{\top}
    \mathbf v_{\mx}^{\mathbf E_{\mx}}(\mz)
    \right], \\
    \tilde W_{\mx,h,1}(\mz)
    &=
    \tilde{\sigma}_h(\mx)^{-1}
    \mathcal L_{\mx,h}(\mz)
    \left[
    1-
    \tilde{\mathbf a}_h(\mx,\mathbf E_{\mx})^{\top}
    \mathbf v_{\mx}^{\mathbf E_{\mx}}(\mz)
    \right].
\end{align*}
By \Cref{lemma:D.uniform_population_moment_orders} and \Cref{lemma:D.uniform_empirical_moments},
\begin{align*}
    \sup_{\mx\in\mathcal K}
    \left|
    \hat{\sigma}_h(\mx)^{-1}
    -
    \tilde{\sigma}_h(\mx)^{-1}
    \right|
    =
    O_{\P}(h^{-d}r_{n,h}),
    \quad
    \sup_{\mx\in\mathcal K}
    \left|
    \tilde{\sigma}_h(\mx)^{-1}
    \right|
    =
    O(h^{-d}).
\end{align*}
Also, using
\begin{align*}
    \hat{\mathbf a}_h-\tilde{\mathbf a}_h
    =
    \bm{\hat{\mu}}_{h,2}^{-1}
    \left(
    \bm{\hat{\mu}}_{h,1}-\bm{\tilde{\mu}}_{h,1}
    \right)
    +
    \left(
    \bm{\hat{\mu}}_{h,2}^{-1}
    -
    \bm{\tilde{\mu}}_{h,2}^{-1}
    \right)
    \bm{\tilde{\mu}}_{h,1},
\end{align*}
with the arguments $(\mx,\mathbf E_{\mx})$ suppressed only in this display, together with \Cref{lemma:D.uniform_population_moment_orders} and \Cref{lemma:D.uniform_empirical_moments}, gives
\begin{align*}
    \sup_{\mx\in\mathcal K}
    \left\|
    \hat{\mathbf a}_h(\mx,\mathbf E_{\mx})
    -
    \tilde{\mathbf a}_h(\mx,\mathbf E_{\mx})
    \right\|_2
    =
    O_{\P}(h^{-1}r_{n,h}).
\end{align*}
Furthermore,
\begin{align*}
    \sup_{\mx\in\mathcal K}
    \left\|
    \hat{\mathbf a}_h(\mx,\mathbf E_{\mx})
    \right\|_2
    =
    O_{\P}(h^{-1}),
    \quad
    \sup_{\mx\in\mathcal K}
    \left\|
    \tilde{\mathbf a}_h(\mx,\mathbf E_{\mx})
    \right\|_2
    =
    O(h^{-1}).
\end{align*}
On the support of $\mathcal L_{\mx,h}$, we have $\|\mathbf v_{\mx}^{\mathbf E_{\mx}}(\mz)\|_2\leq h$. Therefore
\begin{align*}
    \sup_{\mx\in\mathcal K}
    \sup_{\mz\in\mathcal M}
    \left|
    1-
    \hat{\mathbf a}_h(\mx,\mathbf E_{\mx})^{\top}
    \mathbf v_{\mx}^{\mathbf E_{\mx}}(\mz)
    \right|
    \mathbf 1\left\{\mathcal L_{\mx,h}(\mz)\neq0\right\}
    =
    O_{\P}(1),
\end{align*}
and the corresponding population factor is uniformly $O(1)$. Hence
\begin{align*}
    &\sup_{\mx\in\mathcal K}
    \frac1n
    \sum_{i=1}^n
    \left|
    \hat W_{\mx,h,1}\left(\mX^{(i)}\right)
    -
    \tilde W_{\mx,h,1}\left(\mX^{(i)}\right)
    \right| \\
    &\leq
    O_{\P}(h^{-d}r_{n,h})
    \sup_{\mx\in\mathcal K}
    \frac1n
    \sum_{i=1}^n
    \mathcal L_{\mx,h}\left(\mX^{(i)}\right)
    +
    O(h^{-d})
    O_{\P}(h^{-1}r_{n,h})
    \sup_{\mx\in\mathcal K}
    \frac1n
    \sum_{i=1}^n
    \mathcal L_{\mx,h}\left(\mX^{(i)}\right)
    \left\|
    \mathbf v_{\mx}^{\mathbf E_{\mx}}\left(\mX^{(i)}\right)
    \right\|_2.
\end{align*}
Since
\begin{align*}
    \sup_{\mx\in\mathcal K}
    \frac1n
    \sum_{i=1}^n
    \mathcal L_{\mx,h}\left(\mX^{(i)}\right)
    =
    O_{\P}(h^d)
\end{align*}
and, on the support of $\mathcal L_{\mx,h}$,
\begin{align*}
    \left\|
    \mathbf v_{\mx}^{\mathbf E_{\mx}}\left(\mX^{(i)}\right)
    \right\|_2
    \leq h,
\end{align*}
we also have
\begin{align*}
    \sup_{\mx\in\mathcal K}
    \frac1n
    \sum_{i=1}^n
    \mathcal L_{\mx,h}\left(\mX^{(i)}\right)
    \left\|
    \mathbf v_{\mx}^{\mathbf E_{\mx}}\left(\mX^{(i)}\right)
    \right\|_2
    =
    O_{\P}(h^{d+1}).
\end{align*}
Combining the last three displays gives
\begin{align*}
    \sup_{\mx\in\mathcal K}
    \frac1n
    \sum_{i=1}^n
    \left|
    \hat W_{\mx,h,1}\left(\mX^{(i)}\right)
    -
    \tilde W_{\mx,h,1}\left(\mX^{(i)}\right)
    \right|
    =
    O_{\P}(r_{n,h}),
\end{align*}
which proves \eqref{eq:E.uniform_weight_remainder_average} for $s=1$.

It remains to prove \eqref{eq:E.uniform_weight_remainder_localized_loss}. By Condition~\ref{con:M1}, $D_{\mathbb{M}}$ defined by \eqref{eq:B.diameter_metric_space} is finite. Thus, for $y\in B_{\mathbb M}(\tilde y_h(\mx),\delta)$ and $\omega\in\mathbb M$,
\begin{align*}
    \left|
    d_{\mathbb M}^2(y,\omega)
    -
    d_{\mathbb M}^2\left(\tilde y_h(\mx),\omega\right)
    \right|
    \leq
    2D_{\mathbb{M}}\delta.
\end{align*}
Combining this deterministic bound with \eqref{eq:E.uniform_weight_remainder_average} gives
\begin{align*}
    &\sup_{\mx\in\mathcal K}
    \sup_{y\in B_{\mathbb M}\left(\tilde y_h(\mx),\delta\right)}
    \left|
    \frac{1}{n\delta}
    \sum_{i=1}^n
    \left[
    \hat W_{\mx,h,s}\left(\mX^{(i)}\right)
    -
    \tilde W_{\mx,h,s}\left(\mX^{(i)}\right)
    \right]
    \left[
    d_{\mathbb M}^2\left(y,Y^{(i)}\right)
    -
    d_{\mathbb M}^2\left(\tilde y_h(\mx),Y^{(i)}\right)
    \right]
    \right| \\
    &\leq
    2D_{\mathbb{M}}
    \sup_{\mx\in\mathcal K}
    \frac1n
    \sum_{i=1}^n
    \left|
    \hat W_{\mx,h,s}\left(\mX^{(i)}\right)
    -
    \tilde W_{\mx,h,s}\left(\mX^{(i)}\right)
    \right|
    =
    O_{\P}(r_{n,h}).
\end{align*}
This proves \eqref{eq:E.uniform_weight_remainder_localized_loss} and completes the proof.
\end{proof}

\begin{lemma}[Uniform stochastic minimizer rate around the oracle target] \label{lemma:E.uniform_stochastic_rate}
Assume Conditions~\ref{con:U-K1}, \ref{con:U-B1}, \ref{con:U-D1}, \ref{con:U-D2}, \ref{con:M1}, \ref{con:U-M2}, \ref{con:U-D3}, \ref{con:U-M3}, and~\ref{con:U-M4}. Fix $s\in\{0,1\}$. If $s=1$, assume in addition Condition~\ref{con:U-K2}. Assume that the relevant suprema are measurable. Then
\begin{align}
    \sup_{\mx\in\mathcal K}
    d_{\mathbb M}\left(
    \hat{m}_{h,s}(\mx),
    \tilde{m}_{h,s}(\mx)
    \right)^{\beta_{\oplus,\mathcal K}-1}
    =
    O_{\P}
    \left(
    \left(
    \frac{\log n}{nh^d}
    \right)^{1/2}
    \right).
    \label{eq:E.uniform_stochastic_minimizer_rate}
\end{align}
\end{lemma}

\begin{proof}[Proof of \Cref{lemma:E.uniform_stochastic_rate}]
Fix $s\in\{0,1\}$. If $s=1$, Condition~\ref{con:U-K2} is assumed in addition. Set
\begin{align*}
    p
    &:=
    \beta_{\oplus,\mathcal K}-1,
    \quad
    r_{n,h}
    :=
    \left(
    \frac{\log n}{nh^d}
    \right)^{1/2}.
\end{align*}
By Condition~\ref{con:U-M3}, $p>0$. By \Cref{lemma:D.uniform_oracle_minimizer}, for all sufficiently small $h$,
\begin{align*}
    \sup_{\mx\in\mathcal K}
    d_{\mathbb M}\left(\tilde{m}_{h,s}(\mx),m_{\oplus}(\mx)\right)
    \leq
    \frac{r_{\mathbb M,\mathcal K}}{2},
\end{align*}
where $r_{\mathbb M,\mathcal K}>0$ is the constant in Condition~\ref{con:U-M4}. Hence \Cref{lemma:E.uniform_oracle_fluctuation} may be applied with $\tilde y_h(\mx)=\tilde m_{h,s}(\mx)$. Let $\delta_{\mathrm{fluc}}>0$ be the radius in \Cref{lemma:E.uniform_oracle_fluctuation}. Choose
\begin{align*}
    \delta_0
    \in
    \left(
    0,
    \min\left\{
    \eta_{\oplus,\mathcal K},
    2^{-1/p}\delta_{\mathrm{fluc}}
    \right\}
    \right).
\end{align*}
Define
\begin{align*}
    \Gamma_n
    &:=
    \left\{
    \sup_{\mx\in\mathcal K}
    d_{\mathbb M}\left(
    \hat{m}_{h,s}(\mx),
    \tilde{m}_{h,s}(\mx)
    \right)
    <
    \delta_0
    \right\}.
\end{align*}
By \Cref{lemma:D.uniform_empirical_minimizer}, $\mathbb P(\Gamma_n)\to1$.

On $\Gamma_n$, Condition~\ref{con:U-M3} implies that, for every $\mx\in\mathcal K$,
\begin{align}
    C_{\oplus,\mathcal K}
    d_{\mathbb M}\left(
    \hat{m}_{h,s}(\mx),
    \tilde{m}_{h,s}(\mx)
    \right)^{p+1}
    \leq
    \tilde{M}_{h,s}\left(\mx,\hat{m}_{h,s}(\mx)\right)
    -
    \tilde{M}_{h,s}\left(\mx,\tilde{m}_{h,s}(\mx)\right).
    \label{eq:E.uniform_oracle_margin_for_hat}
\end{align}
Since $\hat{m}_{h,s}(\mx)$ minimizes $\hat{M}_{h,s}(\mx,\cdot)$,
\begin{align*}
    \hat{M}_{h,s}\left(\mx,\hat{m}_{h,s}(\mx)\right)
    -
    \hat{M}_{h,s}\left(\mx,\tilde{m}_{h,s}(\mx)\right)
    \leq
    0.
\end{align*}
Recalling that
\begin{align*}
    \hat{T}_{h,s}(\mx,y)
    &:=
    \hat{M}_{h,s}(\mx,y)-\tilde{M}_{h,s}(\mx,y),
\end{align*}
we obtain, for every $\mx\in\mathcal K$,
\begin{align}
    \tilde{M}_{h,s}\left(\mx,\hat{m}_{h,s}(\mx)\right)
    -
    \tilde{M}_{h,s}\left(\mx,\tilde{m}_{h,s}(\mx)\right)
    \leq
    \left|
    \hat{T}_{h,s}\left(\mx,\hat{m}_{h,s}(\mx)\right)
    -
    \hat{T}_{h,s}\left(\mx,\tilde{m}_{h,s}(\mx)\right)
    \right|.
    \label{eq:E.basic_uniform_stochastic_ineq}
\end{align}

By \Cref{lemma:E.uniform_weight_remainder}, for every $\varepsilon>0$ there exists $L<\infty$ such that, for all sufficiently large $n$,
\begin{align*}
    \mathbb P(\Omega_{n,L})
    \geq
    1-\varepsilon,
\end{align*}
where
\begin{align*}
    \Omega_{n,L}
    &:=
    \left\{
    \sup_{\mx\in\mathcal K}
    \sup_{y\in\mathbb M}
    \frac{
    \left|
    \left[
    \hat{T}_{h,s}(\mx,y)
    -
    \hat{T}_{h,s}\left(\mx,\tilde{m}_{h,s}(\mx)\right)
    \right]
    -
    \left[
    \hat{S}_{h,s}(\mx,y)
    -
    \hat{S}_{h,s}\left(\mx,\tilde{m}_{h,s}(\mx)\right)
    \right]
    \right|
    }{
    d_{\mathbb M}\left(y,\tilde{m}_{h,s}(\mx)\right)
    }
    \leq
    Lr_{n,h}
    \right\},
\end{align*}
with the convention that the ratio is zero when $y=\tilde{m}_{h,s}(\mx)$. Indeed, this event follows from \eqref{eq:E.uniform_weight_remainder_average} and the deterministic bound
\begin{align*}
    \left|
    d_{\mathbb M}^2(y,\omega)
    -
    d_{\mathbb M}^2\left(\tilde{m}_{h,s}(\mx),\omega\right)
    \right|
    \leq
    2D_{\mathbb M}
    d_{\mathbb M}\left(y,\tilde{m}_{h,s}(\mx)\right),
    \quad y,\omega\in\mathbb M.
\end{align*}

Choose $A>0$ so large that
\begin{align}
    Lr_{n,h}
    \leq
    \frac{C_{\oplus,\mathcal K}}{2}t^p
    \label{eq:E.uniform_A_large_condition}
\end{align}
whenever $t^p\geq Ar_{n,h}$. It is enough to take $A\geq 2L/C_{\oplus,\mathcal K}$.

For $k=0,1,2,\ldots$, define
\begin{align*}
    \mathcal A_{n,k}
    &:=
    \left\{
    2^k Ar_{n,h}
    <
    \sup_{\mx\in\mathcal K}
    d_{\mathbb M}\left(
    \hat{m}_{h,s}(\mx),
    \tilde{m}_{h,s}(\mx)
    \right)^p
    \leq
    2^{k+1} Ar_{n,h}
    \right\},
\end{align*}
and set
\begin{align*}
    \rho_{n,k}
    :=
    \left(
    2^{k+1}Ar_{n,h}
    \right)^{1/p}, \quad
    \ell_{n,k}
    :=
    \left(
    2^kAr_{n,h}
    \right)^{1/p}.
\end{align*}
On $\mathcal A_{n,k}\cap\Gamma_n$, we have $\rho_{n,k}\leq\delta_{\mathrm{fluc}}$. Indeed,
\begin{align*}
    \rho_{n,k}
    \leq
    2^{1/p}
    \sup_{\mx\in\mathcal K}
    d_{\mathbb M}\left(
    \hat{m}_{h,s}(\mx),
    \tilde{m}_{h,s}(\mx)
    \right)
    <
    2^{1/p}\delta_0
    \leq
    \delta_{\mathrm{fluc}}.
\end{align*}
Moreover, for such $k$, the simplified bound in \Cref{lemma:E.uniform_oracle_fluctuation} is applicable with $\delta=\rho_{n,k}$. To see this, note that $\rho_{n,k}\geq (Ar_{n,h})^{1/p}$, and since $h<1$ for all sufficiently small $h$,
\begin{align*}
    \log\left(1/r_{n,h}\right)
    =
    \frac12\log\left(\frac{nh^d}{\log n}\right)
    \leq
    \frac12\log n
\end{align*}
whenever $r_{n,h}<1$. Hence $\log(1/\rho_{n,k})=O(\log n)$ uniformly over the relevant peeling shells.

Fix $k\geq0$. On $\mathcal A_{n,k}$, there exists $\mx_{n,k}\in\mathcal K$ such that
\begin{align*}
    d_{\mathbb M}\left(
    \hat{m}_{h,s}(\mx_{n,k}),
    \tilde{m}_{h,s}(\mx_{n,k})
    \right)
    >
    \ell_{n,k}.
\end{align*}
The upper bound defining $\mathcal A_{n,k}$ also gives
\begin{align*}
    d_{\mathbb M}\left(
    \hat{m}_{h,s}(\mx_{n,k}),
    \tilde{m}_{h,s}(\mx_{n,k})
    \right)
    \leq
    \rho_{n,k}.
\end{align*}
Therefore, on $\mathcal A_{n,k}\cap\Gamma_n\cap\Omega_{n,L}$, applying \eqref{eq:E.uniform_oracle_margin_for_hat}, \eqref{eq:E.basic_uniform_stochastic_ineq}, and \eqref{eq:E.uniform_A_large_condition} at $\mx=\mx_{n,k}$ yields
\begin{align*}
    &
    \left|
    \hat{S}_{h,s}
    \left(
    \mx_{n,k},
    \hat{m}_{h,s}(\mx_{n,k})
    \right)
    -
    \hat{S}_{h,s}
    \left(
    \mx_{n,k},
    \tilde{m}_{h,s}(\mx_{n,k})
    \right)
    \right| \\
    &\geq
    C_{\oplus,\mathcal K}
    d_{\mathbb M}\left(
    \hat{m}_{h,s}(\mx_{n,k}),
    \tilde{m}_{h,s}(\mx_{n,k})
    \right)^{p+1}
    -
    Lr_{n,h}
    d_{\mathbb M}\left(
    \hat{m}_{h,s}(\mx_{n,k}),
    \tilde{m}_{h,s}(\mx_{n,k})
    \right) \\
    &\geq
    \frac{C_{\oplus,\mathcal K}}{2}
    d_{\mathbb M}\left(
    \hat{m}_{h,s}(\mx_{n,k}),
    \tilde{m}_{h,s}(\mx_{n,k})
    \right)^{p+1} \\
    &\geq
    \frac{C_{\oplus,\mathcal K}}{2}
    \ell_{n,k}^{p+1}.
\end{align*}
Since $\hat{m}_{h,s}(\mx_{n,k})\in B_{\mathbb M}(\tilde{m}_{h,s}(\mx_{n,k}),\rho_{n,k})$, this implies
\begin{align*}
    &
    \sup_{\mx\in\mathcal K}
    \sup_{y\in B_{\mathbb M}\left(\tilde{m}_{h,s}(\mx),\rho_{n,k}\right)}
    \left|
    \hat{S}_{h,s}(\mx,y)
    -
    \hat{S}_{h,s}\left(\mx,\tilde{m}_{h,s}(\mx)\right)
    \right| \\
    &\geq
    \frac{C_{\oplus,\mathcal K}}{2}
    \ell_{n,k}^{p+1}.
\end{align*}
By Markov's inequality and \Cref{lemma:E.uniform_oracle_fluctuation},
\begin{align*}
    \mathbb P\left(
    \mathcal A_{n,k}\cap\Gamma_n\cap\Omega_{n,L}
    \right)
    &\leq
    \frac{
    2
    \E\left[
    \sup_{\mx\in\mathcal K}
    \sup_{y\in B_{\mathbb M}\left(\tilde{m}_{h,s}(\mx),\rho_{n,k}\right)}
    \left|
    \hat{S}_{h,s}(\mx,y)
    -
    \hat{S}_{h,s}\left(\mx,\tilde{m}_{h,s}(\mx)\right)
    \right|
    \right]
    }{
    C_{\oplus,\mathcal K}\ell_{n,k}^{p+1}
    } \\
    &\leq
    C
    \frac{
    \rho_{n,k}r_{n,h}
    }{
    \ell_{n,k}^{p+1}
    } \\
    &\leq
    \frac{C}{A}2^{-k},
\end{align*}
where $C<\infty$ does not depend on $n$, $h$, $A$, or $k$.

Consequently,
\begin{align*}
    &\mathbb P
    \left[
    \sup_{\mx\in\mathcal K}
    d_{\mathbb M}\left(
    \hat{m}_{h,s}(\mx),
    \tilde{m}_{h,s}(\mx)
    \right)^p
    >
    Ar_{n,h}
    \right] \\
    &\leq
    \mathbb P(\Gamma_n^c)
    +
    \mathbb P(\Omega_{n,L}^c)
    +
    \sum_{k=0}^{\infty}
    \mathbb P\left(
    \mathcal A_{n,k}\cap\Gamma_n\cap\Omega_{n,L}
    \right) \\
    &\leq
    \mathbb P(\Gamma_n^c)
    +
    \mathbb P(\Omega_{n,L}^c)
    +
    \frac{C}{A}.
\end{align*}
Since $\mathbb P(\Gamma_n^c)\to0$, $L$ can be chosen so that $\mathbb P(\Omega_{n,L}^c)$ is arbitrarily small, and then $A$ can be chosen large enough, we conclude that
\begin{align*}
    \sup_{\mx\in\mathcal K}
    d_{\mathbb M}\left(
    \hat{m}_{h,s}(\mx),
    \tilde{m}_{h,s}(\mx)
    \right)^p
    =
    O_{\P}(r_{n,h}).
\end{align*}
Substituting $p=\beta_{\oplus,\mathcal K}-1$ and the definition of $r_{n,h}$ proves \eqref{eq:E.uniform_stochastic_minimizer_rate}.
\end{proof}

\begin{proof}[Proof of \Cref{thm:uniform_rate}]
We prove the local constant and local linear assertions simultaneously. Fix $s\in\{0,1\}$. If $s=1$, assume in addition Condition~\ref{con:U-K2}. Set
\begin{align*}
    p
    &:=
    \beta_{\oplus,\mathcal K}-1.
\end{align*}
By Condition~\ref{con:U-M3}, $p>0$. From \Cref{lemma:E.uniform_population_bias},
\begin{align*}
    \sup_{\mx\in\mathcal K}
    d_{\mathbb M}\left(
    \tilde{m}_{h,s}(\mx),
    m_{\oplus}(\mx)
    \right)^p
    =
    O(h^2).
\end{align*}
Since $t\mapsto t^{1/p}$ is increasing on $[0,\infty)$, this implies
\begin{align*}
    \sup_{\mx\in\mathcal K}
    d_{\mathbb M}\left(
    \tilde{m}_{h,s}(\mx),
    m_{\oplus}(\mx)
    \right)
    =
    O\left(h^{2/p}\right)
    =
    O\left(h^{2/(\beta_{\oplus,\mathcal K}-1)}\right).
\end{align*}
Similarly, by \Cref{lemma:E.uniform_stochastic_rate},
\begin{align*}
    \sup_{\mx\in\mathcal K}
    d_{\mathbb M}\left(
    \hat{m}_{h,s}(\mx),
    \tilde{m}_{h,s}(\mx)
    \right)^p
    =
    O_{\P}
    \left(
    \left(
    \frac{\log n}{nh^d}
    \right)^{1/2}
    \right),
\end{align*}
and hence
\begin{align*}
    \sup_{\mx\in\mathcal K}
    d_{\mathbb M}\left(
    \hat{m}_{h,s}(\mx),
    \tilde{m}_{h,s}(\mx)
    \right)
    =
    O_{\P}
    \left(
    \left(
    \frac{\log n}{nh^d}
    \right)^{1/(2p)}
    \right)
    =
    O_{\P}
    \left(
    \left(
    \frac{\log n}{nh^d}
    \right)^{1/(2\beta_{\oplus,\mathcal K}-2)}
    \right).
\end{align*}
The triangle inequality gives
\begin{align*}
    &
    \sup_{\mx\in\mathcal K}
    d_{\mathbb M}\left(
    \hat{m}_{h,s}(\mx),
    m_{\oplus}(\mx)
    \right) \\
    &\leq
    \sup_{\mx\in\mathcal K}
    d_{\mathbb M}\left(
    \hat{m}_{h,s}(\mx),
    \tilde{m}_{h,s}(\mx)
    \right)
    +
    \sup_{\mx\in\mathcal K}
    d_{\mathbb M}\left(
    \tilde{m}_{h,s}(\mx),
    m_{\oplus}(\mx)
    \right) \\
    &=
    O\left(h^{2/(\beta_{\oplus,\mathcal K}-1)}\right)
    +
    O_{\P}
    \left(
    \left(
    \frac{\log n}{nh^d}
    \right)^{1/(2\beta_{\oplus,\mathcal K}-2)}
    \right).
\end{align*}
This proves the asserted rate for $s=0$ under Condition~\ref{con:U-K1}, and for $s=1$ under Conditions~\ref{con:U-K1} and~\ref{con:U-K2}.
\end{proof}

\section{Verification of Auxiliary Conditions}
\label{app:condition_verification}
\setcounter{equation}{0}
\renewcommand{\theequation}{F.\arabic{equation}}
\renewcommand{\theHequation}{F.\arabic{equation}}

This appendix records sufficient conditions under which the uniform kernel-complexity assumptions in Conditions~\ref{con:U-K1} and~\ref{con:U-K2} hold. We first give an elementary verification for Euclidean predictor spaces. We then provide a manifold-level sufficient condition based on finitely definable local geometry on the compact normal tube used in the uniform theory. The latter condition applies, in particular, when the Riemannian metric and the relevant local frames are real analytic. It covers standard predictor manifolds including spheres, finite products of spheres, flat tori, and compact evaluation regions of the SPD cone under the affine-invariant Riemannian metric.

The results below are sufficient-condition results. The main uniform theory continues to treat Conditions~\ref{con:U-K1} and~\ref{con:U-K2} as high-level empirical-process assumptions and does not require the predictor manifold to be real analytic.

\subsection{Euclidean verification}

In this subsection, vectors in $\mathbb R^d$ are written in boldface. Let $\mathbf e_1,\ldots,\mathbf e_d$ denote the standard basis of $\mathbb R^d$. For $\mathbf x,\mathbf z\in\mathbb R^d$, we write
\begin{align*}
    x_j:=\mathbf e_j^\top\mathbf x,
    \quad
    z_j:=\mathbf e_j^\top\mathbf z,
    \quad j=1,\ldots,d.
\end{align*}
The Euclidean norm is denoted by $\|\cdot\|_2$.

\begin{lemma}[Euclidean normalized coordinate multiplier classes] \label{lemma:F.euclidean_multiplier_classes}
Let $\mathcal K\subset\mathbb R^d$ be compact and let $h_0\in(0,\infty)$. For $r,s\in\{1,\ldots,d\}$, define
\begin{align*}
    \mathcal G_{1,r}
    &:=
    \left\{
    \mathbf z\mapsto
    \frac{z_r-x_r}{h}
    \mathds 1
    \left(
    \|\mathbf z-\mathbf x\|_2\leq h
    \right):
    \mathbf x\in\mathcal K,\ 
    0<h<h_0
    \right\}, \\
    \mathcal G_{2,r,s}
    &:=
    \left\{
    \mathbf z\mapsto
    \frac{(z_r-x_r)(z_s-x_s)}{h^2}
    \mathds 1
    \left(
    \|\mathbf z-\mathbf x\|_2\leq h
    \right):
    \mathbf x\in\mathcal K,\ 
    0<h<h_0
    \right\}.
\end{align*}
Then $\mathcal G_{1,r}$ and $\mathcal G_{2,r,s}$ are uniformly bounded VC-subgraph classes. Consequently, they are of VC type, uniformly over $r,s\in\{1,\ldots,d\}$.
\end{lemma}

\begin{proof}[Proof of \Cref{lemma:F.euclidean_multiplier_classes}]
Fix $r,s\in\{1,\ldots,d\}$. We first verify the envelope bound. If $\|\mathbf z-\mathbf x\|_2\leq h$, then
\begin{align*}
    |z_r-x_r|
    \leq
    \|\mathbf z-\mathbf x\|_2
    \leq h,
    \qquad
    |z_s-x_s|
    \leq
    \|\mathbf z-\mathbf x\|_2
    \leq h.
\end{align*}
Therefore,
\begin{align*}
    \left|
    \frac{z_r-x_r}{h}
    \mathds 1\{\|\mathbf z-\mathbf x\|_2\leq h\}
    \right|
    \leq 1
\end{align*}
and
\begin{align*}
    \left|
    \frac{(z_r-x_r)(z_s-x_s)}{h^2}
    \mathds 1\{\|\mathbf z-\mathbf x\|_2\leq h\}
    \right|
    \leq 1.
\end{align*}
Thus both classes have envelope one.

It remains to verify the VC-subgraph property. Write $\lambda:=h^{-1}$. Since $0<h<h_0$, the scale parameter satisfies $\lambda>h_0^{-1}$. For fixed $(\mathbf x,\lambda)$, define
\begin{align*}
    B(\mathbf x,\lambda)
    &:=
    \left\{
    \mathbf z\in\mathbb R^d:
    \lambda^2\|\mathbf z-\mathbf x\|_2^2\leq1
    \right\}.
\end{align*}
This is exactly the Euclidean ball $\{\mathbf z:\|\mathbf z-\mathbf x\|_2\leq h\}$.

Consider first a function in $\mathcal G_{1,r}$, written in the form
\begin{align*}
    g_{\mathbf x,\lambda}^{(1,r)}(\mathbf z)
    &:=
    \lambda(z_r-x_r)
    \mathds 1\{B(\mathbf x,\lambda)\}.
\end{align*}
Its subgraph is
\begin{align*}
    \operatorname{subgraph}
    \left(g_{\mathbf x,\lambda}^{(1,r)}\right)
    &:=
    \left\{
    (\mathbf z,t)\in\mathbb R^d\times\mathbb R:
    t<g_{\mathbf x,\lambda}^{(1,r)}(\mathbf z)
    \right\}.
\end{align*}
Because $g_{\mathbf x,\lambda}^{(1,r)}(\mathbf z)=\lambda(z_r-x_r)$ on $B(\mathbf x,\lambda)$ and $g_{\mathbf x,\lambda}^{(1,r)}(\mathbf z)=0$ outside $B(\mathbf x,\lambda)$, this subgraph can be decomposed as
\begin{align*}
    \operatorname{subgraph}
    \left(g_{\mathbf x,\lambda}^{(1,r)}\right)
    &=
    \left\{
    (\mathbf z,t):
    \lambda^2\|\mathbf z-\mathbf x\|_2^2\leq1,\ 
    t<\lambda(z_r-x_r)
    \right\} \\
    &\quad\cup
    \left\{
    (\mathbf z,t):
    \lambda^2\|\mathbf z-\mathbf x\|_2^2>1,\ 
    t<0
    \right\}.
\end{align*}
Each set in this union is described by finitely many polynomial inequalities in the variables $(\mathbf z,t)$ and the parameters $(\mathbf x,\lambda)$. For instance,
\begin{align*}
    \lambda^2\|\mathbf z-\mathbf x\|_2^2\leq1
\end{align*}
is equivalent to
\begin{align*}
    \lambda^2\sum_{j=1}^d(z_j-x_j)^2-1\leq0,
\end{align*}
and
\begin{align*}
    t<\lambda(z_r-x_r)
\end{align*}
is equivalent to
\begin{align*}
    t-\lambda(z_r-x_r)<0.
\end{align*}
Hence the subgraphs of the functions in $\mathcal G_{1,r}$ form a semialgebraic family whose number of defining polynomial inequalities and polynomial degrees are bounded only in terms of $d$. In particular, these bounds do not depend on $\mathbf x\in\mathcal K$, $h\in(0,h_0)$, or the coordinate index $r$.

The same argument applies to $\mathcal G_{2,r,s}$. For fixed $(\mathbf x,\lambda)$, write
\begin{align*}
    g_{\mathbf x,\lambda}^{(2,r,s)}(\mathbf z)
    &:=
    \lambda^2(z_r-x_r)(z_s-x_s)
    \mathds 1\{B(\mathbf x,\lambda)\}.
\end{align*}
Its subgraph is
\begin{align*}
    \operatorname{subgraph}
    \left(g_{\mathbf x,\lambda}^{(2,r,s)}\right)
    &=
    \left\{
    (\mathbf z,t):
    \lambda^2\|\mathbf z-\mathbf x\|_2^2\leq1,\ 
    t<\lambda^2(z_r-x_r)(z_s-x_s)
    \right\} \\
    &\quad\cup
    \left\{
    (\mathbf z,t):
    \lambda^2\|\mathbf z-\mathbf x\|_2^2>1,\ 
    t<0
    \right\}.
\end{align*}
For example, the first set in the last display is the intersection of the two polynomial inequalities
\begin{align*}
    \lambda^2\sum_{j=1}^d(z_j-x_j)^2-1\leq0
\end{align*}
and
\begin{align*}
    t-\lambda^2(z_r-x_r)(z_s-x_s)<0.
\end{align*}
The second set is the intersection of
\begin{align*}
    \lambda^2\sum_{j=1}^d(z_j-x_j)^2-1>0
\end{align*}
and
\begin{align*}
    t<0.
\end{align*}
Thus the subgraph is a finite union of finite intersections of polynomial inequalities in $(\mathbf z,t,\mathbf x,\lambda)$. The number of inequalities is fixed, and their degrees are bounded by a constant depending only on $d$. These bounds do not depend on the particular values of $\mathbf x$, $h$, $r$, or $s$.

The preceding displays show that the subgraphs of the functions in $\mathcal G_{1,r}$ and $\mathcal G_{2,r,s}$ all belong to a single parametric family of sets with the following form: each set is obtained from a fixed finite number of polynomial inequalities in $(\mathbf z,t,\mathbf x,\lambda)$ by taking finitely many intersections and unions. The number of polynomial inequalities is fixed, and the maximum polynomial degree is bounded by a constant depending only on $d$. Thus, although the parameters $(\mathbf x,\lambda)$ vary with $\mathbf x\in\mathcal K$ and $0<h<h_0$, the algebraic complexity of the sets does not increase. The restriction $\mathbf x\in\mathcal K$ only selects a subclass of the same semialgebraic family indexed by $\mathbf x\in\mathbb R^d$ and $\lambda>h_0^{-1}$. Therefore, no semialgebraic structure is required for the compact set $\mathcal K$ itself.

We now apply the standard VC theorem for semialgebraic classes \cite{van der Vaart and Wellner (1996)}. This theorem states that a class of subsets of a Euclidean space described by a fixed finite Boolean combination of polynomial inequalities, with a uniformly bounded number of inequalities and uniformly bounded polynomial degrees, has finite VC dimension. Intuitively, such a class cannot shatter arbitrarily large finite point sets because all possible membership patterns are generated by polynomial signs of uniformly bounded algebraic complexity. Therefore, the subgraph classes associated with $\mathcal G_{1,r}$ and $\mathcal G_{2,r,s}$ are VC classes. Equivalently, $\mathcal G_{1,r}$ and $\mathcal G_{2,r,s}$ are VC-subgraph classes.

Finally, each function in $\mathcal G_{1,r}$ and $\mathcal G_{2,r,s}$ is bounded in absolute value by one, as shown at the beginning of the proof. A uniformly bounded VC-subgraph class is of VC type. Since the coordinate indices $r$ and $(r,s)$ range over only finitely many possibilities, the envelope and VC-type constants can be chosen uniformly over all $r,s\in\{1,\ldots,d\}$.
\end{proof}

\begin{lemma}[Euclidean implication from U-K1 to U-K2]
\label{lem:app_UK1_implies_UK2_euclidean}
Suppose $\mathcal M=\mathbb R^d$ with its Euclidean metric, let $\mathcal K\subset\mathbb R^d$ be compact, and let $h_0\in(0,\infty)$. Assume that Condition~\ref{con:U-K1} holds for the usual Euclidean radial local-design class generated by $K$ over $\mathbf x\in\mathcal K$ and $0<h<h_0$. Then Condition~\ref{con:U-K2} holds for the corresponding Euclidean multiplier-augmented local-design classes over the same range of $(\mathbf x,h)$.
\end{lemma}

\begin{proof}[Proof of \Cref{lem:app_UK1_implies_UK2_euclidean}]
In the Euclidean case, we use the single global canonical ordered orthonormal frame. Then
\begin{align*}
    \theta_{\mathbf x}(\mathbf z)
    &\equiv1,
    &
    i(\mathbf x)
    &=\infty,
    &
    \Log_{\mathbf x}(\mathbf z)
    &=\mathbf z-\mathbf x.
\end{align*}
Hence the zeroth-order local-design class in Condition~\ref{con:U-K1} reduces to
\begin{align*}
    \mathcal F_0
    &:=
    \left\{
    \mathbf z\mapsto
    K\left(
    \frac{\|\mathbf z-\mathbf x\|_2}{h}
    \right):
    \mathbf x\in\mathcal K,\ 
    0<h<h_0
    \right\}.
\end{align*}
By Condition~\ref{con:U-K1}, $\mathcal F_0$ is of VC type with bounded envelope $K_\infty:=\|K\|_\infty$.

For $r,s\in\{1,\ldots,d\}$, the first- and second-order Euclidean multiplier-augmented classes appearing in Condition~\ref{con:U-K2} are
\begin{align*}
    \mathcal F_{1,r}
    &:=
    \left\{
    \mathbf z\mapsto
    K\left(
    \frac{\|\mathbf z-\mathbf x\|_2}{h}
    \right)
    \frac{z_r-x_r}{h}:
    \mathbf x\in\mathcal K,\ 
    0<h<h_0
    \right\}, \\
    \mathcal F_{2,r,s}
    &:=
    \left\{
    \mathbf z\mapsto
    K\left(
    \frac{\|\mathbf z-\mathbf x\|_2}{h}
    \right)
    \frac{(z_r-x_r)(z_s-x_s)}{h^2}:
    \mathbf x\in\mathcal K,\ 
    0<h<h_0
    \right\}.
\end{align*}
Because $K$ is supported on $[0,1]$, the factor
$K(\|\mathbf z-\mathbf x\|_2/h)$ vanishes whenever
$\|\mathbf z-\mathbf x\|_2>h$. Therefore inserting the indicator
$\mathds 1\{\|\mathbf z-\mathbf x\|_2\leq h\}$ does not change the functions in $\mathcal F_{1,r}$ or $\mathcal F_{2,r,s}$. Thus $\mathcal F_{1,r}$ is a subclass of the product class
\begin{align*}
    \mathcal F_0\mathcal G_{1,r}
    &:=
    \left\{
    fg:
    f\in\mathcal F_0,\ 
    g\in\mathcal G_{1,r}
    \right\},
\end{align*}
and $\mathcal F_{2,r,s}$ is a subclass of
\begin{align*}
    \mathcal F_0\mathcal G_{2,r,s}
    &:=
    \left\{
    fg:
    f\in\mathcal F_0,\ 
    g\in\mathcal G_{2,r,s}
    \right\}.
\end{align*}
By \Cref{lemma:F.euclidean_multiplier_classes}, $\mathcal G_{1,r}$ and $\mathcal G_{2,r,s}$ are uniformly bounded VC-subgraph classes with envelope one. In particular, they are VC-type classes uniformly over $r$ and $(r,s)$.

It remains only to justify that the product classes above are of VC type. Let $\mathcal G$ denote either $\mathcal G_{1,r}$ or $\mathcal G_{2,r,s}$. Fix an arbitrary probability measure $Q$ on $\mathbb R^d$. Since $\mathcal F_0$ is of VC type with envelope $K_\infty$, for any $\epsilon>0$ it admits an $L_2(Q)$-net with radius $\epsilon K_\infty/4$ and polynomial cardinality in $\epsilon^{-1}$. Since $\mathcal G$ is of VC type with envelope one, it admits an $L_2(Q)$-net with radius $\epsilon/4$ and polynomial cardinality in $\epsilon^{-1}$. If the covering centers are not elements of the original class, replace each nonempty covering ball by one representative element of the class contained in that ball. This increases the covering radius by at most a factor of two and does not change the polynomial order of the covering number. Hence we may choose $f_m\in\mathcal F_0$ and $g_l\in\mathcal G$ such that
\begin{align*}
    \|f-f_m\|_{L_2(Q)}
    &\leq
    \frac{\epsilon K_\infty}{2},
    &
    \|g-g_l\|_{L_2(Q)}
    &\leq
    \frac{\epsilon}{2}.
\end{align*}
Since $|g|\leq1$ and $|f_m|\leq K_\infty$, we have
\begin{align*}
    \left\|
    fg-f_mg_l
    \right\|_{L_2(Q)}
    &\leq
    \left\|
    (f-f_m)g
    \right\|_{L_2(Q)}
    +
    \left\|
    f_m(g-g_l)
    \right\|_{L_2(Q)} \\
    &\leq
    \left\|
    f-f_m
    \right\|_{L_2(Q)}
    +
    K_\infty
    \left\|
    g-g_l
    \right\|_{L_2(Q)} \\
    &\leq
    \epsilon K_\infty.
\end{align*}
Thus the products of the net elements form an $L_2(Q)$-net for
$\mathcal F_0\mathcal G$ with envelope $K_\infty$ and with polynomial covering cardinality. Hence $\mathcal F_0\mathcal G$ is of VC type.

Applying this argument with $\mathcal G=\mathcal G_{1,r}$ and
$\mathcal G=\mathcal G_{2,r,s}$ shows that
$\mathcal F_0\mathcal G_{1,r}$ and $\mathcal F_0\mathcal G_{2,r,s}$ are VC-type classes. Their subclasses $\mathcal F_{1,r}$ and $\mathcal F_{2,r,s}$ are therefore also of VC type. Since there are only finitely many coordinate indices $r$ and $(r,s)$, the covering-number constants can be chosen uniformly over all $r,s\in\{1,\ldots,d\}$. This proves Condition~\ref{con:U-K2}.
\end{proof}

\begin{lemma}[Standard Euclidean kernels satisfying Conditions~\ref{con:U-K1} and~\ref{con:U-K2}]
\label{lem:app_standard_kernels_UK}
Suppose $\mathcal M=\mathbb R^d$ with its Euclidean metric, let $\mathcal K\subset\mathbb R^d$ be compact, and let $h_0\in(0,\infty)$. Suppose that $K$ satisfies Condition~\ref{con:P-K1}. If either the Euclidean radial translate-dilate class
\begin{align*}
    \left\{
    \mathbf z\mapsto
    K\left(
    \frac{\|\mathbf z-\mathbf x\|_2}{h}
    \right):
    \mathbf x\in\mathcal K,\ 
    0<h<h_0
    \right\}
\end{align*}
is of VC type or $K$ is piecewise polynomial with finitely many pieces on $[0,1]$, then Conditions~\ref{con:U-K1} and~\ref{con:U-K2} hold over $\mathbf x\in\mathcal K$ and $0<h<h_0$. In particular, the conclusion holds for the uniform, triangular, Epanechnikov, biweight, and triweight kernels,
\begin{align*}
    K_{\mathrm{unif}}(t)
    &:=
    \mathds 1
    \left(
    0\leq t\leq1
    \right), \\
    K_{\mathrm{tri}}(t)
    &:=
    (1-t)
    \mathds 1
    \left(
    0\leq t\leq1
    \right), \\
    K_{\mathrm{Epa}}(t)
    &:=
    (1-t^2)
    \mathds 1
    \left(
    0\leq t\leq1
    \right), \\
    K_{\mathrm{biw}}(t)
    &:=
    (1-t^2)^2
    \mathds 1
    \left(
    0\leq t\leq1
    \right), \\
    K_{\mathrm{triw}}(t)
    &:=
    (1-t^2)^3
    \mathds 1
    \left(
    0\leq t\leq1
    \right).
\end{align*}
\end{lemma}

\begin{proof}[Proof of \Cref{lem:app_standard_kernels_UK}]
In the Euclidean case, we use the single global canonical ordered orthonormal frame. We first verify Condition~\ref{con:U-K1}. Let
\begin{align*}
    \mathcal F_0
    &:=
    \left\{
    \mathbf z\mapsto
    K\left(
    \frac{\|\mathbf z-\mathbf x\|_2}{h}
    \right):
    \mathbf x\in\mathcal K,\ 
    0<h<h_0
    \right\}.
\end{align*}
If $\mathcal F_0$ is assumed to be of VC type, then Condition~\ref{con:U-K1} holds directly, since Condition~\ref{con:P-K1} gives the bounded envelope $\|K\|_\infty$.

Suppose instead that $K$ is piecewise polynomial with finitely many pieces on $[0,1]$. Choose a finite partition
\begin{align*}
    0=a_0<a_1<\cdots<a_L=1
\end{align*}
such that $K$ agrees with a polynomial $p_\ell$ on each open interval piece $(a_{\ell-1},a_\ell)$. Put $\lambda=h^{-1}$ and introduce an auxiliary radial variable $r\geq0$ satisfying
\begin{align*}
    r^2
    =
    \lambda^2\|\mathbf z-\mathbf x\|_2^2.
\end{align*}
For each breakpoint $a_\ell$, $\ell=0,\ldots,L$, the case $r=a_\ell$ is described by the polynomial equality $r=a_\ell$, together with the subgraph inequality $t<K(a_\ell)$. Since the number of breakpoints is finite and the values $K(a_\ell)$ are fixed constants, adding these breakpoint cases does not change the fact that the subgraph family has uniformly bounded semialgebraic complexity.

For a fixed polynomial piece, the part of the subgraph corresponding to $a_{\ell-1}<r<a_\ell$ is described by the conditions
\begin{align*}
    a_{\ell-1}
    <
    r
    <
    a_\ell,
    \qquad
    r^2
    =
    \lambda^2\sum_{j=1}^d(z_j-x_j)^2,
    \qquad
    t<p_\ell(r).
\end{align*}
These are polynomial equalities and inequalities in $(\mathbf z,t,\mathbf x,\lambda,r)$. The part outside the support of $K$ is described by
\begin{align*}
    r>1,
    \qquad
    r^2
    =
    \lambda^2\sum_{j=1}^d(z_j-x_j)^2,
    \qquad
    t<0,
\end{align*}
again using only polynomial equalities and inequalities. Taking the finite union over all polynomial pieces, together with the breakpoint cases and the outside-support case, gives the subgraph of each function in $\mathcal F_0$ after projecting out the auxiliary variable $r$.

By the Tarski--Seidenberg theorem, projections of semialgebraic sets are semialgebraic. In the present argument, this means that after we describe the subgraph using the auxiliary radial variable $r$, we may eliminate $r$ without leaving the class of semialgebraic sets. Therefore, the subgraphs of the functions in $\mathcal F_0$ form a semialgebraic family. Moreover, the number of polynomial inequalities and their degrees are bounded by constants depending only on $d$ and on the finite piecewise-polynomial representation of $K$, not on $\mathbf x\in\mathbb R^d$ or $h\in(0,h_0)$. Restricting the location parameter to $\mathbf x\in\mathcal K$ only takes a subclass of this semialgebraic family, so no semialgebraic assumption on $\mathcal K$ is needed. By the standard VC theorem for semialgebraic classes \cite{van der Vaart and Wellner (1996)}, this subgraph family has finite VC dimension. Hence $\mathcal F_0$ is a VC-subgraph class. Since $K$ is bounded by Condition~\ref{con:P-K1}, $\mathcal F_0$ is therefore of VC type, and Condition~\ref{con:U-K1} holds.

By \Cref{lem:app_UK1_implies_UK2_euclidean}, Condition~\ref{con:U-K1} implies Condition~\ref{con:U-K2} in the Euclidean predictor setting. Hence both Conditions~\ref{con:U-K1} and~\ref{con:U-K2} hold for the kernels covered by the lemma.

Finally, each displayed kernel is bounded, nonnegative, supported on $[0,1]$, and piecewise polynomial with finitely many pieces. In particular, these standard kernels are of finite total variation on $[0,1]$. The verification above, however, uses their finite piecewise-polynomial structure, which gives a direct semialgebraic, and hence VC-type, argument for the induced kernel classes. Each kernel also satisfies Condition~\ref{con:P-K1}. Therefore, the uniform, triangular, Epanechnikov, biweight, and triweight kernels satisfy Conditions~\ref{con:U-K1} and~\ref{con:U-K2}.
\end{proof}

\subsection{Finite localization and definable function classes}

We next formulate a manifold-level sufficient condition. The use of definability below is only a verification device. It is not imposed in the main asymptotic theory.

\begin{definition}[Finite definability on a compact manifold region]
\label{def:F.finite_definability}
Let $\mathfrak R$ be a fixed o-minimal expansion of the real field. Let $\mathcal M_1$ and $\mathcal M_2$ be finite-dimensional smooth manifolds, let $\mathcal C\subset\mathcal M_1\times\mathcal M_2$ be contained in a compact subset of $\mathcal M_1\times\mathcal M_2$, and let $H:\mathcal C\to\mathbb R^m$. We say that $H$ is \emph{finitely $\mathfrak R$-definable on $\mathcal C$} if there exist finitely many product-coordinate charts
\begin{align*}
    (\mathcal U^\ell,\varphi^\ell)
    \quad\text{on }\mathcal M_1,
    \qquad
    (\mathcal V^\ell,\psi^\ell)
    \quad\text{on }\mathcal M_2,
    \qquad
    \ell=1,\ldots,L,
\end{align*}
whose product domains cover $\mathcal C$, and, for every $\ell$, there exist $\mathfrak R$-definable open sets
\begin{align*}
    W_1^\ell
    &\subset
    \mathbb R^{\dim\mathcal M_1},
    &
    W_2^\ell
    &\subset
    \mathbb R^{\dim\mathcal M_2},
\end{align*}
such that
\begin{align*}
    (\varphi^\ell\times\psi^\ell)
    \left\{
    \mathcal C\cap(\mathcal U^\ell\times\mathcal V^\ell)
    \right\}
    \subset
    W_1^\ell\times W_2^\ell,
\end{align*}
and an $\mathfrak R$-definable function
\begin{align*}
    \widetilde H^\ell:
    W_1^\ell\times W_2^\ell
    \to
    \mathbb R^m
\end{align*}
such that the coordinate representation
\begin{align*}
    H\circ
    \left[
    (\varphi^\ell)^{-1}\times(\psi^\ell)^{-1}
    \right]
\end{align*}
agrees with $\widetilde H^\ell$ on
\begin{align*}
    (\varphi^\ell\times\psi^\ell)
    \left\{
    \mathcal C\cap(\mathcal U^\ell\times\mathcal V^\ell)
    \right\}.
\end{align*}
\end{definition}

\begin{lemma}[Uniformly definable subgraph families]
\label{lemma:F.definable_subgraph}
Let $\mathfrak R$ be a fixed o-minimal expansion of the real field. Let $\Xi\subset\mathbb R^q$ be an $\mathfrak R$-definable parameter set and let
\begin{align*}
    \mathcal H
    &:=
    \left\{
    \mathbf z\mapsto H(\mathbf z;\eta):
    \eta\in\Xi
    \right\}
\end{align*}
be a uniformly bounded class of real-valued functions on a Euclidean set $D\subset\mathbb R^p$. Suppose that the subgraph family is uniformly $\mathfrak R$-definable in the sense that there exists an $\mathfrak R$-definable set
\begin{align*}
    \mathcal S
    \subset
    \mathbb R^p\times\mathbb R\times\Xi
\end{align*}
such that, for every $\eta\in\Xi$,
\begin{align*}
    \left\{
    (\mathbf z,t)\in D\times\mathbb R:
    t<H(\mathbf z;\eta)
    \right\}
    =
    \left\{
    (\mathbf z,t)\in D\times\mathbb R:
    (\mathbf z,t,\eta)\in\mathcal S
    \right\}.
\end{align*}
Then $\mathcal H$ is a VC-subgraph class and hence is of VC type.
\end{lemma}

\begin{proof}[Proof of \Cref{lemma:F.definable_subgraph}]
The assumption means that all subgraphs of functions in $\mathcal H$, when restricted to $D\times\mathbb R$, arise as fibers of a single definable set $\mathcal S$ with respect to the parameter $\eta$. The unrestricted fibers
\begin{align*}
    \left\{
    (\mathbf z,t)\in\mathbb R^p\times\mathbb R:
    (\mathbf z,t,\eta)\in\mathcal S
    \right\},
    \quad \eta\in\Xi,
\end{align*}
form a definable family in an o-minimal structure. Such families have finite combinatorial complexity: in particular, their shatter functions grow at most polynomially in the number of points, and hence they have finite VC dimension; see \cite{Laskowski (1992), Johnson and Laskowski (2010)}. Restricting a VC class to the subset $D\times\mathbb R$ preserves the VC property. Therefore, the restricted subgraph family
\begin{align*}
    \left\{
    \left\{
    (\mathbf z,t)\in D\times\mathbb R:
    t<H(\mathbf z;\eta)
    \right\}:
    \eta\in\Xi
    \right\}
\end{align*}
has finite VC dimension. Hence $\mathcal H$ is a VC-subgraph class. Since $\mathcal H$ is uniformly bounded, the standard entropy bound for bounded VC-subgraph classes implies that $\mathcal H$ is of VC type; see \cite{van der Vaart and Wellner (1996)}.
\end{proof}

\begin{lemma}[Finite Boolean patching of VC classes]
\label{lemma:F.finite_boolean_patching}
Let $\mathscr A_1,\ldots,\mathscr A_L$ be VC classes of subsets of a common set $\mathcal S$. Let $\mathfrak B$ be a fixed Boolean expression in $L$ set arguments, formed using finitely many unions, intersections, and complements. Then
\begin{align*}
    \left\{
    \mathfrak B(A_1,\ldots,A_L):
    A_l\in\mathscr A_l,\ 
    l=1,\ldots,L
    \right\}
\end{align*}
is a VC class.
\end{lemma}

\begin{proof}[Proof of \Cref{lemma:F.finite_boolean_patching}]
Let $\mathcal P=\{s_1,\ldots,s_m\}\subset\mathcal S$ be an arbitrary finite set. For a class $\mathscr A$ of subsets of $\mathcal S$, write
\begin{align*}
    \Pi_{\mathscr A}(m)
    &:=
    \max_{\mathcal P:\,|\mathcal P|=m}
    \left|
    \left\{
    A\cap\mathcal P:
    A\in\mathscr A
    \right\}
    \right|
\end{align*}
for its trace number on $m$ points. Once the traces $A_l\cap\mathcal P$, $l=1,\ldots,L$, are fixed, the trace of $\mathfrak B(A_1,\ldots,A_L)$ on $\mathcal P$ is completely determined, because Boolean operations commute with restriction to $\mathcal P$. Hence the number of distinct traces generated by the displayed class on $\mathcal P$ is bounded by
\begin{align*}
    \prod_{l=1}^L \Pi_{\mathscr A_l}(m).
\end{align*}
Since each $\mathscr A_l$ is a VC class, the Sauer--Shelah lemma implies that $\Pi_{\mathscr A_l}(m)$ is bounded by a polynomial in $m$. Since $L$ and the Boolean expression $\mathfrak B$ are fixed, the product above is also bounded by a polynomial in $m$.

A class that shatters arbitrarily large finite sets would have trace number $2^m$ for arbitrarily large $m$. For all sufficiently large $m$, the polynomial bound is strictly smaller than $2^m$. Therefore the displayed Boolean-patched class cannot shatter all sufficiently large finite sets, and hence it has finite VC dimension.
\end{proof}

\begin{definition}[Tame local geometry on the uniform normal tube]
\label{def:F.tame_local_geometry}
Let $\mathcal K\subset\mathcal M$ be compact, fix $\rho\in(0,i(\mathcal K))$, and consider the finite smooth ordered-orthonormal-frame cover in \eqref{eq:U.fixed_frame_cover}. For each $\alpha\in\{1,\ldots,N_{\mathcal K}\}$, define
\begin{align*}
    \mathcal D^\alpha_\rho
    &:=
    \left\{
    (\mx,\mz)\in\mathcal M\times\mathcal M:
    \mx\in\mathcal K\cap\mathcal O^\alpha,\ 
    d_{\mathcal M}(\mx,\mz)<\rho
    \right\}.
\end{align*}
Since $\rho<i(\mathcal K)$, the logarithmic map $\Log_{\mx}(\mz)$ and the volume-density function $\theta_{\mx}(\mz)$ are well-defined on $\mathcal D^\alpha_\rho$. Moreover, by \Cref{lemma:A.uniform_normal_neighborhoods}, $\mathcal K^\rho$ is compact. Hence $\mathcal D^\alpha_\rho$ is contained in the compact set $\mathcal K\times\mathcal K^\rho$, where
\begin{align*}
    \mathcal K^\rho
    &:=
    \left\{
    \mz\in\mathcal M:
    d_{\mathcal M}(\mz,\mathcal K)\leq\rho
    \right\}.
\end{align*}
For $(\mx,\mz)\in\mathcal D^\alpha_\rho$, define
\begin{align*}
    q(\mx,\mz)
    &:=
    d_{\mathcal M}^2(\mx,\mz), \\
    b_{\alpha,r}(\mx,\mz)
    &:=
    \left[
    \bm{\Phi}_{\mathbf E^\alpha_{\mx}}
    \left(
    \Log_{\mx}(\mz)
    \right)
    \right]_r,
    \quad
    r=1,\ldots,d.
\end{align*}
We say that the fixed frame cover has \emph{tame local geometry} on the uniform normal tube if there exists a fixed o-minimal expansion $\mathfrak R$ of the real field such that, for every $\alpha$, the vector-valued map
\begin{align*}
    (\mx,\mz)
    \mapsto
    \left(
    q(\mx,\mz),
    \theta_{\mx}(\mz)^{-1},
    b_{\alpha,1}(\mx,\mz),
    \ldots,
    b_{\alpha,d}(\mx,\mz)
    \right)
\end{align*}
is finitely $\mathfrak R$-definable on $\mathcal D^\alpha_\rho$ in the sense of \Cref{def:F.finite_definability}.
\end{definition}

\subsection{Verification on manifolds with tame local geometry}

\begin{proposition}[Piecewise-polynomial kernels under tame local geometry]
\label{prop:F.tame_UK_verification}
Let $\mathcal K\subset\mathcal M$ be compact, fix
$\rho\in(0,i(\mathcal K))$, and suppose that the fixed finite local-frame
cover has tame local geometry on the uniform normal tube. Suppose that $K$
satisfies Condition~\ref{con:P-K1} and is piecewise polynomial with finitely
many pieces on $[0,1]$. Then Conditions~\ref{con:U-K1}
and~\ref{con:U-K2} hold for every $h_0\in(0,\rho)$.
\end{proposition}

\begin{proof}[Proof of \Cref{prop:F.tame_UK_verification}]
Fix $\alpha\in\{1,\ldots,N_{\mathcal K}\}$ and $h_0\in(0,\rho)$. Since $K$ is supported on $[0,1]$, a nonzero kernel value implies
\begin{align*}
    d_{\mathcal M}(\mx,\mz)
    \leq h
    <h_0
    <\rho.
\end{align*}
Since $\mx\in\mathcal K$ and $\rho<i(\mathcal K)$, we also have $\rho<i(\mx)$. Thus, on the support of every kernel window considered here, the logarithmic map and the volume-density function are well-defined. In particular, the indicator of $B_{\mathcal M}(\mx,i(\mx))$ in the definition of $\mathcal L_{\mx,h}$ is redundant on these classes, and no complexity condition on the injectivity-radius function is required.

For this fixed frame patch, consider the following zero-extended local-design classes. For $\mx\in\mathcal K\cap\mathcal O^\alpha$ and $0<h<h_0$, define
\begin{align*}
    F_{\alpha,0;\mx,h}(\mz)
    &:=
    \begin{cases}
    \displaystyle
    \theta_{\mx}(\mz)^{-1}
    K\left(
    \frac{d_{\mathcal M}(\mx,\mz)}{h}
    \right),
    & d_{\mathcal M}(\mx,\mz)<\rho, \\
    0,
    & d_{\mathcal M}(\mx,\mz)\geq\rho,
    \end{cases} \\
    F_{\alpha,1,r;\mx,h}(\mz)
    &:=
    \begin{cases}
    \displaystyle
    \theta_{\mx}(\mz)^{-1}
    K\left(
    \frac{d_{\mathcal M}(\mx,\mz)}{h}
    \right)
    \frac{b_{\alpha,r}(\mx,\mz)}{h},
    & d_{\mathcal M}(\mx,\mz)<\rho, \\
    0,
    & d_{\mathcal M}(\mx,\mz)\geq\rho,
    \end{cases} \\
    F_{\alpha,2,r,s;\mx,h}(\mz)
    &:=
    \begin{cases}
    \displaystyle
    \theta_{\mx}(\mz)^{-1}
    K\left(
    \frac{d_{\mathcal M}(\mx,\mz)}{h}
    \right)
    \frac{b_{\alpha,r}(\mx,\mz)b_{\alpha,s}(\mx,\mz)}{h^2},
    & d_{\mathcal M}(\mx,\mz)<\rho, \\
    0,
    & d_{\mathcal M}(\mx,\mz)\geq\rho.
    \end{cases}
\end{align*}
Then set
\begin{align*}
    \mathcal F_{\alpha,0}
    &:=
    \left\{
    F_{\alpha,0;\mx,h}:
    \mx\in\mathcal K\cap\mathcal O^\alpha,\ 
    0<h<h_0
    \right\}, \\
    \mathcal F_{\alpha,1,r}
    &:=
    \left\{
    F_{\alpha,1,r;\mx,h}:
    \mx\in\mathcal K\cap\mathcal O^\alpha,\ 
    0<h<h_0
    \right\}, \\
    \mathcal F_{\alpha,2,r,s}
    &:=
    \left\{
    F_{\alpha,2,r,s;\mx,h}:
    \mx\in\mathcal K\cap\mathcal O^\alpha,\ 
    0<h<h_0
    \right\}.
\end{align*}
On the support of the kernel factor, these zero-extended functions agree with the original local-design functions because $h<h_0<\rho$. Moreover, if $d_{\mathcal M}(\mx,\mz)\geq\rho$, then
\begin{align*}
    \frac{d_{\mathcal M}(\mx,\mz)}{h}
    >
    1,
\end{align*}
so the compact support of $K$ implies that the kernel factor is zero. Thus the zero-extended functions are measurable representatives of the original local-design functions, with no need to evaluate $\Log_{\mx}(\mz)$, $\theta_{\mx}(\mz)$, or $b_{\alpha,r}(\mx,\mz)$ outside the uniform normal tube. Consequently, VC-type bounds for the zero-extended classes transfer to the original classes in Conditions~\ref{con:U-K1} and~\ref{con:U-K2}.

We first verify the VC-subgraph property. Fix one of the finitely many product-coordinate blocks appearing in the finite-definability representation of the vector-valued map in \Cref{def:F.tame_local_geometry}. In this coordinate block, the functions
\begin{align*}
    q(\mx,\mz),
    \qquad
    \theta_{\mx}(\mz)^{-1},
    \qquad
    b_{\alpha,1}(\mx,\mz),\ldots,b_{\alpha,d}(\mx,\mz)
\end{align*}
agree simultaneously, on the relevant part of $\mathcal D_\rho^\alpha$, with $\mathfrak R$-definable functions on a definable Euclidean neighborhood. Hence, for the purpose of proving a VC bound, we may enlarge the location-parameter range from the possibly nondefinable set corresponding to $\mx\in\mathcal K\cap\mathcal O^\alpha$ to the ambient definable coordinate neighborhood, and later restrict back to the original parameter set. Passing to a subclass cannot increase VC dimension.

Since $h$ ranges over the definable interval $(0,h_0)$, the operations of multiplication, division by $h$, and finite Boolean combinations preserve definability in the same o-minimal structure. Let
\begin{align*}
    0=a_0<a_1<\cdots<a_L=1
\end{align*}
be a finite partition such that $K$ agrees with a polynomial $p_\ell$ on each open interval piece $(a_{\ell-1},a_\ell)$. On the coordinate block introduce an auxiliary radial variable $u\geq0$ satisfying
\begin{align*}
    u^2h^2=q(\mx,\mz).
\end{align*}
For each polynomial piece, the conditions
\begin{align*}
    a_{\ell-1}<u<a_\ell,
    \qquad
    u^2h^2=q(\mx,\mz),
\end{align*}
together with the appropriate subgraph inequality for the zeroth-, first-, or second-order function, are definable in the coordinate variables, the parameter variables, and $u$. For each breakpoint $a_\ell$, $\ell=0,\ldots,L$, the case $u=a_\ell$ is described by the definable equality $u=a_\ell$, together with the corresponding subgraph inequality obtained by replacing the kernel value by the fixed constant $K(a_\ell)$. Since there are only finitely many such breakpoints and the values $K(a_\ell)$ are fixed constants, adding these cases does not affect uniform definability.

On the part of the normal tube where $u>1$, the kernel factor is zero, so the corresponding subgraph condition is simply $t<0$; this outside-support case is again definable using
\begin{align*}
    u>1,
    \qquad
    u^2h^2=q(\mx,\mz).
\end{align*}
These finitely many polynomial-piece, breakpoint, and outside-support cases do not affect uniform definability. Projecting out the auxiliary variable $u$ preserves definability in an o-minimal structure. Thus each local subgraph family is uniformly $\mathfrak R$-definable and is VC by \Cref{lemma:F.definable_subgraph}.

It remains to pass from coordinate blocks to the global zero-extended functions. Pulling a VC class back under a fixed coordinate map preserves VC dimension, since traces on finite point sets are unchanged under a fixed map. The finitely many coordinate-domain membership sets are fixed sets and therefore form finite VC classes. The inside-normal-tube condition $d_{\mathcal M}(\mx,\mz)<\rho$ is, on each coordinate block, represented by the definable inequality
\begin{align*}
    q(\mx,\mz)<\rho^2,
\end{align*}
and its complement gives the zero branch, where the subgraph condition is simply $t<0$. Inactive coordinate blocks may be represented by the empty set, and adjoining the empty set to a VC class preserves the VC property. Therefore the full zero-extended subgraph is obtained from finitely many VC classes by a fixed finite union and finite Boolean operations. The fact that the local pieces share the same parameter \((\mx,h)\) only restricts the resulting Boolean-patched class to a subclass of the class in \Cref{lemma:F.finite_boolean_patching}, and hence cannot increase VC dimension. By \Cref{lemma:F.finite_boolean_patching}, the global zero-extended classes
\begin{align*}
    \mathcal F_{\alpha,0},
    \qquad
    \mathcal F_{\alpha,1,r},
    \qquad
    \mathcal F_{\alpha,2,r,s}
\end{align*}
are VC-subgraph classes.

It remains to verify uniform envelopes. By \Cref{lemma:A.uniform_normal_neighborhoods},
\begin{align*}
    C_\theta
    &:=
    \sup_{\mx\in\mathcal K}
    \sup_{\mz\in B_{\mathcal M}(\mx,\rho)}
    \theta_{\mx}(\mz)^{-1}
    <\infty.
\end{align*}
Therefore,
\begin{align*}
    \left|
    \theta_{\mx}(\mz)^{-1}
    K\left(
    \frac{d_{\mathcal M}(\mx,\mz)}{h}
    \right)
    \right|
    \leq
    C_\theta\|K\|_\infty.
\end{align*}
Moreover, on the kernel support,
\begin{align*}
    \left\|
    \frac{\Log_{\mx}(\mz)}{h}
    \right\|_{\mx}
    =
    \frac{d_{\mathcal M}(\mx,\mz)}{h}
    \leq1.
\end{align*}
Since $\mathbf E^\alpha_{\mx}$ is orthonormal, for every $r,s\in\{1,\ldots,d\}$,
\begin{align*}
    \left|
    \frac{b_{\alpha,r}(\mx,\mz)}{h}
    \right|
    &\leq1, \\
    \left|
    \frac{b_{\alpha,r}(\mx,\mz)b_{\alpha,s}(\mx,\mz)}{h^2}
    \right|
    &\leq1.
\end{align*}
Thus the zeroth-, first-, and second-order classes all have envelopes bounded by $C_\theta\|K\|_\infty$. Since bounded VC-subgraph classes are of VC type, the classes above are VC-type classes.

Finally, the numbers of frame patches and coordinate indices are finite. The VC-type constants may therefore be replaced by their maxima over $\alpha$, $r$, and $s$. This proves Conditions~\ref{con:U-K1} and~\ref{con:U-K2}.
\end{proof}

\begin{proposition}[Real-analytic local geometry implies tame local geometry] \label{prop:F.analytic_implies_tame}
Let $\mathcal K\subset\mathcal M$ be compact and fix $\rho\in(0,i(\mathcal K))$. Suppose that $\mathcal M$ admits a real-analytic manifold structure on a neighborhood of $\mathcal K^\rho$ and that the Riemannian metric is real analytic in this structure. Then the finite local-frame cover in \eqref{eq:U.fixed_frame_cover} can be chosen so that it has tame local geometry on the uniform normal tube with respect to the o-minimal structure $\mathbb R_{\mathrm{an}}$.
\end{proposition}

\begin{proof}[Proof of \Cref{prop:F.analytic_implies_tame}]
Since $\rho<i(\mathcal K)$ and the injectivity-radius function is continuous, there exists an open neighborhood $\mathcal U$ of $\mathcal K$ such that
\begin{align*}
    \rho
    <
    i(\mx),
    \qquad
    \mx\in\mathcal U.
\end{align*}
Thus the exponential map at every $\mx\in\mathcal U$ is a diffeomorphism on the tangent ball of radius $\rho$.

Because the Riemannian metric is real analytic in the given analytic structure, the Christoffel symbols are real analytic in local coordinates. Hence the geodesic equation is a real-analytic ordinary differential equation. By analytic dependence of solutions of analytic ordinary differential equations on initial conditions, the geodesic flow is real analytic on its domain of definition. Consequently, in local analytic coordinates, the map
\begin{align*}
    (\mx,\mathbf u)
    \mapsto
    \left(
    \mx,
    \Exp_{\mx}(\mathbf u)
    \right)
\end{align*}
is real analytic on an open neighborhood of
\begin{align*}
    \left\{
    (\mx,\mathbf u):
    \mx\in\mathcal K,\ 
    \|\mathbf u\|_{\mx}\leq\rho
    \right\}.
\end{align*}
For $\|\mathbf u\|_{\mx}<\rho$, this map is one-to-one and has nonsingular differential because $\rho<i(\mx)$. Therefore, by the real-analytic inverse function theorem, it is a real-analytic diffeomorphism onto the corresponding normal tube. Its inverse is
\begin{align*}
    (\mx,\mz)
    \mapsto
    \left(
    \mx,
    \Log_{\mx}(\mz)
    \right),
\end{align*}
so the moving logarithmic map is real analytic on the uniform normal tube. It follows that
\begin{align*}
    q(\mx,\mz)
    =
    d_{\mathcal M}^2(\mx,\mz)
    =
    \left\|
    \Log_{\mx}(\mz)
    \right\|_{\mx}^2
\end{align*}
is real analytic there.

The differential of the exponential map is also real analytic. Hence the normal-coordinate metric matrix is real analytic, and so is its determinant. Since the volume density is strictly positive on the relevant normal neighborhoods, the reciprocal
\begin{align*}
    (\mx,\mz)
    \mapsto
    \theta_{\mx}(\mz)^{-1}
\end{align*}
is real analytic on the uniform normal tube.

Next choose finitely many relatively compact real-analytic coordinate neighborhoods covering $\mathcal K^\rho$. On each such coordinate neighborhood, the coordinate vector fields form a local analytic frame. Applying the Gram--Schmidt procedure with respect to the analytic metric gives an ordered orthonormal frame whose components are real analytic, after possibly shrinking the coordinate neighborhood. The denominators that arise in the Gram--Schmidt procedure are positive because the metric is positive definite and the coordinate vector fields are linearly independent. Thus the finite frame cover in \eqref{eq:U.fixed_frame_cover} may be chosen to consist of real-analytic ordered orthonormal frames.

For such an analytic frame, the coordinate components
\begin{align*}
    b_{\alpha,r}(\mx,\mz)
    &:=
    \left[
    \bm{\Phi}_{\mathbf E^\alpha_{\mx}}
    \left(
    \Log_{\mx}(\mz)
    \right)
    \right]_r,
    \qquad
    r=1,\ldots,d,
\end{align*}
are real analytic on the corresponding normal-tube domains, because they are obtained by taking analytic frame coordinates of the analytic vector $\Log_{\mx}(\mz)$.

It remains to connect real analyticity with definability in $\mathbb R_{\mathrm{an}}$. We do not use global definability of analytic functions. Instead, cover the compact uniform normal tube by finitely many product-coordinate blocks whose closures are contained in analytic coordinate neighborhoods. On each such block, the functions
\begin{align*}
    q,
    \qquad
    (\mx,\mz)\mapsto\theta_{\mx}(\mz)^{-1},
    \qquad
    b_{\alpha,1},
    \ldots,
    b_{\alpha,d}
\end{align*}
extend real analytically to an open Euclidean neighborhood of the block closure. After choosing a compact coordinate box containing the block and rescaling the box to $[-1,1]^m$, each such extension is a restricted analytic function. Hence it is definable in the o-minimal structure $\mathbb R_{\mathrm{an}}$; see \cite{van den Dries and Miller (1996)}. Therefore these functions are finitely $\mathbb R_{\mathrm{an}}$-definable on the corresponding normal-tube domains in the sense of \Cref{def:F.finite_definability}. Hence the chosen finite frame cover has tame local geometry on the uniform normal tube.
\end{proof}

\begin{corollary}[Standard analytic predictor manifolds]
\label{cor:F.standard_analytic_manifolds}
Let $\mathcal K\subset\mathcal M$ be compact and let $\rho\in(0,i(\mathcal K))$. Suppose that $\mathcal M$ is one of the following Riemannian predictor manifolds:
\begin{enumerate}
    \item Euclidean space $\mathbb R^d$ with its standard metric;
    \item the sphere $\mathbb S^d$ with its standard metric;
    \item a finite product of spheres $\mathbb S^{d_1}\times\cdots\times\mathbb S^{d_L}$ equipped with the product metric;
    \item the standard flat torus $\mathbb T^d:=\mathbb S^1\times\cdots\times\mathbb S^1$ equipped with the product metric;
    \item the SPD cone $\mathcal S_{++}^p$ equipped with the affine-invariant Riemannian metric. Here $T_{\mathbf X}\mathcal S_{++}^p$ is canonically identified with the vector space $\mathcal S^p$ of real symmetric $p\times p$ matrices, and
    \begin{align*}
        g_{\mathbf X}(\mathbf A,\mathbf B)
        &:=
        \tr\left(
        \mathbf X^{-1}\mathbf A
        \mathbf X^{-1}\mathbf B
        \right),
        \quad
        \mathbf X\in\mathcal S_{++}^p,
        \quad
        \mathbf A,\mathbf B\in T_{\mathbf X}\mathcal S_{++}^p.
    \end{align*}
\end{enumerate}
Then the fixed finite local-frame cover can be chosen to have tame local geometry on the uniform normal tube. Consequently, for every $h_0\in(0,\rho)$, every kernel satisfying Condition~\ref{con:P-K1} that is piecewise polynomial with finitely many pieces on $[0,1]$ satisfies Conditions~\ref{con:U-K1} and~\ref{con:U-K2} over the bandwidth range $0<h<h_0$. In particular, the conclusion holds for the uniform, triangular, Epanechnikov, biweight, and triweight kernels displayed in \Cref{lem:app_standard_kernels_UK}.
\end{corollary}

\begin{proof}[Proof of \Cref{cor:F.standard_analytic_manifolds}]
Euclidean spaces and spheres with their standard metrics are real-analytic Riemannian manifolds. Finite products of real-analytic Riemannian manifolds equipped with product metrics are again real analytic, which covers finite products of spheres and flat tori.

It remains only to comment on the SPD cone. The space $\mathcal S_{++}^p$ is an open subset of the finite-dimensional vector space of symmetric matrices. The affine-invariant metric is real analytic because matrix inversion and matrix multiplication are real analytic on $\mathcal S_{++}^p$. Moreover, $\mathcal S_{++}^p$ equipped with the affine-invariant Riemannian metric is complete. Hence, by Hopf--Rinow, the closed metric neighborhood $\mathcal K^\rho$ of the compact set $\mathcal K$ is compact. In particular, there exist constants $0<c_{\mathcal K,\rho}\leq C_{\mathcal K,\rho}<\infty$ such that
\begin{align*}
    c_{\mathcal K,\rho}
    \leq
    \lambda_{\min}(\mathbf X)
    \leq
    \lambda_{\max}(\mathbf X)
    \leq
    C_{\mathcal K,\rho},
    \qquad
    \mathbf X\in\mathcal K^\rho.
\end{align*}
Thus $\mathcal K^\rho$ stays a positive distance away from the boundary of the SPD cone, and the affine-invariant metric is real analytic on an open neighborhood of $\mathcal K^\rho$.

Therefore, in each of the listed cases, the assumptions of \Cref{prop:F.analytic_implies_tame} hold. Hence the fixed finite local-frame cover can be chosen to have tame local geometry on the uniform normal tube. The kernel-complexity conclusion then follows from \Cref{prop:F.tame_UK_verification}. 
\end{proof}

\begin{remark}[Relation to the torus-specific procedure]
The inclusion of the standard flat torus in \Cref{cor:F.standard_analytic_manifolds} verifies the empirical-process conditions for the radial, scalar-bandwidth estimator defined in \Cref{sec:estimation}. It does not identify that estimator with the torus-specific procedures of \cite{Im and Jeon (2026)}, which use a product directional kernel, a vector of coordinate-specific bandwidths, and an asymptotic analysis that imposes no bounded-ratio restriction on the bandwidth components and allows kernel profiles without compact support. Consequently, the torus-specific estimator and its anisotropic theory are not recovered as a direct corollary of the present manifold-level verification.
\end{remark}

\begin{remark}[Scope of the manifold-level verification] \label{rem:F.scope_tame_verification} 
The preceding results verify Conditions~\ref{con:U-K1} and~\ref{con:U-K2} jointly. They do not assert that Condition~\ref{con:U-K2} follows from Condition~\ref{con:U-K1} on an arbitrary Riemannian manifold. The Euclidean implication in \Cref{lem:app_UK1_implies_UK2_euclidean} relies on the fixed global linear coordinates and is retained as a separate elementary result.

The real-analytic assumption in \Cref{prop:F.analytic_implies_tame} is a transparent sufficient condition, not a necessary condition. The same proof applies whenever the squared distance, reciprocal volume density, and local-frame coordinates of the logarithmic map are finitely definable in a common o-minimal expansion on the compact uniform normal tube. Conversely, smoothness alone does not automatically imply the required VC-type entropy bounds. For predictor manifolds outside the tame or analytic class treated here, Conditions~\ref{con:U-K1} and~\ref{con:U-K2} remain explicit high-level empirical-process assumptions.
\end{remark}

\end{document}